\documentclass[letterpaper,notoc]{JHEP3}
\usepackage{psfrag,amsmath,amsthm,amssymb,subfigure,array,verbatim}
\usepackage{mathrsfs}  
\usepackage{multirow}  
\usepackage{url}
\usepackage{ae}  
\usepackage[all]{xy}  
\usepackage{wrapfig}  

\ifpdf     
    \usepackage[pdftex]{graphicx}
    \usepackage{epstopdf}  
\else  
    \usepackage[dvips]{graphicx}
\fi

\newcommand{\be}{\begin{equation}}
\newcommand{\ee}{\end{equation}}
\newcommand{\bea}{\begin{eqnarray}}
\newcommand{\eea}{\end{eqnarray}}

\newcommand{\ra}[1]{ \stackrel{\scriptscriptstyle #1}{ \rightarrow  }}  

\newcommand{\rt}{\rightarrow}
\newcommand{\bbar}{ {b\bar{b}} }
\newcommand{\mbb}{ {m_{b\bar{b}}} }

\newcommand{\Higgswindow}{$90\,\mathrm{GeV} < \mbb < 124\,\mathrm{GeV}$}  

\newcommand{\eS}{ \varepsilon_{S}}
\newcommand{\eB}{ \varepsilon_{B}}
\newcommand{\e}{ \varepsilon}
\newcommand{\vp}{ \vec{p} }

\newcommand{\rSB}{\eS/\eB}
\newcommand{\rsig}{{\text{\sc SIC}}}
\newcommand{\ropt}{\overline{\text{\sc SIC}}}
\newcommand{\DZero}{{D\O}}
\newcommand{\zh}{ZH}
\newcommand{\wh}{WH}
\newcommand{\zbb}{Z b\bar{b}}

\def\lsim{\mathrel{\rlap{\lower4pt\hbox{\hskip1pt$\sim$}}
    \raise1pt\hbox{$<$}}}
\def\gsim{\mathrel{\rlap{\lower4pt\hbox{\hskip1pt$\sim$}}
    \raise1pt\hbox{$>$}}}

\title{Multivariate discrimination and the Higgs+W/Z search}
\author{Jason Gallicchio, John Huth, Michael Kagan, Matthew D. Schwartz\\
Jefferson Physical Laboratory, Harvard University, Cambridge, MA 02138}
\author{Kevin Black, Brock Tweedie \\
Physics Department, Boston Univeristy, Boston, MA 02215}
\date{\today}

\abstract{
A systematic method for optimizing multivariate
discriminants is developed and applied to the important example of a light Higgs boson search at the Tevatron and the LHC.
The Significance Improvement Characteristic (SIC),
defined as the signal efficiency of a cut or multivariate discriminant divided by the square root of the background efficiency,
is shown to be an extremely powerful visualization tool.
SIC curves
demonstrate numerical instabilities in the multivariate
discriminants, show convergence as the number of variables is increased, and display the sensitivity to the optimal
cut values.
For our application,
we concentrate on Higgs boson production in association with a $W$ or $Z$ boson with $H\to \bbar$
and compare to the irreducible standard model
background, $Z/W + \bbar$.
We explore thousands of experimentally motivated, physically motivated, and unmotivated single variable discriminants.
Along with the standard kinematic variables, a number of new ones, such as twist,
are described which should have applicability to many processes.
We find that some single variables, such as the pull angle, are weak discriminants,
but when combined with others they provide important marginal improvement.
We also find that multiple Higgs boson-candidate mass measures, such as from mild and aggressively trimmed jets,
when combined may provide additional discriminating power.
Comparing the significance improvement from our variables to those used
in recent CDF and \DZero~searches, we find that a 10-20\% improvement
in significance against $Z/W+b \bar b$ is possible. Our analysis also
suggests that the $H+W/Z$ channel with $H\to \bbar$ is also viable at the LHC, without requiring a hard cut on the $W/Z$
transverse momentum.
}

\begin{document}


\section{Introduction}
Search strategies for new physics often depend on being able
to find a small signal on top of a large background.  In many
cases, there are an enormous number of possible discriminants
which one would ideally like to combine to maximize search
sensitivity.
One approach is to choose, by some means, a small set of well-understood and
fairly uncorrelated variables and feed them into
a multivariate discriminant such as an Artificial
Neural Network (ANN) or a Boosted Decision Tree (BDT).
This approach has been applied productively in the light Higgs boson
searches at CDF~\cite{CDF} and \DZero~\cite{D0,Abazov:2010zk}.
While these multivariate techniques can often improve
discrimination power, it is almost impossible to
follow their inner workings in full detail.
A major concern is that they
can pick up on unphysical features of the Monte Carlo samples
used to train them,
rather than real differences in the signal and background processes.
At the same time, it is also unclear how dependent the absolute performance
and robustness are on the initial choice of variables.
While we generally expect that there are important multidimensional correlations
that a computer can find better than a person, we would like to be able
to capitalize on this fact in a more systematic manner.
The goal of this paper is to develop such a systematic programme, showing how to
produce combined discriminants that are more trustworthy and better optimized.
Kinematic observables based on well-understood, hard, perturbative physics
become our 4-5 most powerful discriminants.  Beyond that, it is useful
to examine properties of the jets themselves affected by perturbative QCD emissions.

The main example we consider is the production of a Higgs boson at the 14 TeV LHC in association
with a vector boson ($Z$ or $W$), with the Higgs decaying to $\bbar$.
This process was studied by both ATLAS and CMS.  ATLAS, for example,
concluded that the most promising channel, $WH\to\ell\nu b\bar b$, would
not provide enough significance for discovery~\cite{AtlasTDR}.
More recently, $WH$ and $ZH$ were revived with the observation that putting
a hard cut on a single variable can significantly enhance signal-to-background ratio~\cite{Butterworth:2008iy,atlas}.
In particular, imposing a cut $p_T > 200$\,GeV on the reconstructed $Z$ reduces the
signal by factor of 20, but reduces the background by a much larger factor of 320.
This $p_T$ cut (used in conjunction with jet substructure
methods to optimize mass resolution and background shaping), reinstated $WH$ and $ZH$ as possible Higgs
discovery modes at the LHC.
It is, at the outset, unclear whether a multivariate approach would pick up
on the fact that a hard cut on $p_T$ could make $ZH$ or $WH$ viable. It is also unclear
whether the hard $p_T$ cut is optimal, or whether better use could be made of the 95\% of
$ZH$ signal events which this cut throws out.
We seek to optimize multivariate searches for the standard model Higgs boson at the
Tevatron and LHC.

With this motivation, our goal is to analyze
systematically the entire phase space of $ZH$ production with
$H\to b \bar{b}$ to find the kinematic regions that maximize
signal significance.
We will concentrate on separating $ZH \to \ell^+\ell^- b \bar{b}$ from its irreducible background
in the standard model, for example, from $pp \to \zbb\to\ell^+\ell^- b \bar{b}$.
Here, $\ell$ is an electron or muon.
We will also require the $b$'s to appear in separate jets.
This dijet reconstruction approach continues to work well at high Higgs  boson $p_T$ (up to about 400 GeV).
We note that our results are complementary to any further improvements achievable using substructure-based techniques.
We will only briefly discuss the reducible backgrounds, such as mis-tagged $b$-jets.
All of these other backgrounds account for less than half of the total background at CDF \ref{cdfnote10235}.
We will also show results for the $WH$ channel as compared to its irreducible $\ell \nu \bbar$ background.
For both $ZH$ and $WH$, the reducible backgrounds are clearly important to establish the final search reach, but
restricting to irreducible backgrounds more concisely illustrates our main observations.

Our first step is to develop and study a very broad range of single-variable discriminants.
We consider initially thousands of discriminants, including kinematic observables corresponding to the ``hard scattering''
(e.g. $\Delta \eta$ between the $b$ jets, $\hat{s}$, etc.)
along with observables that distinguish signal from background
due to differences in the QCD radiation patterns (e.g. color
connections, subjet multiplicities, etc.).  Some of these variables, such as
twist and pull, are more generally applicable.
We also  at varying the jet sizes
and jet algorithm, and the effect of trimming~\cite{Krohn:2009th}. Many of the variables
are very similar, but enough are sufficiently independent to be considered separately.

Once the input variables are cataloged, we establish a criterion to
evaluate their usefulness. The ultimate measure is, of
course, how much integrated luminosity the collider would need
to find (or reject) a hypothesized signal to a certain
statistical significance, say $5\sigma$. Even with this
criterion, there are multiple measures of significance ---
should we compare the events in a single optimized signal bin
to the Monte Carlo prediction for signal and background? Should
we look for an excess in the signal region compared to the
sidebands? Should we fit curves and compare the $\chi^2$ for
various simulated distributions? How do we treat the
experimental systematic uncertainties? In fact, it is nearly
impossible to get an accurate measure of the final search reach
from a theoretical study without
collaboration-approval-dependent full detector simulation.
Nevertheless, the relative importance of different
discriminants should be roughly independent of the final search
strategy. We therefore argue that the Significance Improvement Characteristic (SIC), defined as the
signal efficiency divided by the square root of background efficiency resulting from a cut on a given discriminant,
\begin{equation}
  \rsig\equiv\frac{\eS}{\sqrt{\eB}},
\end{equation}
is a practical and useful measure.
We will show that this quantity, viewed as a function of
$\eS$ allows us to efficiently explore the convergence and limitations of the multivariate combinations.
One can also use the maximum of $\rsig$, $\ropt$, to rank variables and as a quantitative measure of final efficiency.

This paper is organized as follows:
Section~\ref{sec:cross} describes the selection cuts we use for the signal and background samples,
and the resulting cross sections.
We only consider irreducible backgrounds, so the numbers present do not directly translate into a discovery potential.
In Section~\ref{sec:vars}, we describe our single variable discriminants. This section includes variables
which are useful at the hard parton level, such as jet $p_T$'s, physically motivated variables, such as
helicity angles, variables dependent on the radiation pattern, such as pull~\cite{Gallicchio:2010sw},
and some useful variables that do not have an obvious physical motivation as well. Section~\ref{sec:eff} discusses some efficiency measures and motivates
the  $\rsig$ curves. Section~\ref{sec:bdt} gives a brief introduction to some of the multivariate measures we consider, and
explains some features of boosted decision trees (BDTs), our method of choice.
Section~\ref{sec:multi} describes improvements when variables are combined and our algorithm for finding
the optimal set. We show that in the case of associated Higgs boson production the $\rsig$ curves
for Boosted Decision Trees continue to increase in performance until the
of 6$^\mathrm{th}$ or 7$^\mathrm{th}$ variable is added. The addition of more variables does not provide additional
discrimination --- it does not improve the SIC.
The efficiencies used in these sections
are all based on samples constrained to lie within a
fixed Higgs mass window using a particular jet algorithm.
We justify this window and algorithm in Section~\ref{sec:jetting}. Section~\ref{sec:jetting} also shows that
additional improvement may result from combining multiple measures of the $\bbar$ invariant mass, from different jet algorithms.
For the final discriminants the mass window is removed and $\mbb$ is included in the multivariate
analysis.  Section~\ref{sec:money} shows the final discriminant combinations for the Tevatron and the LHC.
A summary and discussion is presented in Section~\ref{sec:conc}.

\section{Event Generation \label{sec:cross}}
\label{sec:Cross Sections}
The bulk of this paper will refer to a reference sample
of events generated initially with {\sc madgraph v4.4.26}~\cite{Alwall:2007st}: $pp\to
ZH \to \ell^+ \ell^- \bbar$ for signal and $pp \to Z\bbar\to \ell^+\ell^- \bbar $ for
background at $\sqrt{S}$=14\,TeV. These are then showered through {\sc pythia v8.140}~\cite{Sjostrand:2007gs}.
Jets are reconstructed using {\sc fastjet v2.4.2}~\cite{Cacciari:2005hq}, and these (along with the leptons) serve as our
``detector-level'' objects.
The multivariate analysis is done using
the {\sc tmva v4.0.4} package~\cite{tmva} that comes with {\sc root v5.27.02}~\cite{root}.
Our reference Higgs boson mass is $m_H = 120$\,GeV throughout: above the LEP limit of 115\,GeV
but below 130\,GeV where $b \bar b$ decay no longer dominates.
Masses below 115\,GeV are excluded by LEP, and decay to $WW^*$ starts to dominate above around 135\,GeV
\cite{Djouadi:2005gi}.
We will also consider $ZH$ events at the Tevatron $p\bar{p}$ with $\sqrt{S} = 1.96$\,TeV,
and $WH$ events at the Tevatron and LHC.
The $WH$ events will be compared to their  $W \bbar$ irreducible backgrounds. Reducible backgrounds such as
$Wjj$ with false $b$-tags or $t\bar{t}$ will not be considered, for simplicity.

Generator-level cuts are described in Table \ref{tab:cuts}.
It is important that the cuts applied at the hard parton level, in {\sc madgraph},
not be as tight as the cuts used for the final jets.
We found that a factor of 2 margin was wide enough while maintaining acceptable generation efficiency.
We did not apply a cut on $\mbb$ in the generated samples.
Once this sample is generated and showered, we require two $b$-tagged jets.  Our operative definition of $b$-tagging
matches $B$-hadrons from the intermediate event record to final-state jets with $\Delta R_{jB}$ smaller than the jet clustering
radius. We then cut on the $b$-jet $p_T$ and rapidity.

\begin{table}[h!]
\label{table:gencuts}
\begin{center}
\begin{tabular}{|c|c|}
\hline hard-parton level cuts & detector level cuts \\
\hline
$p_T^b > 7$ GeV & $p_T^b > 15$ GeV \\
$p_T^\mu > 3$ GeV& $p_T^\mu > 6$ GeV \\
$p_T^e > 3$ GeV & $p_T^e > 20$ GeV (LHC), $10$ GeV (Tevatron) \\
$|\eta_b| < 5$ and  $|\eta_\ell| < 5$  &  $|\eta_b| < 2.5$ and  $|\eta_\ell| < 2.5$ \\
\hline
\end{tabular}
\caption{Cuts applied for event generation.\label{tab:cuts}}
\end{center}
\end{table}

We generated 3 million signal events and 30 million background events.
After $b$-tagging and detector cuts were applied, we were left with
around 2M signal and 4M background events.
Within our fiducial Higgs Mass Window of \Higgswindow, we ended up with
1.5M signal and 0.6M background simulated events.
This specific window will be justified later as the one that maximizes significance.

\begin{table}[t]
\label{table:Cross Sections}
\begin{center}
\begin{tabular}{|c|c|c|c|c|}
\hline
\                                        &   \multicolumn{2}{c|}{LHC (14 TeV)}        & \multicolumn{2}{c|}{Tevatron (1.96 TeV)}     \\ \hline
Integrated Luminosity,  $\int\!\mathscr{L}$     &  \multicolumn{2}{c|}{30 $fb^{-1}$}  & \multicolumn{2}{c|}{10 $fb^{-1}$}            \\ \hline
\                                        & $pp \rt ZH$  & $pp \rt Z b \bar b$  & $p\bar p \rt ZH$ & $p\bar p \rt Z b \bar b$         \\ \hline
$\sigma$ times Branching Ratio           & 33.4 fb      & 57,200 fb            & 3.63 fb     & 1250 fb              \\ \hline
After Generator-Level Cuts               & 31.5 fb      & 26,000 fb            & 3.40 fb     &  570 fb              \\ \hline
Two $b$ Tags \% (of Gen-Level)           & 57\%         & 25\%                 & 81\%        & 25\%                 \\ \hline
Higgs Mass Window \% (of Gen-Level)      & 40\%         &  4\%                 & 52\%        &  3\%                 \\ \hline
Initiated by $gg$ (as opposed to $q \bar q$) & 0\%      & 90\%                 &  0\%        & 27\%                 \\ \hline
Xsec (in Higgs Mass Window)             & 12.3 fb      & 1100 fb              & 1.8 fb      & 14.9 fb              \\ \hline
Events (Xsec $\times\int\!\mathscr{L}$)  & 370          & 33,700               & 18          & 149                  \\ \hline
Starting $B/S$                           &   \multicolumn{2}{c|}{91.1}         &     \multicolumn{2}{c|}{8.2}       \\ \hline
Starting $S/\sqrt{B}$                    &   \multicolumn{2}{c|}{2.02}         &     \multicolumn{2}{c|}{1.47}      \\ \hline
\end{tabular}
\caption{
Cross Sections for LHC and Tevatron signal and
background. The lowest 6 rows refer to a Higgs mass-window cut, \Higgswindow ,
where the mass is computed from the hardest two $b$-tagged $R\!=\!0.5$ anti-$k_T$ jets. The significances here will
be the baseline references from which we compute fractional improvements.
This applies even for samples without an $\mbb$ cut where different $\mbb$ measures are used as part of a multivariate
discriminant.
}
\end{center}
\end{table}

The overall cross sections and efficiencies for these cuts are shown in Table~\ref{table:Cross Sections}.
The row in this table labeled ``Higgs Window\,\%'' refers the percent of events with ~(a window we find later to be optimal, see Section~\ref{sec:jetting} below).
For this table, jets are found using the anti-$k_T$ algorithm with $R\!=\!0.5$ and $\mbb$ is the invariant mass of the two
$b$-tagged jets in the event.
The ``Xsec''  row, and the rows below, provide cross sections and a normalization for the
initial significance. These are all after the $\mbb$ mass window cuts, but with no other discriminating variable applied.
Our improvements will be compared to this significance.
We will later treat the mass window in a more sophisticated way,
combining multiple mass measures and jet algorithms.
Here find it pedagogically useful to have a standard set of reference efficiencies.

The row in the table labeled ``Initiated by $gg$'', which is also for events within the $\mbb$ window,
shows an important distinction between the Tevatron and the LHC.
Note that the signal is all $q\bar{q}$ initiated at both machines. Since
the $Z\bbar$ background at the LHC is mostly $gg$ initiated it
will be easier to distinguish from signal than the backgrounds at
the Tevatron, which are mostly $q\bar{q}$-initiated, and therefore more similar to  the signal.

\section{Single Variable Discriminants \label{sec:vars}}
In this section we catalog all of the variables we use as potential discriminants between signal and background.
The language will refer to the $ZH\to \ell^+ \ell^- \bbar$ process, although the same variables can be used for the $WH$ sample
with $\ell \nu$ replacing $\ell^+ \ell^-$. The variables will be split roughly into two classes, {\bf kinematic} and {\bf radiation}.
{\it Kinematic} variables are those
which are meaningful at the hard parton-level, such as $\mbb$. They are expressible in terms of the 4-momenta of the
$\ell^+$, $\ell^-$, $b$, and $\bar{b}$.  Although the kinetic variables can be defined at
the parton level, we will measure them using the $b$-jet momenta.
{\it Radiation} variables, such as the number of charged hadrons in a jet,
are those which result mainly from QCD radiation.
In the Monte Carlo, these variables are populated due to the parton shower.

To begin, consider how many independent degrees of
freedom there are at the hard parton level.
We will treat showering and initial state radiation later.
The underlying hard process we are
interested in is $pp \to \zh \to \ell^+ \ell^- \bbar$. The final state
is characterized by the 4-momenta of the four final state
particles, which is 12 degrees of freedom including the
mass-shell constraints of the $b$'s and leptons.
One approach to constructing discriminants is to simply throw
the 12 degrees of freedom, $p_b^x, p_b^y ,p_b^z$, {\it etc.},  into a multivariate analysis and
hope for the best. However, it makes more sense to consider
physically motivated variables.
Azimuthal rotation invariance, and
the overall $\vp_T=0$ constraint reduce the physical degrees of freedom to 9,
and $Z$ and $H$ invariant mass constraints reduce the number down to 7.
Since there are multiple ways
physics can motivate the choice of variables,
this will result in many more than 7 variables.
We use the standard  coordinate system with $\pm \hat{z}$ pointing in the beam direction, $y$ is the rapidity, $\eta$ is the pseudorapidity, and $\phi$ is the azimuthal angle.

We first consider variables which are natural from the experimental point
of view. These include things like transverse
momenta, invariant masses, rapidity differences,
angular separations, {\it etc.}. These will be cataloged in Section~\ref{sec:hardlab}. This can be
thought of as a type of bottom-up parametrization.
The alternative is a top-down parametrization, motivated by the physical process.
One can start with variables that characterize the $ZH$ production, such as
$\hat{s}$ and the production polar angle $\theta^\star$.  Then when the Higgs and $Z$ decay, one can think about various
angles constructed from their decay products in their rest frame. This approach will lead to variables
discussed in Sections~\ref{sec:twist} and \ref{sec:helicity}.
Combining the bottom-up and top-down ways of thinking leads to an even larger set of possible discriminants, discussed in
Section~\ref{sec:menu}. Sections~\ref{sec:showered} to~\ref{sec:missing} describe some of the showered variables.


\subsection{Lab-frame Kinematical Variables \label{sec:hardlab}}
First, consider variables which are natural to define and measure in the lab-frame.
Using only the 4-momenta of the $b$ jets in the lab frame, we have
\begin{itemize}
\item $p_T$ of the two $b$'s:  $p_T^{b1}$ for the
    higher-$p_T$, and $p_T^{b2}$ for the lower-$p_T$ one
\item $\Delta\eta_{\bbar} =  \eta_b - \eta_{\bar{b}} $
\item $\Delta\phi_{\bbar} =  \phi_b - \phi_{\bar{b}} $ \qquad (properly wrapped to have a range between $-\pi$ and $\pi$)
\item $\Delta R_{\bbar} = \sqrt{ (\Delta\eta_{\bbar})^2 +
    (\Delta\phi_{\bbar})^2}$
\end{itemize}
The same variables are also considered for the lepton system,
with $\ell^+$ and $\ell^-$ replacing the $\bbar$. There, one
can also treat the positively and negatively charged leptons
separately, although we found that sorting by $p_T$, as with the $b$'s,
tends to work better.

There are variables involving the reconstructed Higgs boson
with four-momenta $p_H^\mu = p_b^\mu + p_{\bar{b}}^\mu$, and
reconstructed $Z$, with $p_Z^\mu = p_{\ell^+}^\mu +
p_{\ell^-}^\mu$:
\begin{itemize}
\item $p_T^Z$ and $p_T^H$, the $p_T$ of the $Z$ and Higgs boson. At the parton-level these are the same, but they end up slightly
    different due to jet reconstruction.
\item $p_T^{CM}$: Center of Mass $p_T$, the magnitude of the vectorial sum of the 2 $b$-jet and 2 lepton $p_T$'s. This is zero for the
    parton-level process, but non-zero after showering and jet reconstruction.
    $p_T^{CM}$ is not included in our analysis, as discussed below in the missing $E_T$ section.
\end{itemize}

\begin{figure}[t]
\begin{center}
\includegraphics[width=0.3\textwidth]{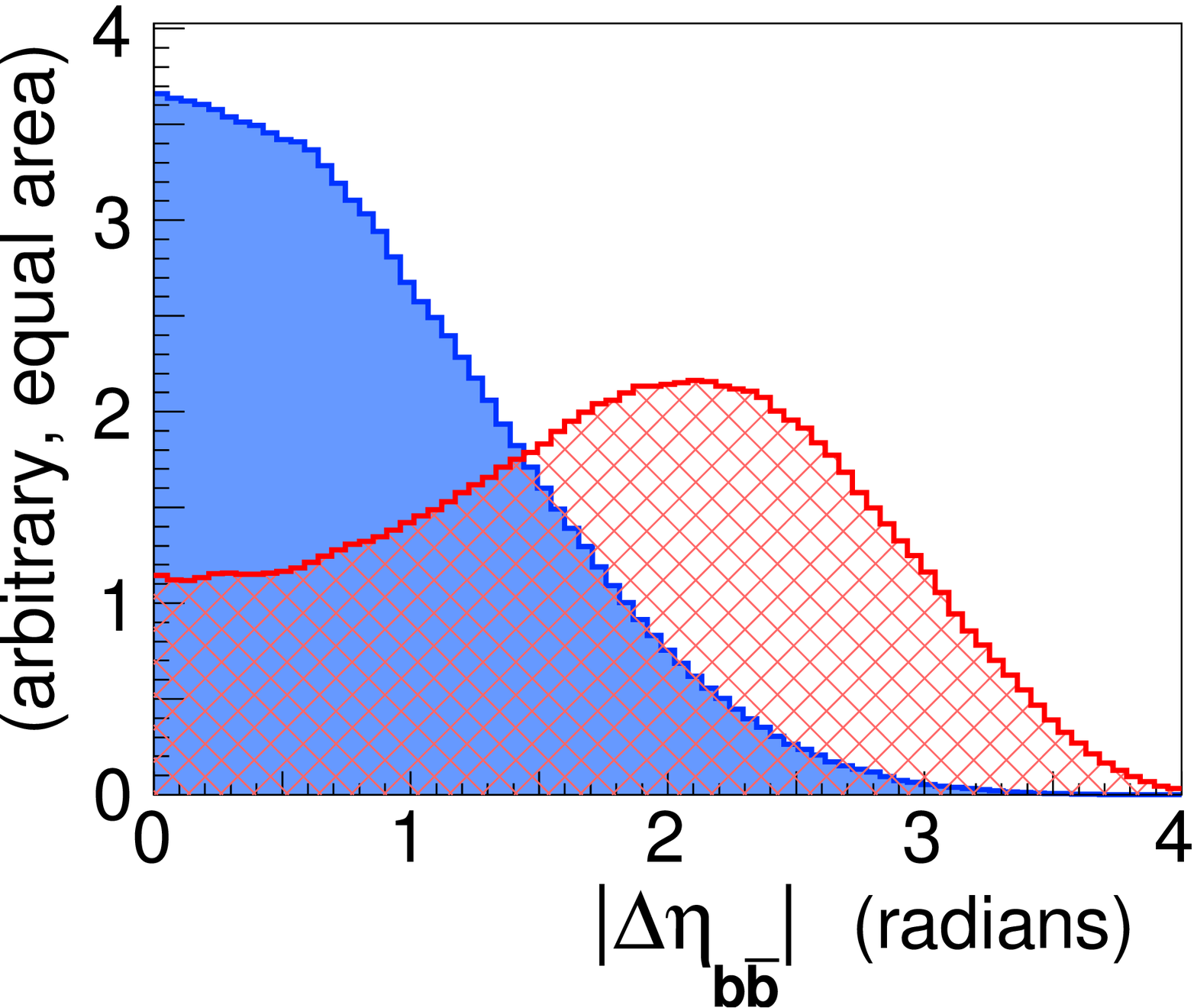}
\includegraphics[width=0.3\textwidth]{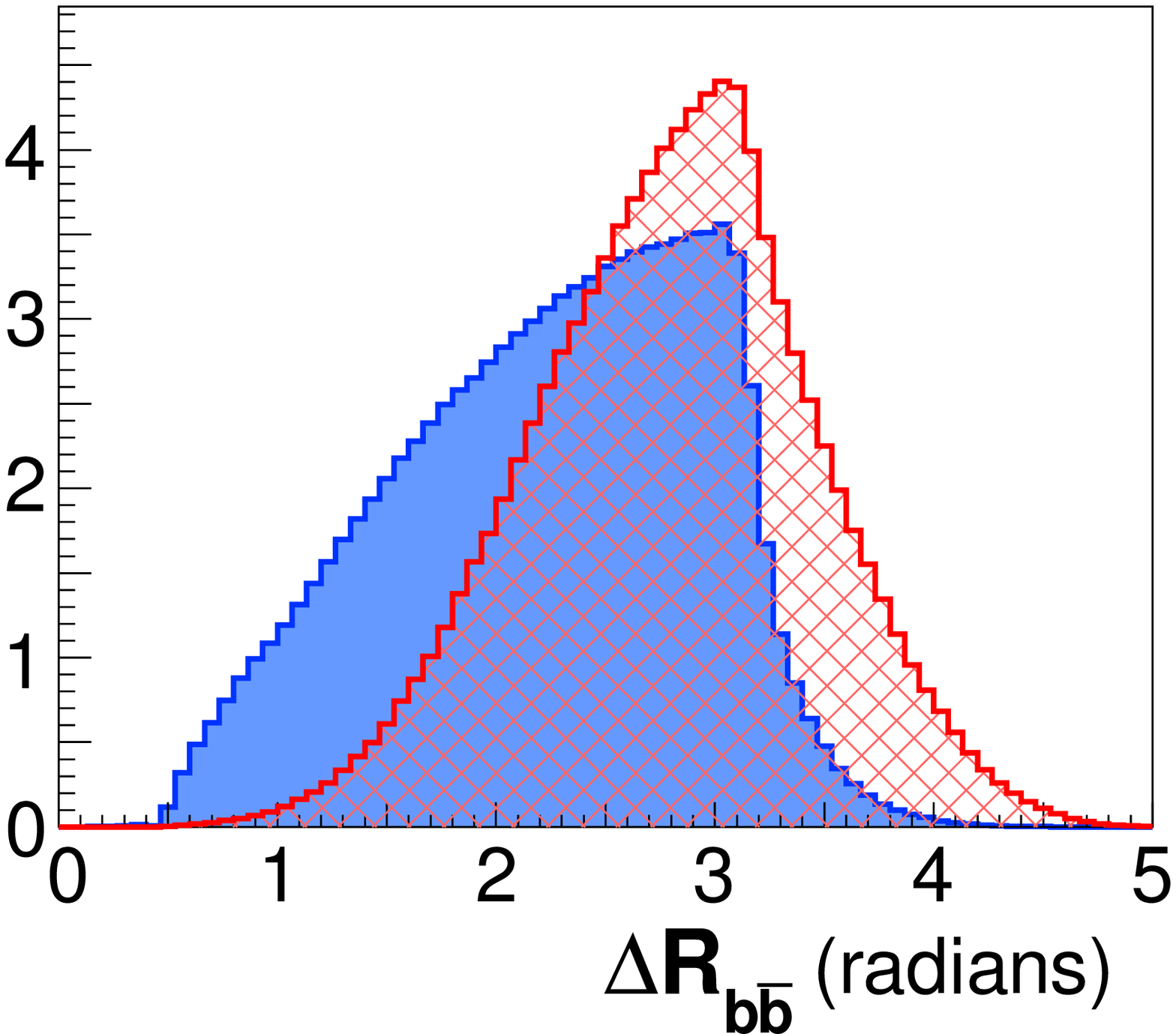}
\includegraphics[width=0.3\textwidth]{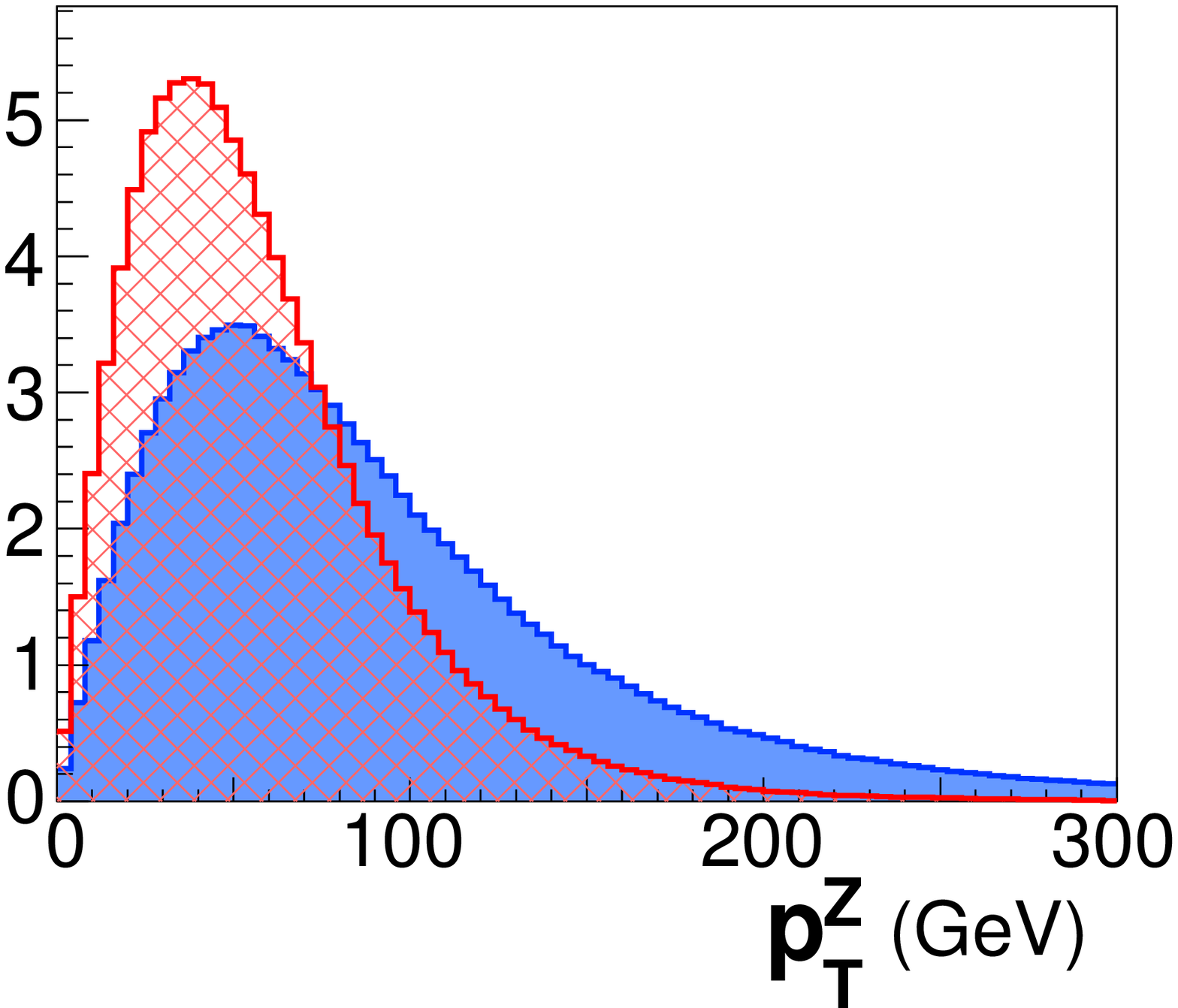}
\includegraphics[width=0.3\textwidth]{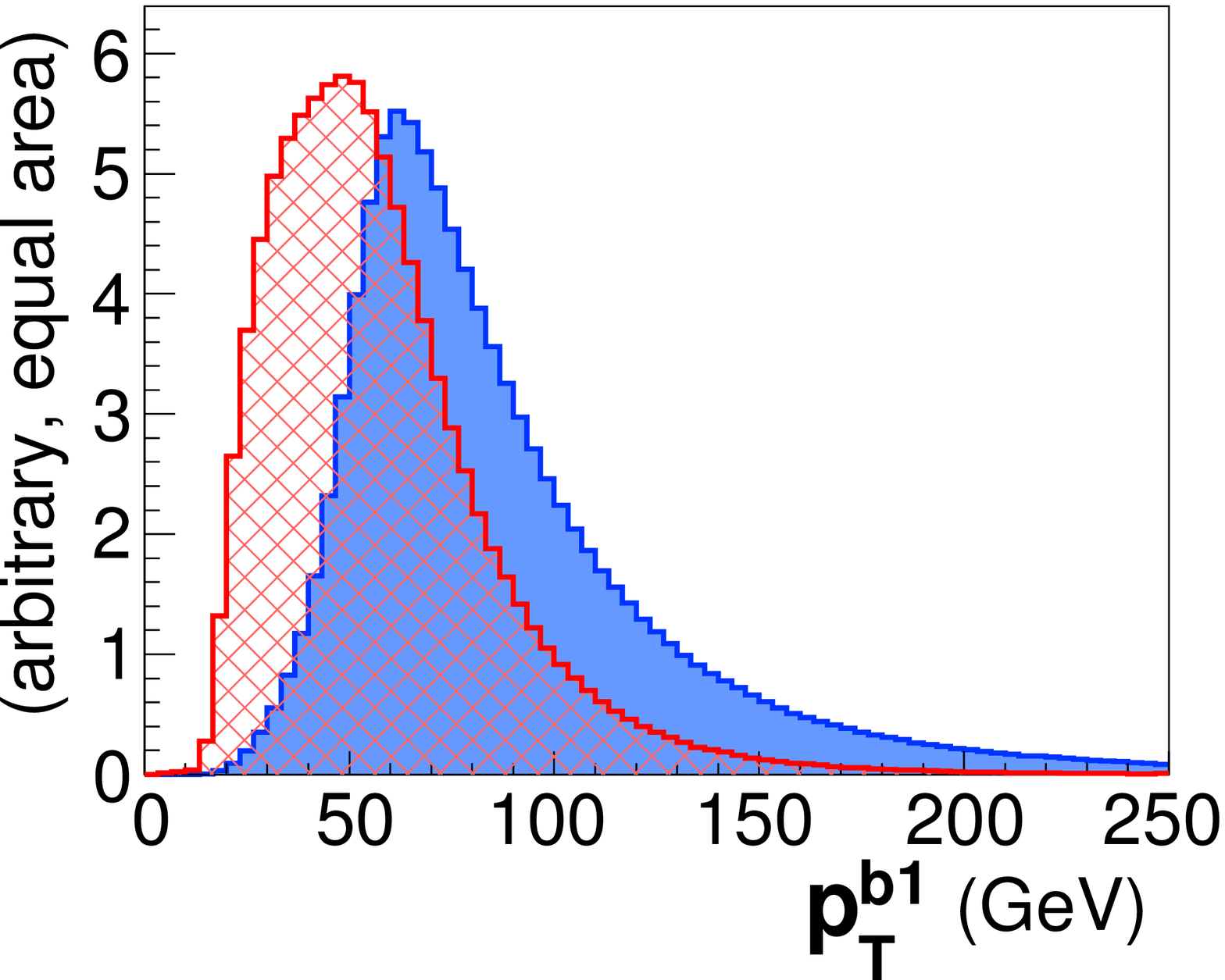}
\includegraphics[width=0.3\textwidth]{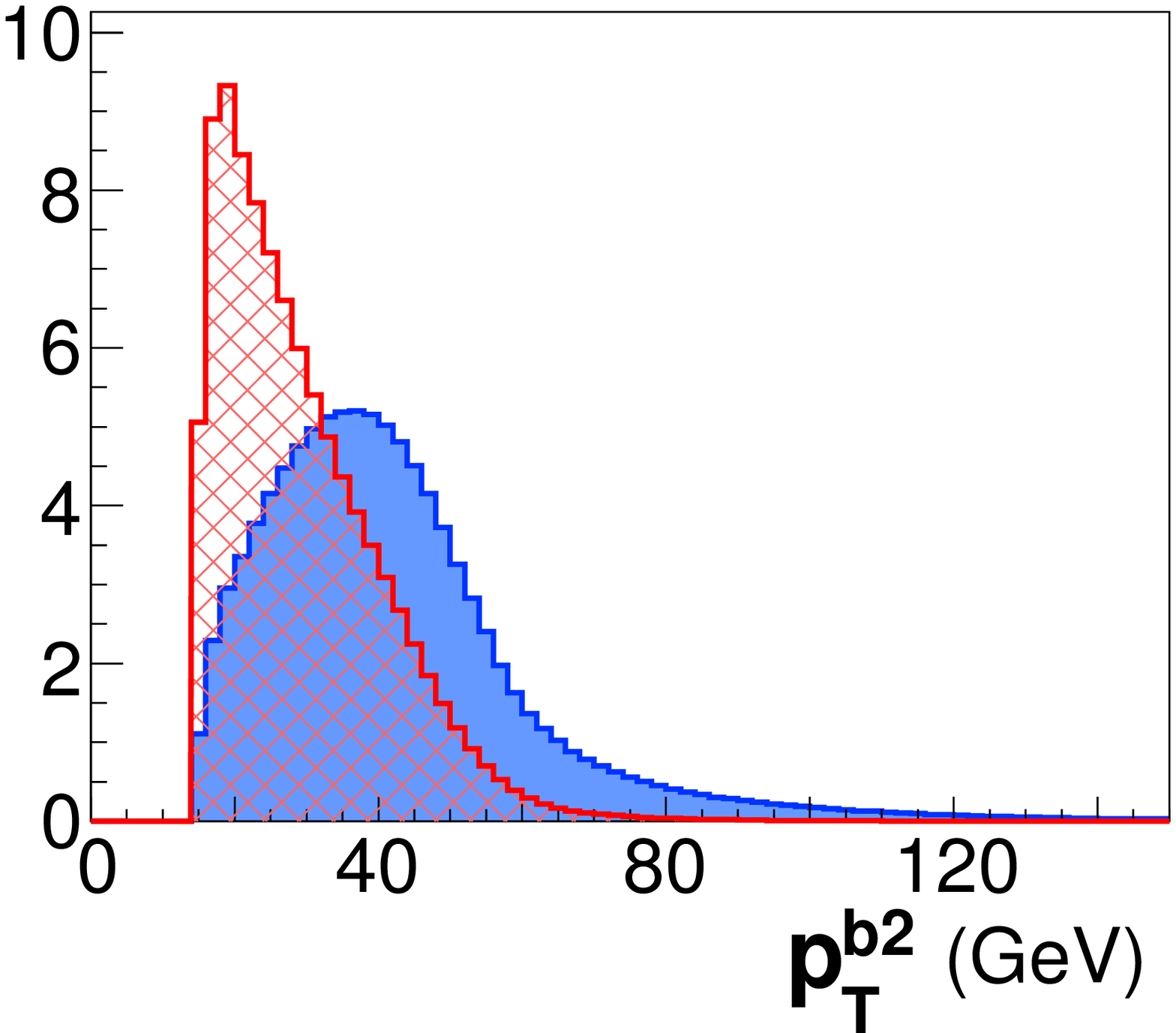}
\includegraphics[width=0.3\textwidth]{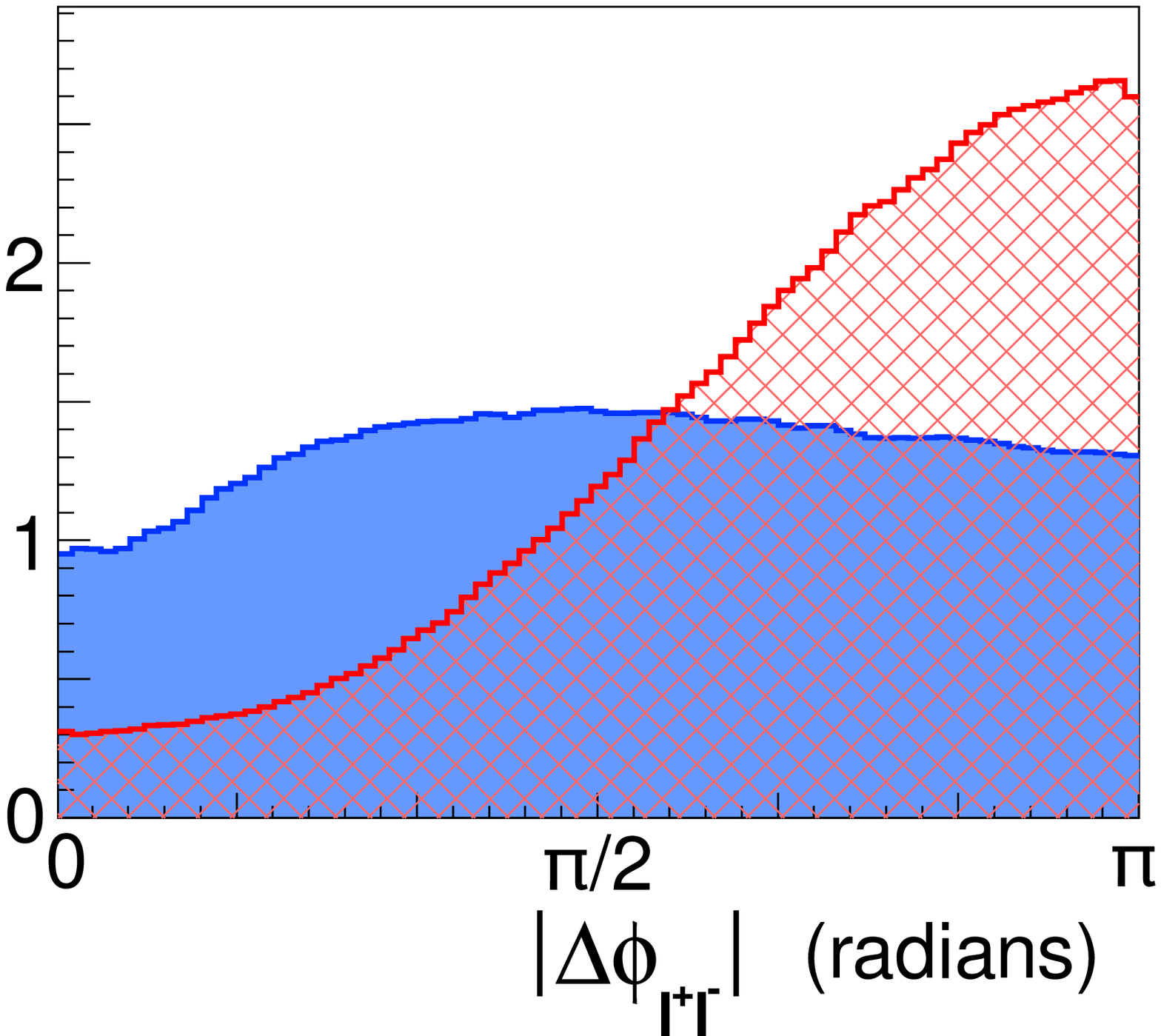}
\caption{Some lab-frame kinematic variables for $\zh$ signal (solid blue)
and $\zbb$ background (hashed red) at the LHC.
Events satisfy selection cuts and the Higgs mass window cut, \Higgswindow . Horizontal
axes are in radians or GeV as appropriate,
and vertical axes are in arbitrary units with signal and background normalized to the same area.}
\label{fig:lab}
\end{center}
\end{figure}

We show a set of these variables in Figure~\ref{fig:lab}.
Some of these, such as $\Delta \eta_{b \bar b}$, look to provide very good discriminants. All of the variables
are shown for anti-$k_T$ $R\!=\!0.5$ jets \cite{Cacciari:2008gp}.
Iterative jet algorithms, including anti-$k_T$ jets are defined by
first assigning all energy depositions into their own protojet.
At each stage of the clustering, calculate the
\emph{distance between} each pair of protojets, defined by
\begin{equation}
d_{ij} = min( k_{t_i}^{2p},k_{t_j}^{2p})\frac{\Delta_{ij}^{2}}{R^{2}} ,
\end{equation}
along with the \emph{beam distance} of each protojet, defined by
\begin{equation}
d_{iB} = k_{t_i}^{2p} ,
\end{equation}
where $\Delta_{ij}^{2} = (y_{i} - y_{j})^{2} + (\phi_{i} - \phi_{j})^{2}$
and $k_{ti}$, $y_{i}$, and $\phi_{i}$ are the transverse momentum,
rapidity, and azimuth of protojets $i$ and $j$.
The parameter $p = 1$ for the $k_T$ algorithm,
$p = 0$ for the Cambridge/Achten algorithm,
and $p = -1$ for the anti-$k_{T}$ algorithm.
When one of the distances between protojets is the smallest,
those protojets are combined into another protojet by adding their 4-vectors.
When one of the beam distances
is the smallest, that protojet gets promoted to a real jet and removed from further
consideration.  Either way, all distances are computed again for the new set
of protojets, and the process repeats until all protojets have been promoted.

Some other variables inspired by previous Higgs boson searches include:
\begin{itemize}
\item acoplanarity = $\left| \pi - |\Delta \phi_{b \bar b}| \right| + \left| \pi - \Sigma \theta _{b \bar b} \right|$.    Also for $\ell^+ \ell^-$.
\item $m_T^{b \bar b} = \mbb^2 + p_x^2 + p_y^2$: transverse
    mass of $\bbar$ system.  Also for $\ell^+ \ell^-$.
\item $m_{\ell \ell b}$: invariant mass of the $Z$ combined with a $b$-jet.  The max or min over the two $b$'s can also be used.
\item $p_T^{\mathrm{imbalance}} = |\vec p_T^{b1} + \vec p_T^{b2} + \vec p_T^{\ell} - \vec p_T^{\nu}$|, for the $WH$ search.
\end{itemize}



\subsection{Twist  \label{sec:twist} }
\begin{figure}[ht]
\begin{center}
\psfrag{qqHZbbll}{\!\!\!\!\!\!\!Higgs Boson Signal}
\psfrag{ggZbbll}{Background Initiated by $gg$}
\psfrag{b/bbar dEta}{$\Delta \eta_{b \bar b}$}
\psfrag{b/bbar dPhi}{$\Delta \phi_{b \bar b}$}
\includegraphics[height=2in]{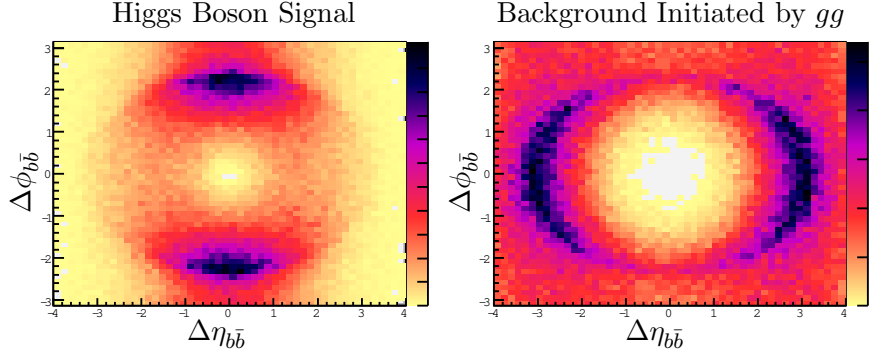}
\caption{$\Delta \eta_{b \bar b}$ vs $\Delta \phi_{b \bar b}$ for
the Higgs boson signal (left) and the $gg$ initiated $Z b\bar b$ background dominant at the LHC (right).
This is at the hard parton level, and for $p_T^H>50$ GeV.  The difference is less dramatic for
lower $p_T$ or for the Tevatron, where the $q \bar q$-initiated background dominates.
Absolute values could also be taken for $b$ indistinguishable from $\bar b$.
}
\label{fig:b_bbar_dPhi_vs_b_bbar_dEta}
\end{center}
\end{figure}

\begin{figure}[ht]
\begin{center}
\vspace{-0.5cm}
\begin{tabular}{cc}
 Signal-Like Twist $\tau=\pi/2$ &
 Background-Like Twist $\tau=0$ \\
 \includegraphics[width=0.35\textwidth]{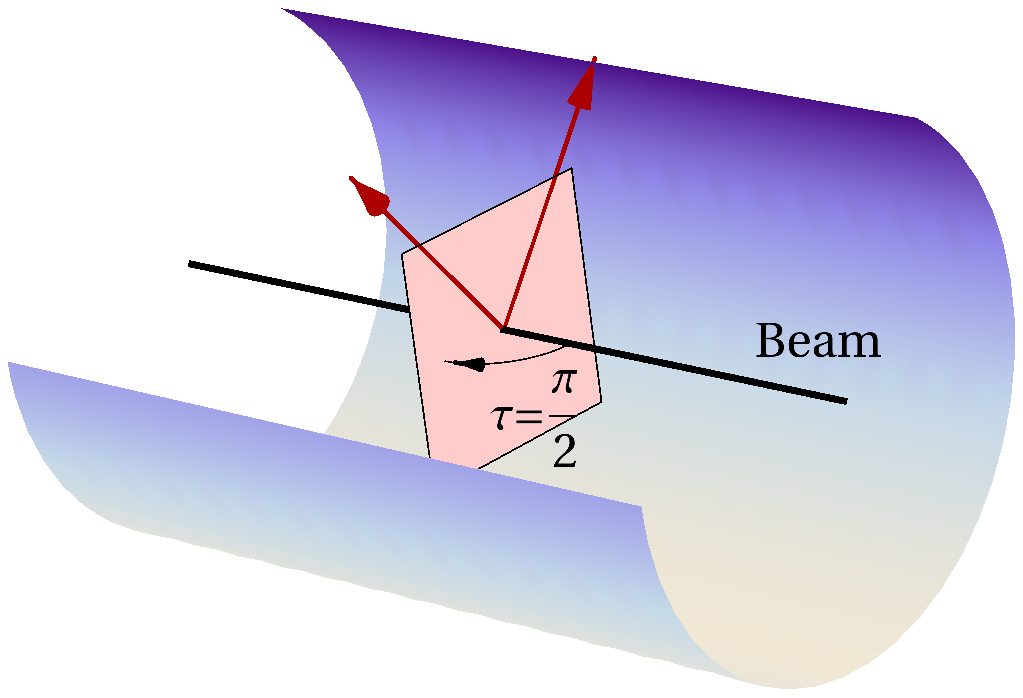} &
 \includegraphics[width=0.35\textwidth]{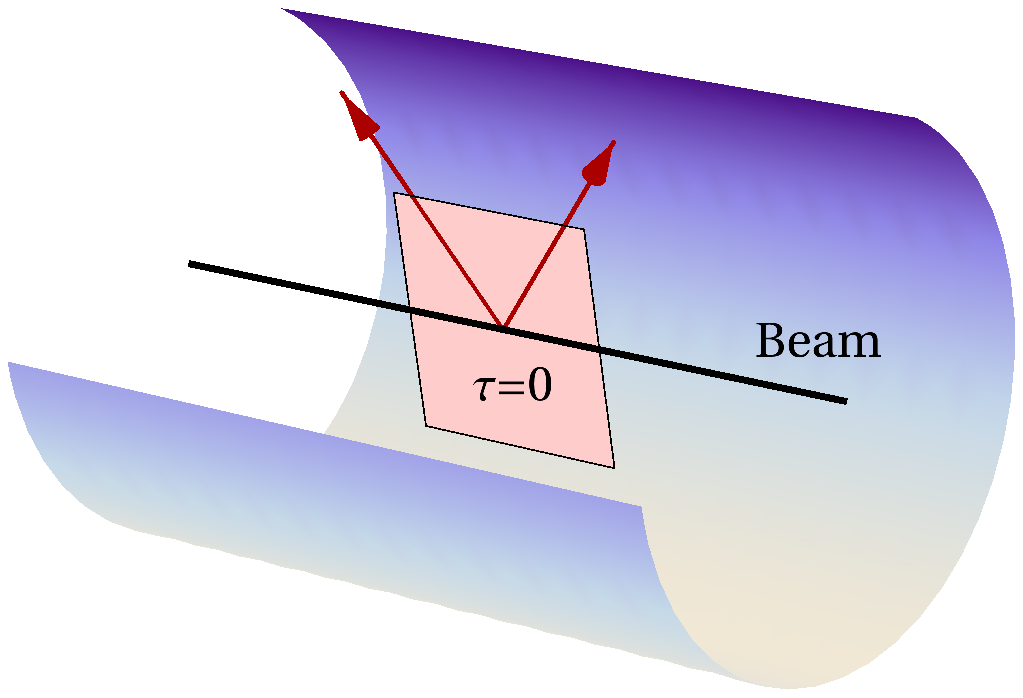}\\
\end{tabular}
\vspace{-0.5cm}
\caption{
Twist angle $\tau$ in 3D with the $b$ and $\bar b$
emerging from the interaction point.
The twist angle is defined to be boost invariant and
does not exactly correspond to the physical rotation angle of a plane.
The case shown, however, has no longitudinal boost.
}
\label{fig:twist2D}
\end{center}
\end{figure}

The strength of the $\bbar$ angle discriminants $\Delta \eta$, $\Delta \phi$ and $\Delta R$, motivates us to explore
the $\bbar$ system more carefully.
Figure~\ref{fig:b_bbar_dPhi_vs_b_bbar_dEta} shows the two-dimensional distribution of parton-level events in the
$(\Delta \eta, \Delta \phi)$ plane for the signal and background.  It
is clear from these plots that the polar angle, which
we call {\bf twist}, would be a good discriminant.  If we think of
$(\Delta \eta, \Delta \phi)$ as 2D Cartesian
coordinates, then polar coordinate combinations
are the familiar $\Delta R = \sqrt{\Delta \eta^2 + \Delta \phi^2}$, and the twist angle:
\begin{equation}
\tau \equiv \tan^{-1}\frac{\Delta \phi}{\Delta \eta} \, .
\end{equation}
Twist is a longitudinal-boost-invariant version of the rotation of the $H/b/\bar{b}$ plane with
respect to the $beam/H$ plane.
This is illustrated in Figure~\ref{fig:twist2D}.
The twist angle is zero when
the particles are separated \emph{along} the cylinder in $\eta$, and $\pi/2$ when
separated \emph{around} the cylinder in $\phi$.

The shape of the signal and background twist distributions in Figure~\ref{fig:b_bbar_dPhi_vs_b_bbar_dEta}
can be understood as follows.
For low Higgs $p_T$, $p_T^H \ll m_H$, the signal lives in bands clustered along $|\Delta\phi| \sim \pi$,
and hence $\tau \sim \pi/2$.
For $p_T \gsim m_H$,
the signal lives in a ring of $\Delta R \sim 2m_H/p_T$, and twist is a powerful variable orthogonal to the radial direction.
For higher $p_T$, the signal becomes somewhat more uniformly distributed in twist, but still retains its $\tau=\pi/2$ preference.
This is all purely a consequence of the spherically-symmetric Higgs boson decay boosted transverse to the beam.
In contrast, for the background, in particular the $gg$-initiated background,
there is a bias for one of the $b$'s to have large rapidity, which leads to low twist. This stems from $t$-channel type singularities
in the $Zb\bar b$ production matrix elements (in the limit of massless $b$'s.)
At higher $p_T$, the background retains much of its $\tau=0$ preference.
The distributions in Figure~\ref{fig:b_bbar_dPhi_vs_b_bbar_dEta} reflect an admixture
of the high and low $p_T$ twist behaviors, biased towards the low $p_T$ behavior, which is
where most of the events lie.

\begin{figure}[t]
\begin{center}
\begin{tabular}{cc}
\multicolumn{2}{c}{Madgraph hard parton level with no cuts} \\
\includegraphics[height=1.5in]{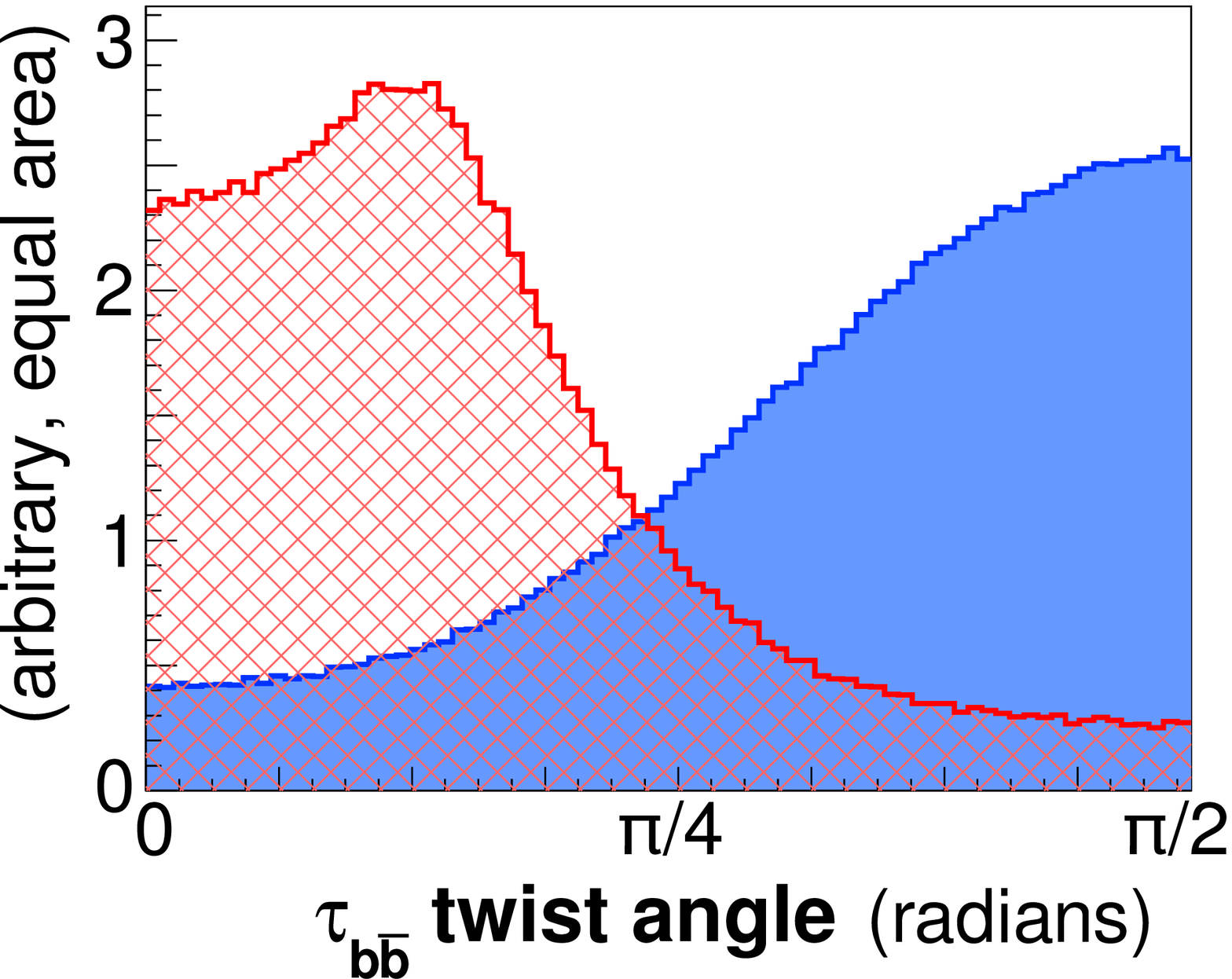} &   
\includegraphics[height=1.5in]{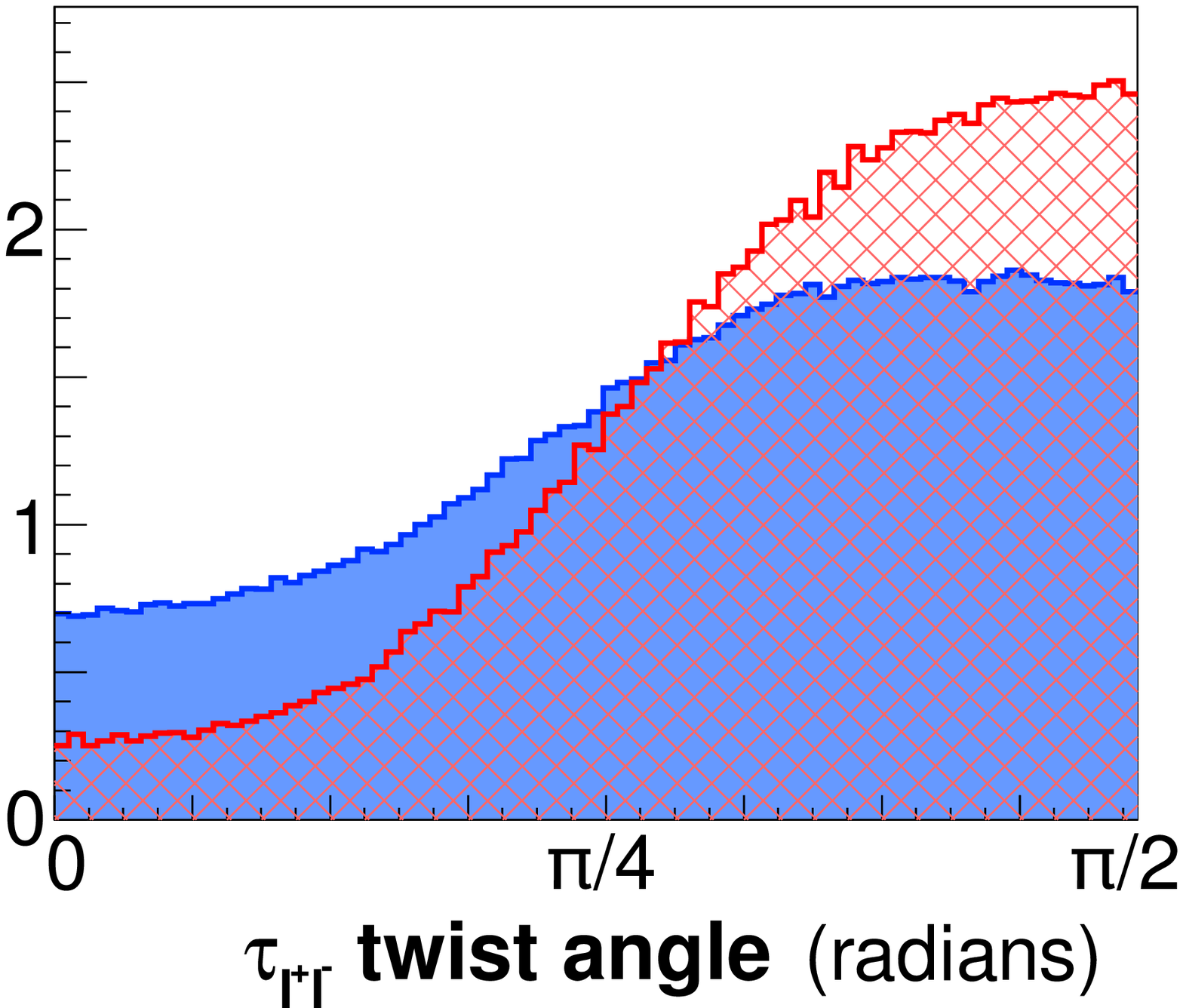} \\
\hline
\multicolumn{2}{c}{Showered and jetted, with detector cuts} \\
\includegraphics[height=1.5in]{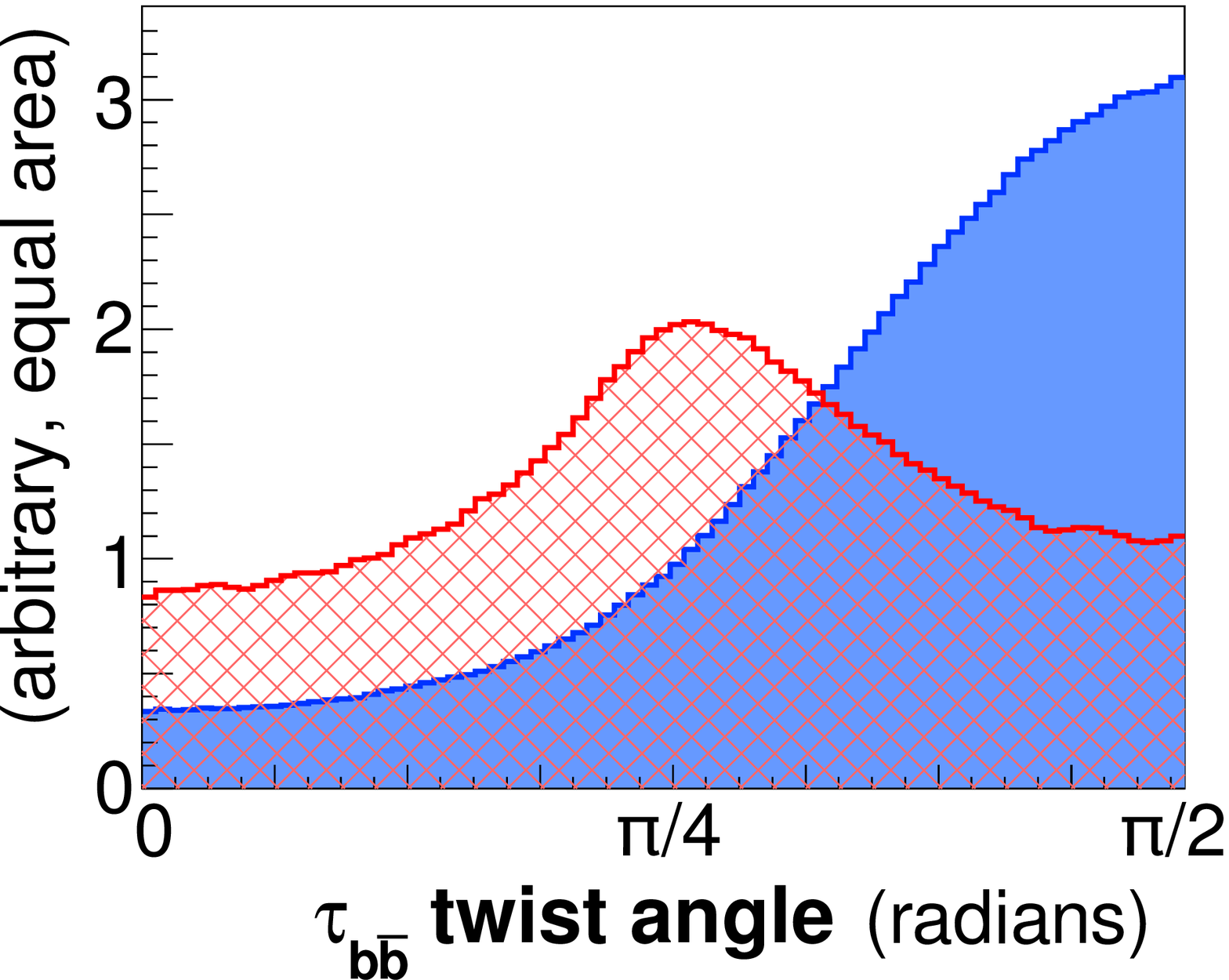} &
\includegraphics[height=1.5in]{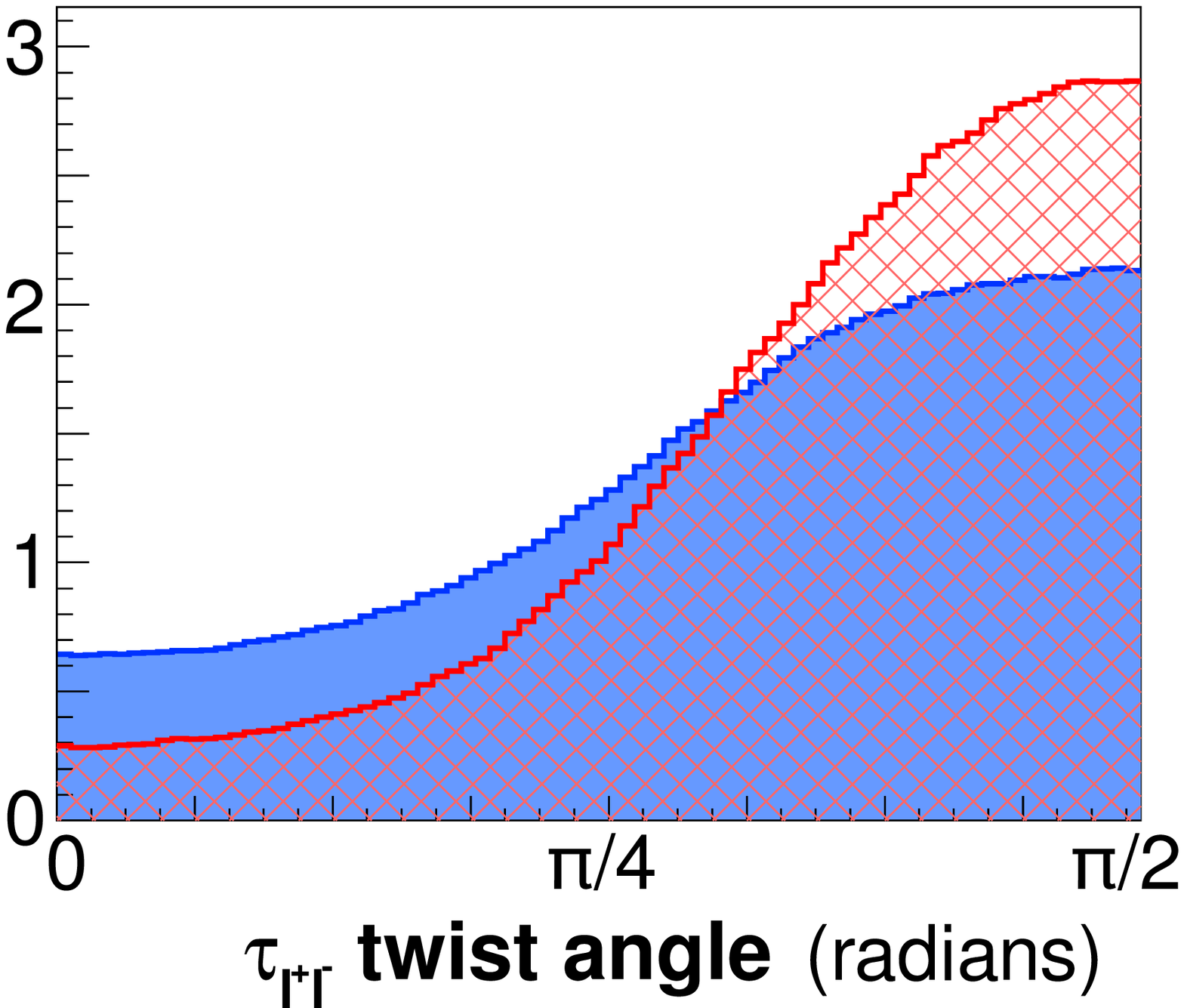} \\
\end{tabular}
\caption{
Twist angle distributions for $\zh$ signal (solid blue)
and $Z\bbar$ background (hashed red), for the LHC
with no $p_T^Z$ cut.
Madgraph hard parton-level with no cuts (top)
and showered jet-level with detector cuts (bottom).
Both are shown only in the Higgs mass-window, \Higgswindow .
Vertical axes are in arbitrary units with signal and background normalized to the same area.
}
\label{fig:twist_redblue}
\end{center}
\end{figure}

We show in Figure~\ref{fig:twist_redblue} the twist distributions for signal and background for the $\bbar$ system
and the $\ell^+\ell^-$ system.  The top panels show the twist distributions at the {\sc madgraph} level before showering and cuts, and the bottom
panels after jets are reconstructed and detector level cuts have been been applied.
We can see that
the discriminating power of twist for the $\bbar$ system is reduced by requiring relatively central jets with a minimal $p_T$.
However, the background's bias towards the
beam is still evident, and twist still provides a useful discriminant which we will incorporate into our
multivariate analysis.  Because of its physical
motivation, we also suspect the twist angle could have much wider
applicability than to the $ZH$ and $WH$ searches we consider here.

\subsection{Helicity and Azilicity Angles \label{sec:helicity}}

Next, we begin to consider variables motivated by the on-shell decays of the Higgs and the $Z$.
In their rest frames, each decay
is parametrized only by two angles $\theta$ and $\phi$. Since the
Higgs boson is a scalar, its decay products are spherically symmetric
and therefore distributed  uniformly in $\phi$ and
$\cos \theta$. Specifically, in the rest frame of the scalar
Higgs, the $b$ and $\bar b$ quarks travel in opposite
directions with a fixed energy ($m_H/2$, up to $b$ mass
corrections).
The rest frame of any fake Higgs boson formed by two $b$-jets will
also have two oppositely-going $b$'s in the $\bbar$
center-of-mass frame. If we impose that $\mbb$ be close to the
Higgs mass, in this frame, the background $b$'s will have
energies close to $m_H/2$ just like the signal, but they will
not be distributed in a spherically symmetric way.

\begin{figure}[t]
\begin{center}
\psfrag{b}{$b$}
\psfrag{bb}{$\bar b$}
\psfrag{e}{$\ell$}
\psfrag{H}{$H$ boost direction}
\psfrag{Z}{$Z$ direction}
\psfrag{BM}{beam}
\psfrag{AZ}{$\phi_{a}$}
\psfrag{TH}{$\theta_{h}$}
\includegraphics[height=2in]{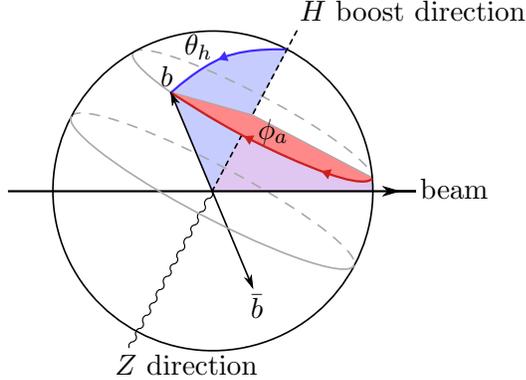}
\caption{Helicity angle $\theta_h$ and azilicity angle $\phi_{a}$
for $H \rightarrow b \bar b$.
Angles can also be defined for the leptons on the $Z \rightarrow \ell^+ \ell^-$ side
of the event.}
\label{fig:angles_higgs}
\end{center}
\end{figure}

Parameterizing the $H$ decay with two polar angles requires a
convention for the axes. Our convention is motivated by the
helicity angle used in $W$ and top studies~\cite{Chwalek:2007pc}.  The angles
are constructed in the rest frame of the Higgs boson, and refer to directions defined
by the $Z$ and beam 3-momenta observed in this frame.   We choose ``latitude'' to be measured
by the {\bf helicity} angle, $\theta$, with the south pole ($\theta=\pi$) defined
to be the direction aligned with the $Z$ (equivalently, the direction along which the $ZH$ parent
CM system moves).  Given this angle, we still have the freedom
to chose any direction perpendicular to the axis as the
$\phi=0$ direction.  We chose $\phi=0$ to be pointing toward the beam which was moving along $+\hat z$ in
lab frame coordinates.
The QCD background will favor $\phi=0$ and $\phi=\pi$.
With this convention, we call longitude the \textbf{azilicity} angle, since it
is the azimuthal $\phi$ angle on the sphere whose north pole is defined when the helicity angle vanishes.
Azilicity is equivalently the angle between the Higgs boson decay plane and the $ZH$ production plane
(constructed with reference to the beams) as viewed in the event's center of momentum frame.
Since $b$ and $\bar b$ are indistinguishable, this angle can be chosen to go between $0$ and $\pi/2$.
A cartoon of these angles is shown in Figure~\ref{fig:angles_higgs}.

\begin{figure}[t]
\begin{center}
\begin{tabular}{cc}
\multicolumn{2}{c}{Madgraph hard partons with no cuts:} \\
\includegraphics[height=1.5in]{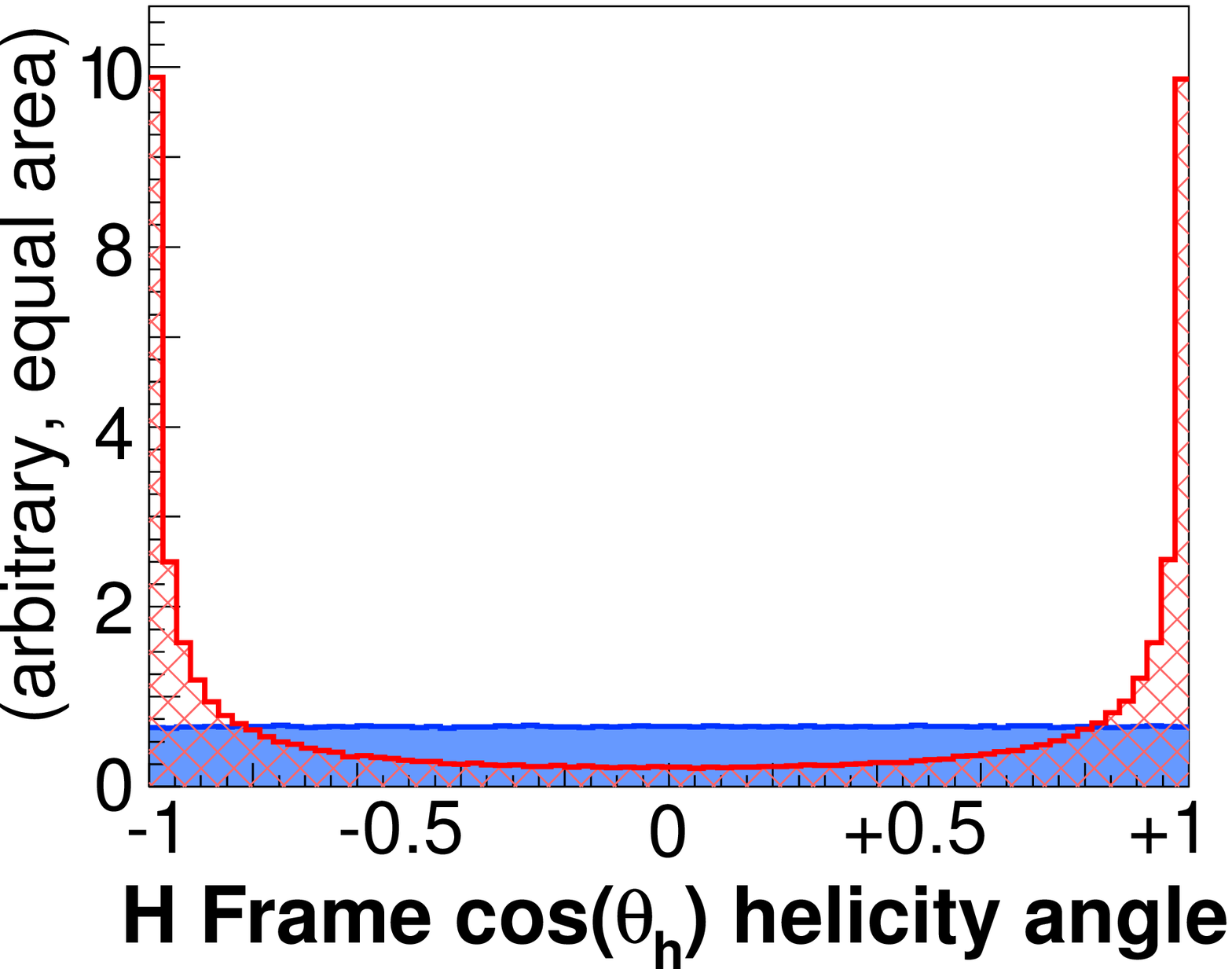} &
\includegraphics[height=1.5in]{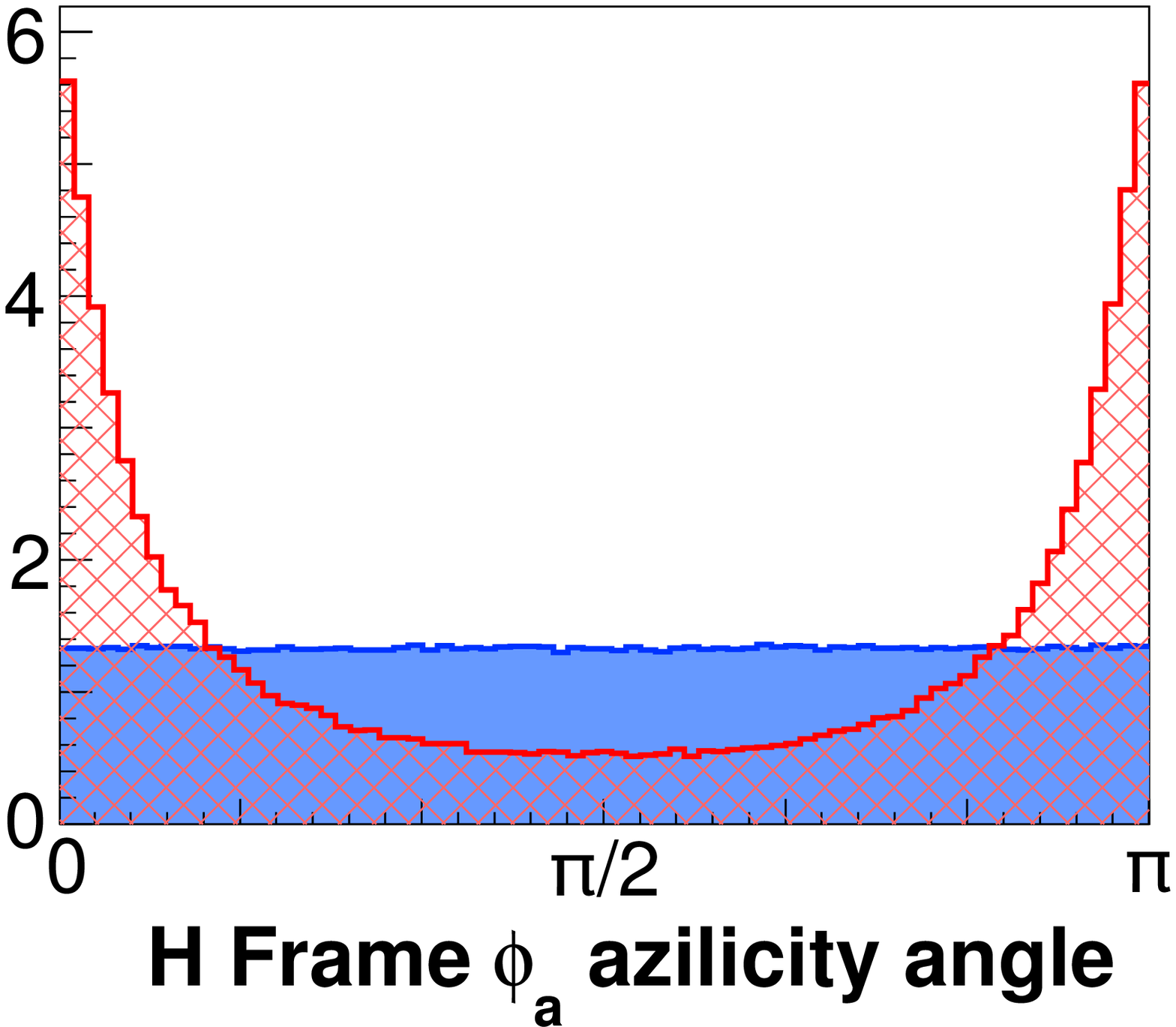} \\
\hline
\multicolumn{2}{c}{Showered and reconstructed, with detector cuts:} \\
\includegraphics[height=1.5in]{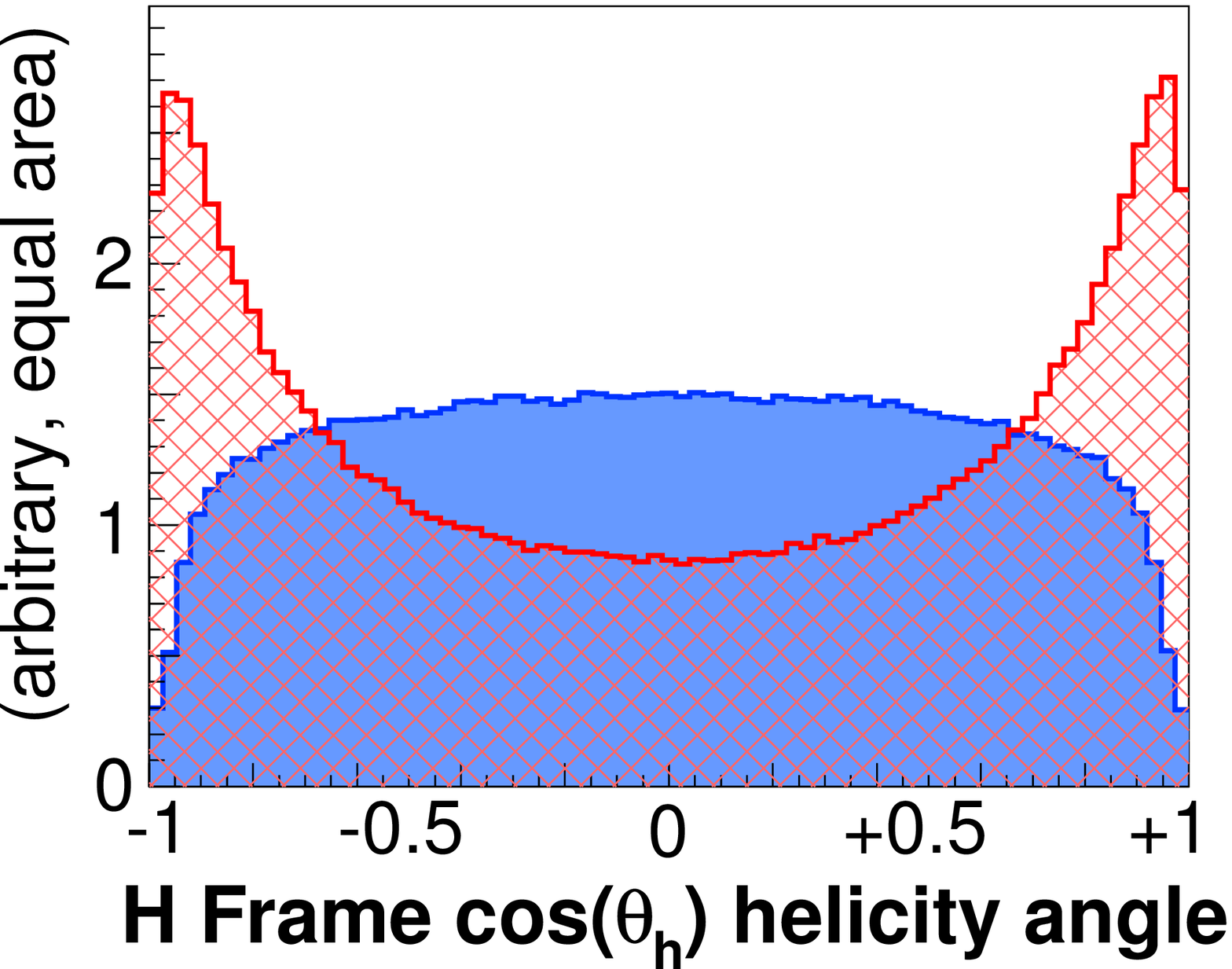} &
\includegraphics[height=1.5in]{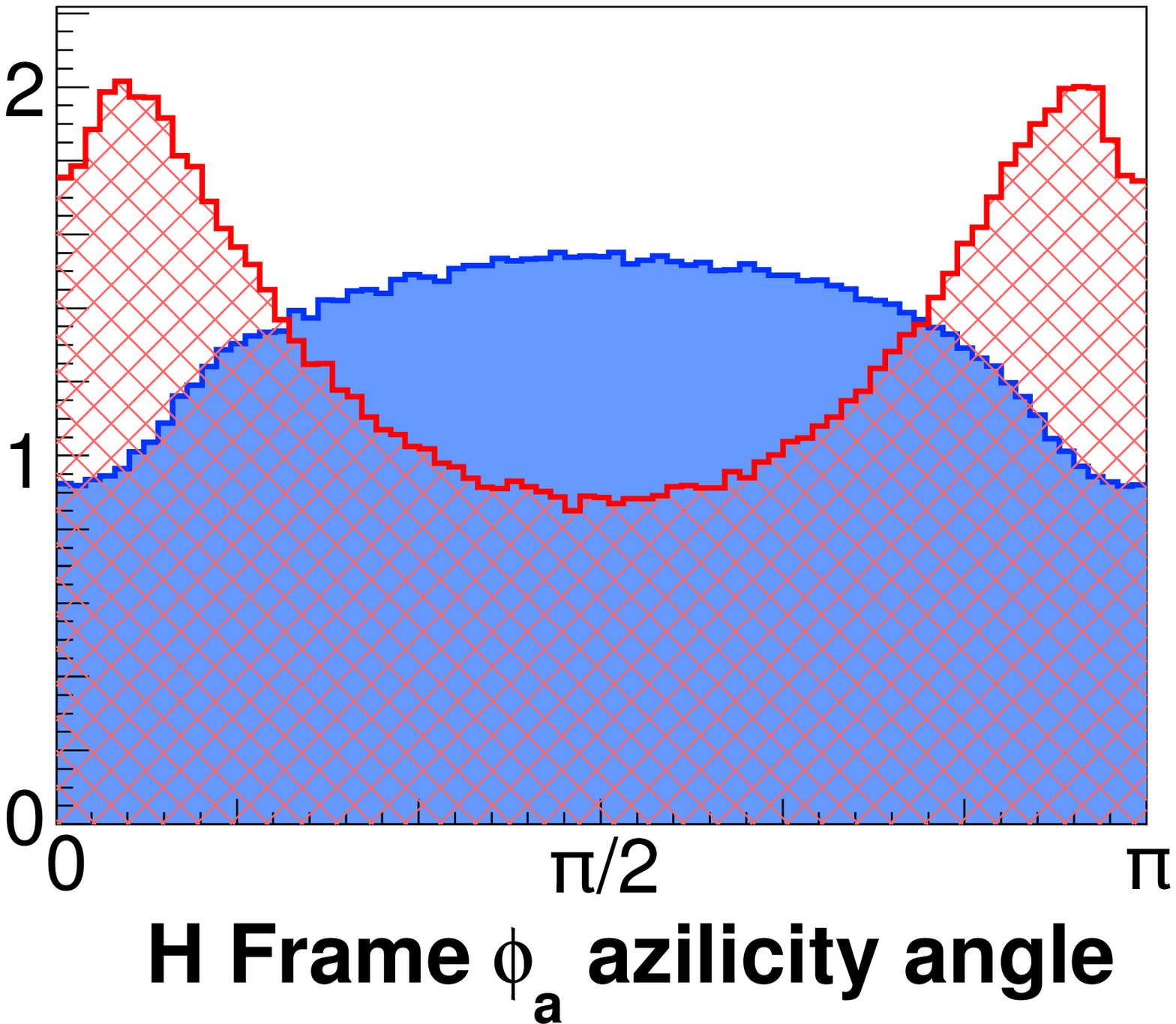} \\
\end{tabular}
\caption{Helicity angle $\theta$ and azilicity angle $\phi$ for the $b$ in
the Higgs boson rest frame, for $\zh$ signal (solid blue)
and $Z\bbar$ background (hashed red) at the LHC.
Madgraph hard parton-level with no cuts (top)
and showered jet-level with detector cuts (bottom).
Both are shown only in the Higgs mass-window, \Higgswindow .
}
\label{fig:H_Frame_b}
\end{center}
\end{figure}

The helicity and azilicity angles offer the promise of very strong discrimination power, because they
are directly tied to physical features of the signal. Indeed, in the top row of Figure~\ref{fig:H_Frame_b},
we can see that, at the parton-level with no cuts, there are strong singularities at both $\theta=0$ and $\phi=0$ for
the background, while the signal distributions are flat, as expected.  Detector observability cuts
remove the most singular background contributions, but still lead to distributions that show some remaining
discriminating power.

\subsection{Kinematic variable construction \label{sec:menu}}

We have described a number of variables constructed out of the 4-momenta of the jets and leptons in various frames.
There are large combination of variables which can be formed from the measurable kinematic variables.
However, using physically motivated
variables can help the automated process in the right direction.
By searching for useful combinations up front, a
neural network, for example, does not have to ``discover'' how
to take an invariant mass or boost to the Higgs boson rest frame.

Thus, we now consider various unintuitive combinations of variables.
The procedure is:
\begin{itemize}
\item Pick a particle:  high-$p_T$ $b$-jet, low-$p_T$
    $b$-jet, high-$p_T$ lepton, low-$p_T$ lepton, Higgs,
    $Z$.
\item Optionally transform to a boosted frame:  Higgs, $Z$,
    System Center of Mass (CM).
\item  Optionally rotate the polar axis
    to point along the initial direction of the particle
    whose frame you are in (as for helicity and azilicity angles).
\item Pick a kinematic property:  $p_T$, $\eta$, $\phi$,
    $\cos(\theta)$, {\it  etc.}.
\item Optionally pick a second particle to form a sum or
    difference, sometimes with a coordinate transformation
    as in $\Delta R$ and twist $\tau$, and sometimes with a
    more complicated combination as in invariant-mass.
\item For vector quantities, optionally take the magnitude of vector sums, $| \vp_1 \pm \vp_2|$ or
scalar sums,  $| \vec{p}_1| \pm |\vec{p}_2|$.
\end{itemize}

\begin{figure}[t]
\begin{center}
\includegraphics[width=0.3\textwidth]{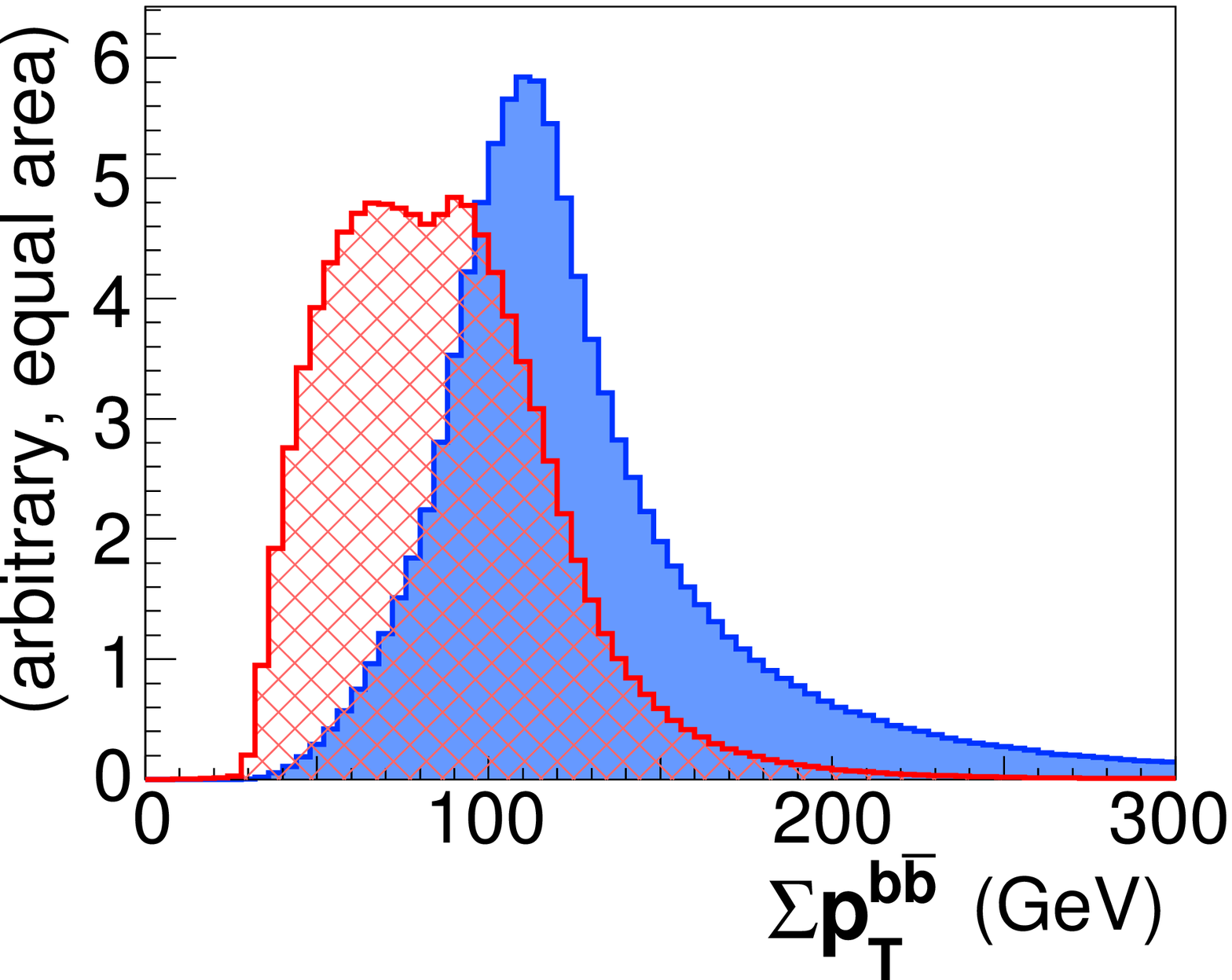}
\includegraphics[width=0.3\textwidth]{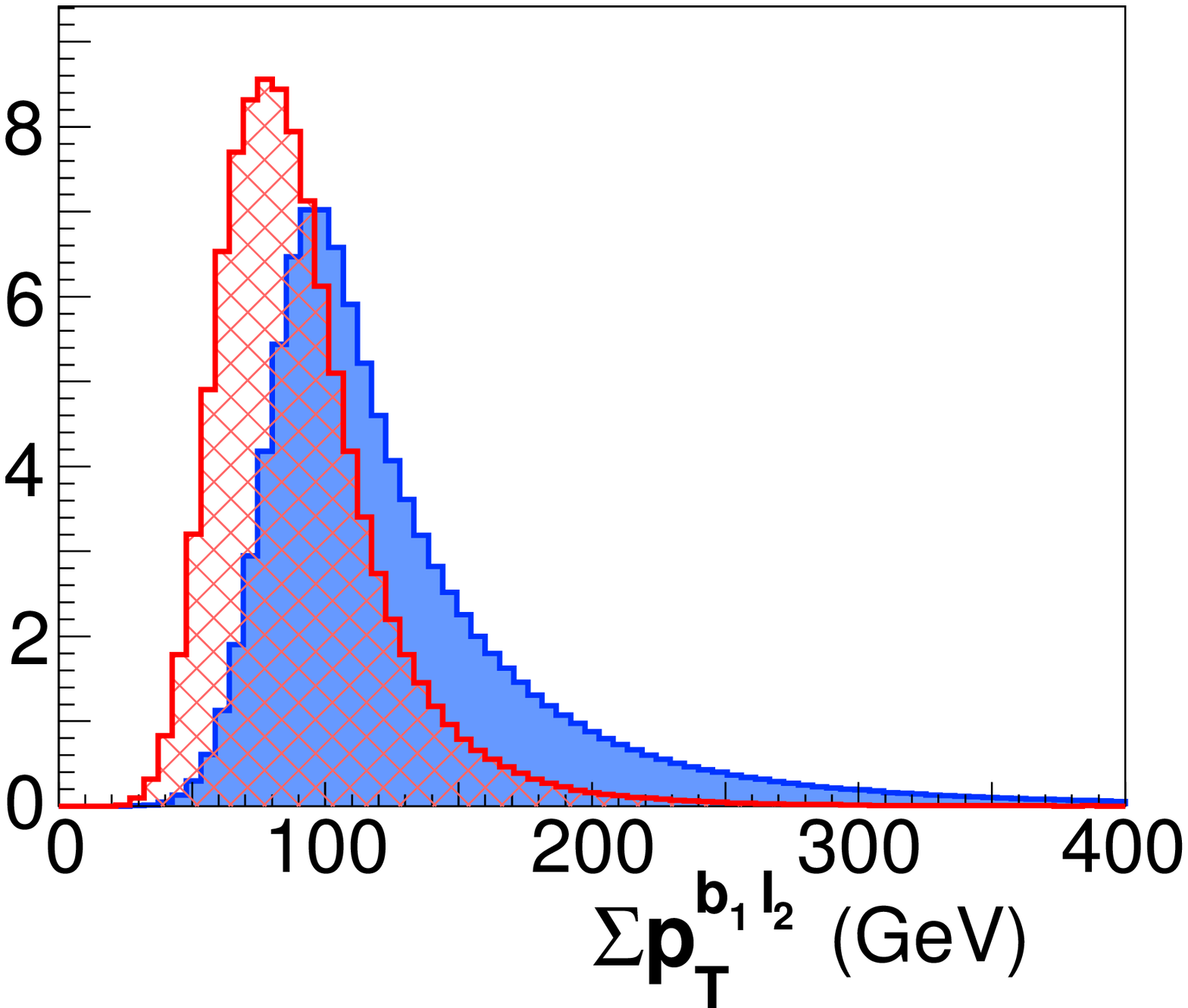}
\includegraphics[width=0.3\textwidth]{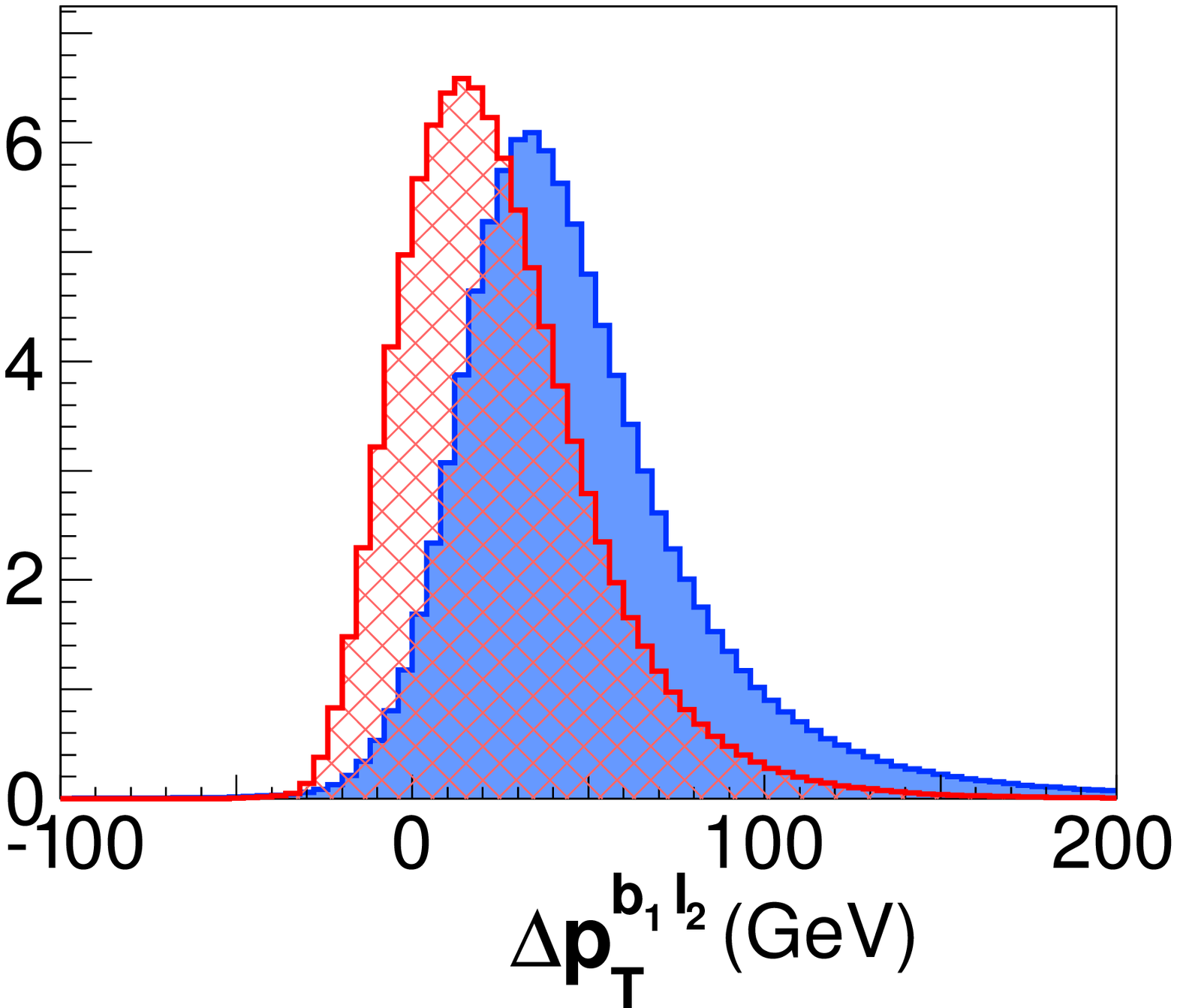}
\\
\includegraphics[width=0.3\textwidth]{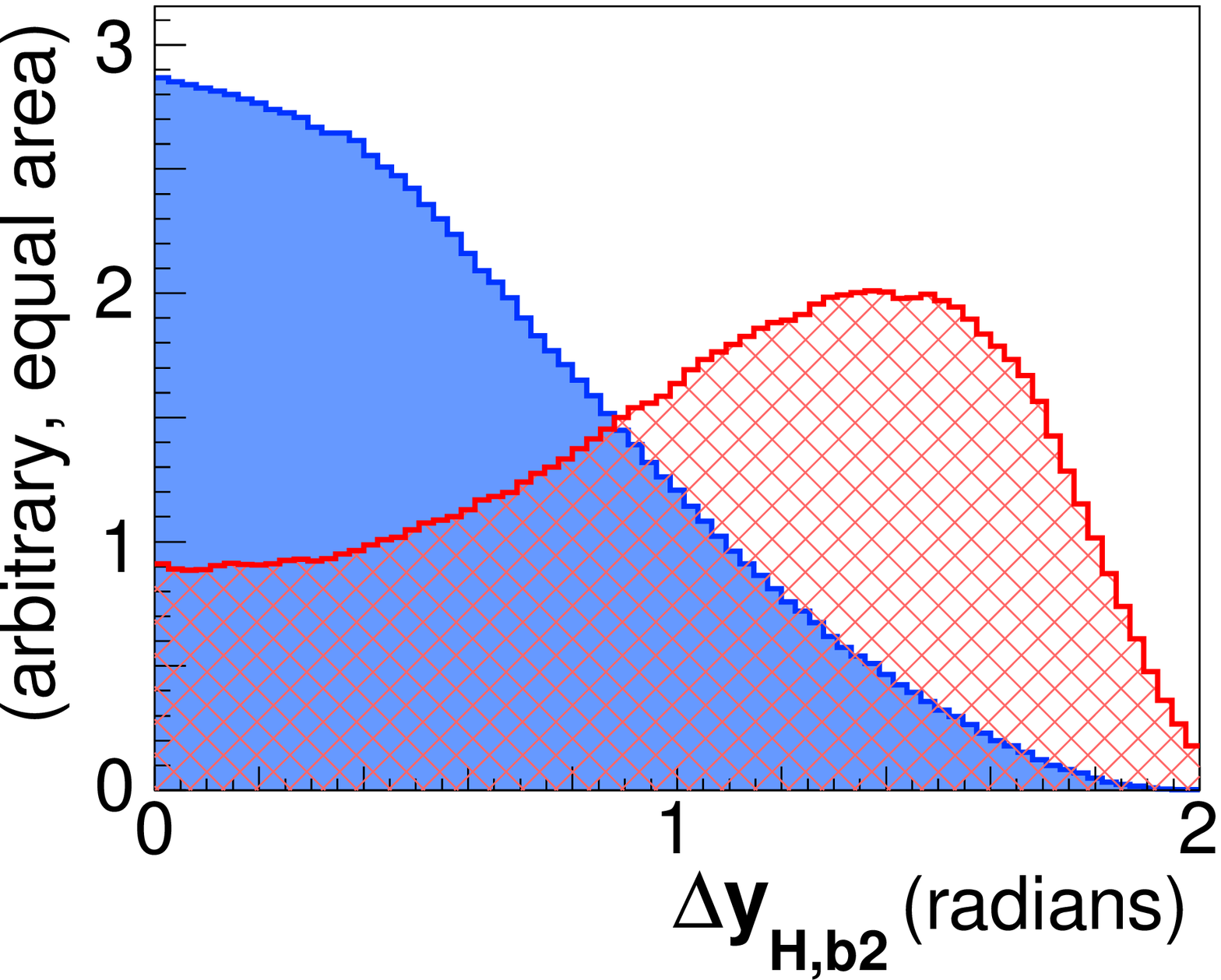}
\includegraphics[width=0.3\textwidth]{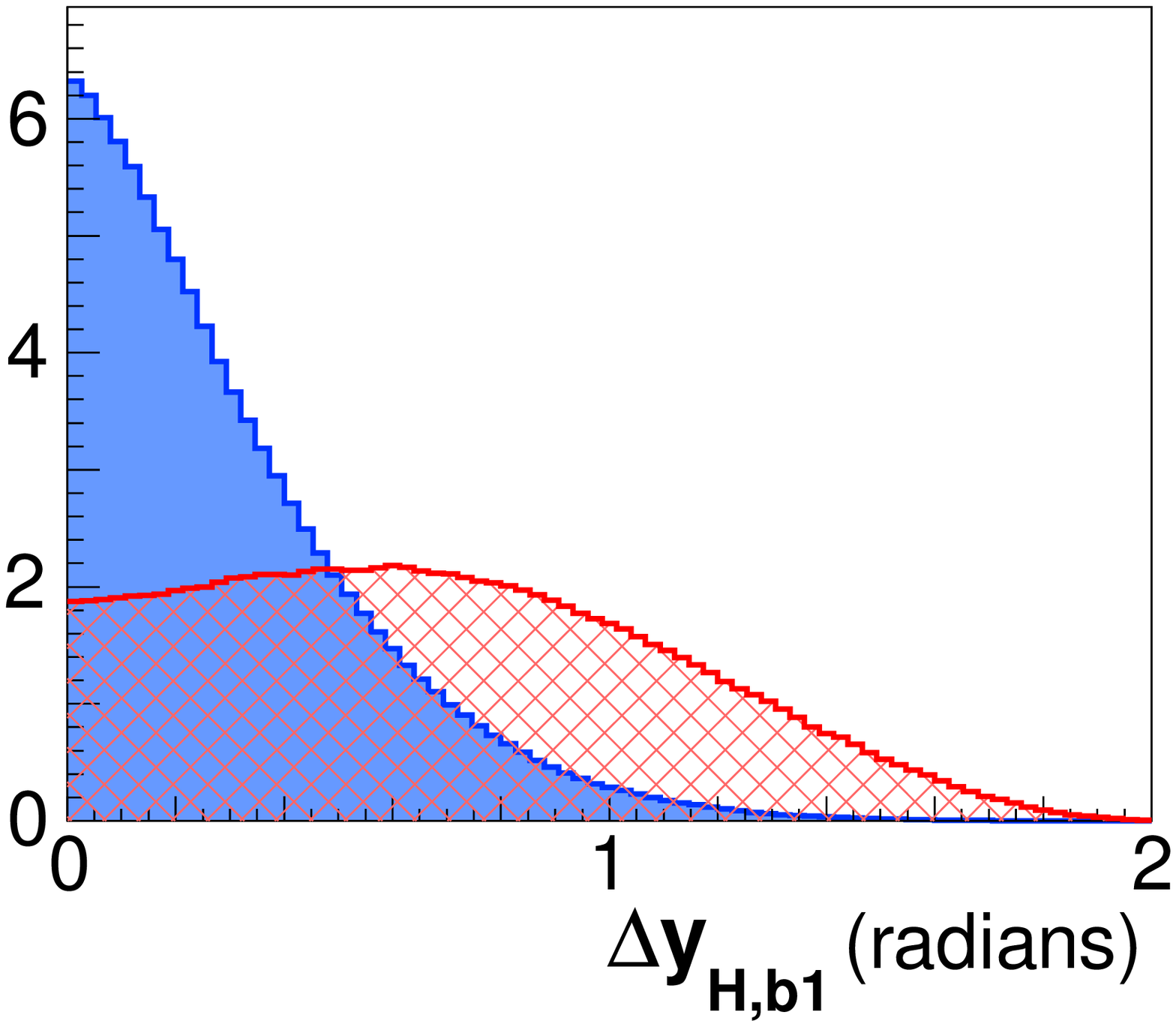}
\includegraphics[width=0.3\textwidth]{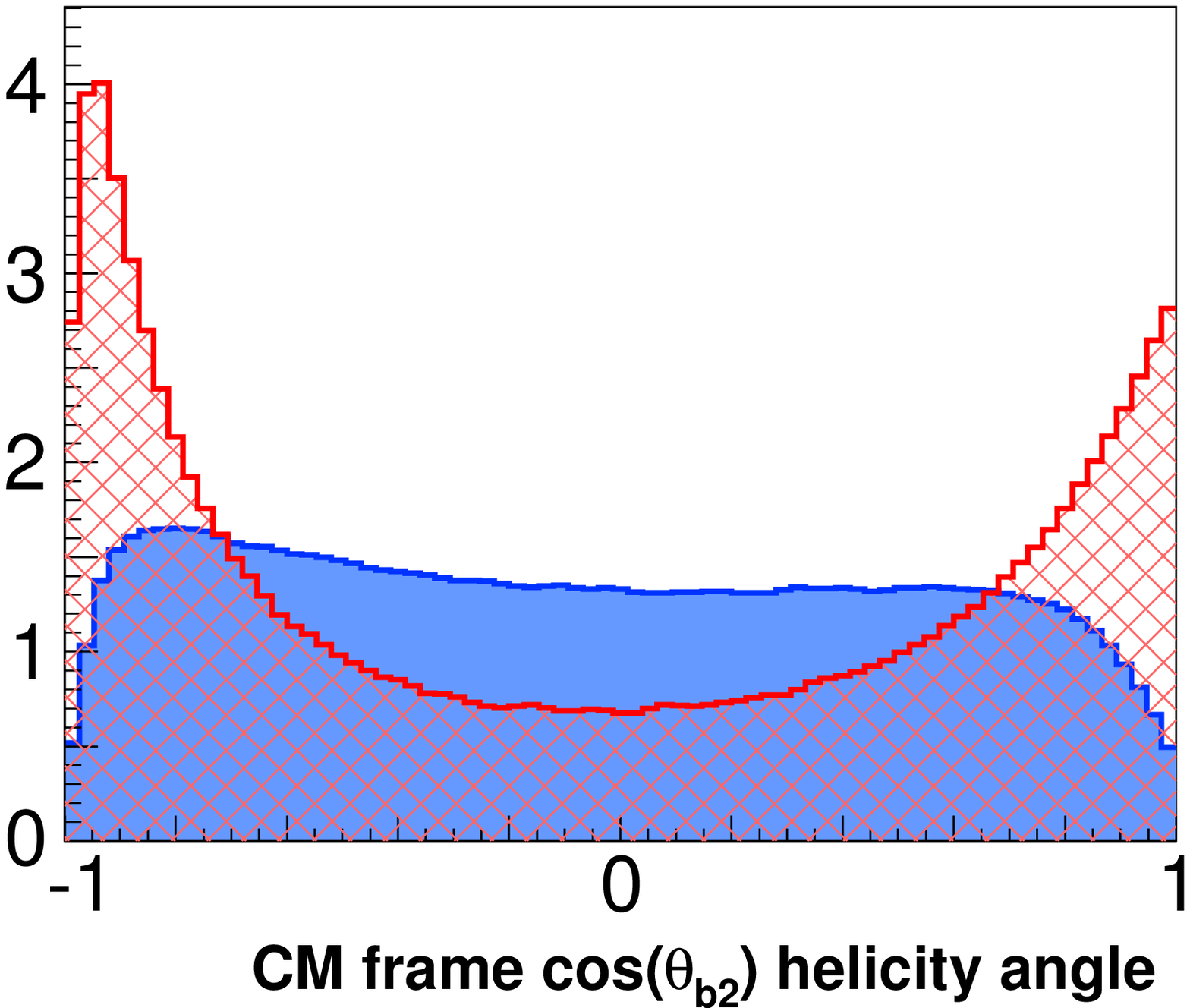}
\\
\includegraphics[width=0.3\textwidth]{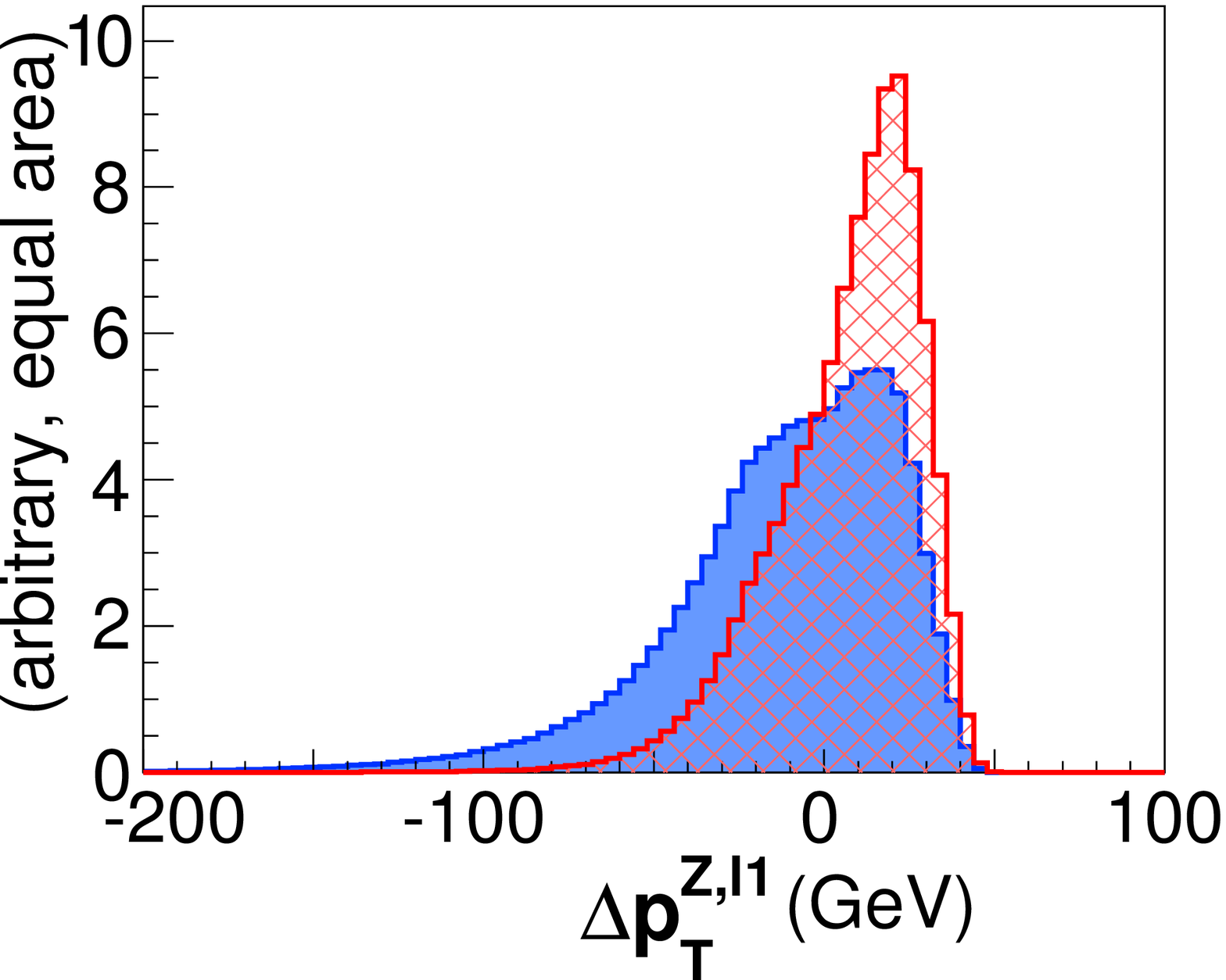}
\includegraphics[width=0.3\textwidth]{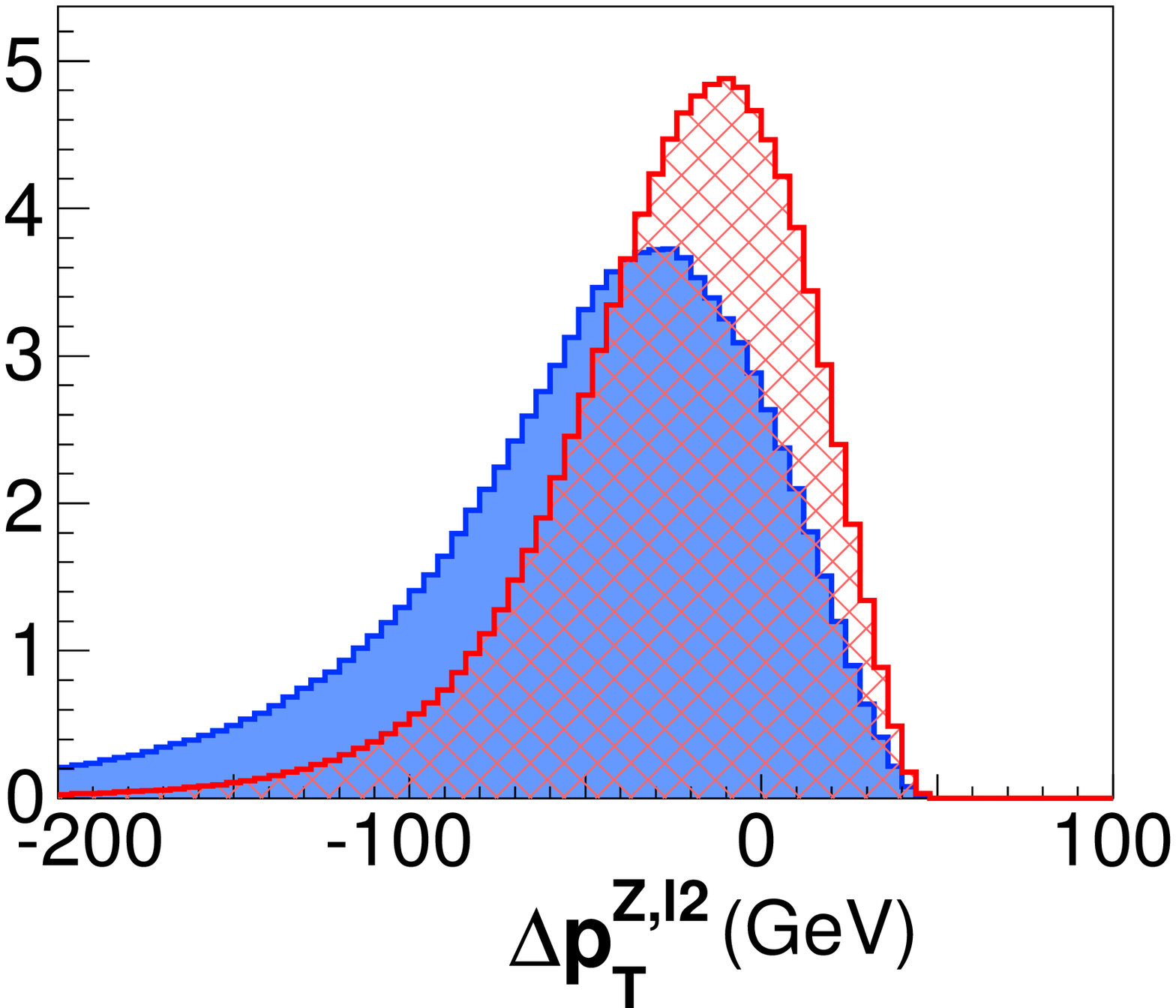}
\includegraphics[width=0.3\textwidth]{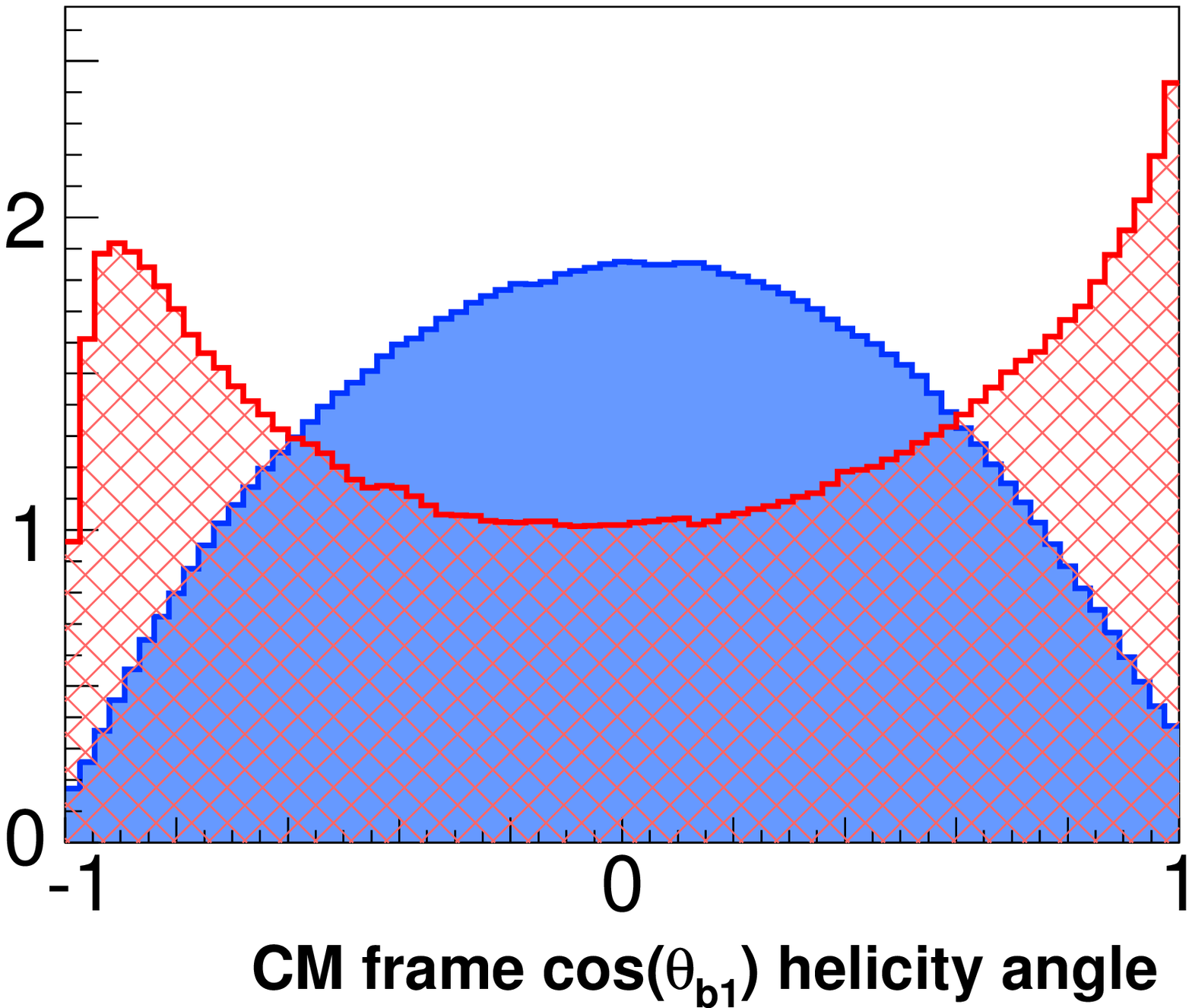}
\\
\includegraphics[width=0.3\textwidth]{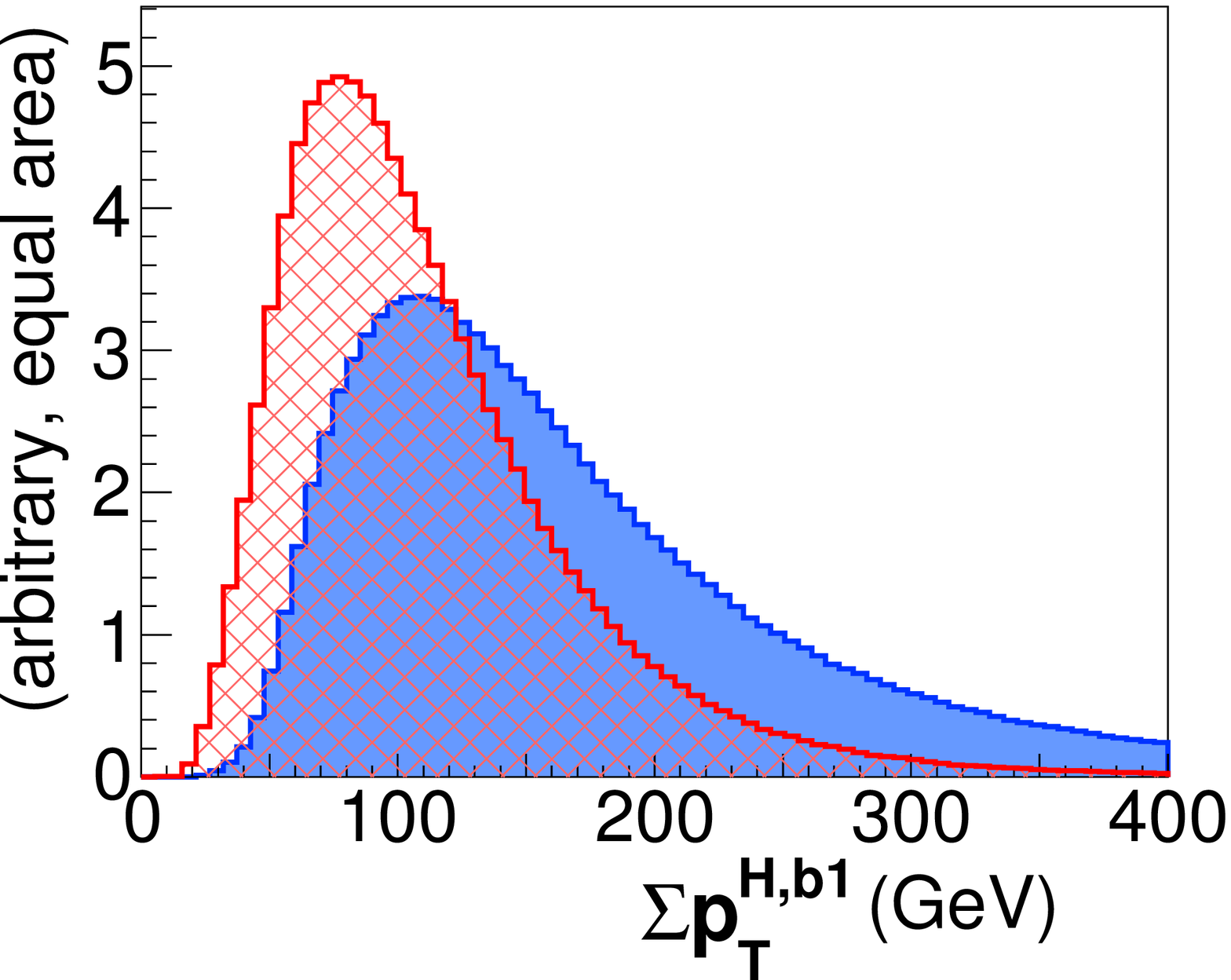}
\includegraphics[width=0.3\textwidth]{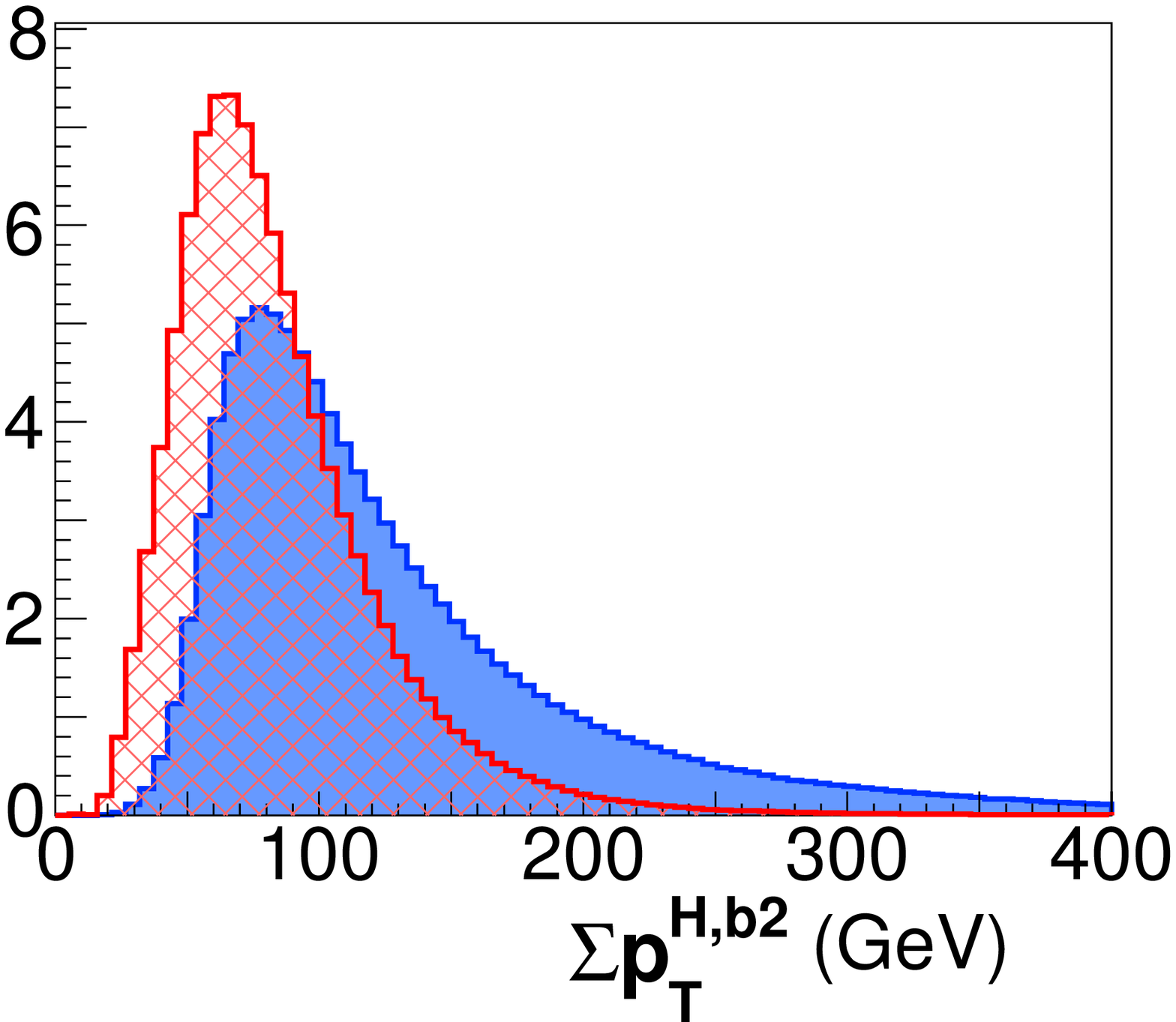}
\includegraphics[width=0.3\textwidth]{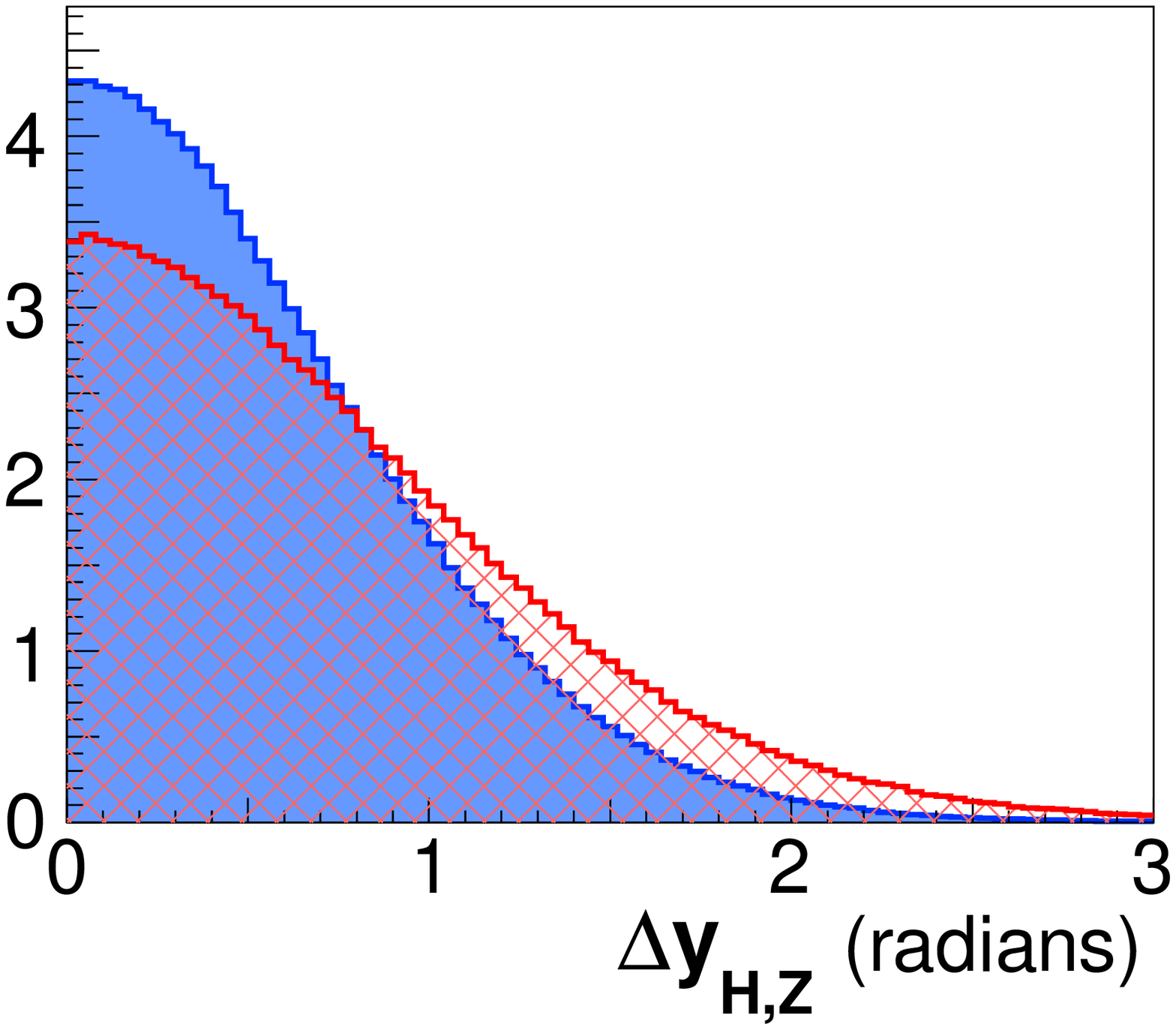}
\caption{
A selection of various variable distributions for $\zh$ signal
(solid blue) and $\zbb$ background (hashed red) at the LHC.
Events satisfy selection cuts and the Higgs mass-window cut, \Higgswindow . Horizontal
axes are in radians or GeV as appropriate,
and vertical axes are in arbitrary units with signal and background normalized to the same area.
}
\label{fig:china}
\end{center}
\end{figure}

Some stranger kinematic variables that prove to be useful
in the multi-variable analysis include:
\begin{itemize}
\item $\Sigma p_T^{b \bar b}= |\vp_T^{\,b1}| + |\vp_T^{\,b2}|$:  Sum of magnitudes of $p_T$'s of the
    two b-jets.
\item $\Sigma p_T^{b1,\ell 2}=|\vp_T^{\,b1}|+|\vp_T^{\,l2}|$:  Sum of magnitudes of $p_T$'s of the
    higher-$p_T$ $b$-jet and the lower-$p_T$ lepton.
\item $\Delta p_T^{Z,\ell 1}=|\vp_T^{\,Z}|-|\vp_T^{\,l1}|$:  Difference in magnitude of $p_T$ between
    $Z$ and higher-$p_T$ lepton.
\item $\Delta p_T^{b1,\ell 2}=|\vp_T^{\,b1}|-|\vp_T^{\,l2}|$:  Difference in magnitude of $p_T$ between
    higher-$p_T$ $b$-jet and lower-$p_T$ lepton.
\item $\Delta \eta_{b1,\ell 2}$:  Difference in $\eta$ between the
    higher-$p_T$ $b$-jet and the lower-$p_T$ lepton.
\item $\Delta y_{H,b1}$ and $\Delta y_{H,b2}$:  Difference
    in rapidity between $H$ and higher-$p_T$ or
    lower-$p_T$ $b$-jet.
\item $\cos(\theta^*_{b2})$:  Center of Mass frame
    $\cos(\theta)$ of the lower-$p_T$ $b$-jet. Same for higher-$p_T$ $b$-jet.
\end{itemize}
We show the distributions for a number of these Menu-Method variables in Figure~\ref{fig:china}.

\subsection{Radiation Variables \label{sec:showered}}
\begin{figure}[t]
\begin{center}
\includegraphics[width=0.3\textwidth]{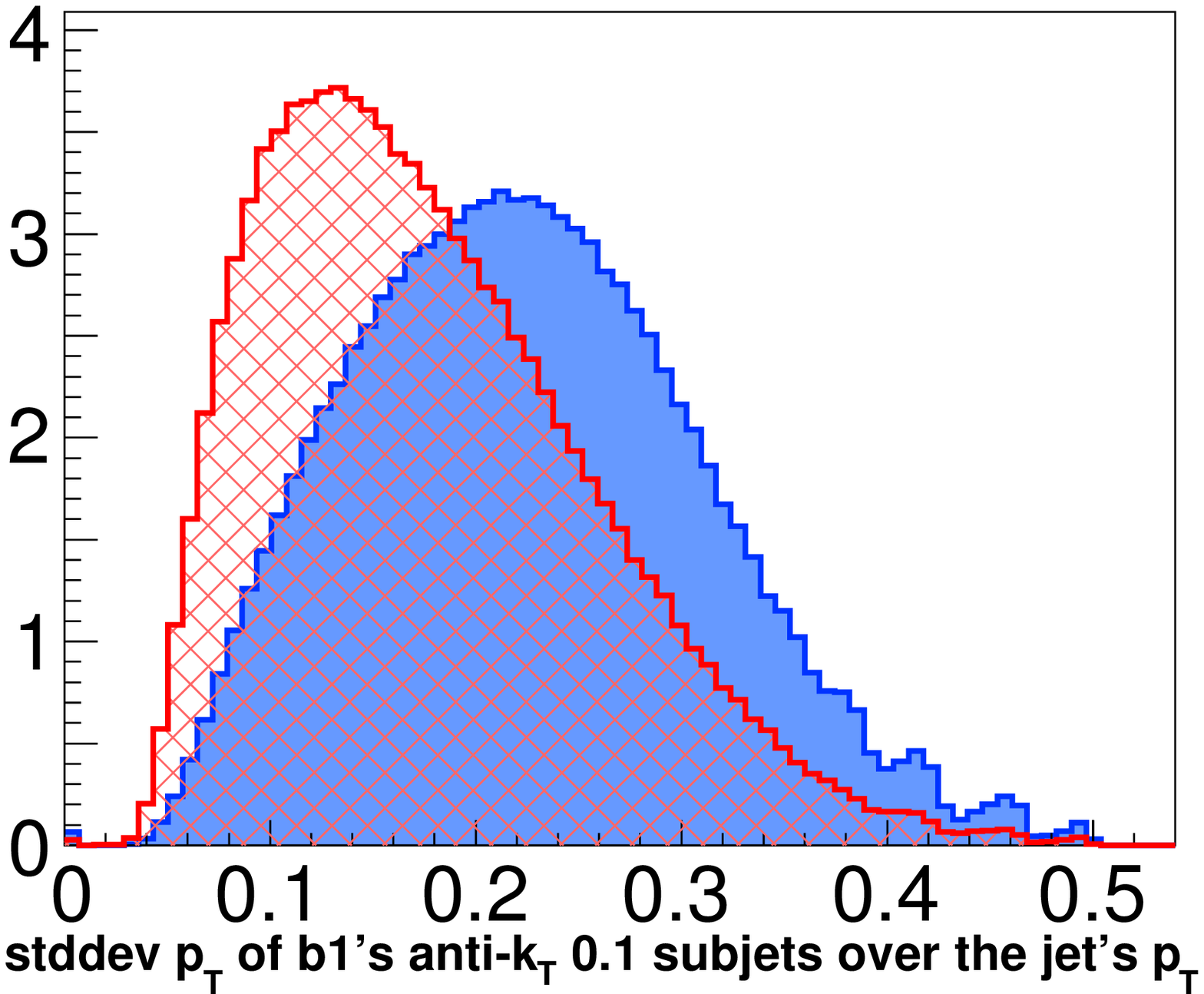}
\includegraphics[width=0.3\textwidth]{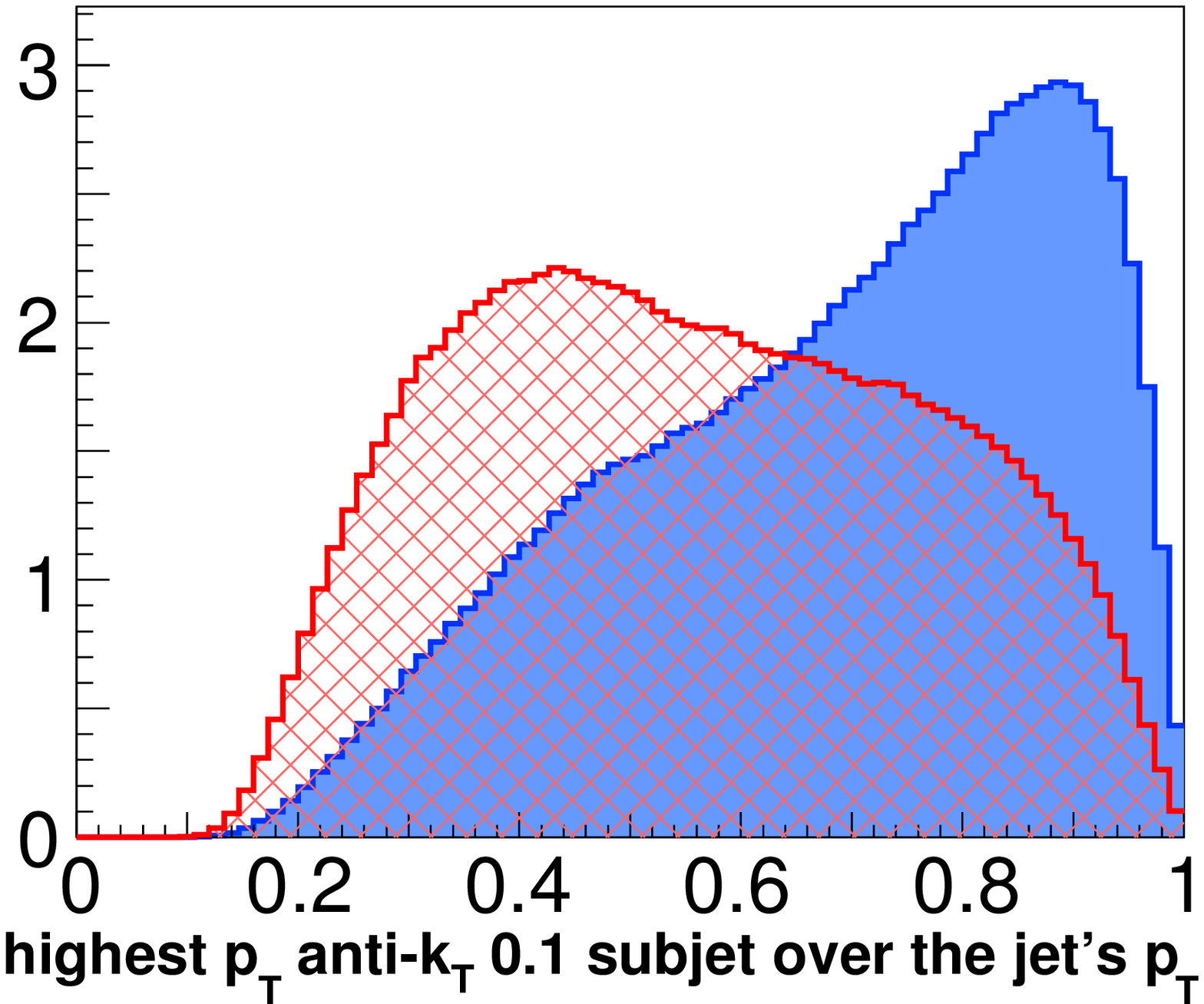}
\includegraphics[width=0.3\textwidth]{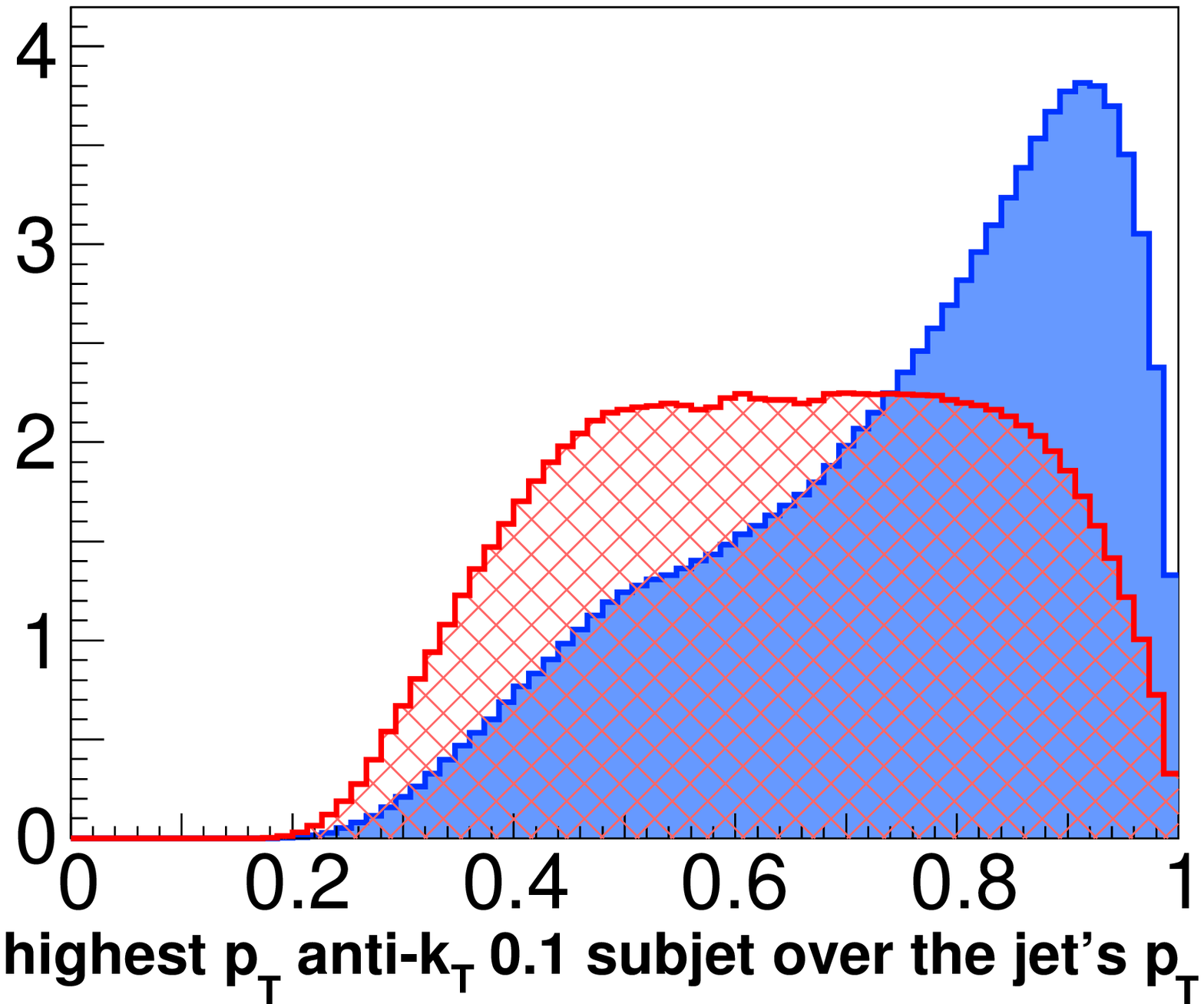}
\\
\includegraphics[width=0.3\textwidth]{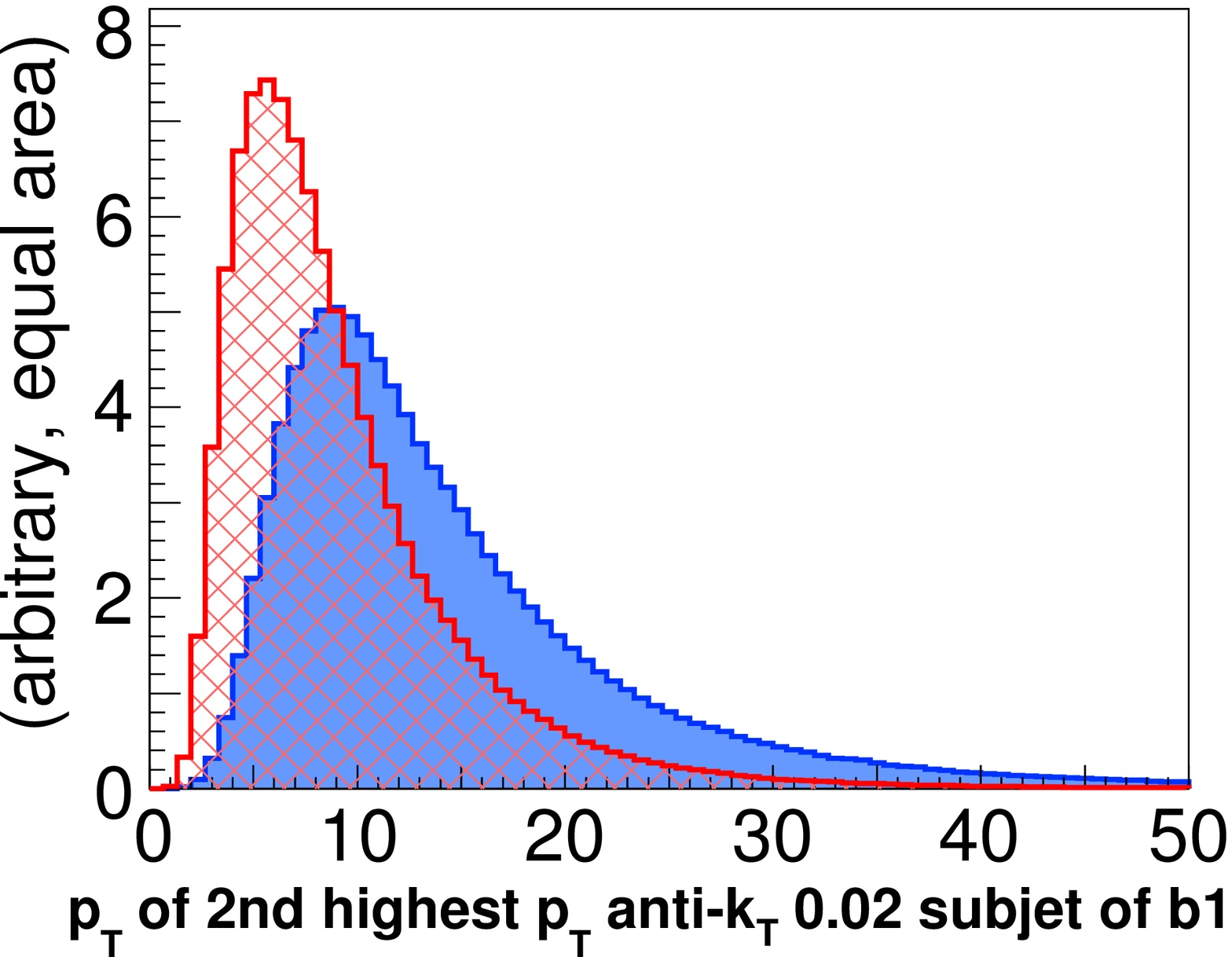}
\includegraphics[width=0.3\textwidth]{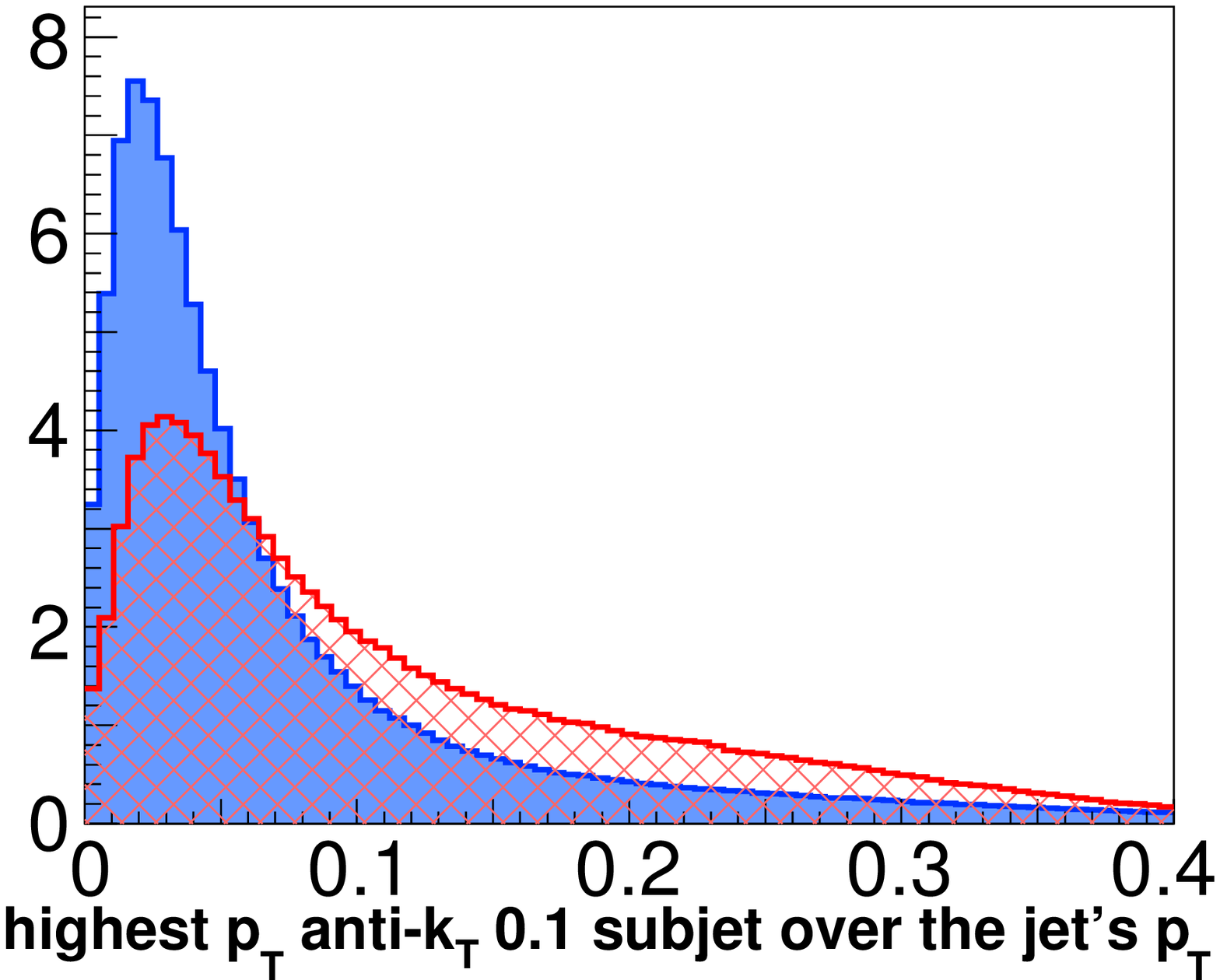}
\includegraphics[width=0.3\textwidth]{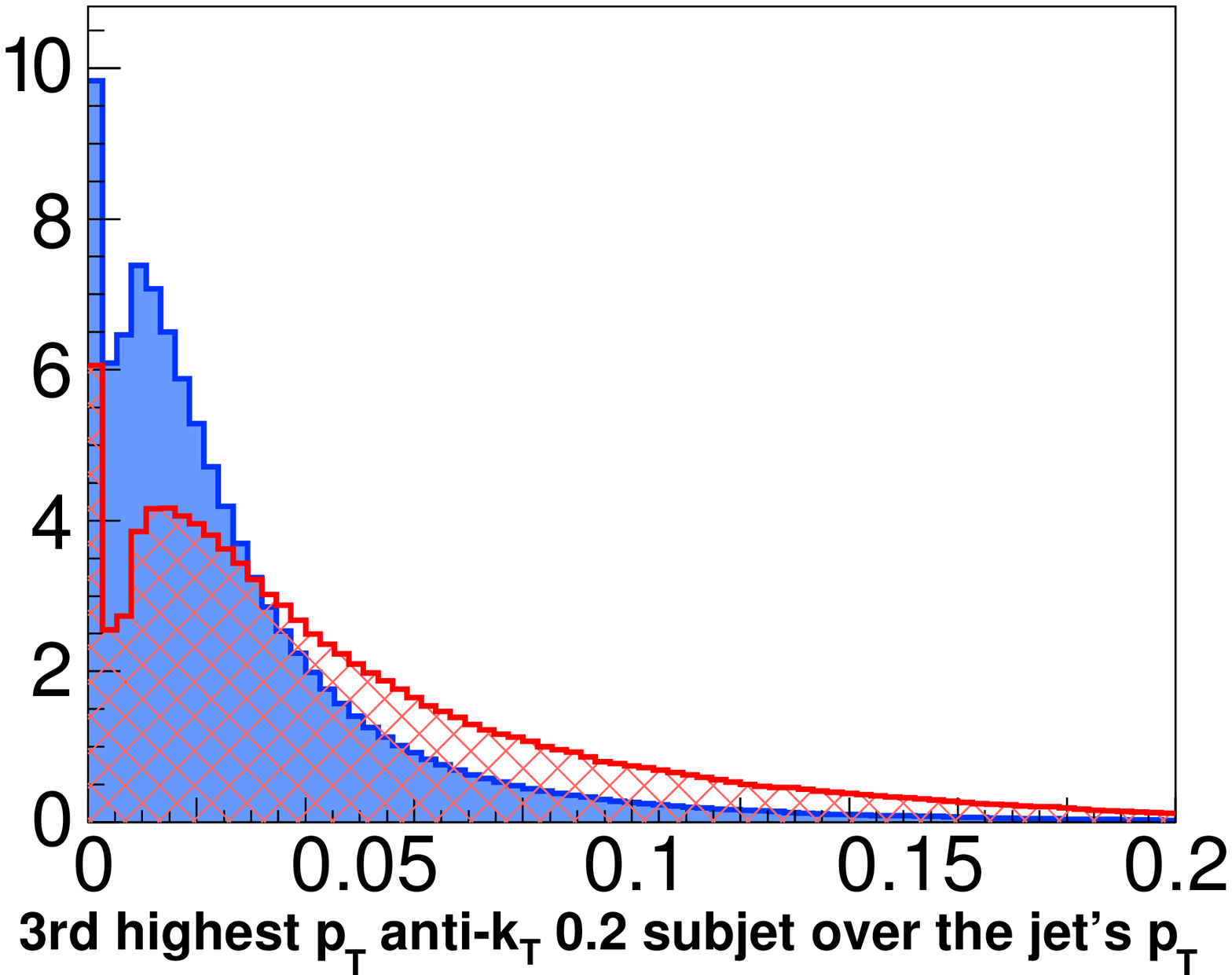}
\caption{
Some subjet variable distributions for $\zh$ signal (solid blue)
and $\zbb$ background (hashed red) athe LHC.
Events satisfy selection cuts and the Higgs mass-window cut, \Higgswindow . Horizontal
axes are in radians or GeV as appropriate,
and vertical axes are in arbitrary units with signal and background normalized to the same area.
}
\label{fig:showeredvariables}
\end{center}
\end{figure}

The above variables are constructed out of the $b$-jet momenta and the lepton momenta. These
are what we have been calling the kinematic variables.
In addition, there are what we call the radiation variables, which are dependent on the radiation pattern of the event.
The radiation variables generally have fixed or meaningless values at the hard parton level, so they are almost
entirely complementary to the hard variables. Some examples include:
\begin{itemize}
\item Mass of each $b$-jet and the jet mass-to-$p_T$ ratio, where the jet's 4-vector
    is the sum of its components' 4-vectors (the ``E-Scheme'').
\item Rapidity $y$ in addition to pseudorapidity $\eta$  of each massive $b$-jet.
\item Subjet multiplicity for each $b$-jet, with different subjet algorithms and sizes.
\item Average $p_T$ of the small subjets within each $b$-jet.
\item The $p_T$ of hardest, $2^{\text{nd}}$  hardest, and $3^{\text{rd}}$ hardest subjets within each $b$-jet.
\item Radial moments (``girth'') of each $b$-jet (see Section~\ref{sec:girth} below).
\item Jet Angularities (see Section~\ref{sec:girth} below).
\item Planar Flow of each $b$-jet. This was not found to be useful. See~\cite{Almeida:2008yp} for the definition.
\item Pull of each $b$-jet (see Section~\ref{sec:pull} below).
\item Extra jets in the event. Extra jets were not found useful. However, we have not
attempted the analysis with a proper matched sample, so we cannot make a strong statement about extra jets. We suspect that this
is an important issue for removing the $t\bar{t}$ contamination, but not so much for the irreducible backgrounds we consider here.
\end{itemize}
Some of the more powerful of these variables are shown in Figure~\ref{fig:showeredvariables}.

\subsection{Radial Moments: Girth and Jet Angularities\label{sec:girth}}

\begin{figure}[ht]
\begin{center}
\includegraphics[width=0.3\textwidth]{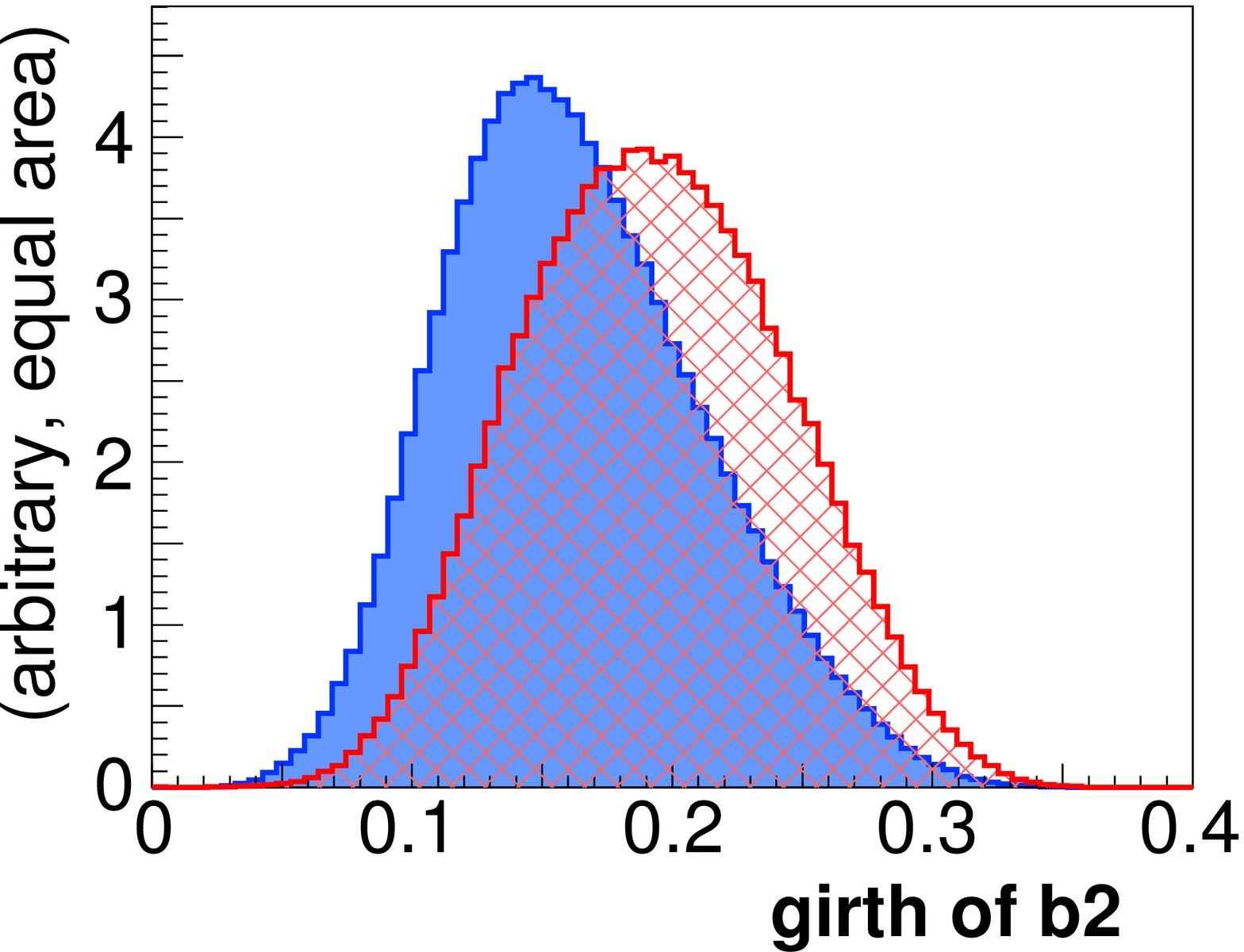}
\includegraphics[width=0.3\textwidth]{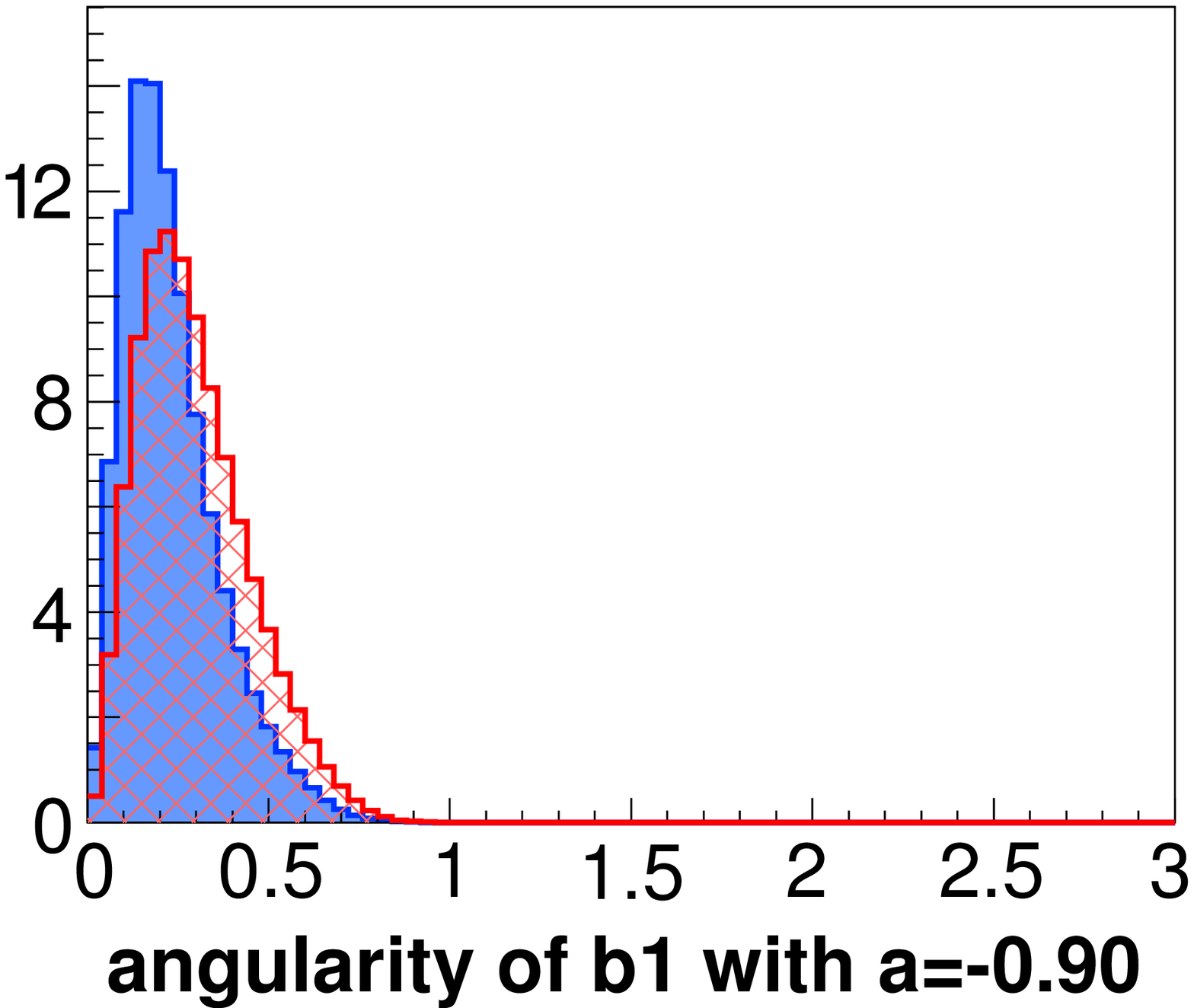}
\includegraphics[width=0.3\textwidth]{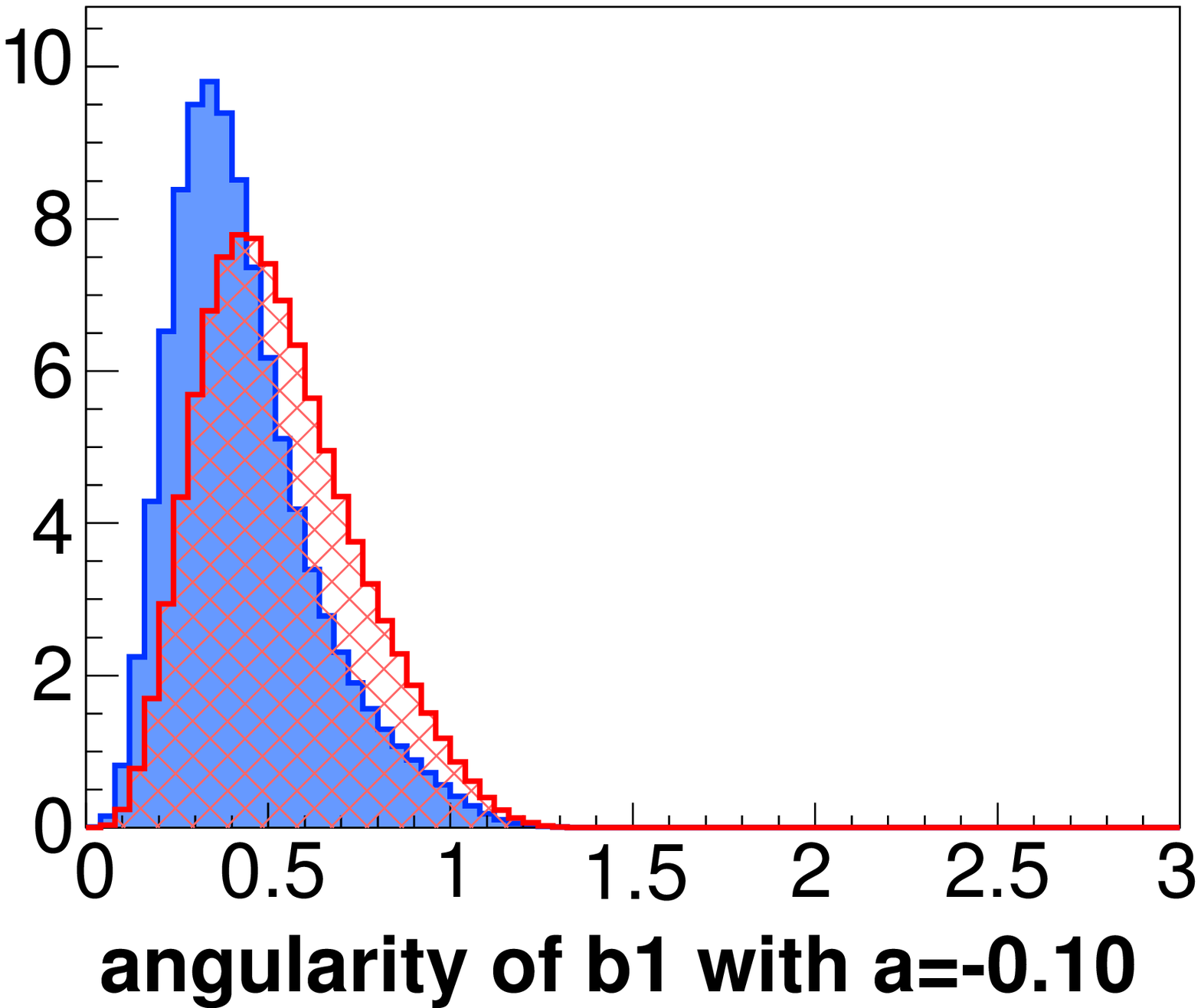}
\caption{
Girth and angularity distributions for $\zh$ signal (solid blue)
and $Z\bbar$ background (hashed red) at the LHC.
Events satisfy selection cuts and the Higgs mass-window cut, \Higgswindow . Horizontal
axes are in radians or GeV as appropriate,
and vertical axes are in arbitrary units with signal and background normalized to the same area.
}
\label{fig:girth_and_angularity}
\end{center}
\end{figure}

\begin{figure}[t]
\begin{center}
\includegraphics[width=0.5\textwidth]{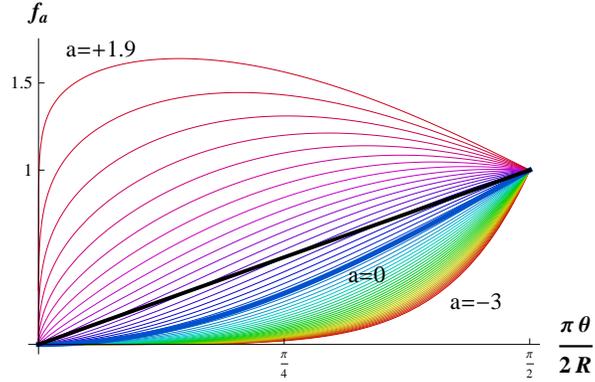}
\caption{Profiles for different choices of the angularity $a$ parameter spaced at 0.1 intervals
(in rainbow) and linear radial-moment ``girth'' (in black)}
\label{fig:angularities}
\end{center}
\end{figure}

The distribution of particles within a jet can be can be useful for
distinguishing jets initiated by different flavors of quark or by a gluon.
Even in our signal and irreducible background with two $b$-tagged jets,
this distribution has proven useful.

One infrared safe way of characterizing the jets is to integrate (sum)
the energy or $p_T$ distribution against a radially symmetric profile.
Different choices of profile and overall normalization lead to different
observables.
Distances $r_i$ of each particle or cell are calculated in $(y,\phi)$ space
with respect the location of the jet.
The jet location $(y,\phi)$ is defined by the anti-$k_{T}$ algorithm `E-scheme'
as the 4-vector sum of all inputs (particles or calorimeter towers.)
It is important to use
rapidity (rather than pseudorapidity) for the jet location because the jet is massive in this scheme.
A radial moment sums these distances (or a function of these distances),
weighted by a quantity like $p_T$, then normalized to the total $p_T$ of the jet.
For example, the linear radial moment, {\bf girth}, is defined as~\cite{Gallicchio:2010sw}
\begin{equation}
\textrm{Girth}:\qquad  g =  \sum_{i \in \mathrm{jet}} \frac{ p_T^i \, |r_i|  }{ p_T^{jet} } \, .
\end{equation}
The girth distribution is shown in Figure~\ref{fig:girth_and_angularity}.

Jet Angularities are also radial moments, but their ``radial distances'' are rescaled
into the angular coordinates appropriate for $e^+e^-$ event shapes.
They are defined by~\cite{Almeida:2008yp}
\begin{equation}
\textrm{Jet Angularities}:~~~~ A_a =
\frac{1}{ m_{jet} } \sum_{i \in \mathrm{jet}}  E_i \ f_a( \frac{\pi |r_i|}{2R} ) \,,
\end{equation}
with
\begin{equation}
f_a(\theta) =  \sin^a \! \theta \,
        \left( 1 - \cos \theta \right)^{1-a} \,,
\end{equation}
with $a<2$. The kernel function $f_a(\theta)$ is inspired by full event-shape angularities~\cite{Berger:2003iw},
but modified so that the edge of a jet at $|r_i|\!=\!R$ is mapped to $\pi/2$.
Profiles for different choices of the $a$ parameter are
shown in Figure~\ref{fig:angularities}.
Note that the energies $E_i$ are used in the definition, instead of $p_T$'s, and the angularities are
normalized by the jet mass.
Radial moments like jet angularities and girth are especially interesting because it may
be possible to calculate them accurately in QCD, see for example~\cite{Ellis:2010rw}.


\subsection{Pull \label{sec:pull}}
Pull tries to capture the difference in color structure between the Higgs boson signal
and the QCD background. It was introduced in Ref.~\cite{Gallicchio:2010sw},
and then immediately used in the D\O\ search \cite{D0pull} for $\zh$ with $Z\to \nu \bar{\nu}$.

\begin{figure}[t]
\begin{center}
\includegraphics[width=0.25\hsize]{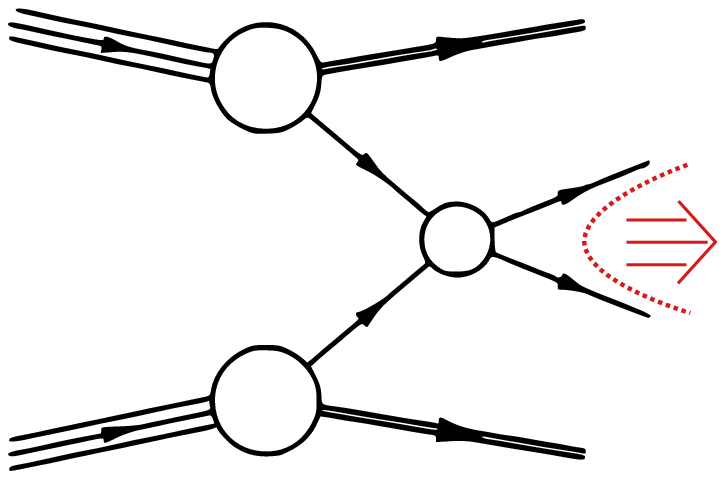}
\includegraphics[width=0.25\hsize]{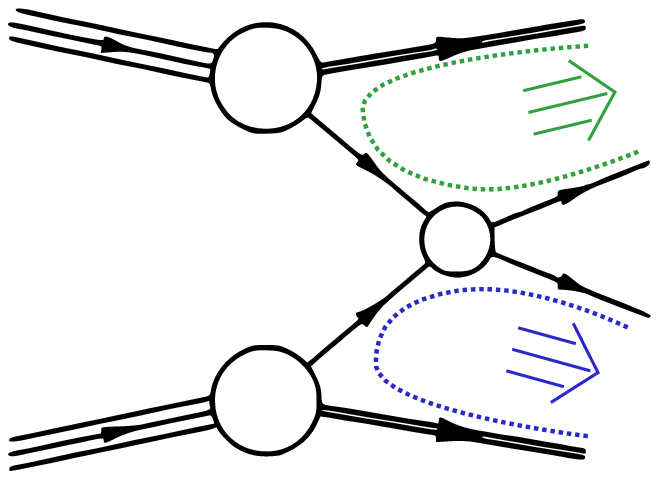}
\end{center}
\caption{
Color connections for a signal-like ($pp\to H\to \bbar$) on the left
and background-like ($pp\to \bbar$) on the right.
Our signal and background each have a colorless $Z$ or $W$ (not shown) radiating
from of one of the hard quark lines and decaying to leptons.
This doesn't affect the color flow.
}
\label{fig:beam_spray_diagram}
\end{figure}

To leading order in the number of colors (up to $1/N_\text{colors}^2\sim10\%$
corrections), quarks can be described as being ``color-connected'' to other quarks
by a ``color string'' Ref.~\cite{Ellis:1991qj}. This approximation governs much of the parton shower.  The color-singlet
Higgs boson decays to two $b$-quarks that are color-connected to each other, while
$b$'s from the background are color-connected to the proton remnants that travel down the
beam pipe.  This is shown schematically in Figure~\ref{fig:beam_spray_diagram}.
This difference is \emph{independent} of the event kinematics, and therefore, if observable, should
be be complementary to kinematical variables and useful in a multi-variable search.

\begin{figure}[t]
\begin{center}
\psfrag{th}{$\theta_p$}
\psfrag{mp}{$\left| \vec p \right|$}
\psfrag{mpi}{\scriptsize $-\pi $}
\psfrag{pi}{\scriptsize $\pi$}
\psfrag{0.04}{\scriptsize $0.04$}
\psfrag{0.02}{\scriptsize $0.02$}
\psfrag{0}{\scriptsize $0$}
\begin{tabular}{cc}
\textbf{\ Signal Accumulated $\mathbf{p_t}$}
&
\textbf{\ Background Accumulated $\mathbf{p_t}$}
\\
\includegraphics[width=0.45\hsize]{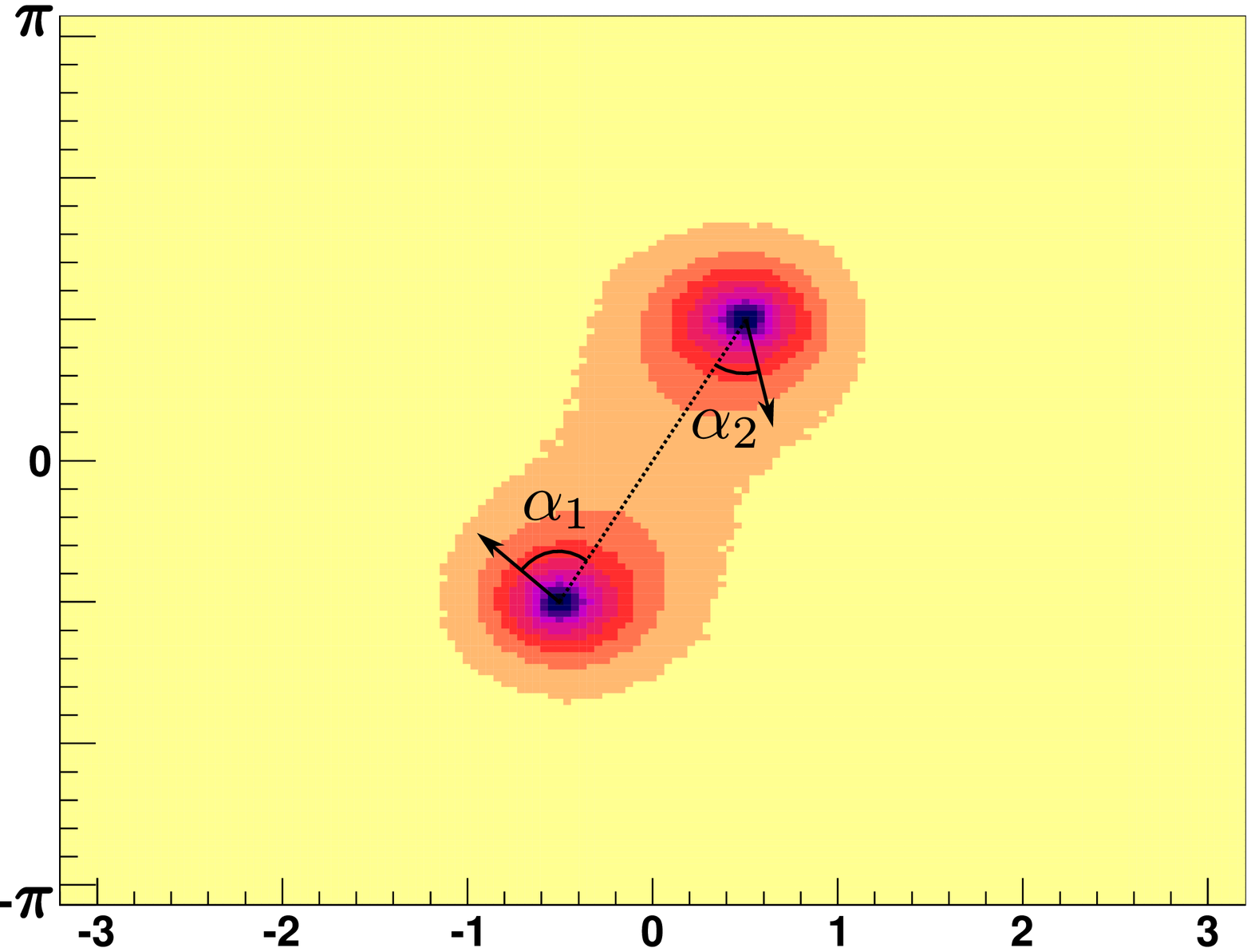}
&
\includegraphics[width=0.45\hsize]{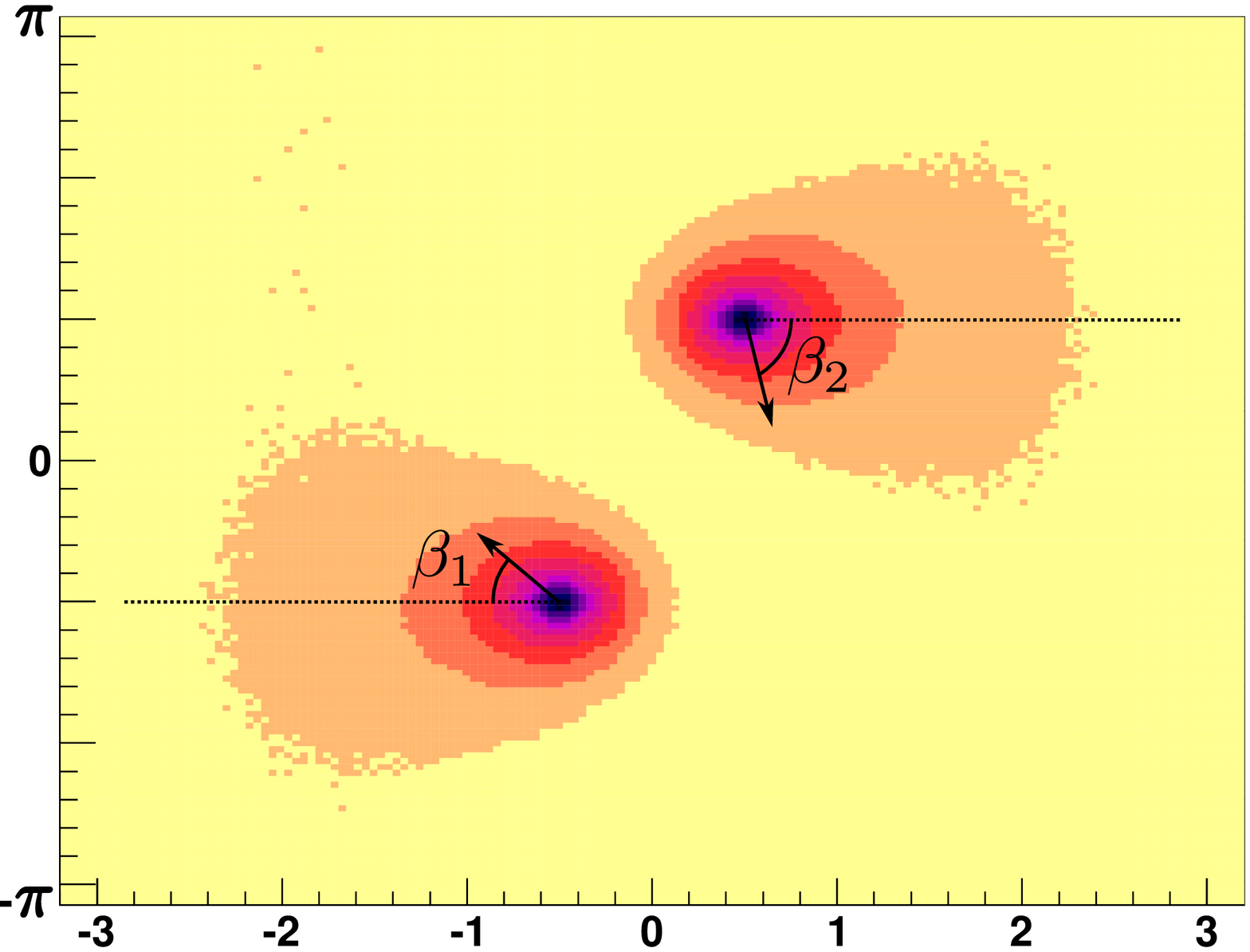}
\end{tabular}
\end{center}
\caption{
A single parton-level signal event showered millions of times (left)
and a single parton-level background event with \emph{identical} kinematics
but different color connections showered millions of times (right).
The color shows the average showered $p_T$ density in
$(\eta,\phi)$ for an ensemble of events with fixed parton-level kinematics.
(Contours are stepped in factors of two.)
Results for $ZH\to Z\bbar$ are displayed on the left, and $gg\to Z\bbar$ on the right.
The underlying color connections are shown with thin lines, and examples of pull vectors (which
would be defined event-by-event) with arrows.  The angles $\alpha_{1,2}$ and $\beta_{1,2}$ are
illustrated as the angles of the pull vectors with respect to the ``signal-like'' and
``background-like'' color connection lines.
}
\label{fig:acc_highres3M_raw_crop}
\end{figure}

The {\bf pull vector} is designed to measure color flow. It is a $p_T$ weighted moment vector
that tends to point toward the color-connected partner of the jet's initiating quark. The pull vector is defined as
\begin{equation}
\textrm{Pull Vector } \vec t = \sum_{i \in \mathrm{jet}} \frac{ p_T^i \, |r_i|  }{ p_T^{jet} }\, \vec r_i
\qquad
\textrm{where } \vec r_i \equiv (y_i - y_\mathrm{jet}, \phi_i - \phi_\mathrm{jet}) \,.
\end{equation}
Without the factor of $|r_i|$, this would be the jet's $p_T$-weighted centroid.
For fixed kinematics, pull has been shown to help separate signal from background~\cite{Gallicchio:2010sw}.
For example, if we fix the $b$'s to have $(\Delta \eta, \Delta \phi)=(1,2)$ with $p_T^H=200$ GeV,
the average $p_T$ measured in the calorimeter is shown for signal and background
in Figure~\ref{fig:acc_highres3M_raw_crop}.
The difference in $p_T$ distributions around the jets holds up even for
individual events. The most effective way to use the pull vector is to calculate a {\bf pull angle}, which is the angle
between the pull vector and some other vector in the event. We will now point out some technical details
that help make pull more effective, and define some example pull angles, such as the ones
drawn in Figure~\ref{fig:acc_highres3M_raw_crop} which are defined in Table~\ref{tab:pulldef}.

\begin{table}
\begin{center}
\begin{tabular}{|m{0.8in}|m{1.6in}|m{1.6in}||m{1.0in}|}
  \hline
  Signal \phantom{xxxx} pull angles &
  $\alpha_1 =$ angle between the high-$p_T$ jet's pull and the direction to the low $p_T$ jet  &
  $\alpha_2 =$ angle between the low-$p_T$ jet's pull and the direction to the high $p_T$ jet & 
  Signal \phantom{xxxx} pull  distance \phantom{x}  $\alpha=\sqrt{\alpha_1^2+\alpha_2^2}$ \\
  \hline
  Background pull angles &
  $\beta_1 =$ angle between the high-$p_T$ jet's pull and the direction to nearest beam &
  $\beta_2 =$ angle between the high-$p_T$ jet's pull and the direction to nearest beam & 
  Background pull distance \phantom{x} $\beta=\sqrt{\beta_1^2 + \beta_2^2}$  \\
  \hline
\end{tabular}
\caption{Definition of useful pull angles.\label{tab:pulldef}.}
\end{center}
\end{table}

We found it important to use anti-$k_T$ jets because the highest energy jets
in the event tend to be circular, whereas the split-merge procedure of a cone-jet
algorithm can nibble away pieces and combine them with an adjacent jet. If that
adjacent jet came from a perturbative QCD radiation
(due to the color-connection in the dipole picture),
these are exactly the pieces
that should be contributing most to the pull.
It's also important that real rapidity $y$ be used, as opposed to pseudo-rapidity $\eta$.
While this is less important for each calorimeter cell, the 4-momenta of a jet can
be massive when constructed by adding together massless calorimeter 4-momenta
(the ``E-Scheme'').  If anti-$k_T$ jets are not available,
the next best option is to use a circular area in the $y/\phi$ plane around the jet's
$y/\phi$ centroid.

\begin{figure}[t]
\begin{center}
\psfrag{pull_toward_otherjet_b_highPt__1_Pt_4v_y}{\tiny{pull of high-$p_T$ $b$ jet: $\alpha_{1}$}}
\psfrag{pull_toward_otherjet_b_lowPt__1_Pt_4v_y}{\tiny{pull of low-$p_T$ $b$ jet: $\alpha_{2}$}}
\psfrag{pull_dist_between_otherjet_b_highPt__1_Pt_4v_y}{\tiny{pull signal-distance: $\alpha$}}
\psfrag{pull_uncrosssed_b_highPt__1_Pt_4v_y}{\tiny{pull of high-$p_T$ $b$ jet: $\beta_{1}$}}
\psfrag{pull_uncrosssed_b_lowPt__1_Pt_4v_y}{\tiny{pull of low-$p_T$ $b$ jet: $\beta_{2}$}}
\psfrag{pull_dist_uncrosssed_otherjet_b_highPt__1_Pt_4v_y}{\tiny{pull backgnd-distance: $\beta$}}
\includegraphics[width=0.3\textwidth]{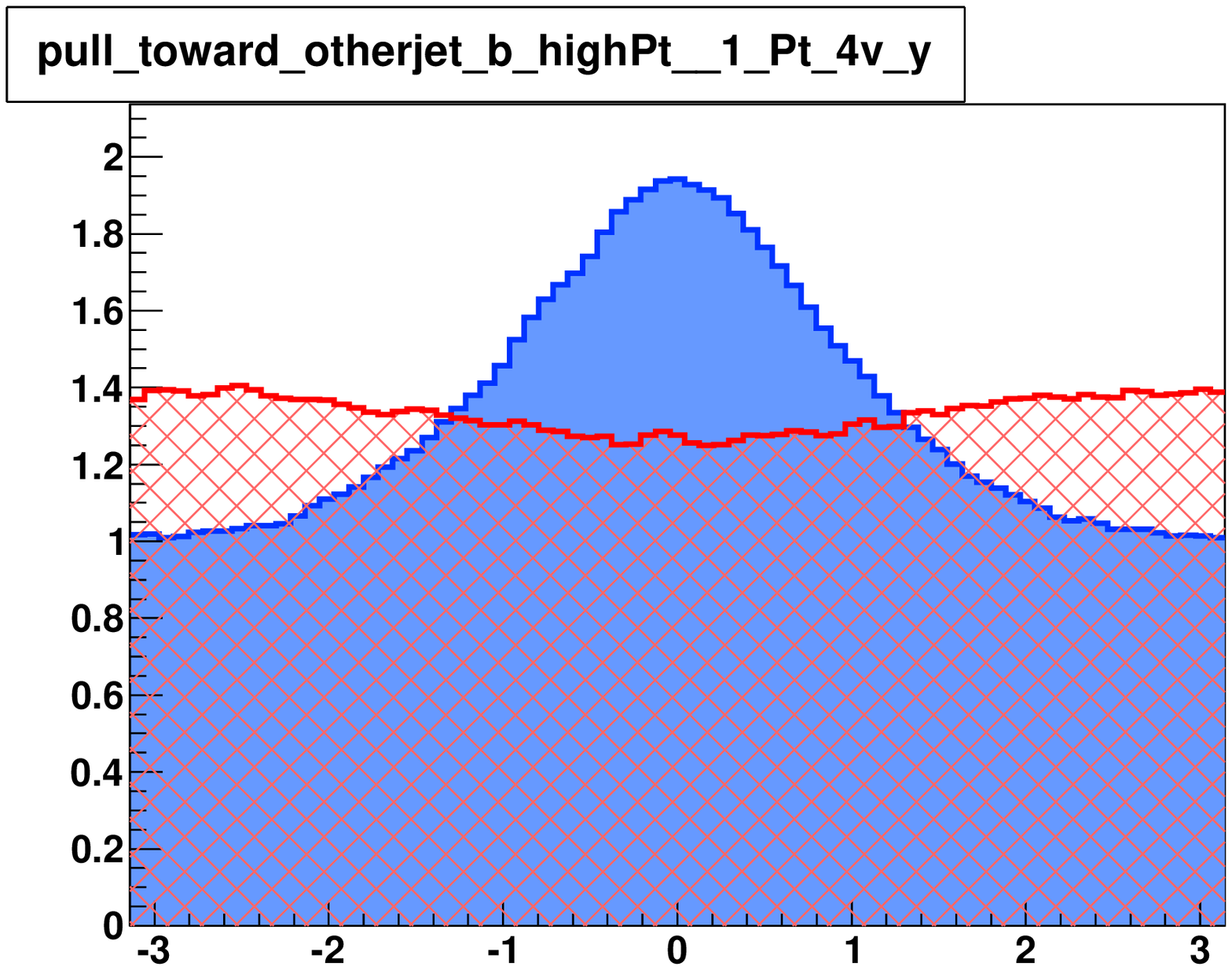}
\includegraphics[width=0.3\textwidth]{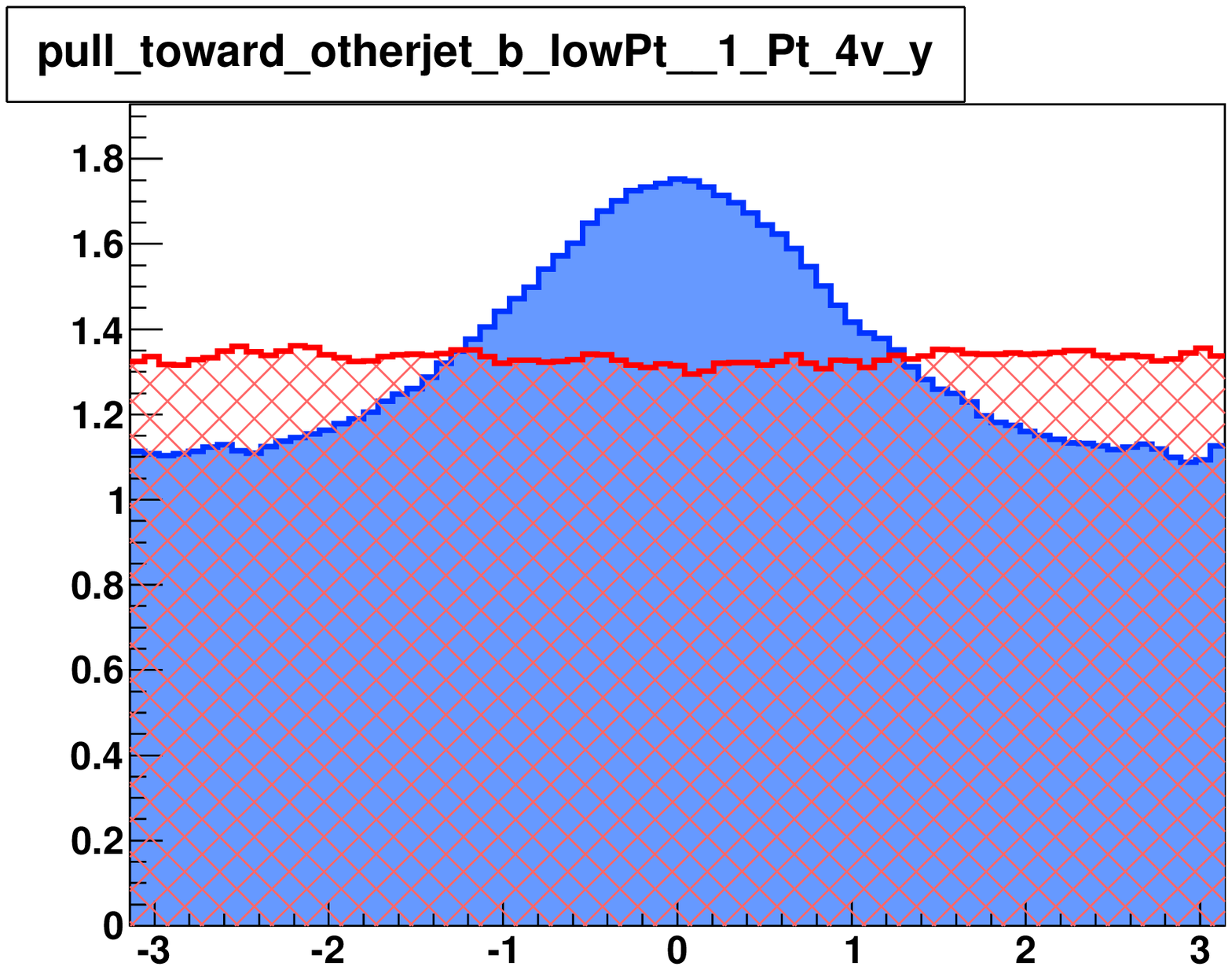}
\includegraphics[width=0.3\textwidth]{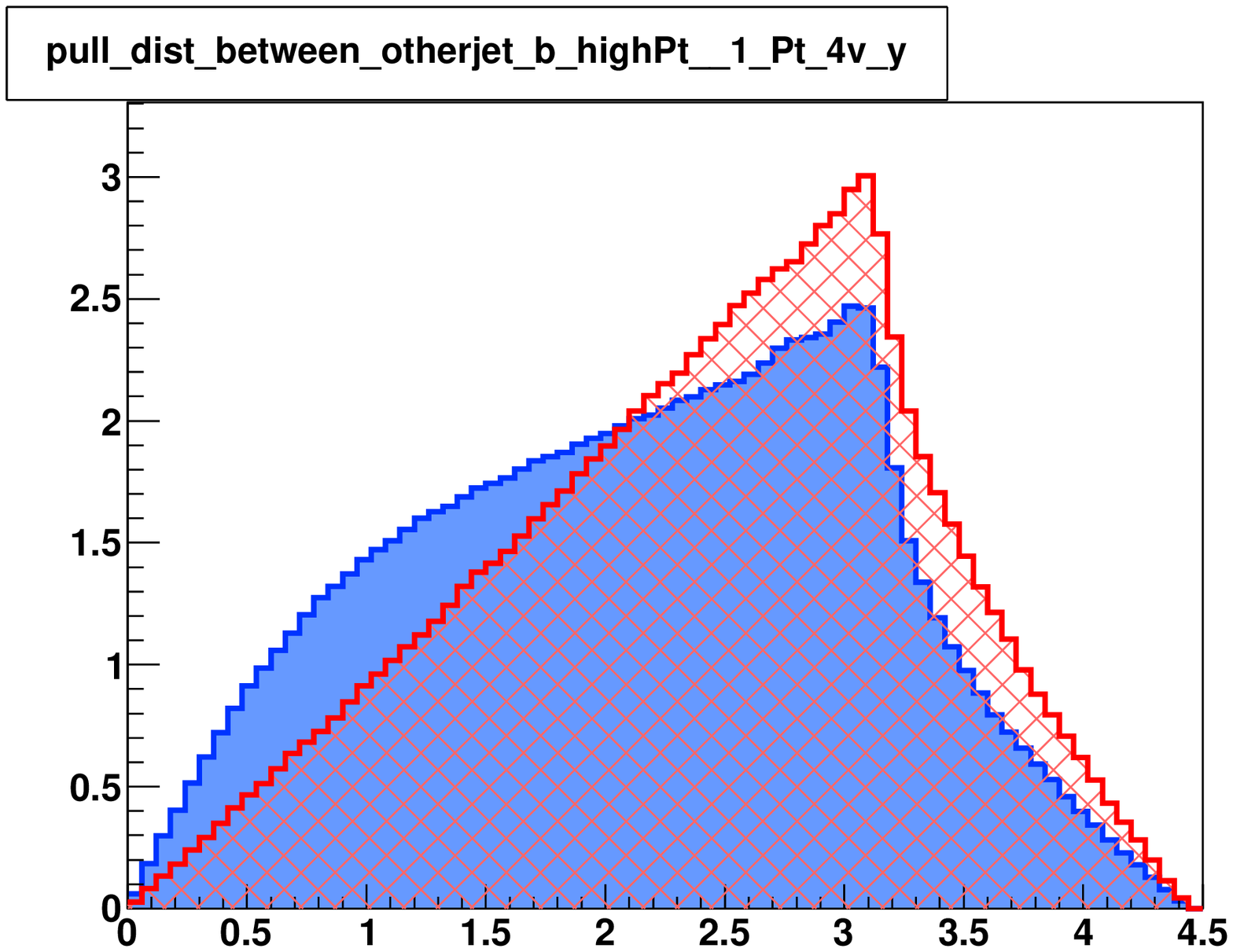}
\\
\includegraphics[width=0.3\textwidth]{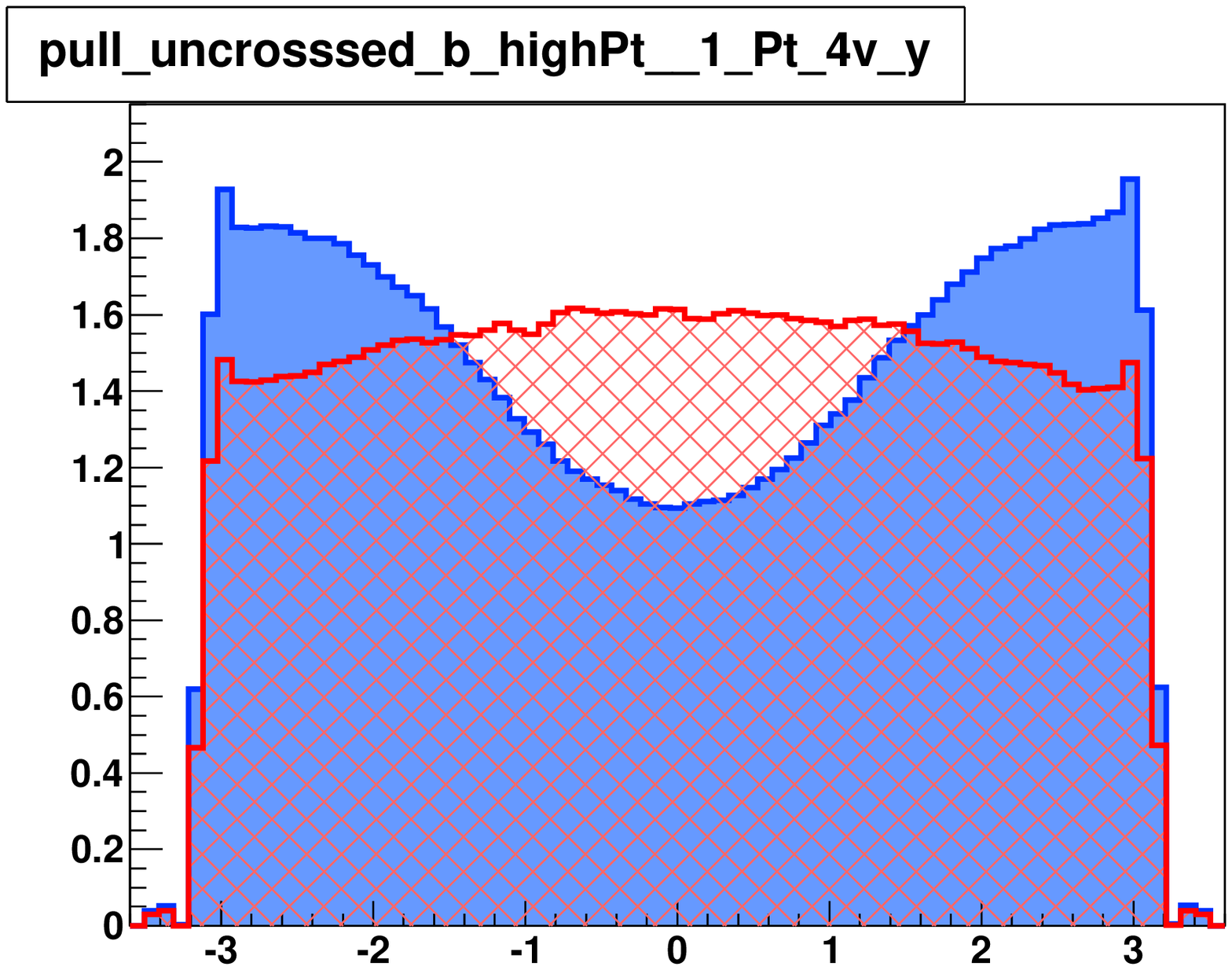}
\includegraphics[width=0.3\textwidth]{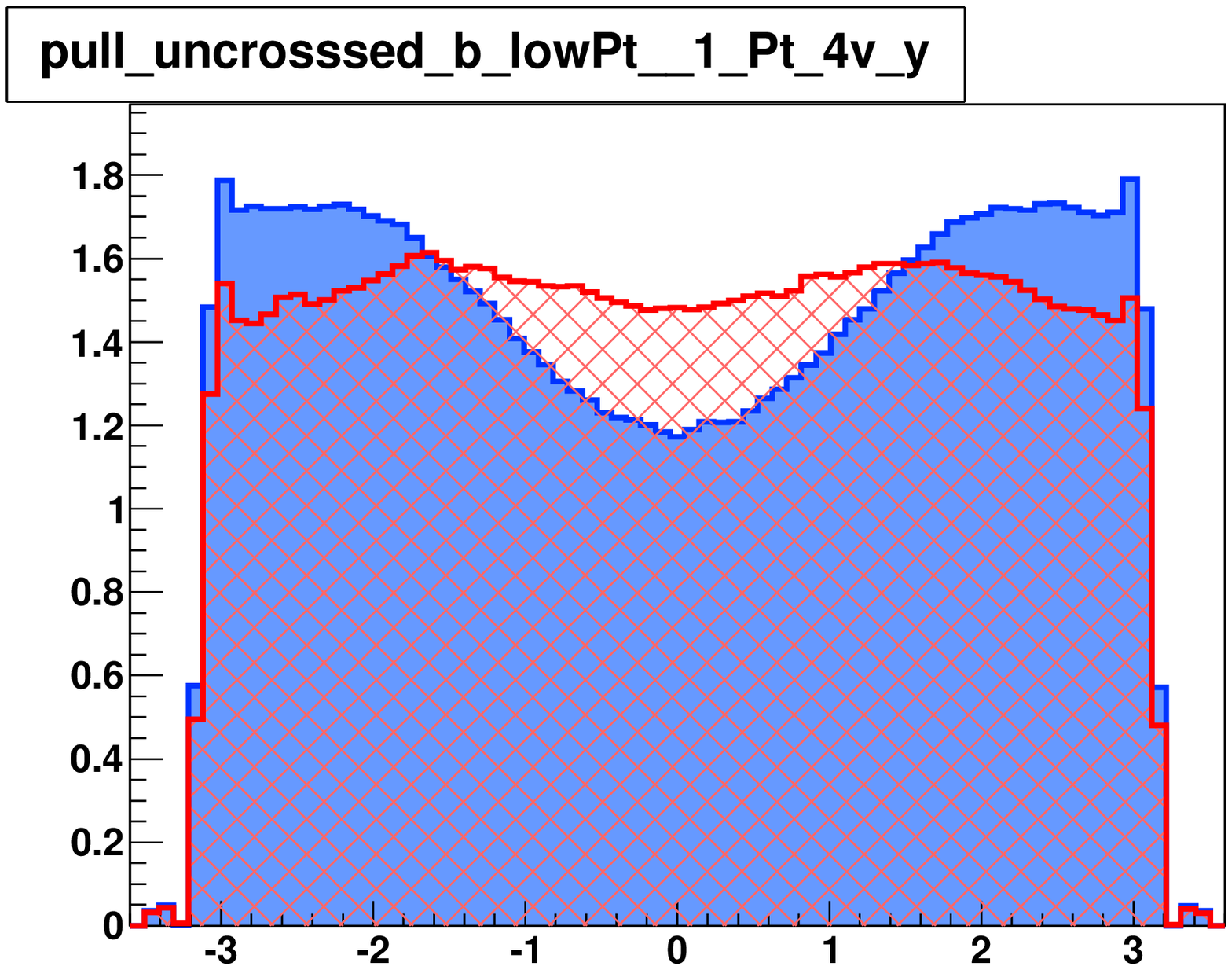}
\includegraphics[width=0.3\textwidth]{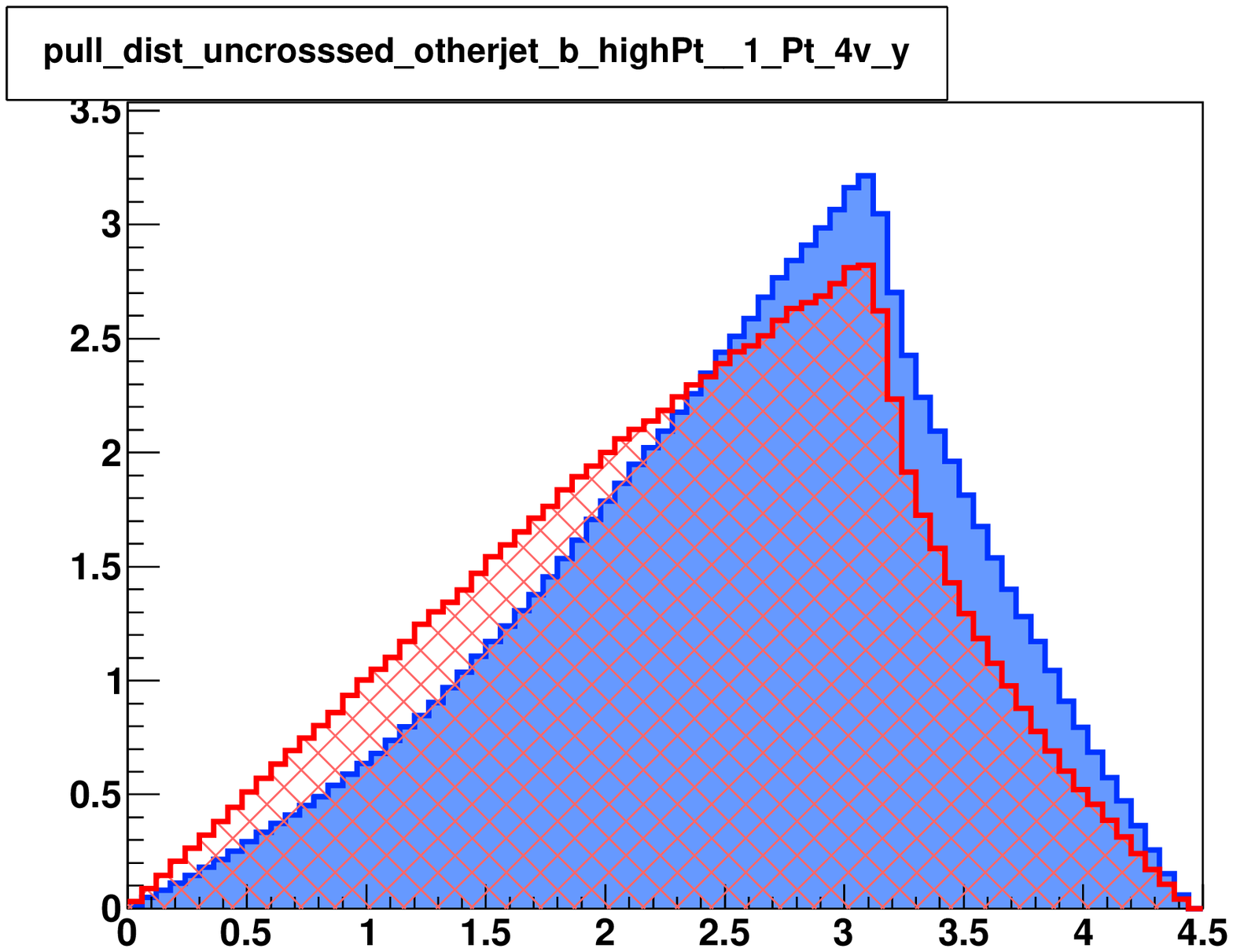}
\\
\caption{
Pull distributions for $\zh$ signal (solid blue) and $\zbb$ background (hashed red) at the LHC.
Events satisfy selection cuts and the Higgs mass-window cut, \Higgswindow . Horizontal
axes are in radians, and vertical axes are in arbitrary
units with signal and background normalized to the same area.
}
\label{fig:pullvariables}
\end{center}
\end{figure}

The 2D distribution of the $y$ and $\phi$ components of the pull vector looks like a Gaussian whose peak
is shifted slightly away from the origin in the direction of the color-connected object.
The magnitude of the vector does not have as much distinguishing power as its angle.
The pull angle of each jet, however, does not have much power without comparing it
to where it ``should'' point: toward the other jet for the signal, and
toward one of the beams (usually the closest) for the background.

The direction toward the other $b$-jet is the \emph{twist angle}, defined in Section~\ref{sec:twist}.
A version of twist that goes from 0 to $2\pi$ is defined as the direction of the lower-$p_T$ jet
from the location of the higher-$p_T$ jet in the $y/\phi$ plane.  The
3D distribution of the pull angles of each jet along with this twist angle contains
all of the useful pull information that can be used to separate signal from background,
but the individual pull angles for a given event are meaningful only with respect to the twist angle.
In an attempt to expose the physics and make the job of a multivariate discriminator
easier, we begin by defining four variables that more directly capture where the pulls
``should'' point for the signal and background, labeled in Figure~\ref{fig:acc_highres3M_raw_crop}
and defined Table~\ref{tab:pulldef}.

Subtracting the pull \emph{vectors} is not useful,
since the magnitude of each vector does not carry much meaning, and
the magnitudes are independent of each other.
Many other attempts to combine the pull angles into a
single variable also sacrifice discrimination power.
The best we have achieved is the ``pull distance,'' the
square root of the sum of squares of the difference angles. These pull angles are defined in Table~\ref{tab:pulldef}.
%


\subsection{Total Energy Variables \label{sec:energy}}
Next, we consider general purpose variables which look at the event as a whole.
Examples include
\begin{itemize}
\item $\hat s$ = CM energy for hard collision, or invariant mass of the reconstructed Higgs boson and Z.
\item $H_T$ = Scalar sum of all $E_T$.
\item $H_z$ = Boost of the center-of-mass system along the beam.
\item $\Sigma p_T$ = Scalar sum of all $p_T$ (which differs from $H_T$ for massive jets).
\item $E_{vis}$  = Scalar sum of all visible energy.
\item Centrality $= \Sigma p_T \  / \ E_{vis}$.
\end{itemize}
For each of these variables, the quantity can be constructed from the complete event,
summing over all {\bf particles} in the event. Particles here can mean topo-clusters or calorimeter cells or
energy deposits coarse-grained into $0.1\times0.1$ cells in $(\eta,\phi)$. In this study
we use the 4-momenta of the stable particles in the event record for what we call the {\it particle} variables.
The same variables can also be constructed using
just the reconstructed {\bf objects} (jets, leptons and photons)
or even just the four {\bf primary  objects} (2 $b$-jets and 2 leptons). We find that using objects
generally works much better than using variables constructed from energy in the complete event.
A particularly
useful variable is centrality constructed from the four primary objects.
Some energy variables are shown in Figure~\ref{fig:shapes}.

\subsection{ Event Shape Variables}
Finally, we look at some event shape variables. The ones we consider
involve eigenvalues of two similar tensors composed of
3-momenta, with the sum over the same sets of particles, reconstructed objects, or four primary objects, as above:

\begin{equation}
\begin{array}{lll}
\textrm{Sphericity Tensor }  \equiv
&\qquad \qquad&
\textrm{Spherocity Tensor } \equiv \\
\dfrac{1}{\Sigma_i |\vec p_i|^2}
\sum_i
\begin{pmatrix}
   p_x p_x & p_x p_y & p_x p_z \\
   p_y p_x & p_y p_y & p_y p_z \\
   p_z p_x & p_z p_y & p_z p_z
\end{pmatrix}
&\qquad \qquad&
\dfrac{1}{\Sigma_i |\vec p_i|}
\sum_i
\left[
\dfrac{1}{|\vec p_i|}
\begin{pmatrix}
   p_x p_x & p_x p_y & p_x p_z \\
   p_y p_x & p_y p_y & p_y p_z \\
   p_z p_x & p_z p_y & p_z p_z
\end{pmatrix}
\right]
\\
\end{array}
\end{equation}
where $i$ labels on the momentum components appearing in the matrix are implicit.

\begin{figure}[t]
\begin{center}
\includegraphics[width=0.3\textwidth]{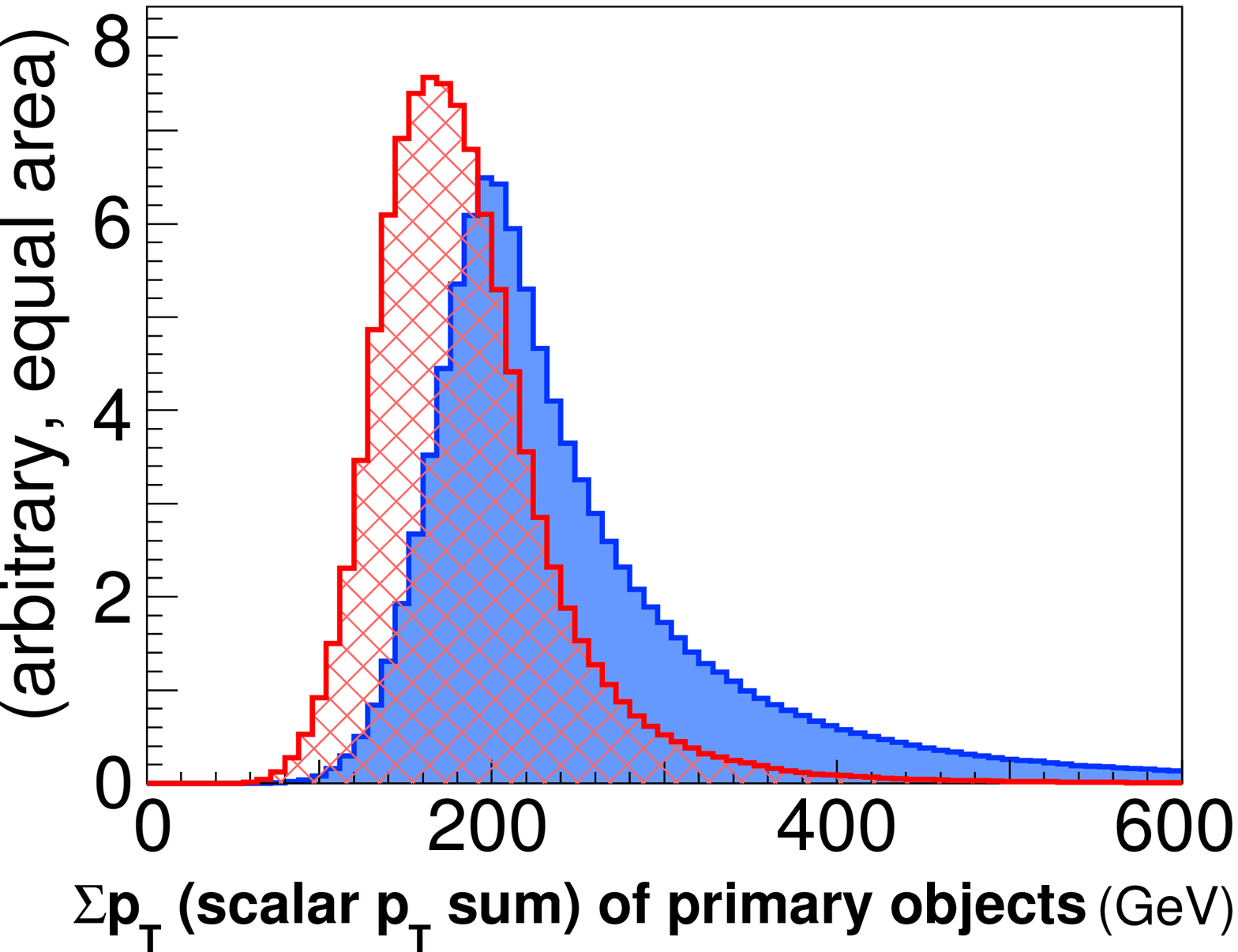}
\includegraphics[width=0.3\textwidth]{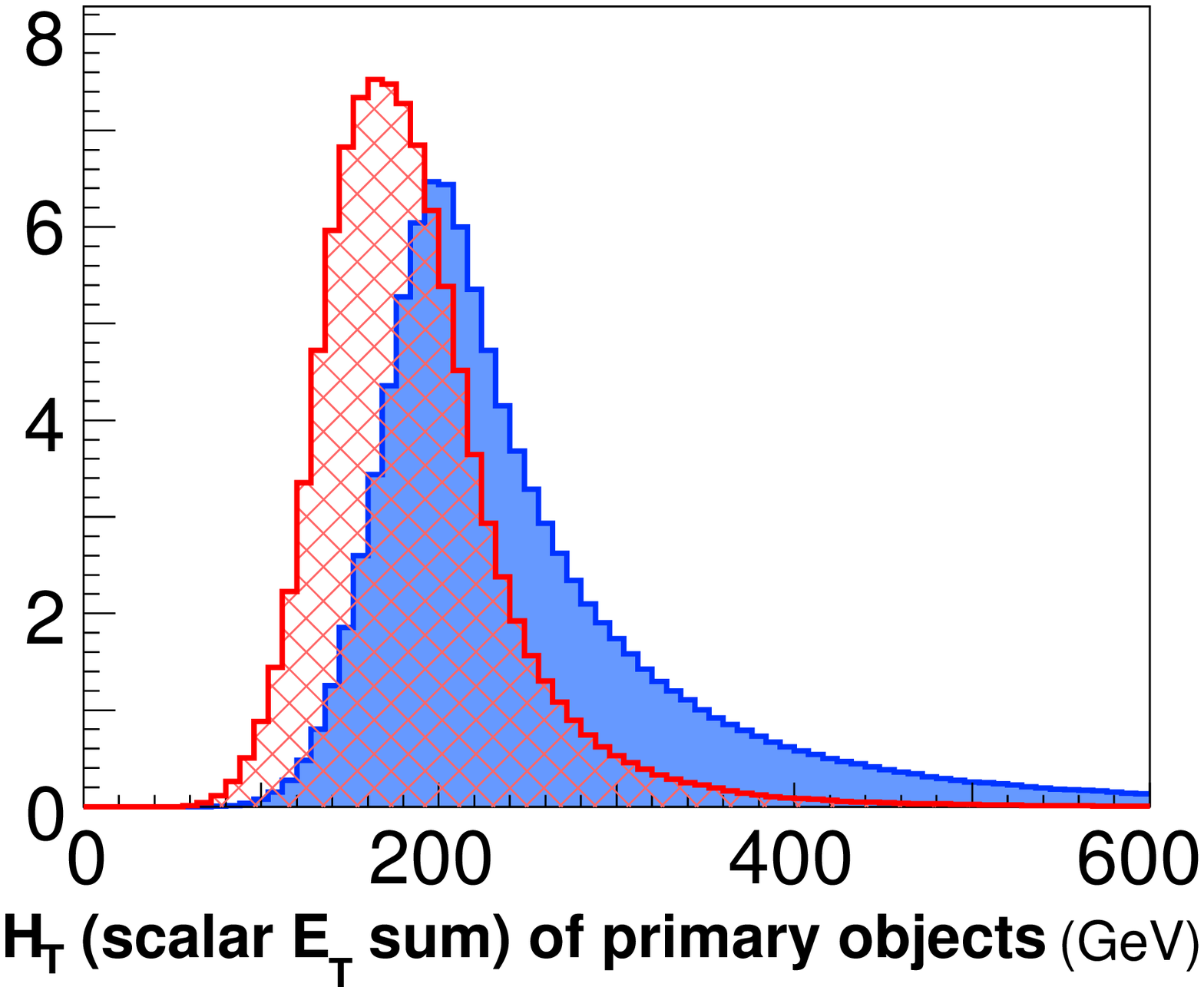}
\includegraphics[width=0.3\textwidth]{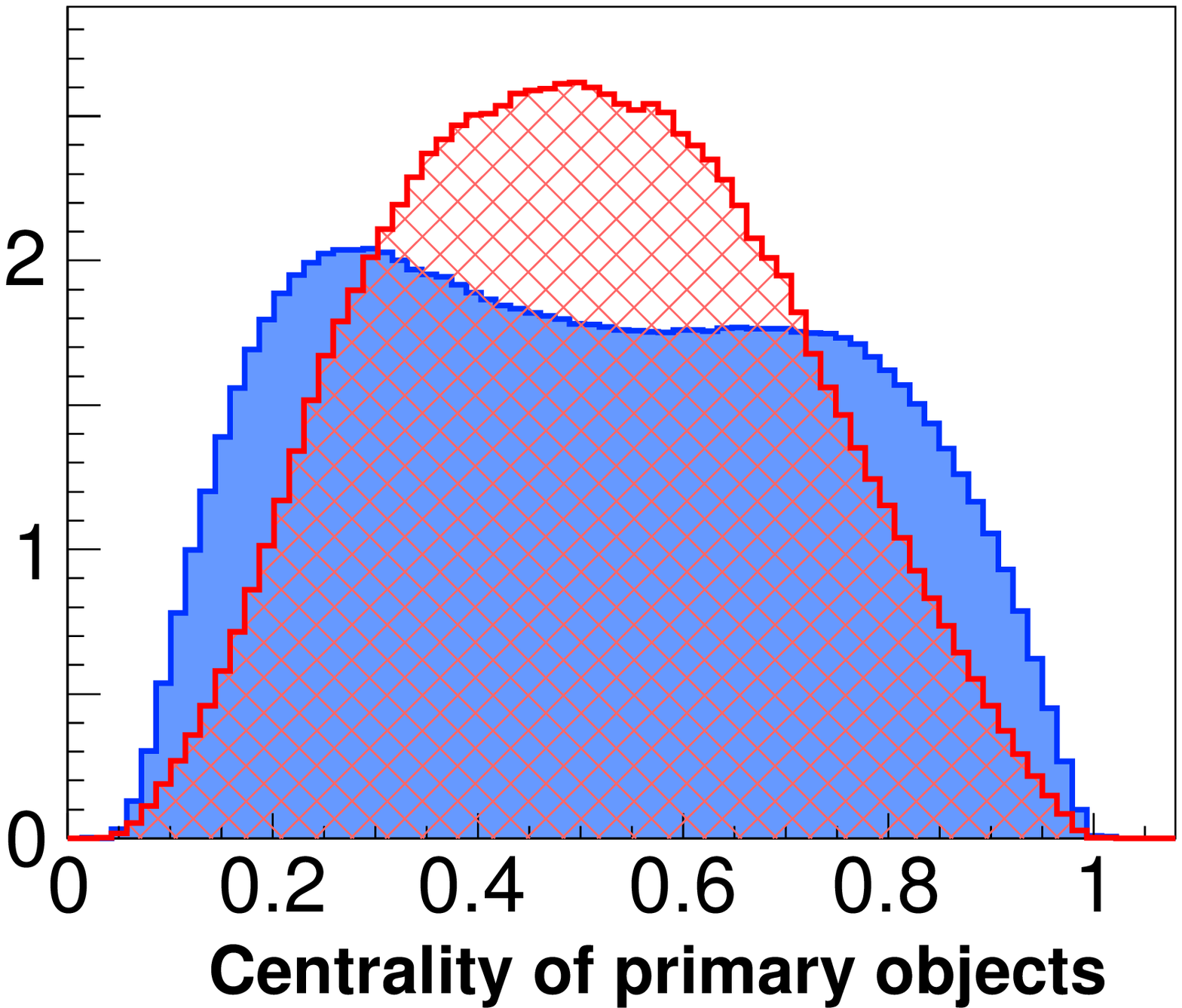}

\includegraphics[width=0.3\textwidth]{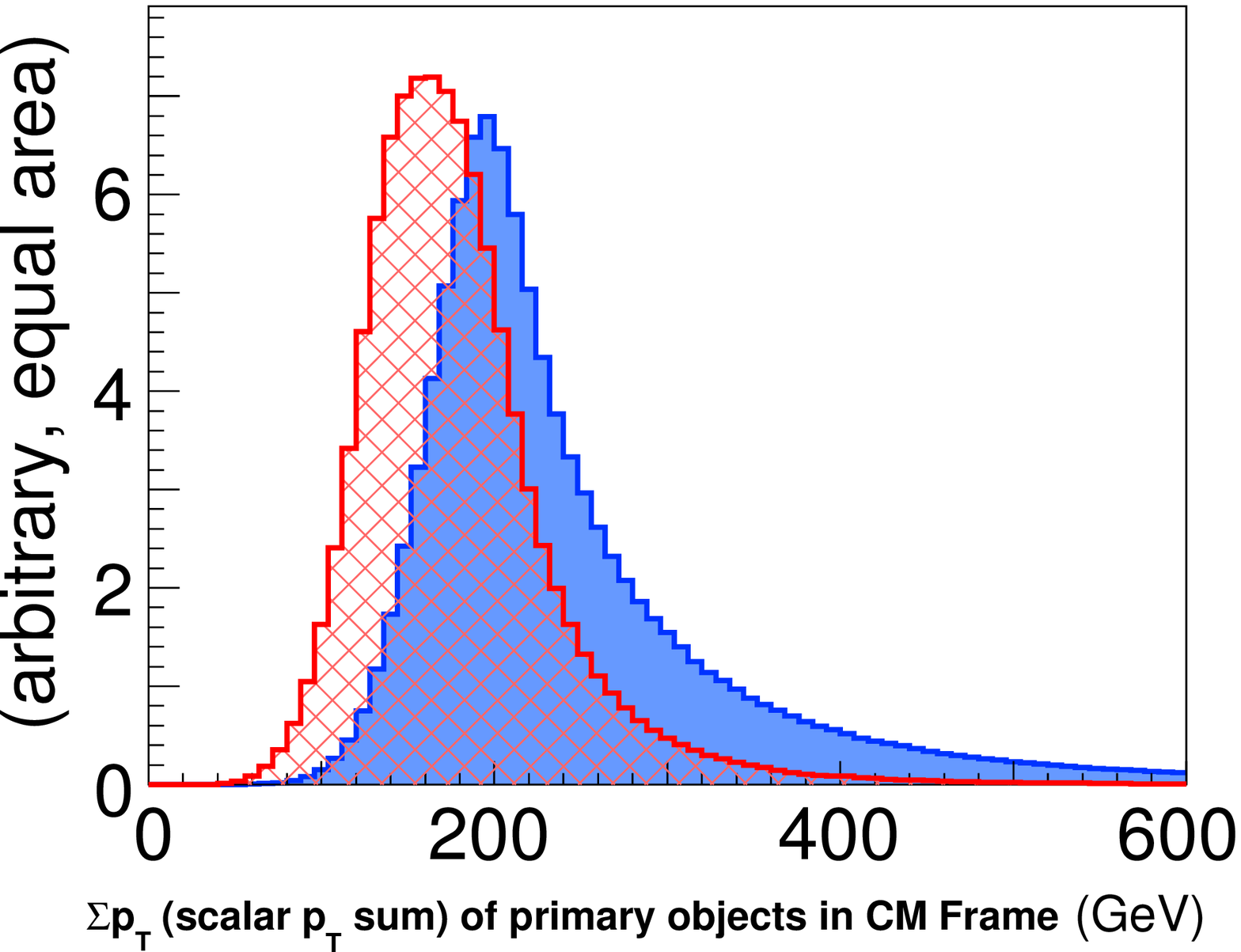}
\includegraphics[width=0.3\textwidth]{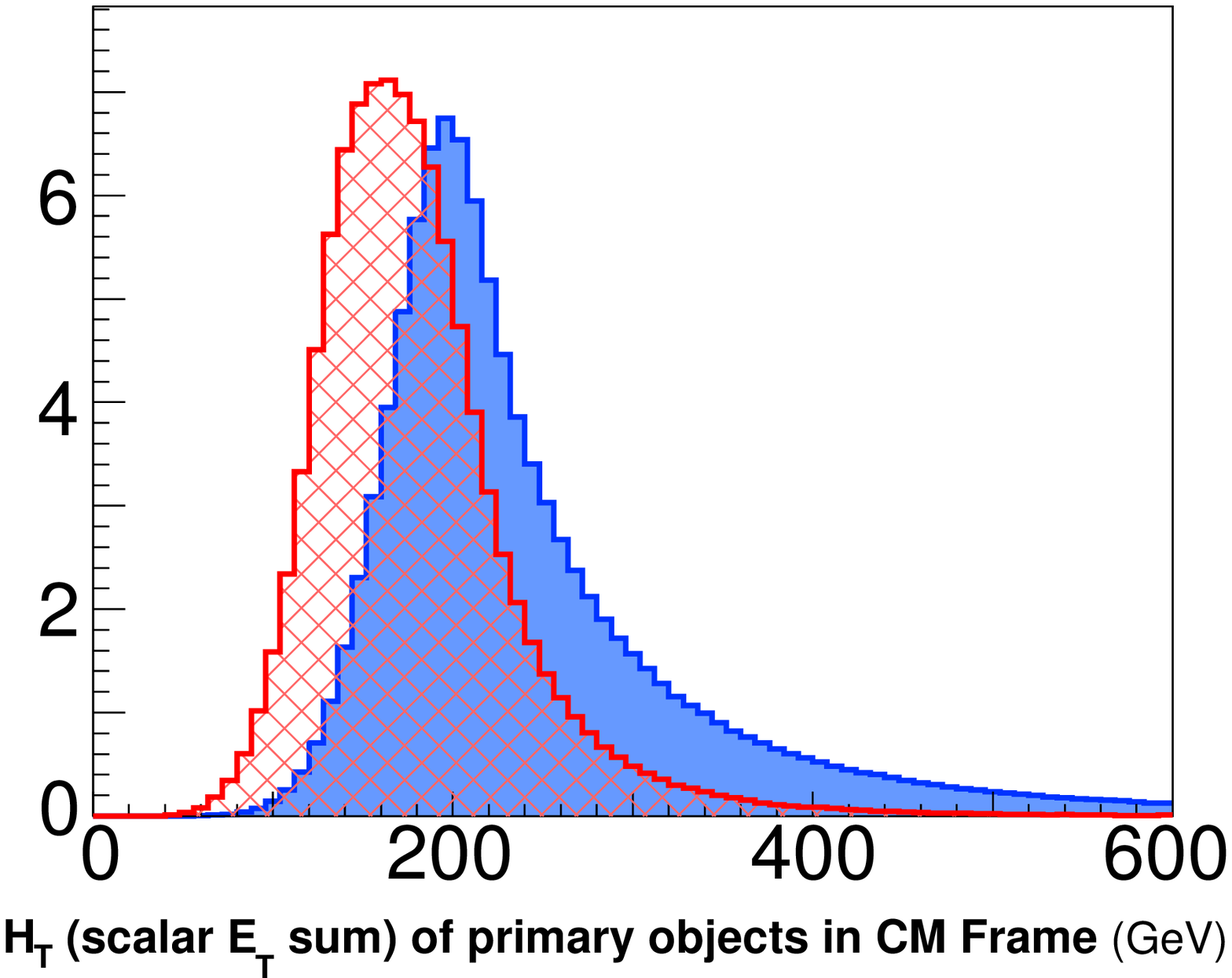}
\includegraphics[width=0.3\textwidth]{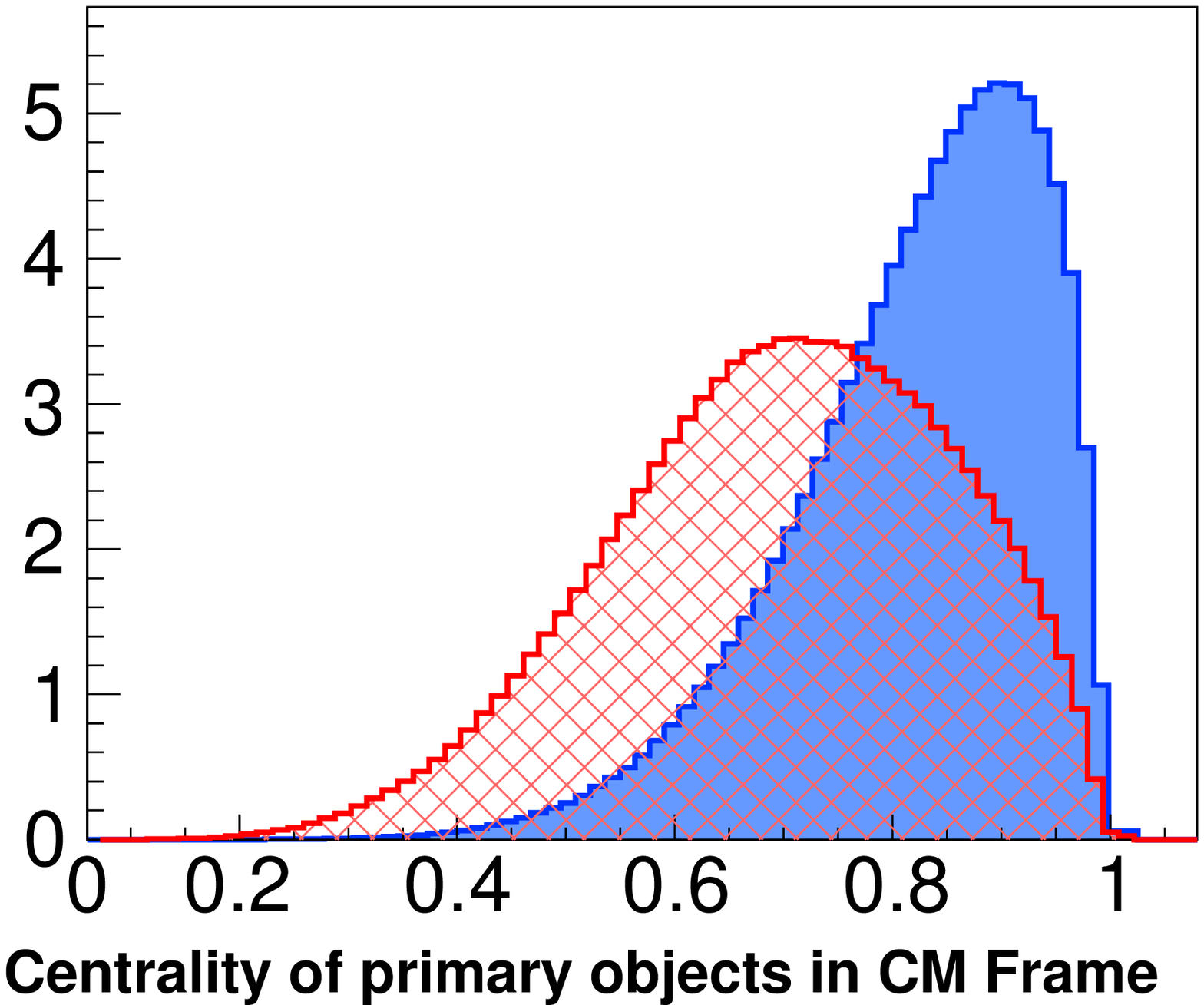}

\includegraphics[width=0.3\textwidth]{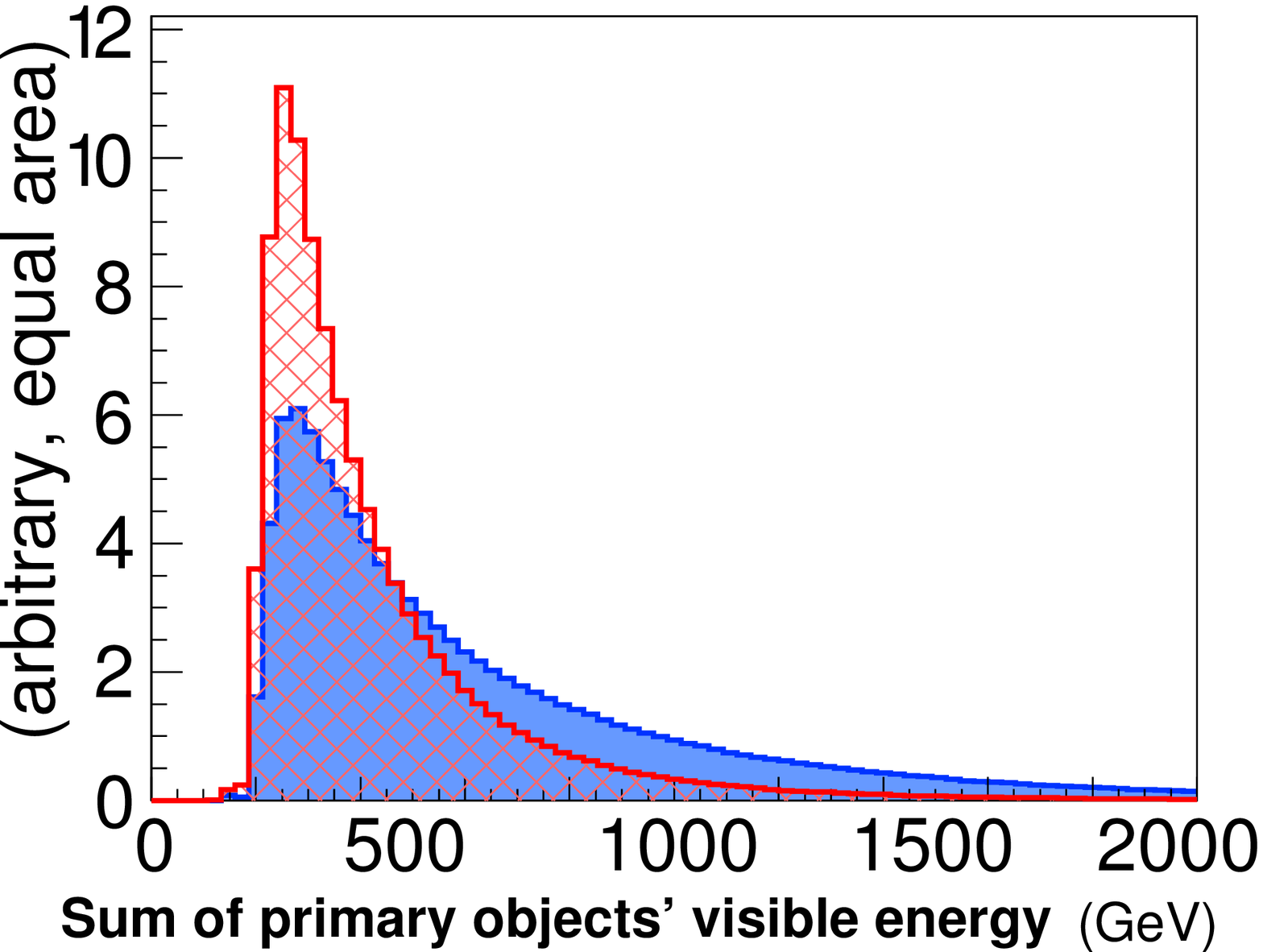}
\includegraphics[width=0.3\textwidth]{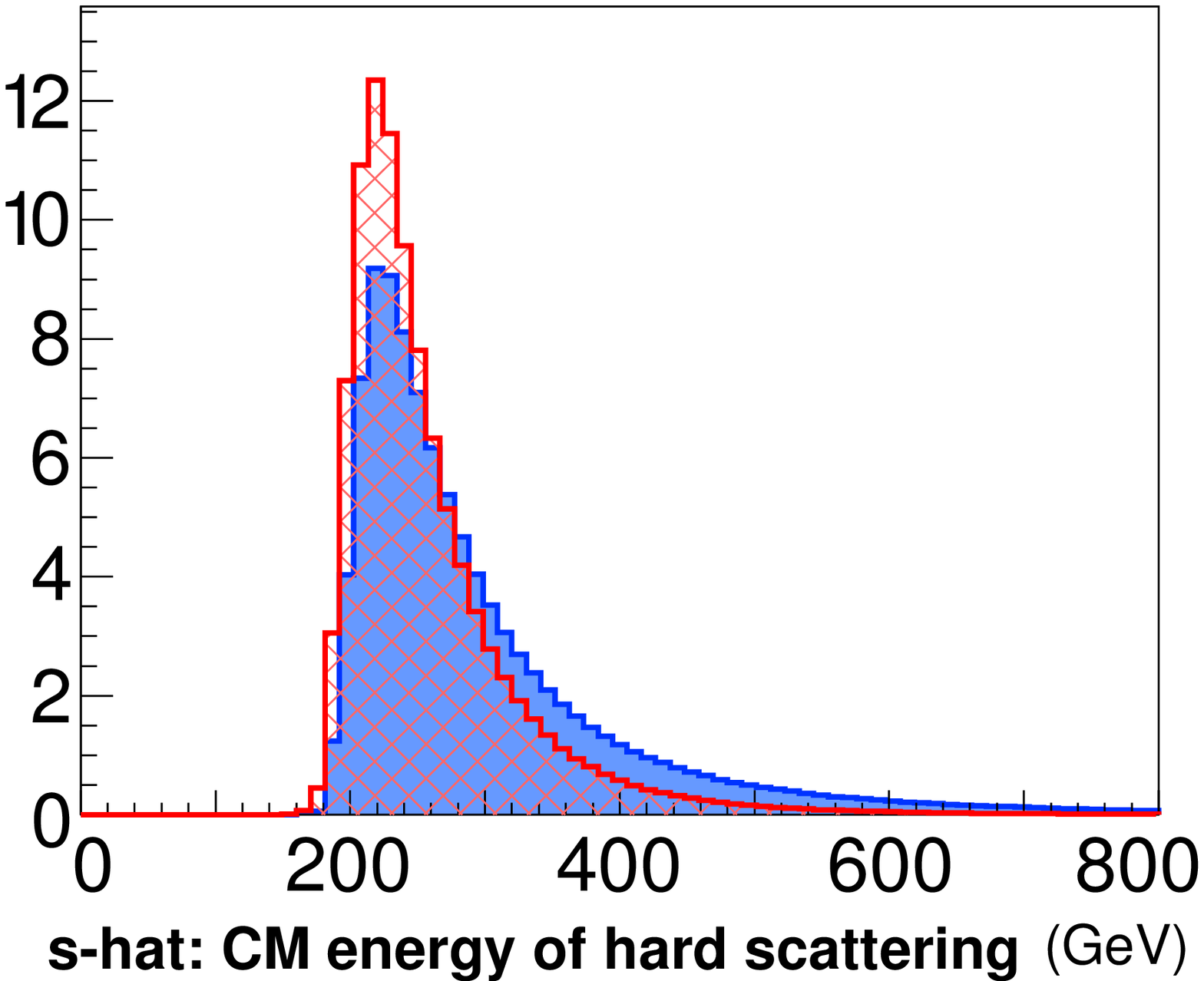}
\includegraphics[width=0.3\textwidth]{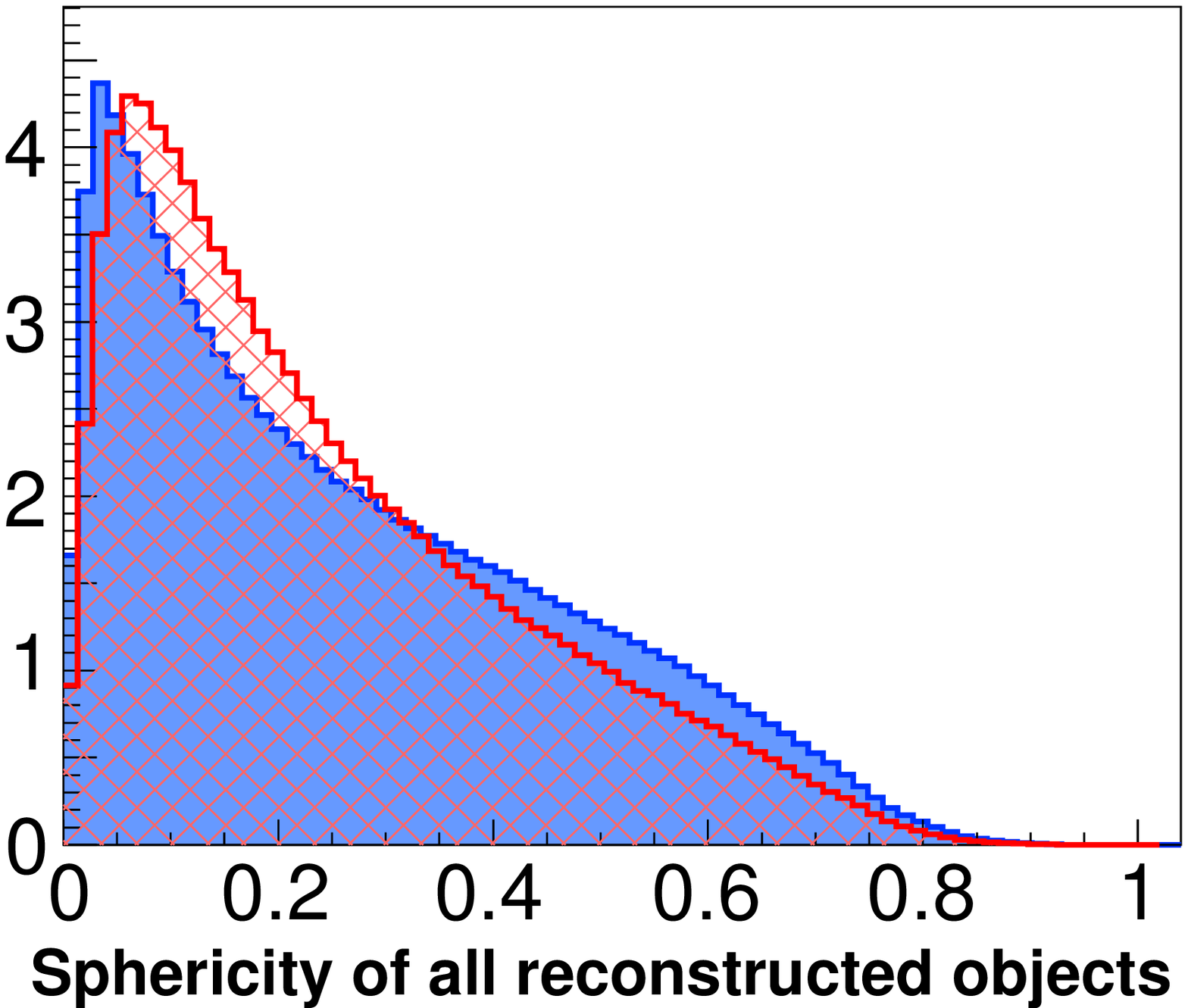}
\caption{
Total energy and event shape variables for $\zh$ signal (solid blue)
and $\zbb$ background (hashed red) at the LHC.
Events satisfy selection cuts and the Higgs mass-window cut, \Higgswindow . Horizontal
axes are in GeV where appropriate,
and vertical axes are in arbitrary units with signal and background normalized to the same area.
}
\label{fig:shapes}
\end{center}
\end{figure}

The eigenvalues of these matrices are computed, then ordered and normalized $\lambda_1 \ge \lambda_2 \ge \lambda_3$ with
$\lambda_1 + \lambda_2 + \lambda_3 = 1$. The event shapes are then defined as \cite{Bjo70}
\begin{itemize}
\item Sphericity and Spherocity:  $S = \frac{3}{2} \, (\lambda_2 + \lambda_3)$ where $0 \le S \le 1$.
    A 2-jet event has $S \approx 0$ while an isotropic one has $S \approx 1$.
\item Aplanarity and Aplanority:  $A = \frac{3}{2} \, \lambda_3$ where $0 \le A \le 1/2$.
    A planar event has $A \approx 0$ while an isotropic one has $A \approx 1/2$.
\item Y variable from Sphericity: $Y = \frac{\sqrt{3}}{2} \, (\lambda_2 - \lambda_3)$.
\item DShape  from Spherocity: $D = 27 \, \lambda_1 \lambda_2  \lambda_3$.
\end{itemize}
Sphericity, Spherocity, and DShape are all highly correlated, Aplanarity and Aplanority are too,
as are the shapes derived from the four primary objects in the lab and CM frame.   Aplanarity seems
to be the most useful shape variables. The others seem useful only to the extent they are correlated with Aplanarity or Centrality.
Some event shapes are shown in Figure~\ref{fig:shapes}.
We also consider the Fox-Wolfram moments of particles, objects, and primary objects. These are defined by projecting against the
Legendre polynomials
\begin{equation}
H_\ell = \frac{1}{E_{vis}} \Sigma_{i,j} |p_i| |p_j| P_{\ell}(\cos \theta_{ij} ) \, .
\end{equation}
The Fox-Wolfram moments were not found particularly useful.

\subsection{Missing $E_T$ variables \label{sec:missing}}
Missing $E_T$ and missing $p_T$ are \emph{not} included in this analysis,
other than the neutrino used to reconstruct the $W$ in the $WH$ channel.
From an experimental point of view, missing $E_T$ is an extremely important
variable. However, its distribution is dominated strongly by experimental mis-measurements and calorimeter resolution.
Without a full detector simulation, missing energy
is inappropriate for our particle-level study and, in fact, causes unphysical instabilities if included
in the multi-variable analysis.

\section{Efficiency measures \label{sec:eff}}
Having cataloged our input variables, we can now explore which
ones have the best distinguishing power. For single variables, we look at the
effect that a simple cut or window (two-sided cut) has on the number of signal and background
events in our fiducial $\mbb$ window. To combine variables, we will use more sophisticated
methods, described in the next section. But for single variables, cuts are as good as one can do.

\begin{figure}[t]
\begin{center}
\psfrag{LHC HZ Signal and Background Efficiencies}{LHC ZH \qquad  ROC Curve : $\varepsilon_B$ vs $\varepsilon_S$}
\psfrag{LHC ZH : Significance : }{}
\psfrag{Signal eff}{\!\!\!\!\!\!\!\!\!\!\!\!\!\!\!\!\!\!\!\!\!\!\!\!\!\!\!\!\!\!\!\!\!\!\!\!\!\!\!\!\!\!\!\!\!\!\! \footnotesize{Higgs Signal Efficiency $\varepsilon_S$}}
\psfrag{Backgr eff}{\!\!\!\!\!\!\!\!\!\!\!\!\!\!\!\!\!\!\!\!\!\!\!\!\!\!\!\!\!\!\!\!\!\!\!\!\!\!\!\!\!\!\!\!\!\!\!\!\!\!\!\!\!\!\!\!\!\!\!\!\! \footnotesize{Background Efficiency $\varepsilon_B$}}
\psfrag{b_highPt_Pt_ak05}{\scriptsize{$p_T^{b1}$}}
\psfrag{b_lowPt_Pt_ak05}{\scriptsize{$p_T^{b2}$}}
\psfrag{Z_Pt_ak05}{\scriptsize{$p_T^Z$}}
\psfrag{bbbar_twist_ak05}{\scriptsize{$\tau_{b \bar b}$}}
\psfrag{bbbar_dEta_ak05}{\scriptsize{$\Delta \eta_{b \bar b}$}}
\psfrag{H_and_b_lowPt_dRap_ak05}{\scriptsize{$\Delta y_{H,b2}$}}
\psfrag{H_and_b_highPt_dRap_ak05}{\scriptsize{$\Delta y_{H,b1}$}}
\psfrag{ZZH_Frame_b_lowPt_costheta_ak05}{\scriptsize{CM $\cos(\theta_{b2})$}}
\psfrag{bbbar_dR_ak05}{\scriptsize{$\Delta R_{b \bar b}$}}
\psfrag{lplm_dPhi_ak05}{\scriptsize{$\Delta \phi_{\ell^+ \ell^-}$}}
\includegraphics[width=0.7\textwidth]{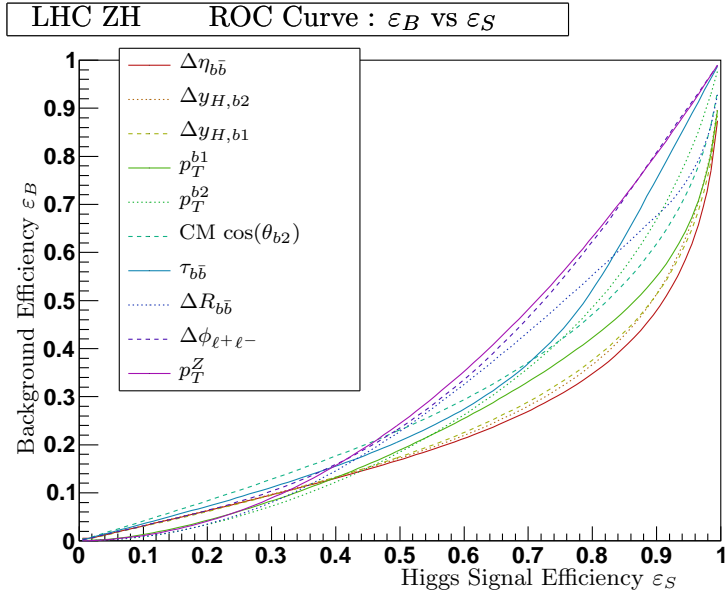}
\caption{ROC curves for some variables, showing the background
efficiency and signal efficiency as parametric functions of the cut value.}
\label{fig:roc}
\end{center}
\end{figure}

Any given cut will keep some fraction $\varepsilon_S$ of signal
events and some other fraction $\varepsilon_B$ of background
events.  These are the signal and background efficiencies of
the cut. For the two sided cuts, there are a number of choices which give
the same $\eS$, so the one which minimizes $\eB$ for a given $\eS$ is chosen.
A standard way to visualize the relationship between
$\varepsilon_S$ and $\varepsilon_B$ is with a ``Receiver Operating
Characteristic'' curve, or ROC curve.  Often the background
rejection $1-\eB$, or the inverse rejection $\frac{1}{\eB}$ is plotted against the signal
efficiency $\eS$. A sample ROC curve for 10 representative single variables is shown in Figure~\ref{fig:roc}.
Lower lines are better, since they give more background rejection for the same signal efficiency,
but since some of the curves cross, the ROC curves do not lead to an immediate observation of which variables are better.
One approach to ranking variables would be to arbitrarily demand a particular signal efficiency and find the variables
that give the best background rejection, but we propose a more useful procedure that does not introduce an arbitrary parameter.

In order to quantify the usefulness of a variable, we need to consider the goal of the signal search.
The signal-to-background ratio ($S/B$) and the significance
($\sigma \sim S/\sqrt{B}$) are the two quantities considered
when trying to see a signal over a large background.
Making a tighter cut will reduce both efficiencies, but not necessarily by the same amount, so $S/B$ may change.
The factor by which
$S/B$ changes is just the ratio of the efficiencies.
\begin{equation}
\frac{S}{B}
\qquad \ra{\text{cut}} \qquad
\frac{\varepsilon_S S}{\varepsilon_S B}
=\left( \frac{\varepsilon_S}{\varepsilon_B}\right) \frac{S}{B} \, .
\end{equation}
The other quantity normally considered is the statistical
significance $\sigma$ which, for large numbers of events,
approaches $S/\sqrt{B}$.
For a convincing discovery,
the expected number of signal events must
exceed the statistical fluctuations of the background:
\begin{equation}
\frac{S}{\sqrt{B}} \gg 1 \,.
\end{equation}
When we cut on our discriminant, the significance changes by
\begin{equation}
\sigma \equiv \frac{S}{\sqrt{B}}
\qquad \ra{\text{cut}} \qquad
\frac{\eS S}{\sqrt{ \eB B}}
= \left(\frac{\eS}{\sqrt{\eB}}\right) \sigma \,.
\end{equation}
It follows that the two quantities we are interested in are
the signal-over-background improvement characteristic
\begin{equation}
\frac{\eS}{\eB} \,,
\end{equation}
and the Significance Improvement Characteristic (SIC)
\begin{equation}
  \rsig \equiv \frac{\eS}{\sqrt{\eB}} \,.
\end{equation}
These quantities tell us the improvement on $S/B$ and $\sigma$ that our discriminant will give.
When systematics dominate, $\rSB$ is more important, and when statistical errors dominate,
$\rsig$ is the more useful measure.
Given that $\rSB$ and $\rsig$ are luminosity independent, they provide good measures of the relative discrimination power
of the various variables. Moreover, plotting these measures, especially $\rsig$ as functions of $\eS$ provides a wonderful
visualization of a variable's effectiveness.

\begin{figure}[t]
\begin{center}
\begin{tabular}{cc}
\psfrag{LHC HZ Signal over Background}{~~~~\tiny{$\text{LHC ZH Signal over Background}$}}
\psfrag{Signal eff}{\!\!\!\!\!\!\!\!\!\!\!\!\!\!\!\!\!\!\!\!\!\!\!\!\!\!\!\!\!\!\!\!\!\!\!\!\!\!\!\!\!\!\!\!\!\!\!\tiny{Higgs Signal Efficiency $\varepsilon_S$}}
\psfrag{b_highPt_Pt_ak05}{\tiny{$p_T^{b1}$}}
\psfrag{b_lowPt_Pt_ak05}{\tiny{$p_T^{b2}$}}
\psfrag{Z_Pt_ak05}{\tiny{$p_T^Z$}}
\psfrag{bbbar_twist_ak05}{\tiny{$\tau_{b \bar b}$}}
\psfrag{bbbar_dEta_ak05}{\tiny{$\Delta \eta_{b \bar b}$}}
\psfrag{H_and_b_lowPt_dRap_ak05}{\tiny{$\Delta y_{H,b2}$}}
\psfrag{H_and_b_highPt_dRap_ak05}{\tiny{$\Delta y_{H,b1}$}}
\psfrag{ZZH_Frame_b_lowPt_costheta_ak05}{\tiny{CM $\cos(\theta_{b2})$}}
\psfrag{bbbar_dR_ak05}{\tiny{$\Delta R_{b \bar b}$}}
\psfrag{lplm_dPhi_ak05}{\tiny{$\Delta \phi_{\ell^+ \ell^-}$}}
\includegraphics[width=0.5\textwidth]{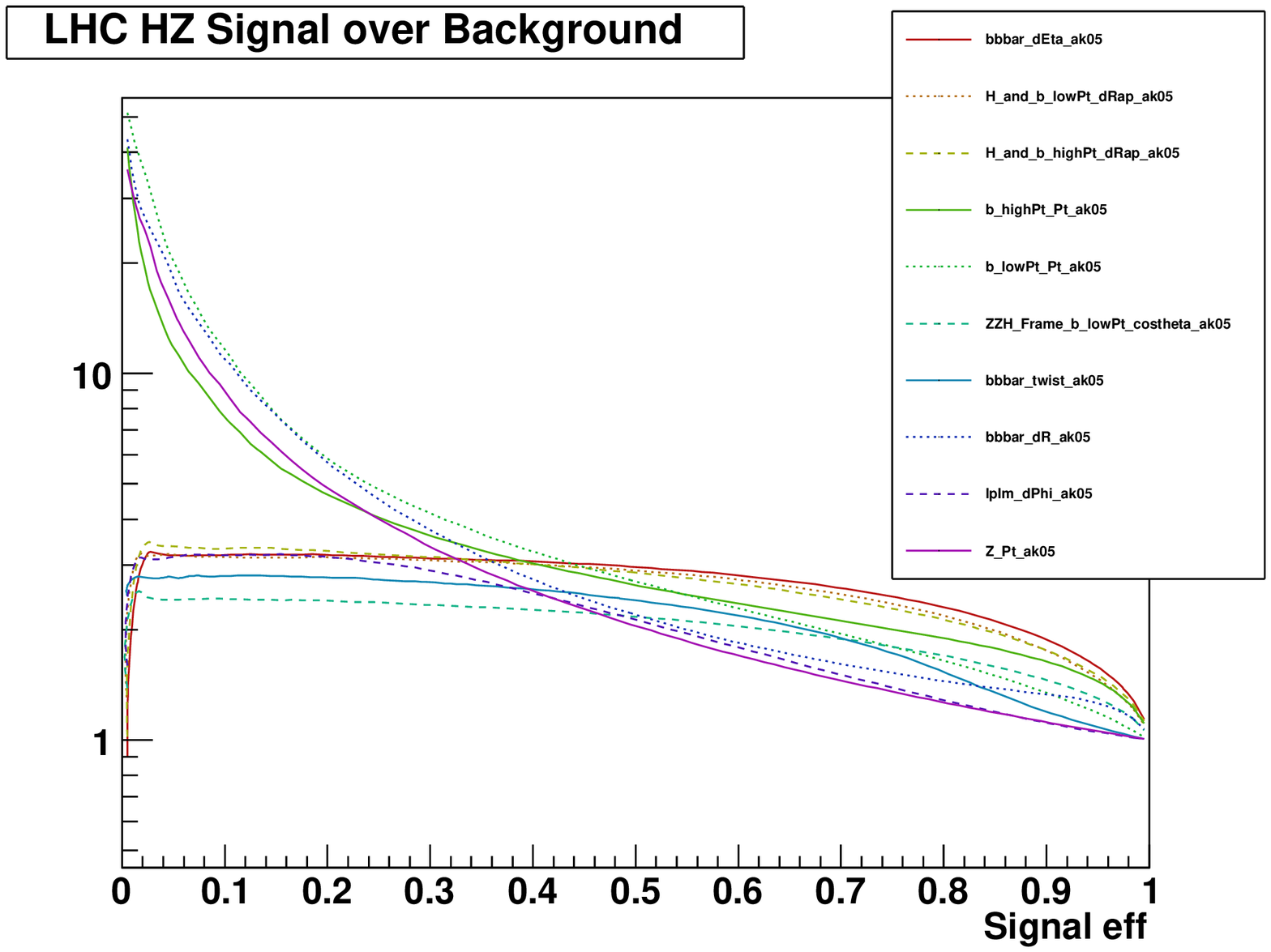}
&
\psfrag{LHC HZ Significance}{\tiny{$S/\sqrt{B}$ Improvement}}
\psfrag{Signal eff}{\!\!\!\!\!\!\!\!\!\!\!\!\!\!\!\!\!\!\!\!\!\!\!\!\!\!\!\!\!\!\!\!\!\!\!\!\!\!\!\!\!\!\!\!\!\!\!\tiny{Higgs Signal Efficiency $\varepsilon_S$}}
\psfrag{b_highPt_Pt_ak05}{\tiny{$p_T^{b1}$}}
\psfrag{b_lowPt_Pt_ak05}{\tiny{$p_T^{b2}$}}
\psfrag{Z_Pt_ak05}{\tiny{$p_T^Z$}}
\psfrag{bbbar_twist_ak05}{\tiny{$\tau_{b \bar b}$}}
\psfrag{bbbar_dEta_ak05}{\tiny{$\Delta \eta_{b \bar b}$}}
\psfrag{H_and_b_lowPt_dRap_ak05}{\tiny{$\Delta y_{H,b2}$}}
\psfrag{H_and_b_highPt_dRap_ak05}{\tiny{$\Delta y_{H,b1}$}}
\psfrag{ZZH_Frame_b_lowPt_costheta_ak05}{\tiny{CM $\cos(\theta_{b2})$}}
\psfrag{bbbar_dR_ak05}{\tiny{$\Delta R_{b \bar b}$}}
\psfrag{lplm_dPhi_ak05}{\tiny{$\Delta \phi_{\ell^+ \ell^-}$}}
\includegraphics[width=0.5\textwidth]{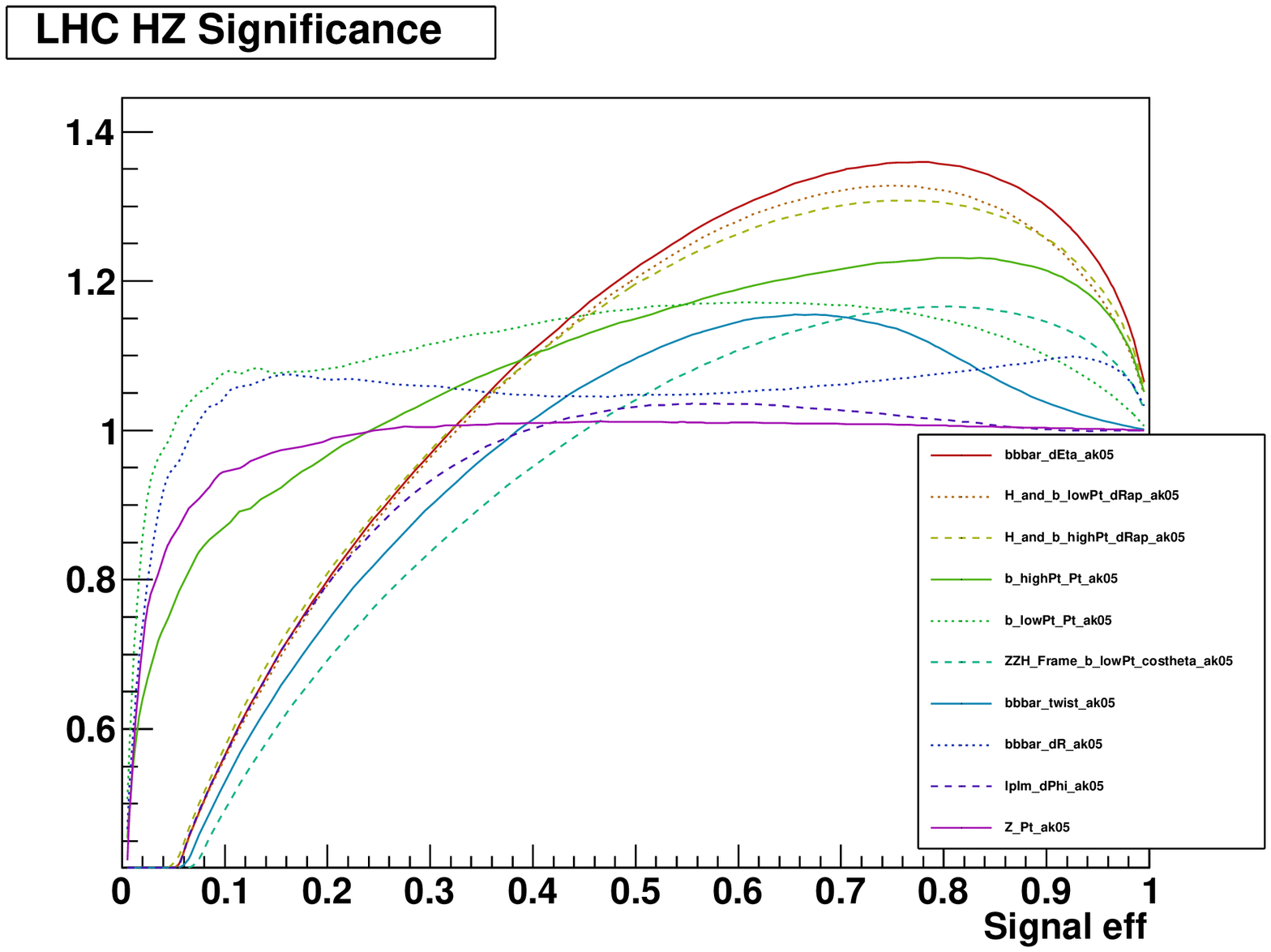}
\end{tabular}
\caption{Improvements in $S/B$ and $\sigma =S/\sqrt{B}$ for the same variables as in the previous figure.} 
\label{fig:soverb}
\end{center}
\end{figure}

We show $\rSB$ and $\rsig$ as a function of $\eS$ in Figure~\ref{fig:soverb}
for the same 10 variables as in the ROC curve in Figure~\ref{fig:roc}.
There are a few salient features of these more powerful
visualizations. First, consider the $S/B$ curves. There are apparently
two classes of variables. For one class (including
$\Delta \eta_\bbar$ and other angle-type variables), the $S/B$ is
essentially flat below $\eS\sim 0.7$ or so. This means that the
variable cannot distinguish signal from background beyond this
point, and cutting harder on these variable is equivalent to
throwing out random events. The other class (including $p_T^Z$ and other variables with long tails)
seem to lead to $S/B$ enhancements which grow
indefinitely towards $\varepsilon_S=0$. This means that if
sufficiently hard cuts are placed, a very large improvement in
$S/B$ can be achieved. This often happens because there is some
limit of the variable in which there are zero background
events.
These qualitative observations notwithstanding,  it
is not at all clear, by looking at the $\rSB$ curves alone, which variable does best, or where to apply
the cut. For this reason, we will focus less on these curves for the multivariate analysis than on $\rsig$.

The $\rsig$ curves provide a much better way of looking at the
information than the $\eS/\eB$ curves. We see that there are still essentially two
classes. For one class (including $\Delta \eta_\bbar$), the
efficiencies have a maximum at some intermediate value of
$\varepsilon_S$. Thus, there is an optimal value of the cut for
these variables to get the maximal enhancement in $S/\sqrt{B}$.
Because of this maximum, we argue that
\begin{equation}
  \ropt \equiv \max  \frac{\eS}{\sqrt{\eB}}
\end{equation}
is a useful characterization of the effectiveness of a
discrimination method. The other class of variables (including
$p_T^Z$) seem to only be able to reduce $S/\sqrt{B}$ and have
$\ropt=1$.

The variables
which can lead to $S/\sqrt{B}$ improvement are often ones which
had flat $S/B$ distributions. Also, the variables which lead to
arbitrarily large $S/B$ enhancements often can only reduce
$S/\sqrt{B}$. Since $p_T^Z$ is one of the second class
of variables, the original efficiencies
from the boosted Higgs boson search~\cite{Butterworth:2008iy} were $\varepsilon_S =
1/20$ and $\varepsilon_B = 1/320$.  We see that
$\varepsilon_S/\varepsilon_B = 16$ while
$\varepsilon_S/\sqrt{\varepsilon_B} = 0.89$.

Starting from these observations, we turn to the multivariate analysis. Among other things, we will find that
some variables which are useless as single variables ({\it i.e.} $\ropt = 1$) start to add marginal efficiency
on top of the ones which are useful to begin with. Moreover, we will see that the optimal set of 10 variables is
very different from the top 10 single variables. This makes perfect sense: less useful variables can become useful after
cuts on more powerful variables are applied, because they have subtle correlations. However, this means we have to carefully
decide on which variables to use, in order avoid throwing out some potentially useful ones which are not useful by themselves.
A systematic procedure for selecting a set of variables is described in Section~\ref{sec:multi}. First, we give a
brief introduction to the multivariate methods.


\section{Multivariate Techniques~\label{sec:bdt}}
In order to make use of all possible discriminating features
between signal and background samples, including complex
non-linear correlations, we will make use of a
multivariate approach.
Many multivariate techniques are efficiently implemented in the {\sc Toolkit for Multivariate Data Analysis with ROOT} (TMVA)
in a way that makes them easy to use and compare.
We refer the reader to the TMVA documentation for more details of the methods~\cite{tmva}.

In this study, we will use mainly Boosted Decision Trees (BDTs) \cite{Freund:1996acm}.
We briefly considered other methods, such as
multilayer perceptron Artificial Neural Networks (ANNs).
We found that BDTs tend to converge faster and run in
shorter time than ANN while giving similar results.
We suspect that ANNs may be optimal for
other applications, such as artificial intelligence, but for high energy physics analyses, they are not
entirely appropriate. The main feature that distinguishes the particle physics applications is that
we are almost always interested in a binary classification: signal vs background. Neural networks seem
better suited for multidimensional outputs, such as in pattern recognition tasks.
(For a more detailed comparison in the context of a particle physics application,
with similar conclusions, see~\cite{Roe:2004na,Roe:2005hm}.)
We also found that a traditional Bayesian likelihood analysis or optimal linear Fisher discriminant is
comparable to the BDT for few variables, but does significantly
worse when multiple variables are combined. Performing a thorough comparison of methods
in the context of collider physics is beyond the scope of the current study. The results
in this paper will all refer to discrimination using the BDT method, which we found optimal in our informal survey
of different methods and different parameters.

A decision tree is a hierarchical set of one-sided cuts used to classify signal (S) versus background (B). For example,
if there are two discriminants $a$ and $b$, a tree might be: if $a>2$ then $\{$if $b<4$ then S$\}$ else $\{$if $b>5$ then S$\}$
else B. One can, in principle, find the single best tree for discriminating S from B within a given Monte Carlo sample of training events,
and given certain constraints on how many cuts we are allowed to apply.
Once this tree is found, one
can then attempt to train another tree to classify correctly the events that the first tree misclassified ({\it i.e.} background
events that end up in regions designated as ``signal-like'' by the tree, and vice-versa).  Then one can
train a third tree to attempt to correct misclassifications of the second tree, and so on.  Typically, this is repeated until a ``forest''
of $O(1000)$ trees has been constructed.   Each successive
tree is trained on the same Monte Carlo sample, but at each stage a {\bf boost} is applied:  misclassified events from the
previous tree are increased in weight, so that the next tree will work harder to better classify them.  After building
up the forest, a weighted vote is taken between all of the trees to form the final discriminant at each point in the multivariate
phase space.
By varying a cut on the weighted vote, {\it i.e.} asking that at least $x\%$ of the trees classify an event as signal for it to pass,
the BDT provides a nearly continuous efficiency measure parametrized by $x$. Thus, varying $x$ can generate the ROC and SIC curves
for our analysis.

For a single variable, a single tree is sufficient. If the variable is monotonic in signal and background,
a one-sided cut can be proven optimal. If the distributions rise and then fall, a two-sided cut is optimal. For more than
one variable, the optimal solution can be calculated exactly, using a Bayesian likelihood. However, to use the Bayesian approach
one needs to know the distributions essentially analytically. This is impossible when many variables are involved,
and one can only sample phase space. In this case, the
best one can do is to know the differential cross sections for events in the vicinity
of the candidate event's kinematics.
For such large under-sampled phase space, the Bayesian likelihood approach is inefficient and multivariate techniques like BDTs
are needed.

\begin{figure}[t]
\begin{center}
  \psfrag{Signal}{\tiny{\!Signal}}
  \psfrag{Background}{\tiny{\!Background}}
  \psfrag{b_highPt__l_highPt_Angle_ak05_b_lowPt__l_highPt_dEta_ak05}{\tiny{\qquad Likelihood}}
  \psfrag{b_highPt__l_highPt_Angle_ak05}{\tiny{$\Delta \theta_{b1,\ell 1}$}}
  \psfrag{b_lowPt__l_highPt_dEta_ak05}{\tiny{$\Delta \eta_{b2,\ell 1}$}}
  \includegraphics[width=0.32\textwidth]{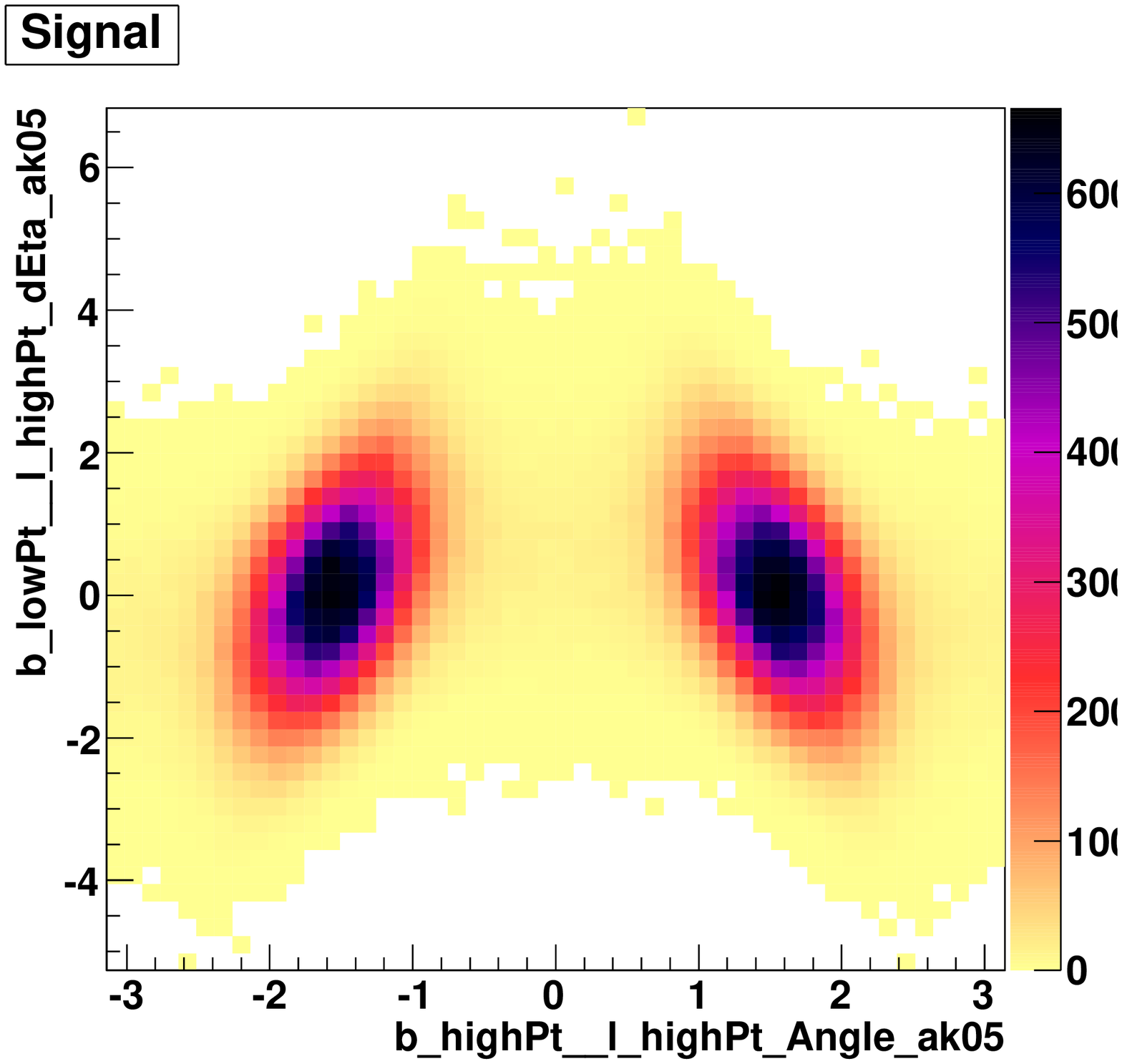}
  \includegraphics[width=0.32\textwidth]{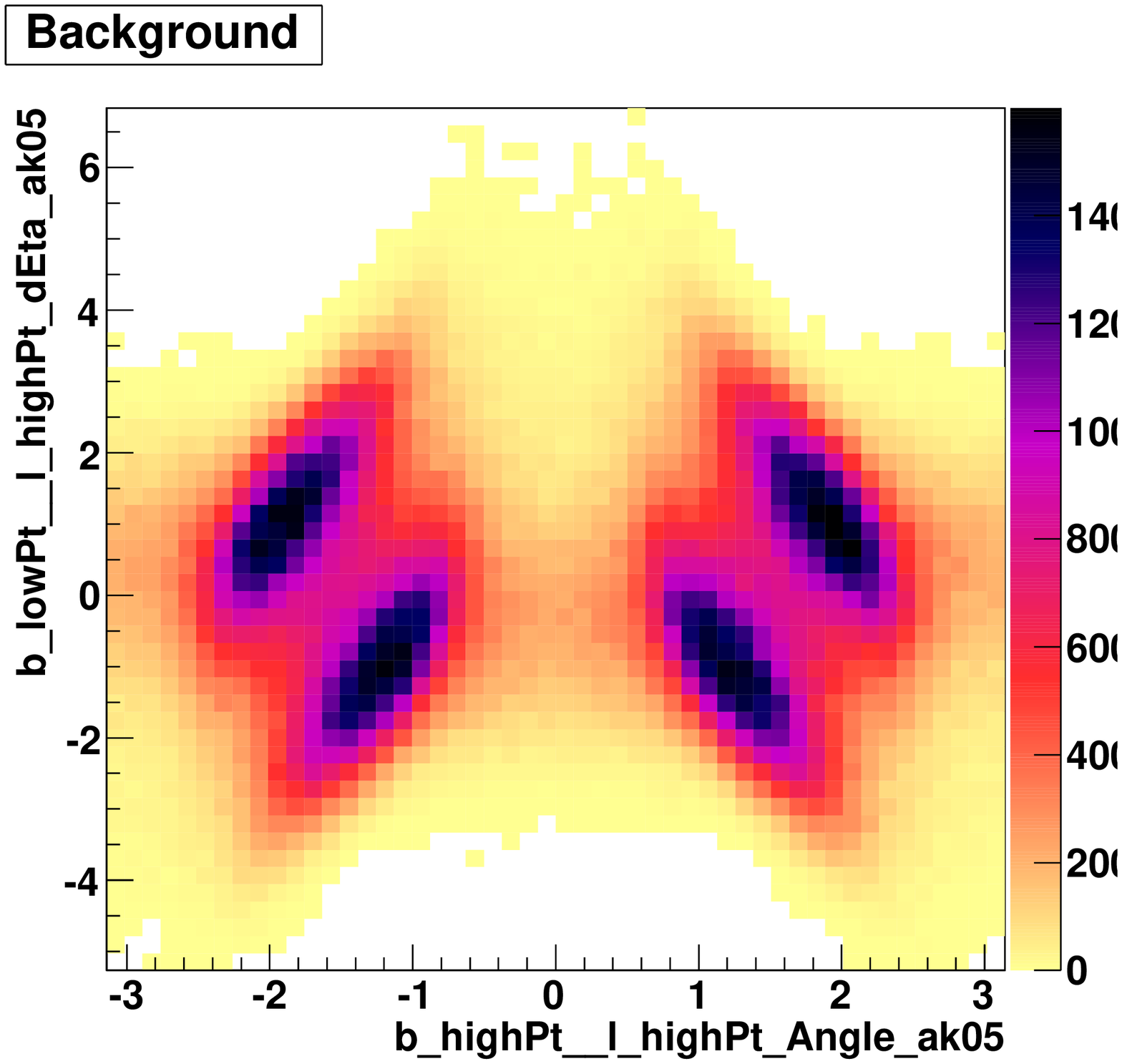}
  \includegraphics[width=0.32\textwidth]{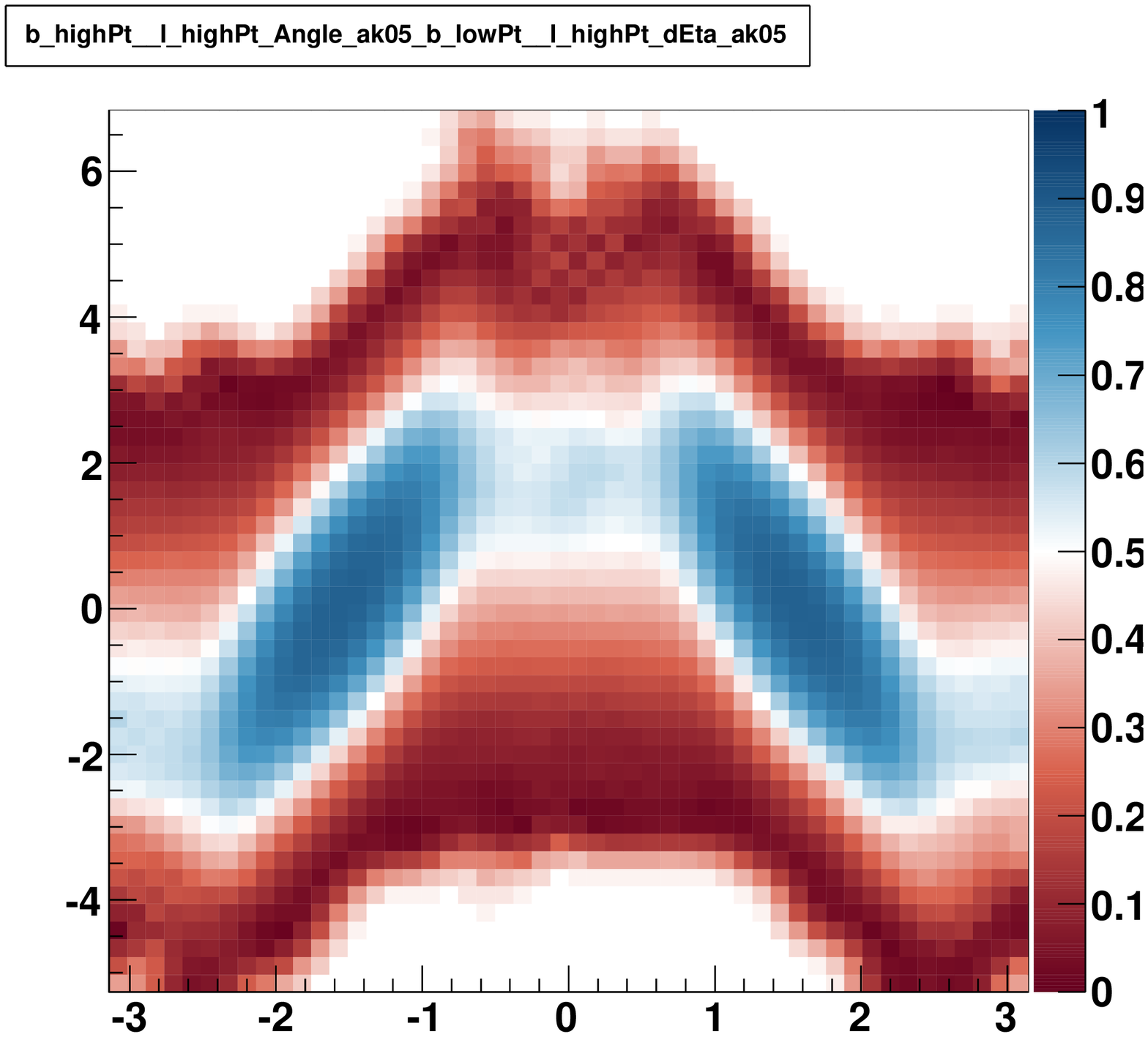}
\caption{
   2D histograms for a particular pair of variables (event counts in each of $50 \times 50$ bins).
   After the histograms are normalized to equal area,
   the counts for each corresponding bin of the signal and background histograms can be combined
   into a likelihood estimate where $L = s/(s+b)$.  (For a different relative normalization, the \emph{shapes} of
   constant probability contours would be identical, but the \emph{values} would change in a monotonic way.
   However, the contour for the probability that yields a particular \emph{signal efficiency} does not change,
   nor does the ROC curve.)
}
\label{fig:exact_likelihood}
\end{center}
\end{figure}

\begin{figure}[ht]
\begin{center}
  \psfrag{b_highPt__l_highPt_Angle_ak05 b_lowPt__l_highPt_dEta_ak05 BDT2}{\tiny{BDT 2}}
  \psfrag{b_highPt__l_highPt_Angle_ak05 b_lowPt__l_highPt_dEta_ak05 BDT8}{\tiny{BDT 8}}
  \psfrag{b_highPt__l_highPt_Angle_ak05 b_lowPt__l_highPt_dEta_ak05 BDT256}{\tiny{BDT 256}}
  \includegraphics[width=0.32\textwidth]{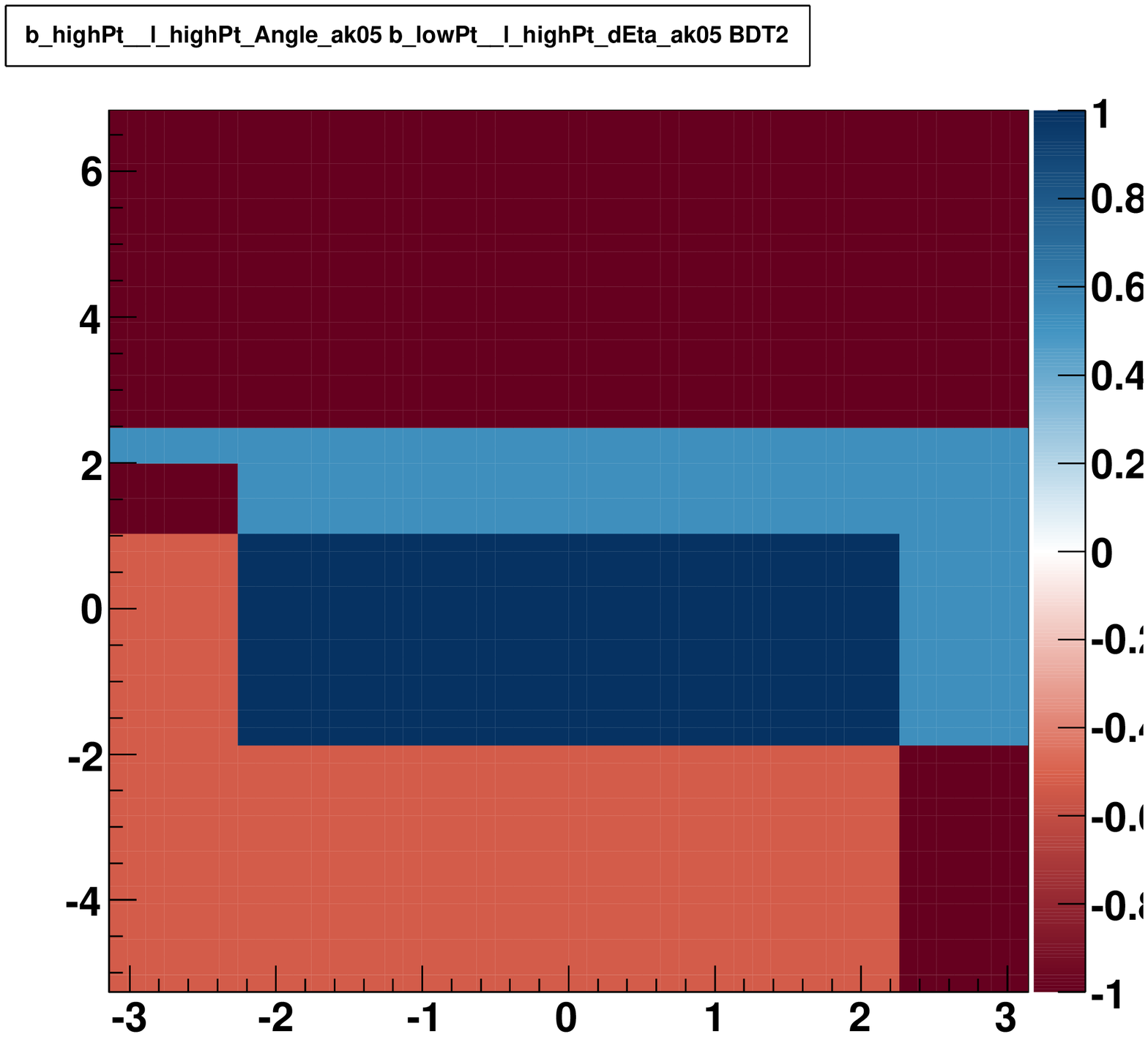}
  \includegraphics[width=0.32\textwidth]{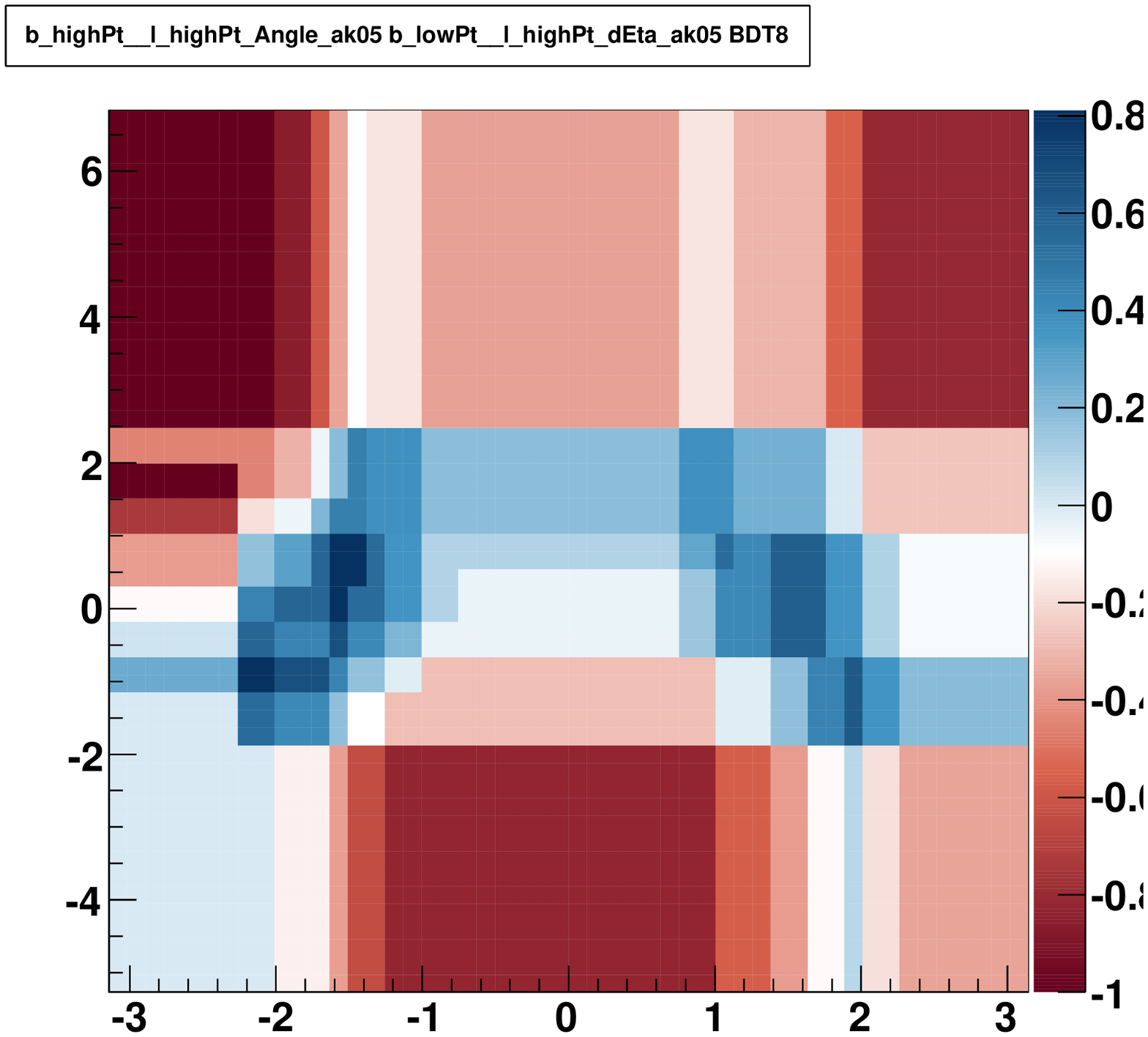}
  \includegraphics[width=0.32\textwidth]{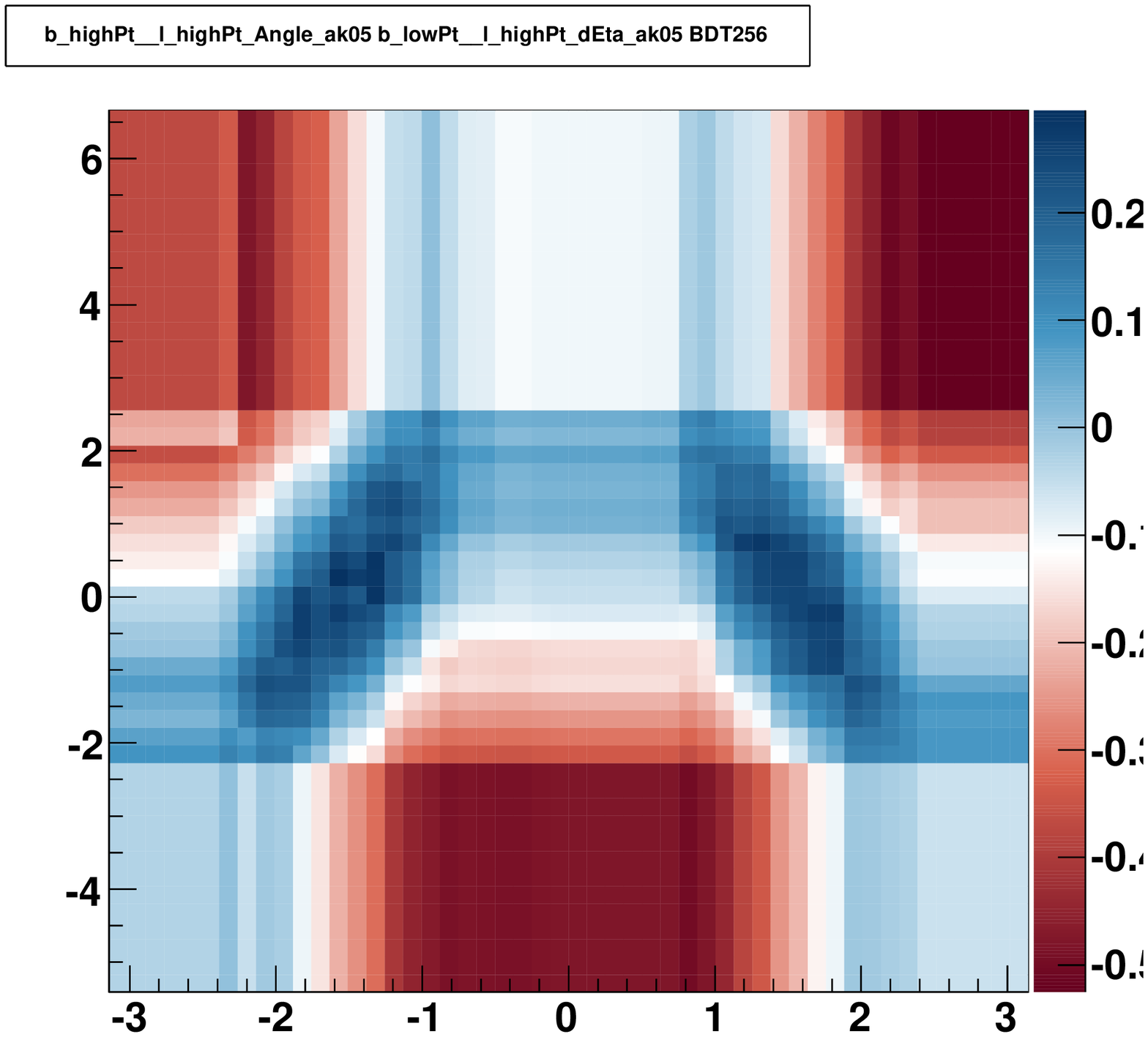}
\caption{
Boosted Decision Tree (BDT) approximations of the above ``exact'' likelihood estimate for 2, 8, and 256 trees.
For regions of no signal and no background events, we defined the probability as 50\% whereas the BDT's
rectangles tend to take the score of the nearest event.
Note that if appropriate absolute values were taken, this particular pair of variables would be linearly correlated and
a suitable linear transformation could be found to decorrelate them and make the job of the BDT's rectangular cuts easier.
Taking absolute values of all symmetric distributions might hide important correlations with additional variables.
}
\label{fig:bdts}
\end{center}
\end{figure}

As an example, consider the signal and background distributions for the two variables $\Delta \theta_{b1,\ell 1}$ and
$\Delta \eta_{b2,\ell 1}$, as shown in Figure~\ref{fig:exact_likelihood}. These two variables are chosen simply because
they have unusual correlations. With our large event samples, we can produce an essentially
smooth 2-dimensional distribution in these variables. Each axis in these figures is sampled into
50 divisions, leading to 2500 bins. From these distributions, we compute the ``exact'' 2D probability density,
as shown in the third panel of this figure. For two variables, the phase space is sampled finely enough that this full likelihood
discriminant is computable. Next, we ask how well a BDT classifier can reproduce this likelihood distribution.
In Figure~\ref{fig:bdts}, we present the result of a BDT using 2, 8 and 256 trees trained on the same 2-dimensional data set.
Even with their rectangular cuts, we see that the BDTs do a good job characterizing the
correlations of the the full 2D probability density.  Certainly 8 and even 256 trees require many
fewer events to train than are needed to sufficiently populate the 2500 bins of the sampled likelihood.
For higher dimensional phase space, we find that
a reasonable number of trees continues to sample the space well,
while a uniform sampling required for a likelihood approach is impossible. We use up to 3000 trees for 10 variables,
although our results barely change beyond 400 trees.

Even within the BDT method, there are many different ways to construct classifiers. For example,
one can train successive trees on misclassified trees from previous runs, as described above. Alternatively, one could train
trees on a random subset of events (the Random Forest approach). For each tree, one can limit the number of branches or the depth
of the tree. And of course the way the subset of relevant input variables for each tree are chosen is critically important as well.
We have not attempted to systematically optimize the BDT method in this paper.

\section{Combining variables\label{sec:multi}}
\begin{figure}[t!]
\begin{center}
\psfrag{LHC HZ Significance}{\qquad \qquad LHC ZH \qquad \qquad  $S/\sqrt{B}$ Improvement}
\psfrag{LHC HZ Significance : }{}
\psfrag{Signal eff}{\!\!\!\!\!\!\!\!\!\!\!\!\!\!\!\!\!\!\!\!\!\!\!\!\!\!\!\!\!\!\!\!\!\!\!\!\!\!\!\!\!\!\!\!\!\!\!Higgs Signal Efficiency $\varepsilon_S$}
\psfrag{bbbar_dEta_ak05}{\tiny{$\Delta \eta_{b \bar b}$}}
\psfrag{bbbar_sumPt_ak05}{\tiny{$\Sigma p_T^{b \bar b}$}}
\psfrag{H_and_b_lowPt_dRap_ak05}{\tiny{$\Delta y_{H,b2}$}}
\psfrag{H_and_b_highPt_dRap_ak05}{\tiny{$\Delta y_{H,b1}$}}
\psfrag{components_b_highPt_ak05_sub_ak020_stddev_pT}{\tiny{subjet $p_T$}}
\psfrag{b_highPt_Pt_ak05}{\tiny{$p_T^{b1}$}}
\psfrag{SumPt_Final4_ak05}{\tiny{$H_T$}}
\psfrag{Ht_Final4_CM_ak05}{\tiny{$H_T$}}
\psfrag{Centrality_Final4_CM_ak05}{\tiny{Centrality}}
\psfrag{H_Frame_b_highPt_costheta_ak05}{\tiny{$\cos(\theta^*_{b1})$}}
\psfrag{ZZH_Frame_b_highPt_costheta_ak05}{\tiny{CM Frame $\cos(\theta^*_{b1})$}}
\psfrag{b_highPt__l_lowPt_sumPt_ak05}{\tiny{$p_T^{b1} + p_T^{\ell 2}$}}
\psfrag{b_lowPt_Pt_ak05}{\tiny{$p_T^{b2}$}}
\psfrag{b_lowPt__l_lowPt_sumPt_ak05}{\tiny{$p_T^{b2} + p_T^{\ell 2}$}}
\includegraphics[width=0.8\textwidth]{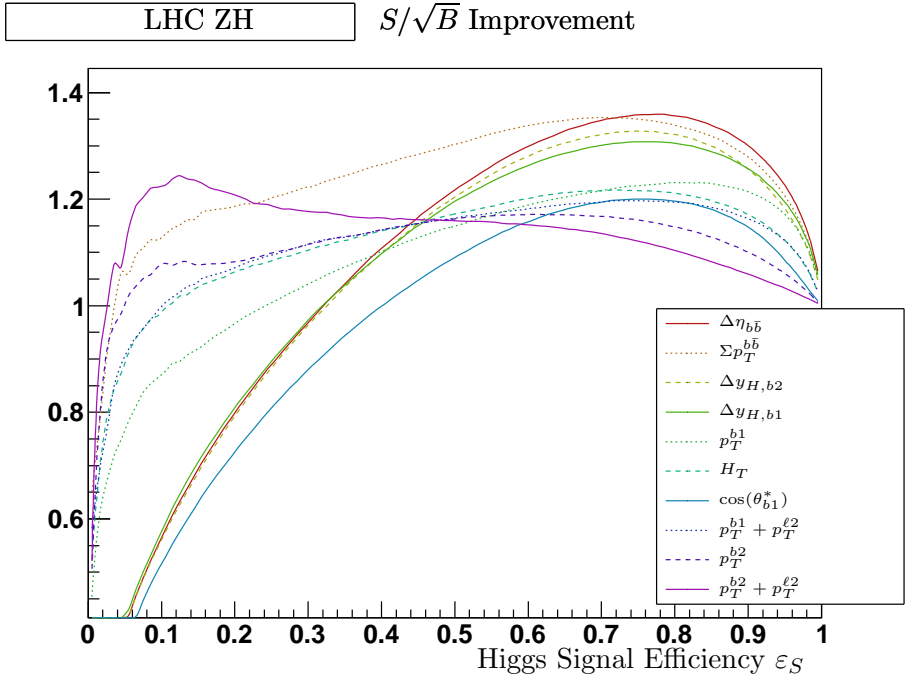}
\caption{SIC curves for the top 10 {\bf individual} variables.}
\label{fig:individuals}
\end{center}
\end{figure}

Having discussed the variables and the boosted decision tree approach to classification, we can now
attempt to construct strong discriminants.
We start by reducing our thousands of variables to approximately 800 reasonably different ones. We then calculate
the significance improvement, and sort the variables
 by their $\ropt= \max(\e_S/\sqrt{\e_B})$ values.
 The top 10 single variables ranked this way are
\begin{equation}
\Delta \eta_{b \bar b},\quad
\Sigma p_T^{b \bar b},\quad
\Delta y_{H,b2},\quad
\Delta y_{H,b1},\quad
p_T^{b1},\quad
H_T,\quad
\cos(\theta^*_{b1}),\quad
|\vp_T^{b1}| + |\vp_T^{\ell 2}|,\quad
|\vp_T^{b2}|,\quad
|\vp_T^{b2}| + |\vp_T^{\ell 2}|
\end{equation}
The corresponding significance improvement curves are shown in Figure~\ref{fig:individuals}.
This is not the most varied possible set. In particular, none of the radiation variables made the list. More unusual variables
will resurface when many variables are combined.

\begin{figure}[t]
\begin{center}
\psfrag{v000000000}{\tiny{$\Delta \eta_{b \bar b}$\ }}     
\psfrag{v000000001}{\tiny{$\Sigma p_T^{b \bar b}$\ }}     
\psfrag{v000000002}{\tiny{$\Delta y_{H,b2}$\ }}     
\psfrag{v000000003}{\tiny{$\Delta y_{H,b1}$\ }}     
\psfrag{v000000004}{\tiny{$p_T^{b1}$\ }}     
\psfrag{v000000005}{\tiny{$H_T$\ }}     
\psfrag{v000000006}{\tiny{$\cos(\theta^*_{b1})$\ }}     
\psfrag{v000000007}{\tiny{\!\!\!\!$p_T^{b1} + p_T^{\ell 2}$\ }}     
\psfrag{v000000008}{\tiny{$p_T^{b2}$\ }}     
\psfrag{v000000009}{\tiny{\!\!\!\!$p_T^{b2} + p_T^{\ell 2}$\ }}     
\caption{Linear correlation coefficients for top 10 {\bf individual} variables for both signal (left) and background (right).
\vspace{1cm}}
\label{fig:top10correlations}
\end{center}
\end{figure}

\begin{figure}[t!]
\begin{center}
\psfrag{LHC HZ Significance}{\qquad \qquad LHC ZH \qquad \qquad  $S/\sqrt{B}$ Improvement}
\psfrag{LHC HZ Significance : }{}
\psfrag{Signal eff}{\!\!\!\!\!\!\!\!\!\!\!\!\!\!\!\!\!\!\!\!\!\!\!\!\!\!\!\!\!\!\!\!\!\!\!\!\!\!\!\!\!\!\!\!\!\!\!Higgs Signal Efficiency $\varepsilon_S$}
\psfrag{bbbar_sumPt_ak05, H_Pt_ak05}{\tiny{$\Sigma p_T^b$, $p_T^H$}}
\psfrag{bbbar_sumPt_ak05, H_and_b_highPt_sumPt_ak05}{\tiny{$\Sigma p_T^b$, $p_T^H \! + \! p_T^{b1}$}}
\psfrag{bbbar_sumPt_ak05, H_and_b_lowPt_dPt_ak05}{\tiny{$\Sigma p_T^b$, $p_T^H \! - \! p_T^{b2}$}}
\psfrag{components_b_highPt_ak05_ak05_sub_kt020_stddev_pT, subjet_topPt_b_lowPt_ak05_ak05_sub_ak020}{\tiny{$\sigma_{p_T}$, $sub$}}
\psfrag{bbbar_dEta_ak05, bbbar_transverse_mass_ak05}{\tiny{$\Delta \eta_{b \bar b}$, $m_T^{b \bar b}$}}
\psfrag{bbbar_sumPt_ak05, H_and_l_highPt_inv_mass_ak05}{\tiny{$\Sigma p_T^b$, $m_{H,\ell 1}$}}
\psfrag{subjet_topPt_b_highPt_ak05_ak05_sub_ak020, subjet_topPt_b_lowPt_ak05_ak05_sub_ak020}{\tiny{Subjets}}
\psfrag{bbbar_sumPt_ak05, H_and_b_lowPt_dRap_ak05}{\tiny{$\Sigma p_T^b$, $\Delta y_{H,b2}$}}
\psfrag{bbbar_dEta_ak05, bbbar_sumPt_ak05}{\tiny{$\Sigma p_T^b$, $\Delta \eta_{b \bar b}$}}
\psfrag{bbbar_dEta_ak05, H_and_Z_dEta_ak05}{\tiny{$\Delta \eta_{b \bar b}$, $\Delta \eta_{H,\!Z}$}}
\psfrag{bbbar_sumPt_ak05, H_and_b_lowPt_dRap_ak05}{\tiny{$\Sigma p_T^{b \bar b}$, $\Delta y_{H,b2}$}}
\psfrag{bbbar_sumPt_ak05, H_and_b_lowPt_dPt_ak05}{\tiny{$\Sigma p_T^{b \bar b}$, $\Delta p_T^{H,b2}$}}
\psfrag{bbbar_sumPt_ak05, ZZH_Frame_l_highPt_Momentum_ak05}{\tiny{$\Sigma p_T^{b \bar b}$, CM $|\vec p|_{\ell 1}$}}
\psfrag{bbbar_dEta_ak05, bbbar_sumPt_ak05}{\tiny{$\Sigma p_T^{b \bar b}$, $\Delta \eta_{b \bar b}$}}
\psfrag{bbbar_transverse_mass_ak05, H_and_b_lowPt_dRap_ak05}{\tiny{$m_T^{b \bar b}$, $\Delta y_{H,b2}$}}
\psfrag{bbbar_sumPt_ak05, H_and_b_highPt_sumPt_ak05}{\tiny{$\Sigma p_T^{b \bar b}$, $\Sigma p_T^{H,b1}$}}
\psfrag{bbbar_dEta_ak05, bbbar_transverse_mass_ak05}{\tiny{$\Delta \eta_{b \bar b}$, $m_T^{b \bar b}$}}
\psfrag{bbbar_sumPt_ak05, H_and_l_lowPt_dPt_ak05}{\tiny{$\Sigma p_T^{b \bar b}$, $\Delta p_T^{H,\ell 2}$}}
\psfrag{bbbar_sumPt_ak05, ZZH_Frame_b_lowPt_costheta_ak05}{\tiny{$\Sigma p_T^{b \bar b}$, CM $\cos(\theta_{b2})$}}
\psfrag{bbbar_dEta_ak05, ZH_dEta_ak05}{\tiny{$\Delta \eta_{b \bar b}$, $\Delta \eta_{Z,H}$}}
\psfrag{components_xxx}{\tiny{subjet $p_T$'s}}
\psfrag{b_highPt_Pt_ak05, components_b_lowPt_ak05_sub_ak030_stddev_pT}{\tiny{$p_T^{b1}$, subjet avg $p_T$}}
\includegraphics[width=0.8\textwidth]{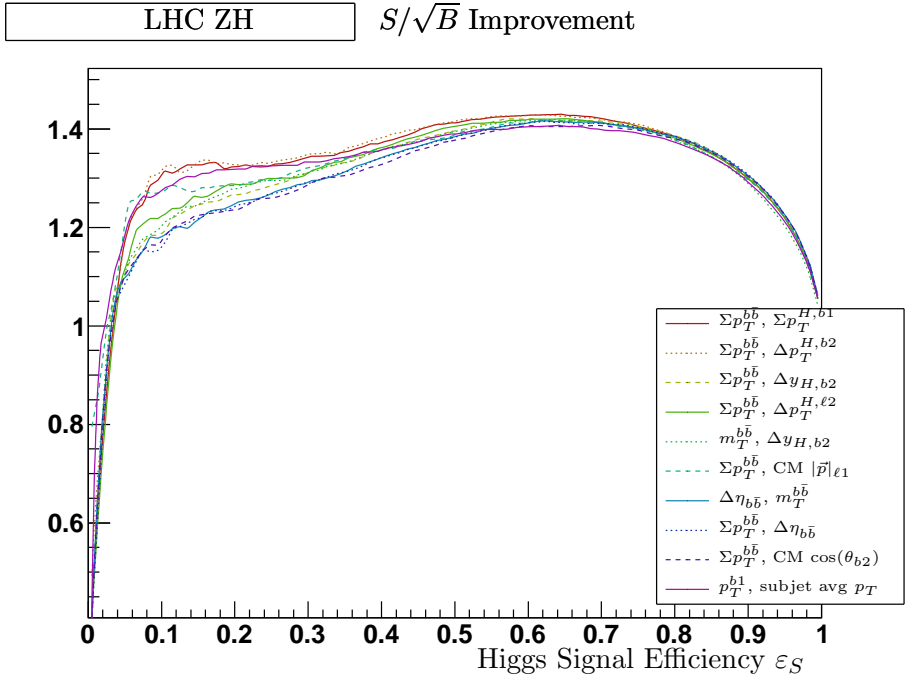}
\caption{SIC curves for the top 10 \textbf{pairs} of variables using Boosted Decision Trees}
\label{fig:pairs}
\end{center}
\end{figure}

\begin{figure}[h]
\begin{center}
\psfrag{TwoVarImp}{\tiny{Two variable improvement percentage}}
\psfrag{Improvement percentage over the better of the two variables}{\tiny{Improvement over the better variable}}
\psfrag{v000000000}{\tiny{$\Delta \eta_{b \bar b}$\ }}     
\psfrag{v000000001}{\tiny{$\Sigma p_T^{b \bar b}$\ }}     
\psfrag{v000000002}{\tiny{$\Delta y_{H,b2}$\ }}     
\psfrag{v000000003}{\tiny{$\Delta y_{H,b1}$\ }}     
\psfrag{v000000004}{\tiny{$p_T^{b1}$\ }}     
\psfrag{v000000005}{\tiny{$H_T$\ }}     
\psfrag{v000000006}{\tiny{$\cos(\theta^*_{b1})$\ }}     
\psfrag{v000000007}{\tiny{\!\!\!\!$p_T^{b1} + p_T^{\ell 2}$\ }}     
\psfrag{v000000008}{\tiny{$p_T^{b2}$\ }}     
\psfrag{v000000009}{\tiny{\!\!\!\!$p_T^{b2} + p_T^{\ell 2}$\ }}     
\includegraphics[width=0.45\textwidth]{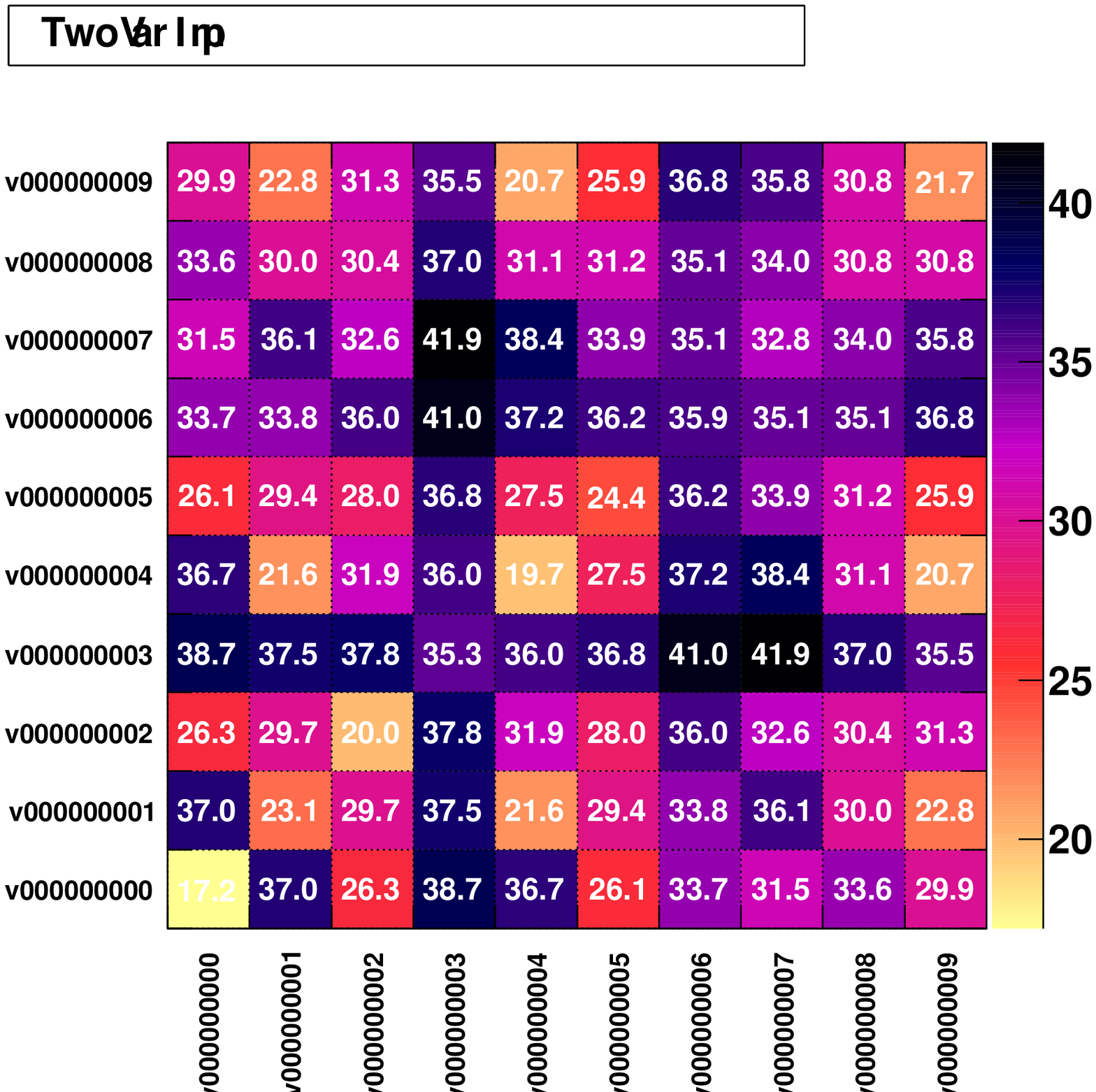}
\includegraphics[width=0.45\textwidth]{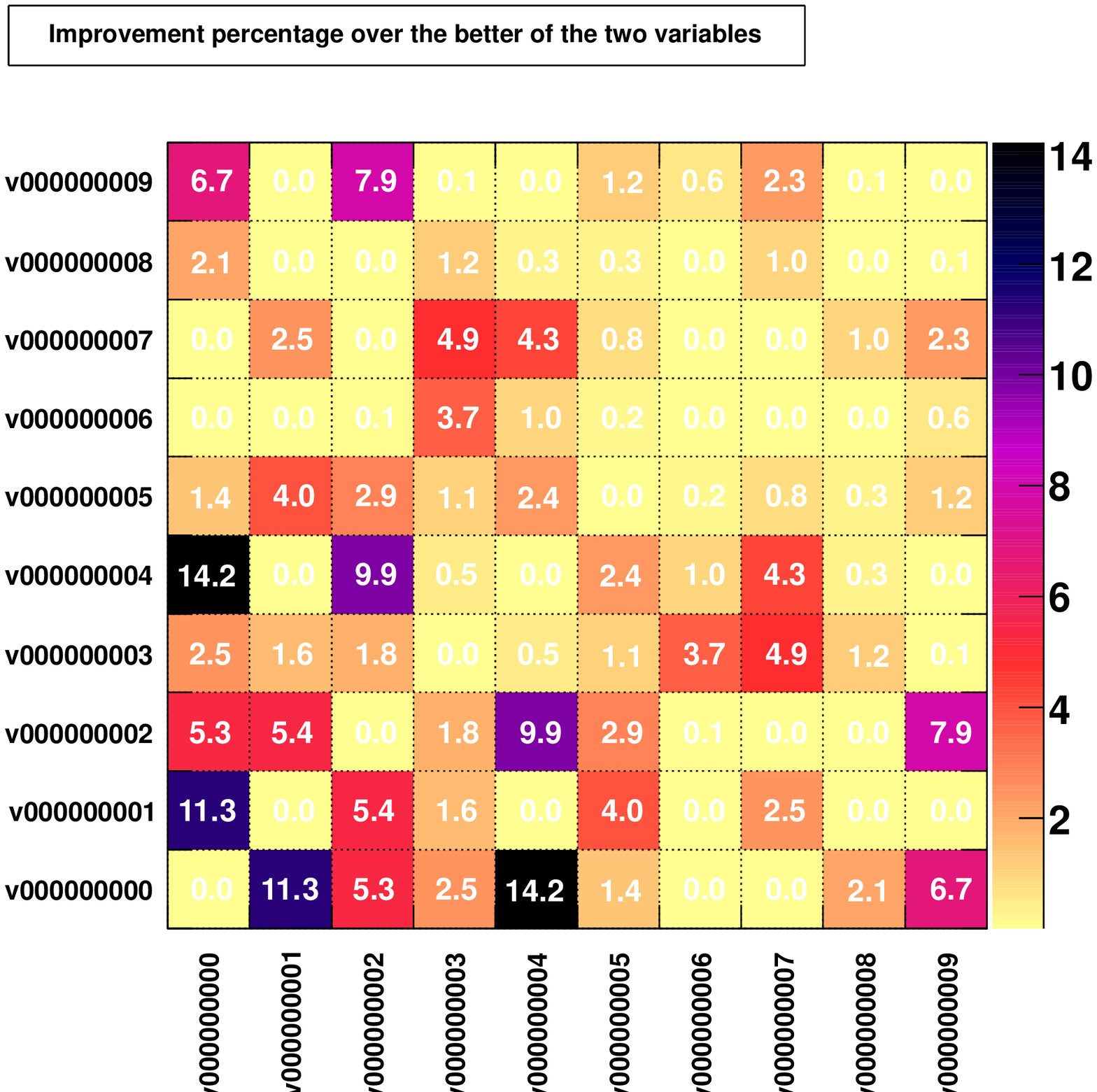}
\caption{Two variable improvement percentage of the top 10 {\bf individual} variables.
The left shows the values of $\ropt$ for the pairs (as a percentage),
with the single variables' scores on the diagonal.
The right shows how much the combination improves over the better of the pair:
$(R_{ij}-\max(R_i,R_j)) \cdot 100\%$.}
\label{fig:pair_improvement}
\end{center}
\end{figure}

\begin{figure}[t]
\begin{center}
\psfrag{v000000000}{\tiny{\!\!\!\!\!\!$|\vp_T^{b1}\!|\!+\!|\vp_T^{b2}|$\ }}     
\psfrag{v000000001}{\tiny{\!\!\!\!\!\!$|\vp_T^{H}\!|\!-\!|\vp_T^{b2}|$\ }}     
\psfrag{v000000002}{\tiny{\!\!\!\!\!\!$|\vp_T^{Z}\!|\!-\!|\vp_T^{b2}|$\ }}     
\psfrag{v000000003}{\tiny{$m_{H,b1}$}}     
\psfrag{v000000004}{\tiny{$E^{\text{obj.}}_\text{vis}$\ }}     
\psfrag{v000000005}{\tiny{$E^{\text{prim.}}_\text{vis}$\ }}     
\psfrag{v000000006}{\tiny{pull $\alpha$\ }}     
\psfrag{v000000007}{\tiny{pull $\beta$\ }}     
\psfrag{v000000008}{\tiny{$A_{b1}^{-0.9}$\ }}     
\psfrag{v000000009}{\tiny{girth b2 }}     
\includegraphics[width=0.48\textwidth]{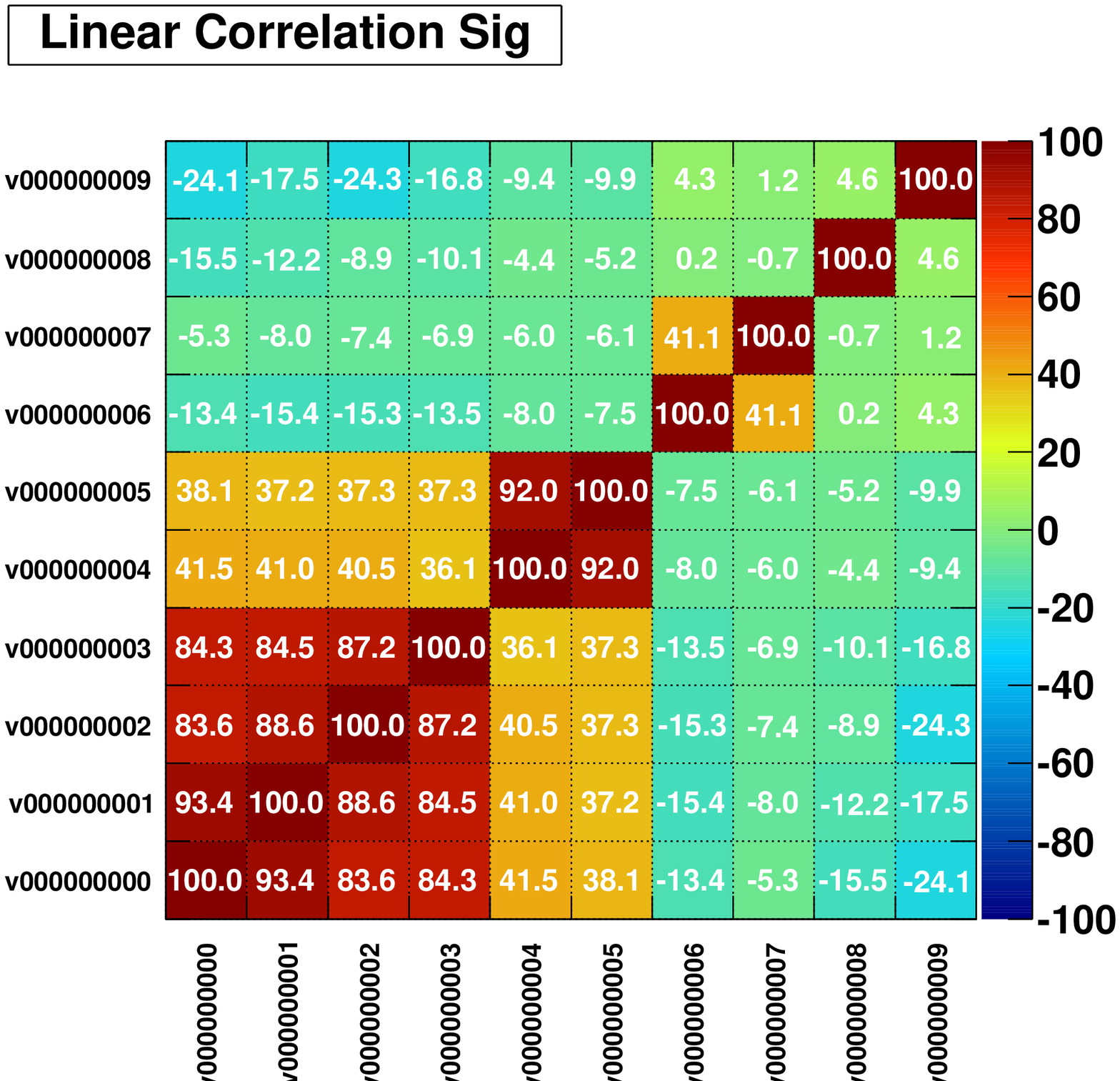}
\includegraphics[width=0.48\textwidth]{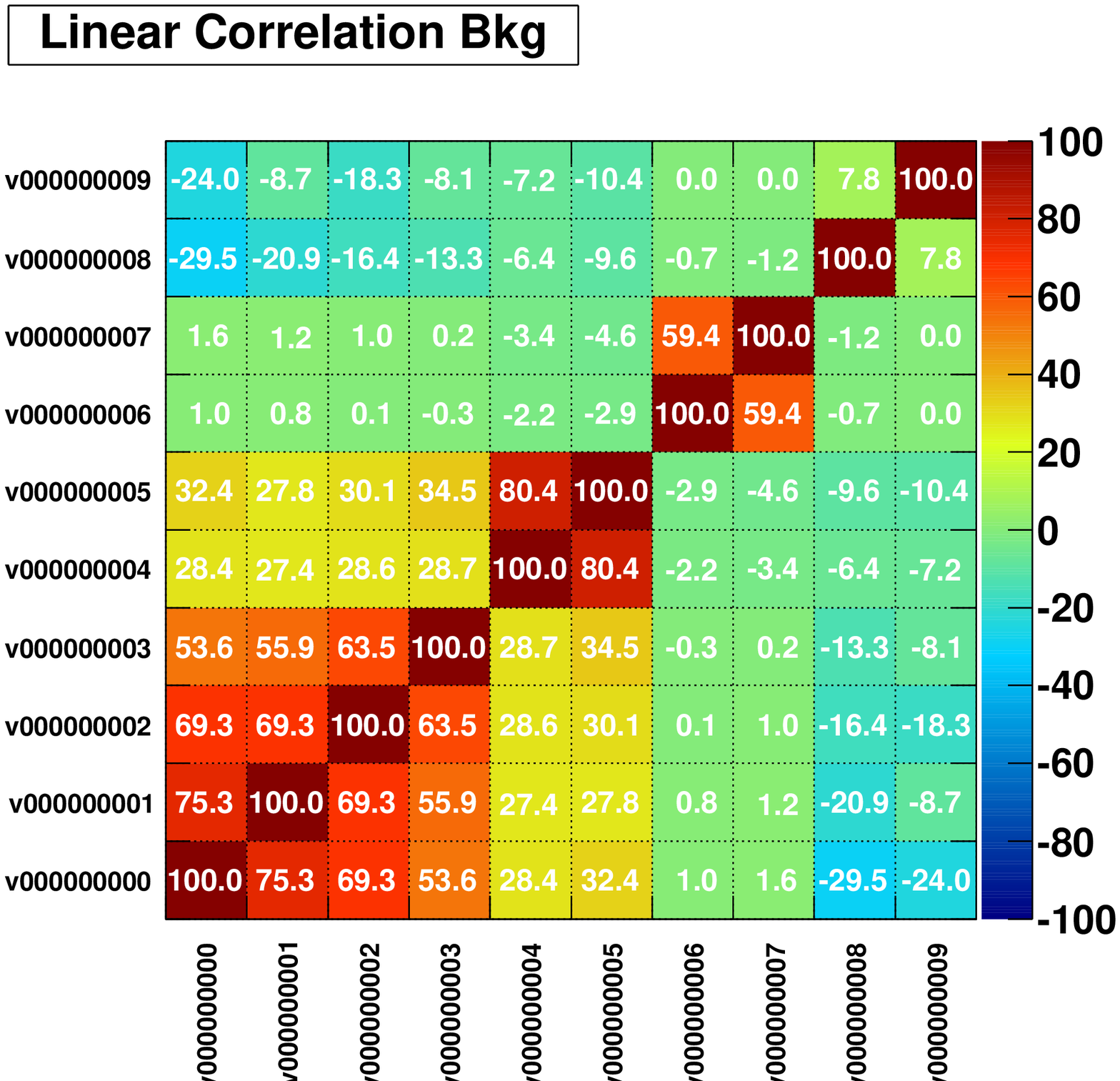}
\caption{Linear correlation coefficients of signal and background (LHC $\zh$ sample) for 10
 variables which are best when combined, as determined
by the algorithm defined in the text.}
\label{fig:correlations}
\end{center}
\end{figure}

To start combining variables, we first look at the linear correlations among some the top variables.
These are shown for signal and background in Figure~\ref{fig:top10correlations}.
These numbers should be interpreted cautiously. Sometimes
variables are highly correlated non-linearly, but may have low
linear correlation coefficients. Also, combining uncorrelated
variables often does not help improve the $S/\sqrt{B}$
at all, whereas combining correlated variables often
does. Nevertheless, we find the linear correlations a useful way to see which variables are measuring similar things.
For example, this matrix shows that since $\Delta y_{H,b2}$ and $\Delta \eta_{b \bar b}$
are 96\% correlated, they have almost exactly the same information, as expected. We will also see that the final set of
variables we choose algorithmically is not nearly as correlated as the top 10 (see Figure~\ref{fig:correlations}).
The $\ropt$ for each pair among these 10 kinematic variables is shown in the matrix in
Figure~\ref{fig:pair_improvement}, along with how much each
variable can improve $\ropt$ over the better of the two.

Next, we consider the top 10 pairs from the complete set of 800 variables. To get these, we would ideally just take every possible pairing and evaluate
the combined $\rsig$. This is marginally possible with pairs, but too computationally intensive for triplets or larger
combinations. Fortuitously, after trying the all possible pairs,
we observed  that nearly identical results follow from simply taking the top 3
single variables and combining pairwise with a reduced
set of 200 of the original variables. This reduced set was selected to contain not more than 99\% correlated information.
We combine the pairs with the BDT method and extract the value of $\ropt$ for each combination.
The significance improvement curves for top 10 pairs are shown in figure~\ref{fig:pairs}.
Note that most of the pairs involve variables not in the original top 10.

This way of building up combinations can be iterated. The algorithm is
\begin{enumerate}
\item Start with the top 3 sets of $n$ variables.
\item Add each of the original 200 variables to the set and compute the maximal significance improvement.
\item Take the best 3 sets of $n+1$ variables for the next iteration
\end{enumerate}
Using this algorithm, we find the the top set of 10 variables for the LHC $\zh$ sample is
\begin{multline}
|\vp_T^{b1}\!|+|\vp_T^{b2}\!|,\quad
|\vp_T^{H} \!|-|\vp_T^{b2}\!|,\quad
|\vp_T^{Z} \!|-|\vp_T^{b2}\!|,\quad
m_{H,b1},\quad
E^{\text{obj.}}_\text{vis},\quad
E^{\text{prim.}}_\text{vis},\quad \\
\text{pull} \alpha,\quad
\text{pull} \beta,\quad
A_{b1}^{-0.9},\quad
\text{girth}_{b2}
\end{multline}
Here  ``obj.'' and ``prim.'' refer to whether these energy variables were constructed from reconstructed objects
in the entire event or just from the primary objects (two leptons and two $b$-jets), as discussed in Section~\ref{sec:energy}. $A_{b1}^{-0.9}$ is
the jet angularity constructed from the hardest $b$-jet with $a=-0.9$.
This set of variables is much less correlated than the top 10 individual variables.
This is shown in Figure~\ref{fig:correlations}, which can be compared to Figure~\ref{fig:top10correlations}.
Also, the set of top 10 combined variables, in contrast to the top 10 individual variables, includes
a number of the showered and event shape discriminants.

\begin{figure}[t]
\begin{center}
\psfrag{Signal eff}{\!\!\!\!\!\!\!\!\!\!\!\!\!\!\!\!\!\!\!\!\!\!\!\!\!\!\!\!\!\!\!\!\!\!\!\!\!\!\!\!\!\!\!\!\!\!\!\footnotesize{Higgs Signal Efficiency $\varepsilon_S$}}
\psfrag{TVT HZ Significance}{\footnotesize{\qquad TVT ZH \qquad \quad $S/\sqrt{B}$ Improvement}}
\psfrag{TVT HW Significance}{\footnotesize{\qquad TVT WH \qquad \quad $S/\sqrt{B}$ Improvement}}
\psfrag{LHC HZ Significance}{\footnotesize{\qquad LHC ZH \qquad \quad $S/\sqrt{B}$ Improvement}}
\psfrag{LHC HW Significance}{\footnotesize{\qquad LHC WH \qquad \quad $S/\sqrt{B}$ Improvement}}
\includegraphics[width=0.48\textwidth]{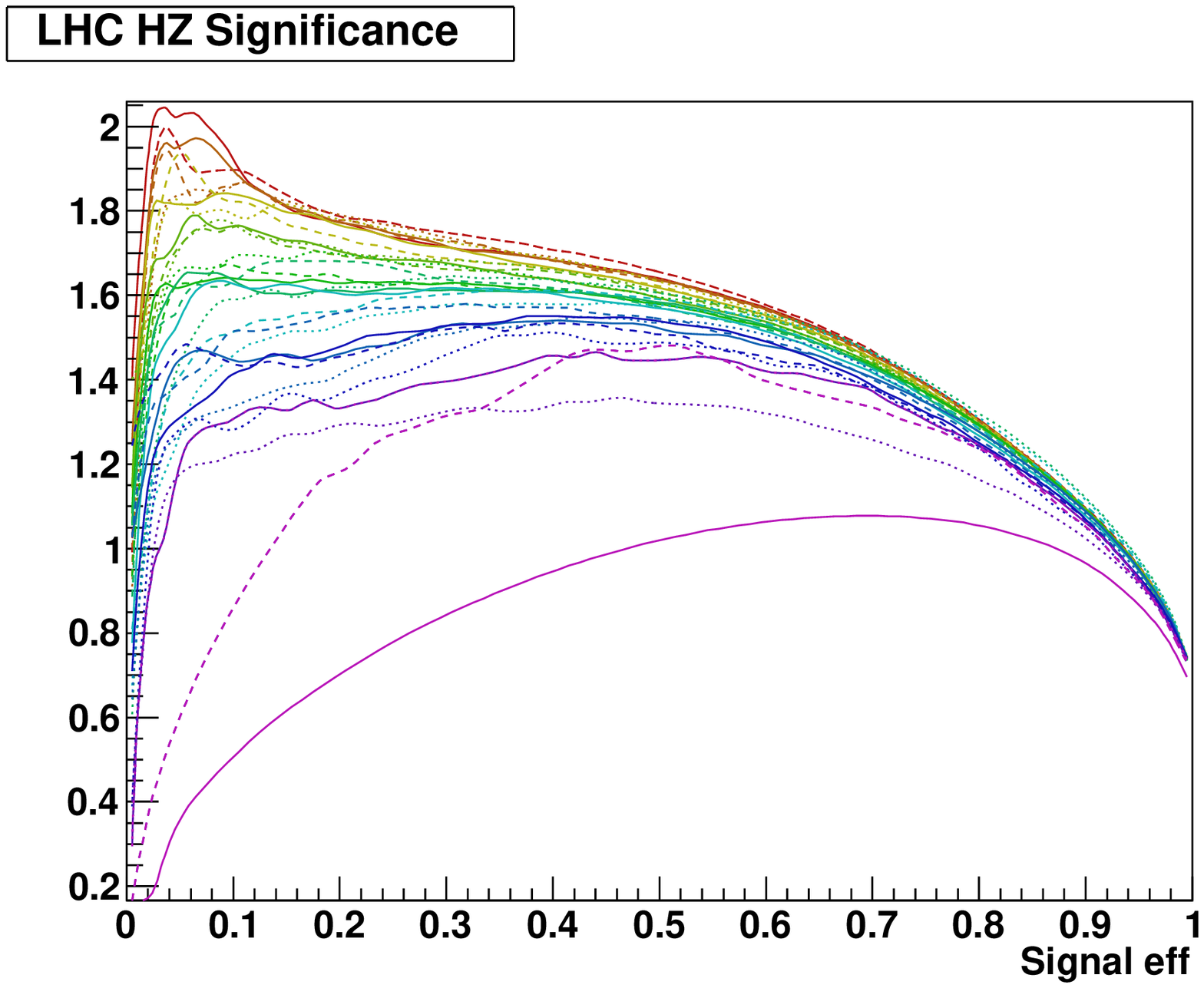}
\includegraphics[width=0.48\textwidth]{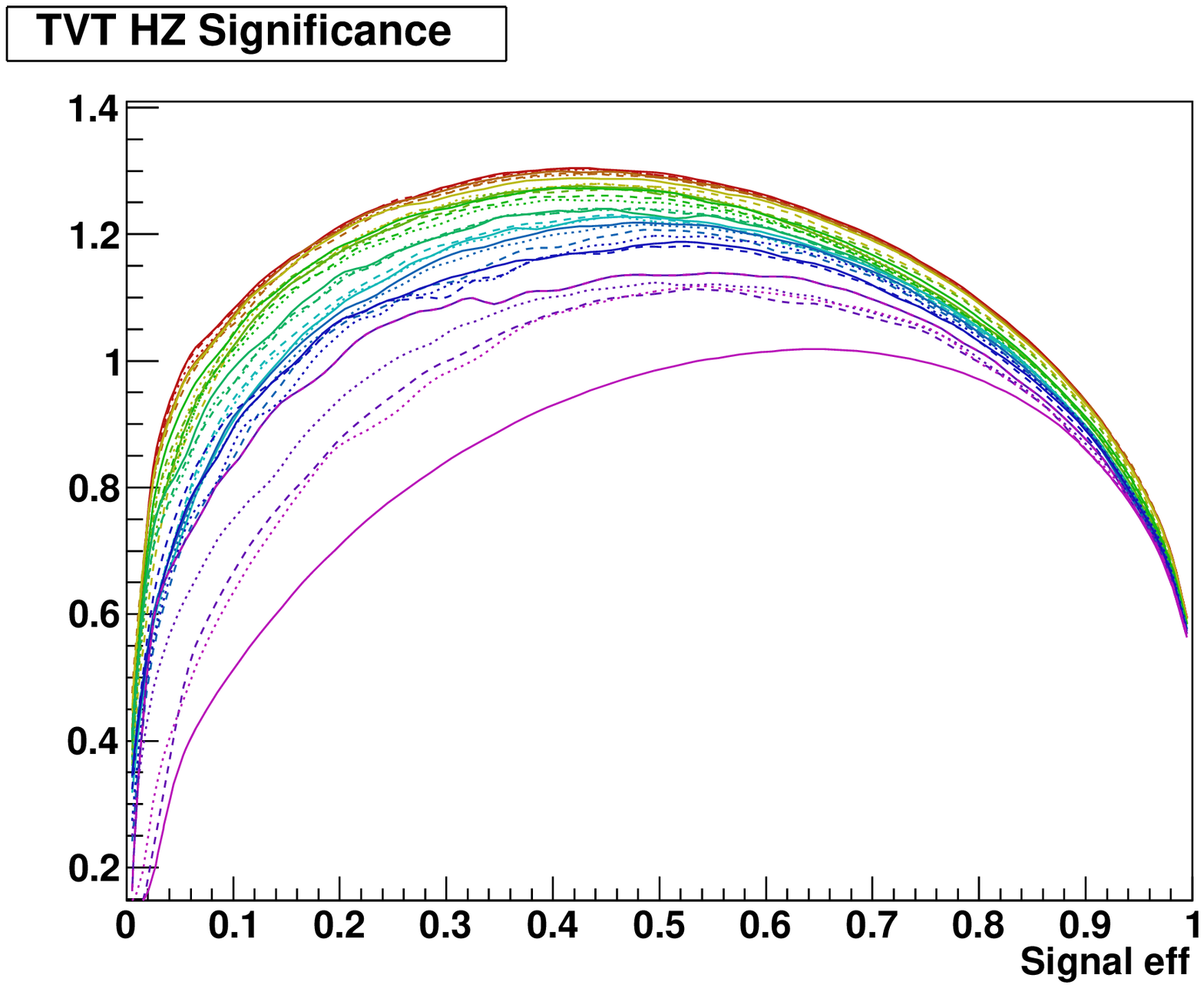}
\includegraphics[width=0.48\textwidth]{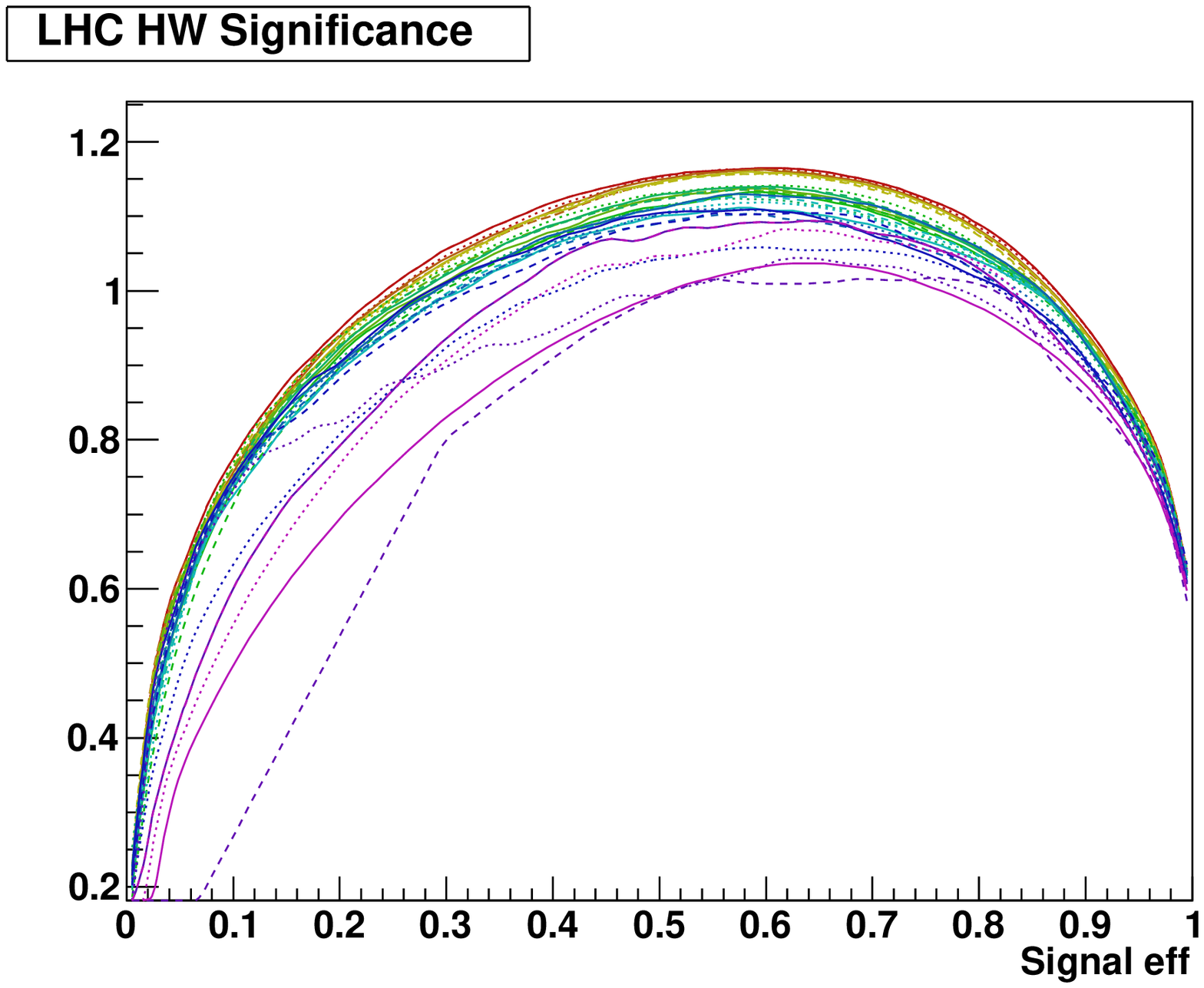}
\includegraphics[width=0.48\textwidth]{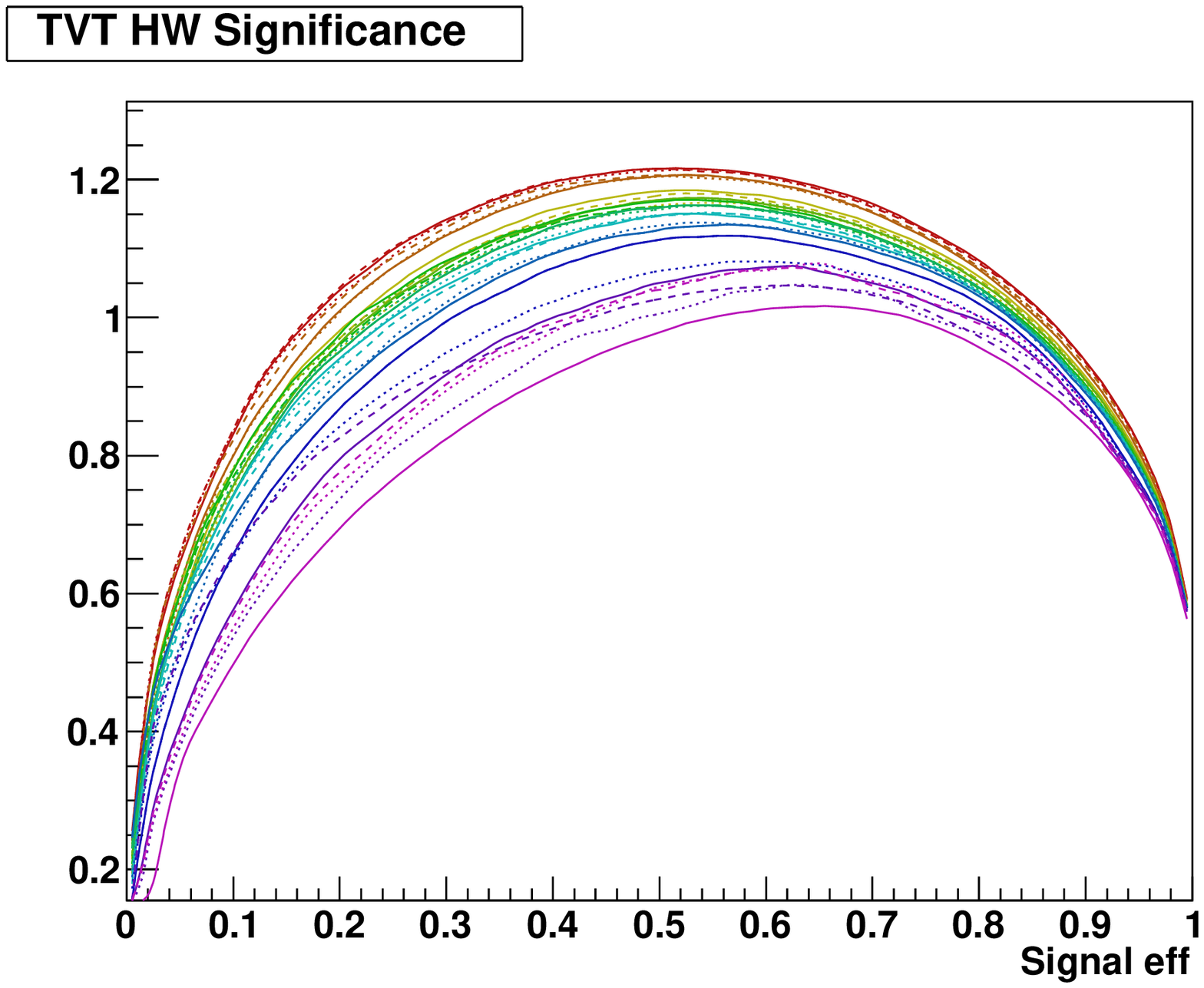}
\caption{Significance improvement characteristics for top combinations of $1 \ldots 11$ variables for $\zh$ and $\wh$
at the LHC and Tevatron.  Each color corresponds to a different number of variables. For a fixed number,
the solid line corresponds to the best set, dashed to the 2nd best,
and dotted to the 3rd best, all of which go on to the next round.
These simples all have \Higgswindow, which, with no additional discriminants,
defines the reference value
$\e_B=\e_S=\text{SIC}=1$.
}
\label{fig:top8}
\end{center}
\end{figure}

We often see convergence towards a final significance when many variables are combined. We show in Figure~\ref{fig:top8} the
significance curves for the top 3 sets of $3,4,5,\cdots,11$ variables. We see convergence at around 8 variables.
The results for the $WH$ sample as compared to its irreducible background, and the Tevatron versions are also shown.
The associated $\rSB$ curves contain equivalent information. But since they all blow up at small $\e_S$,
they are much more difficult to interpret, and therefore not shown.

Note that the significance curves for $\zh$ become poorly estimated at low $\eS$. This is entirely due to lower statistics due to harder cuts.
In fact, these low significances correspond to $\eS\sim 10^{-1}$ and $\eB\sim 10^{-3}$.
It is natural to expect that the multivariate methods will struggle in trying to characterize
an 11 dimensional space to one part in $10^3$.
We find no such instabilities for the $WH$ sample, since the background efficiencies are in general higher for
a given signal efficiency.
This points to one advantage of the differential $\rsig$ visualization technique --
it demonstrates when the statistics of the Monte Carlo sample is becoming a problem.
Indeed, as we were generating the samples, we noted much instability with smaller statistics, which led us
to increase the size of the runs. However, even with 1 million background events, $\eB\sim 10^{-3}$ means
that only a few thousand events are controlling the final efficiency. Efficiencies around $10^{-3}$ seem
to be a practical limitation on this method. To go further, one should put more judicious cuts on the initial sample
so that the tails of the distributions will be more accurately populated.


\section{Optimizing Jet Reconstruction \label{sec:jetting}}
Next, we return to the issue of optimizing the jet size and the $\mbb$ discriminant. For the previous sections, we have fixed the
jet size to anti-$k_T$ with $R\!=\!0.5$ and imposed a fixed window of \Higgswindow. This section justifies that choice,
and describes some more sophisticated options for how to treat $\mbb$.
Ideally, one would like to see a peak in the $\mbb$ distribution by eye to claim a really
satisfying Higgs boson discovery. From a mathematical point of view, statistical significance is much more important than the
aesthetic shape of a curve. Thus, a more powerful approach is, rather than imposing a fixed $\mbb$ window, to include $\mbb$
as a discriminant in the multivariate analysis. This will draw out correlations between $\mbb$ and the other variables. Moreover,
as we will see, including even multiple $\mbb$ measures does much better.

\begin{figure}[t]
\begin{center}
\psfrag{bbbar_inv_mass}{$\mbb$}
\psfrag{bbbar_inv_mass_ak05_signif_in}{\tiny{SIC for different $\mbb$ windows}}
\psfrag{bbbar_inv_mass         r}{\tiny{$\mbb$ for anti-$k_T$ $R\!=\!0.5$ jets}}
\psfrag{ = +2.57\%}{}
  \includegraphics[width=0.45\textwidth]{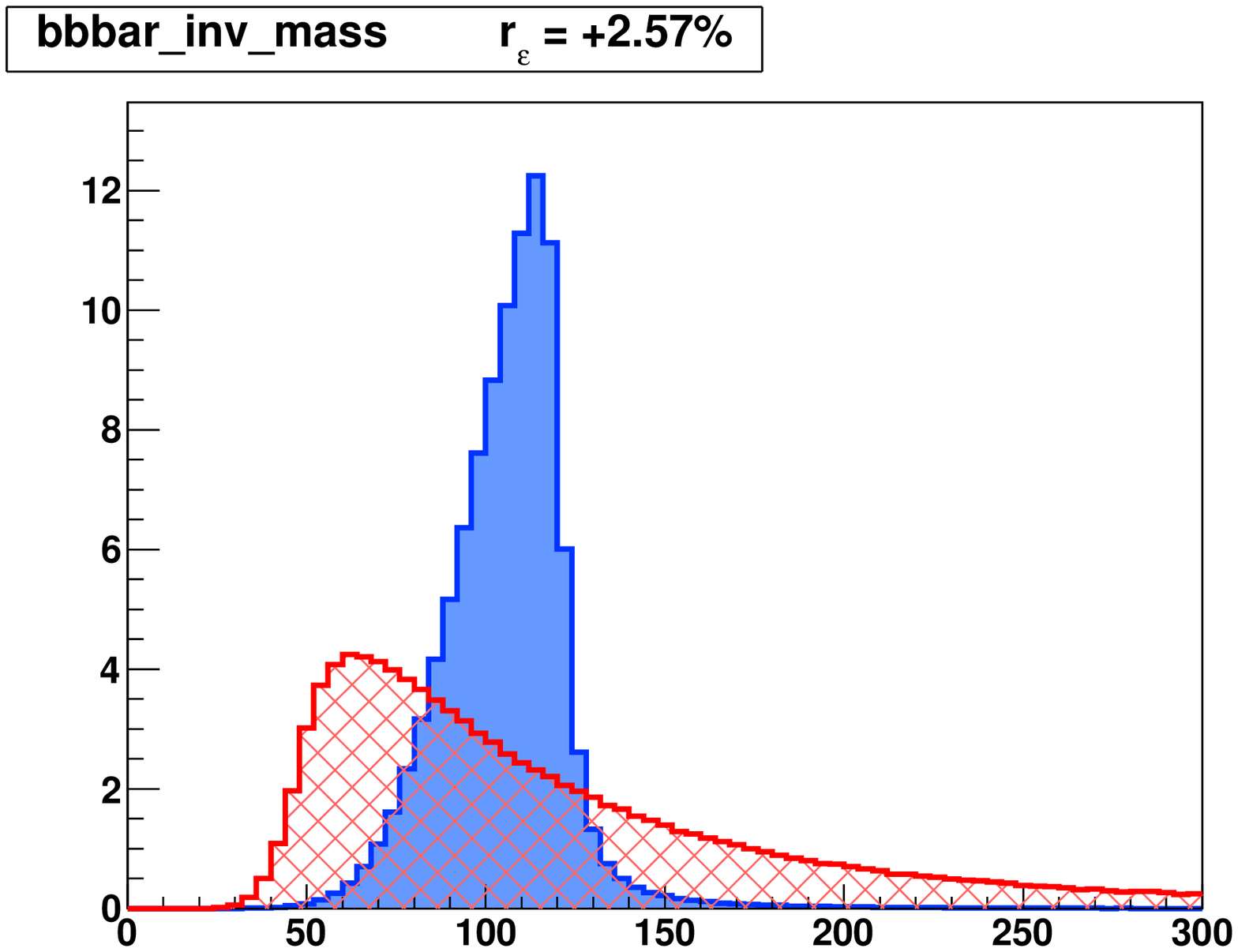}
  \includegraphics[width=0.45\textwidth]{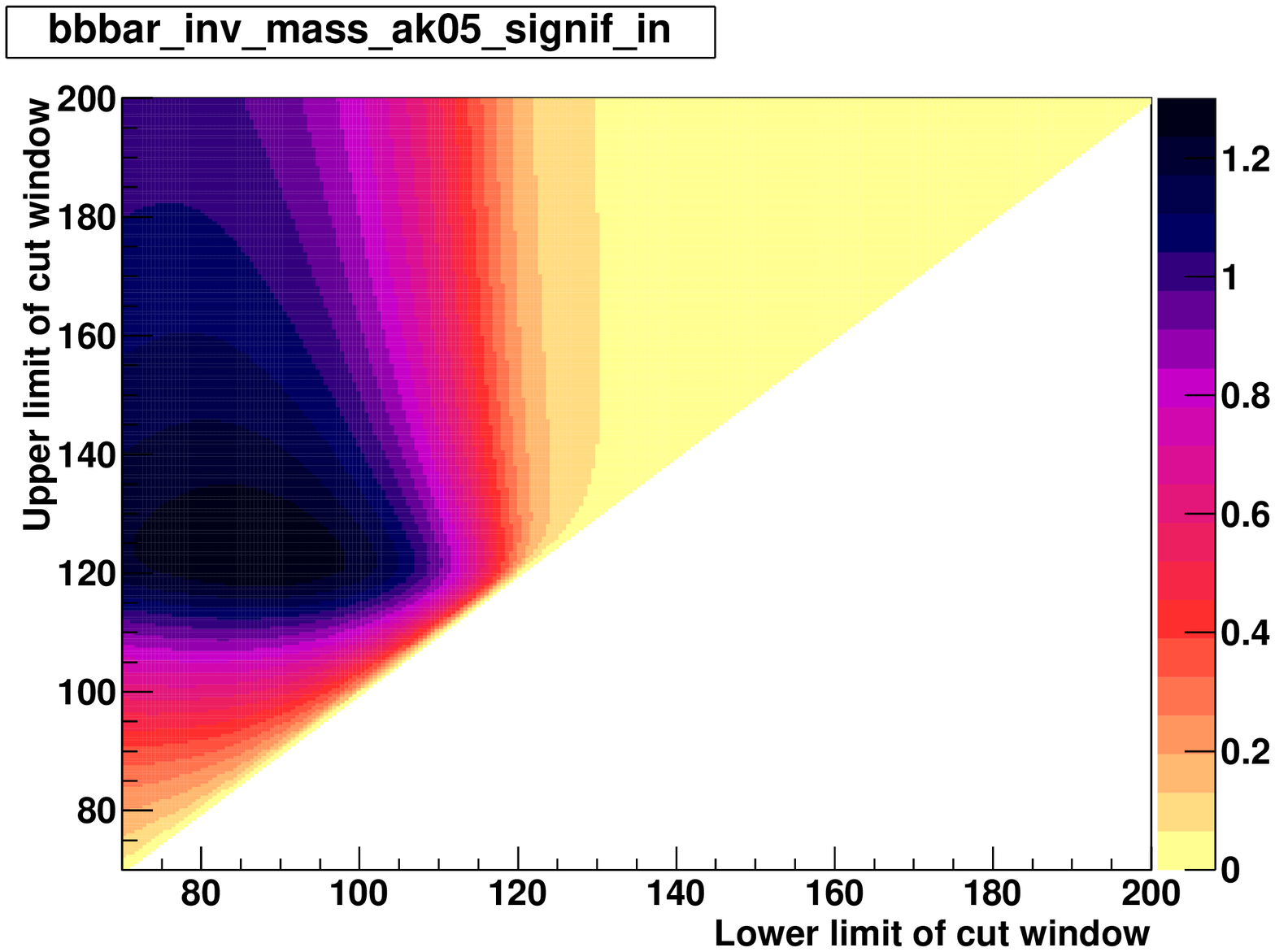}
\caption{
On the left is a representative Higgs Invariant Mass $\mbb$ distribution for
anti-$k_T$ $R\!=\!0.5$ jets for $\zh$ signal (solid blue) and $\zbb$ background (hashed red)
at the LHC with selection cuts.
On the right, the significance improvement can be directly calculated for any two-sided cut.
The maximum here gives our Higgs mass-widow constraint of \Higgswindow.
}
\label{fig:mbb_to_roc}
\end{center}
\end{figure}

To begin, we return to our generator-level sample, still requiring two $b$-tagged jets,
but not imposing an $\mbb$ mass constraint.
We first explain how an optimal window is calculated. Figure~\ref{fig:mbb_to_roc} shows, on the left,
the $\mbb$ distribution for signal and background. For each possible upper and lower edge of the mass window, one can calculate
$\eS$ and $\eB$ for that window. The distribution of $\rsig = \eS/\sqrt{\eB}$ is shown on the right. The optimal window is defined
to be the one that maximizes $\rsig$, which for this figure, anti-$k_T$ $R\!=\!0.5$ jets for $\zh$ at the LHC, is
\Higgswindow. This is the window we have been using in previous sections,
and the window that defined the reference efficiencies ($\rsig=1$).

\begin{figure}[t]
\begin{center}
\psfrag{LHC HZ Significance}{\tiny{$S/\sqrt{B}$ Improvement}}
\psfrag{Higgs boson : Significance : }{}
\psfrag{Signal eff}{\!\!\!\!\!\!\!\!\!\!\!\!\!\!\!\!\!\!\!\!\!\!\!\!\!\!\!\!\!\!\!\!\!\!\!\!\!\!\!\!\!\!\!\!\!\!\!\tiny{Higgs Signal Efficiency $\varepsilon_S$}}
\psfrag{bbbar_inv_mass_ak04}{\tiny{0.4 anti-$k_T$}}
\psfrag{bbbar_inv_mass_ca04}{\tiny{0.4 C/A}}
\psfrag{bbbar_inv_mass_kt04}{\tiny{0.4 $k_T$}}
\psfrag{bbbar_inv_mass_ak05}{\tiny{0.5 anti-$k_T$}}
\psfrag{bbbar_inv_mass_ca05}{\tiny{0.5 C/A}}
\psfrag{bbbar_inv_mass_kt05}{\tiny{0.5 $k_T$}}
\psfrag{bbbar_inv_mass_ak06}{\tiny{0.6 anti-$k_T$}}
\psfrag{bbbar_inv_mass_ca06}{\tiny{0.6 C/A}}
\psfrag{bbbar_inv_mass_kt06}{\tiny{0.6 $k_T$}}
\psfrag{bbbar_inv_mass_ak07}{\tiny{0.7 anti-$k_T$}}
\psfrag{bbbar_inv_mass_ca07}{\tiny{0.7 C/A}}
\psfrag{bbbar_inv_mass_kt07}{\tiny{0.7 $k_T$}}
\begin{tabular}{cc}
\includegraphics[width=0.5\textwidth]{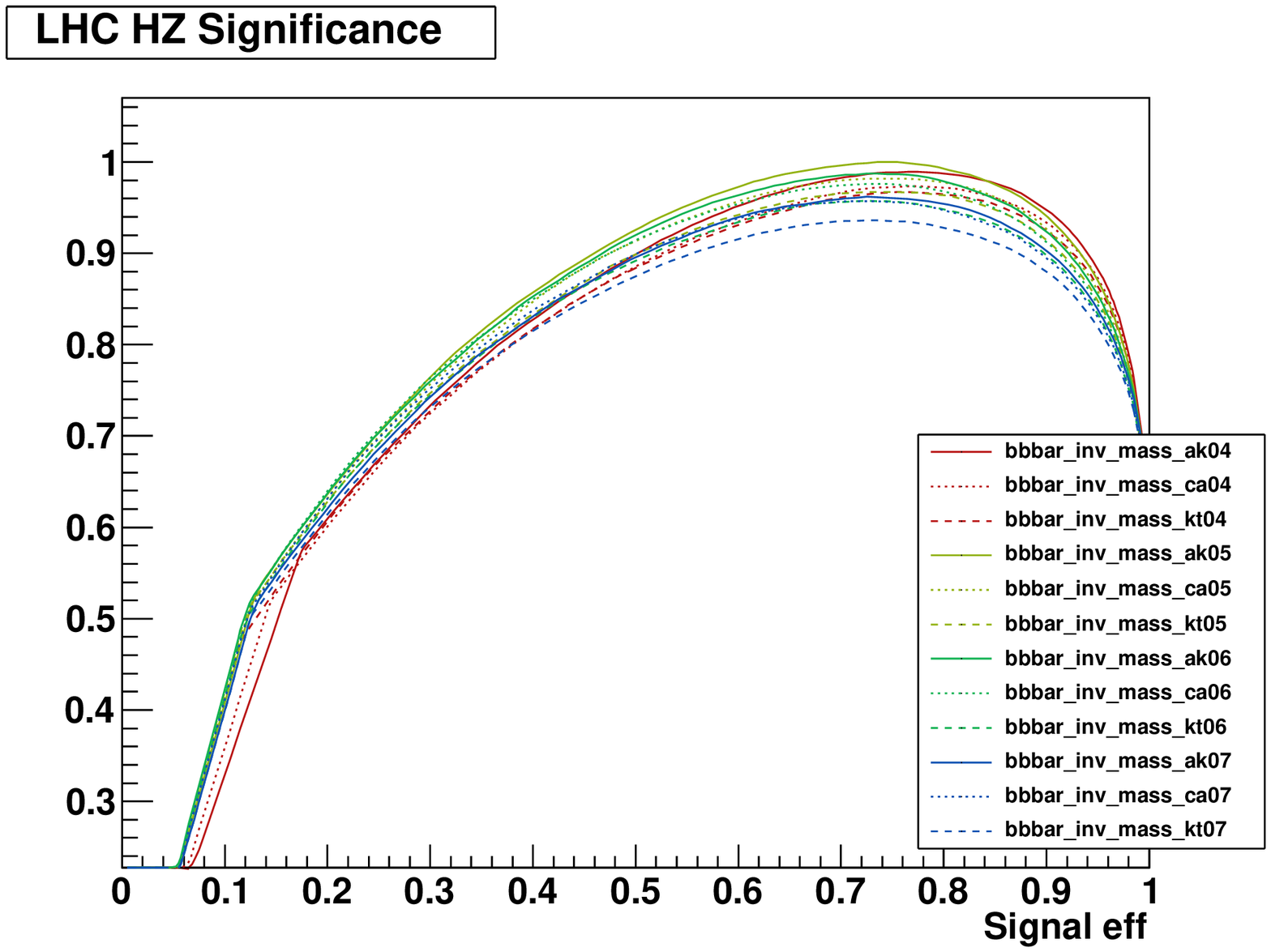}
&
\psfrag{LHC HZ Significance}{\tiny{$S/\sqrt{B}$ Improvement \qquad (zoom)}}
\includegraphics[width=0.5\textwidth]{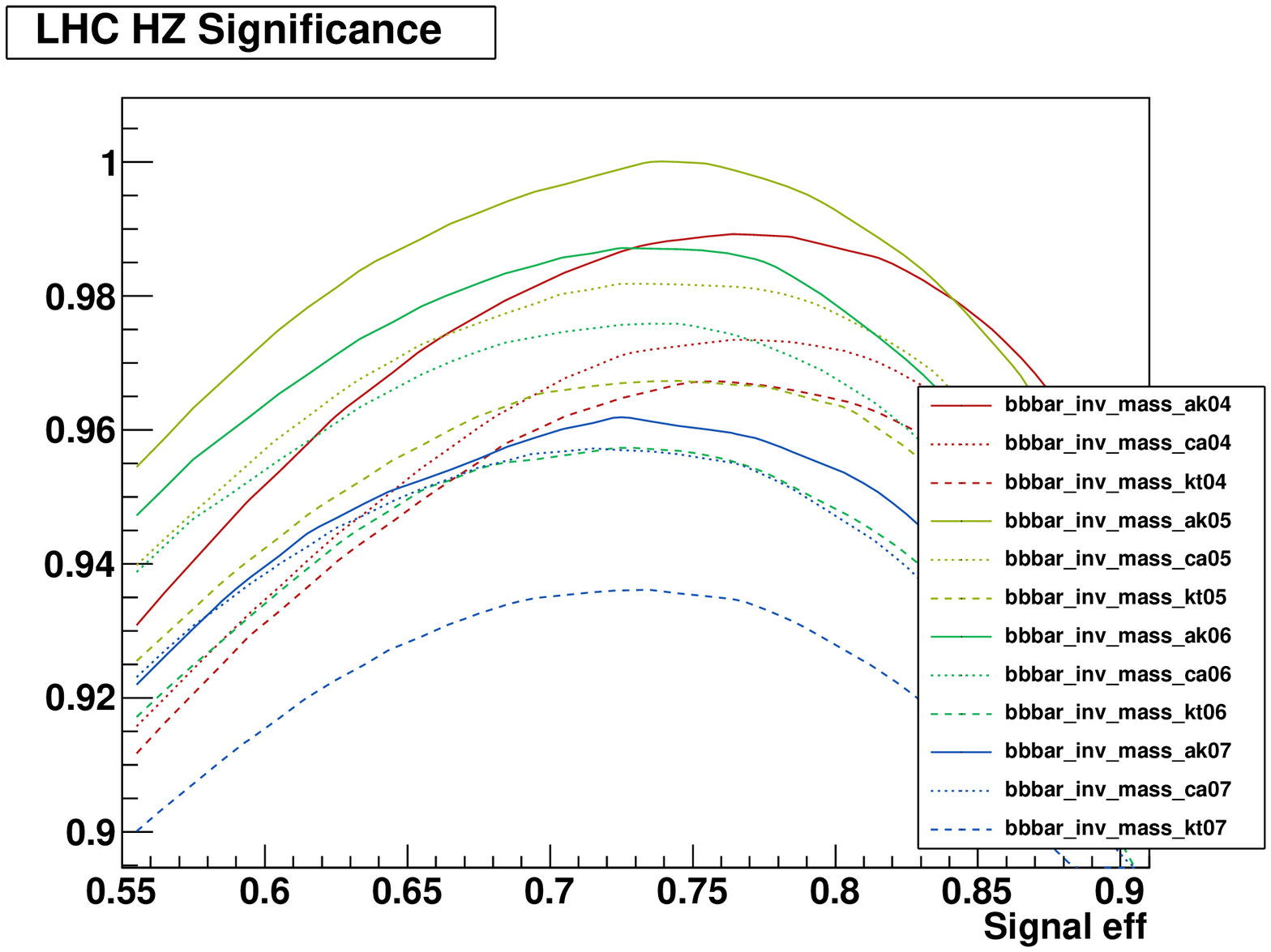}
\end{tabular}
\caption{Significance improvement characteristics varying only the $\mbb$ window for different jet types. SICs are
less than 1 because the reference value is defined using the optimal jet algorithm: anti-$k_T$ with $R\!=\!0.5$ with
optimal mass cut \Higgswindow.
The right panel shows a zoom-in of the peak region in the left panel.
}
\label{fig:masswindow_types}
\end{center}
\end{figure}

Figure~\ref{fig:masswindow_types} shows the $\rsig$ curves for three jet algorithms:
$k_T$~\cite{Catani:1993hr}, anti-$k_T$~\cite{Cacciari:2008gp} and Cambridge/Aachen (C/A)~\cite{Salam:2007xv}
 and for jet sizes from 0.4 to 0.7.
The best anti-$k_T$ jets (solid lines) beat out the
best Cambridge/Aachen (dotted) and $k_T$ (dashed), but they are all quite close.
The optimal seems is anti-$k_T$ $R\!=\!0.5$ for the LHC and anti-$k_T$ $R\!=\!0.7$ for the Tevatron,
but the response of real calorimeters might change this.
The SIC curves are very similar for
$\zh$ and $\wh$ at both machines.

Next, we consider multiple mass measures. This idea was inspired by the work of Soper and Spannowski in Ref.~\cite{Soper:2010xk}.
They considered the effect of combining pruning and trimming for the highly boosted $\zh$ events, and found that the two were somewhat
complimentary. {\it Pruning}~\cite{Ellis:2009me} attempts to find evidence for a heavy particle decaying to boosted collimated hadronic jets, while
{\it trimming}~\cite{Krohn:2009th} is designed to remove contamination from initial-state radiation and the underlying event.
Both trimming and pruning were modifications of the {\it filtering} procedure used for boosted $\zh$ search~\cite{Butterworth:2010ym}
 and for top-tagging~\cite{Kaplan:2008ie}.
For a review, see~\cite{Salam:2009jx} or \cite{Butterworth:2010ym}. Since our $b$'s do not appear in a single fat jet, we restricted our consideration to the trimming algorithm. We wanted to see whether combing mass measures with different amounts of trimming could improve over a single jet type.
\begin{figure}[t]
\begin{center}
\begin{tabular}{cc}
\psfrag{bbbar_inv_mass_ak06_sub_ak020_trimmed_50}{\tiny{$\mbb$ for Aggressive Trimming (GeV)}}
\psfrag{bbbar_inv_mass_ak06_sub_ak020_trimmed_20}{\tiny{$\mbb$ for Aggressive Trimming (GeV)}}
\psfrag{bbbar_inv_mass_ak06_sub_kt005_trimmed_01}{\tiny{$\mbb$ for Mild Trimming (GeV)}}
\psfrag{bbbar_inv_mass_ak06, bbbar_inv_mass_ak06_sub_kt005_trimmed_01}{\tiny{$\mbb$ for No+Mild Trimming (GeV)}}
\includegraphics[width=0.5\textwidth]{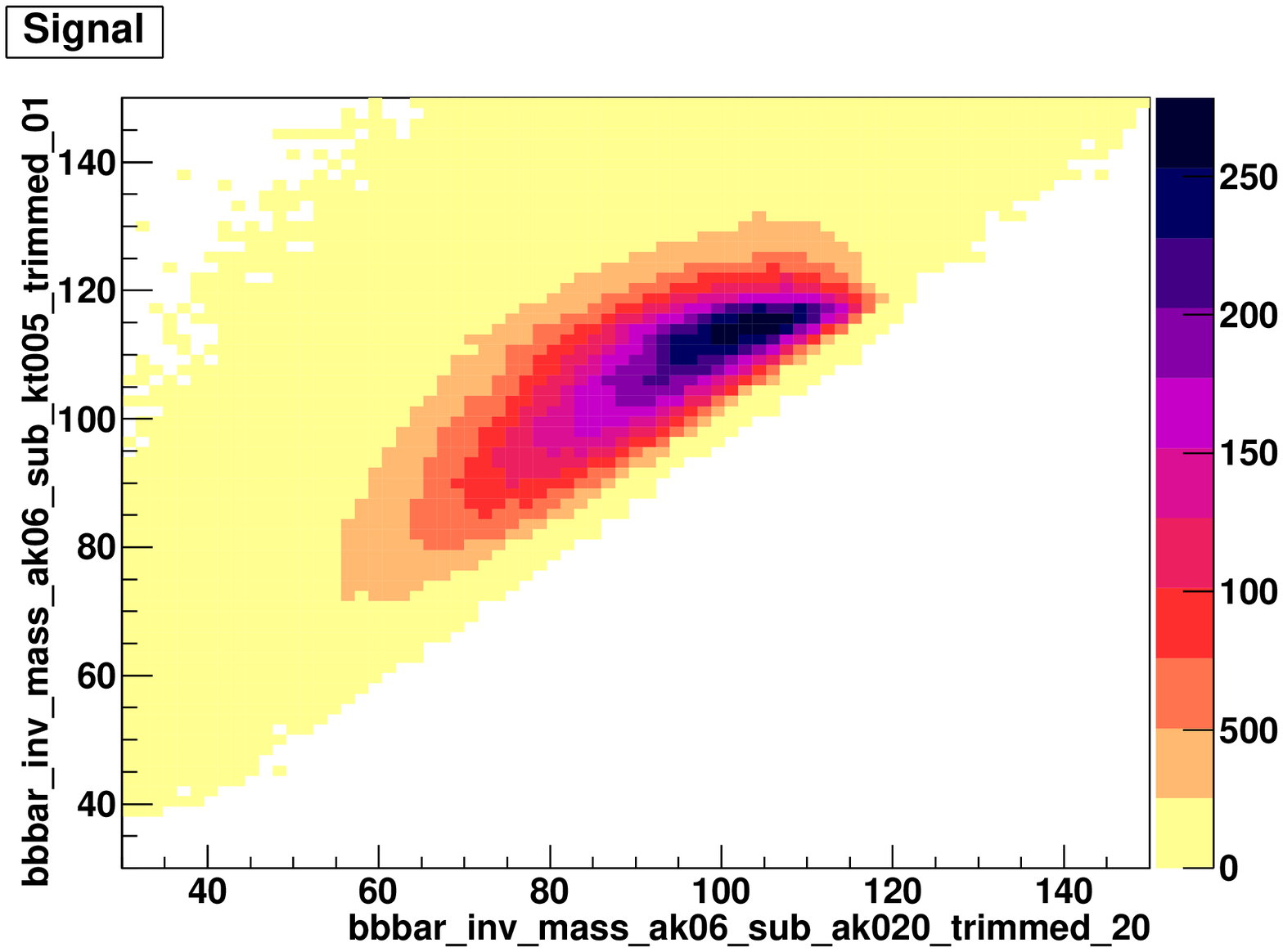}
&
\psfrag{bbbar_inv_mass_ak06_sub_ak020_trimmed_50}{\tiny{$\mbb$ for Aggressive Trimming (GeV)}}
\psfrag{bbbar_inv_mass_ak06_sub_ak020_trimmed_20}{\tiny{$\mbb$ for Aggressive Trimming (GeV)}}
\psfrag{bbbar_inv_mass_ak06_sub_kt005_trimmed_01}{\tiny{$\mbb$ for Mild Trimming (GeV)}}
\psfrag{bbbar_inv_mass_ak06, bbbar_inv_mass_ak06_sub_kt005_trimmed_01}{\tiny{$\mbb$ for No+Mild Trimming (GeV)}}
\includegraphics[width=0.5\textwidth]{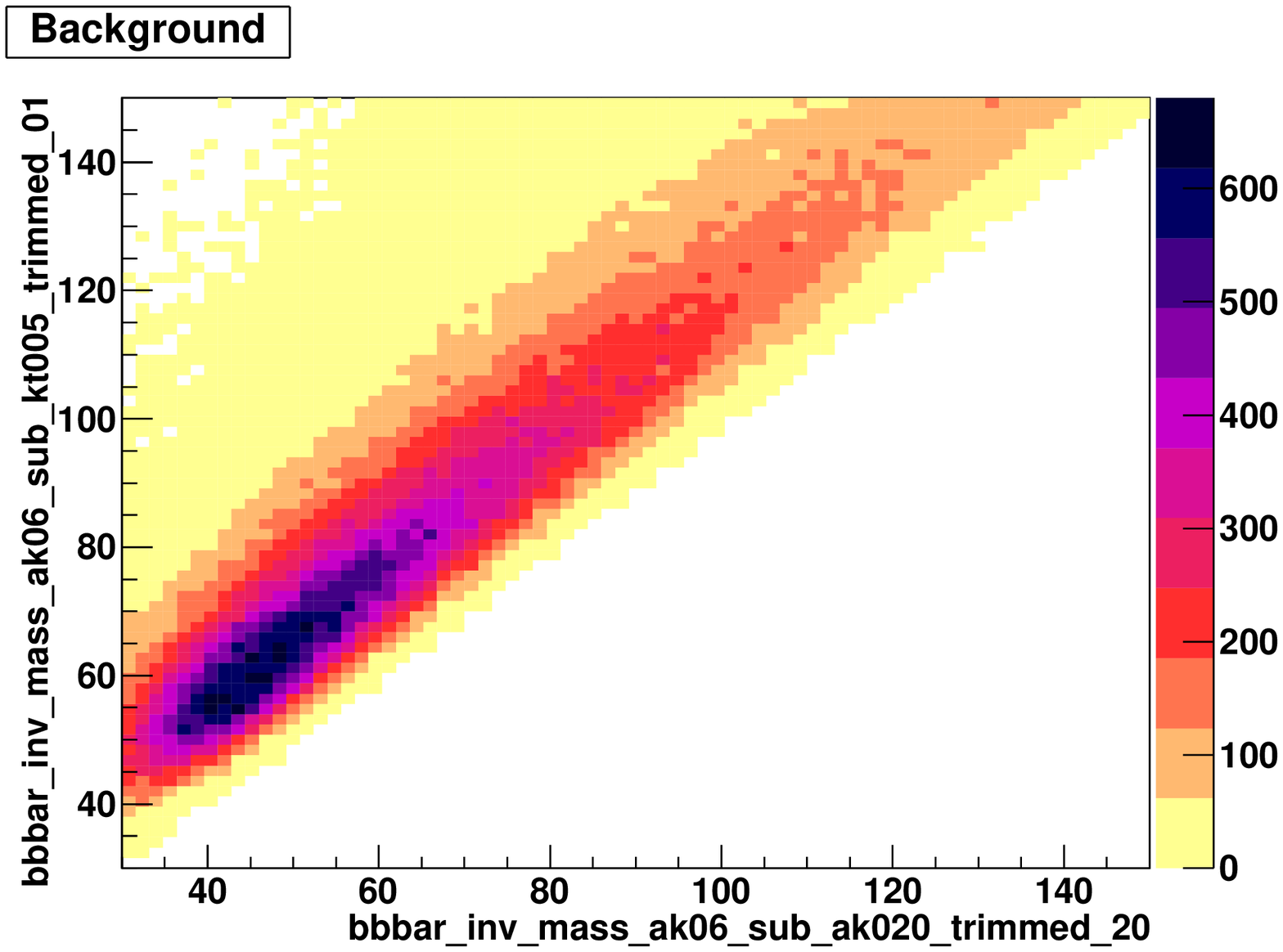}
\end{tabular}
\caption{
2D histograms of $\mbb$ for mild and aggressively trimmed jets show their correlation.
Starting with anti-$k_T$ $R\!=\!0.5$ jets,
aggressive trimming means keeping only 0.2 anti-$k_T$ subjets whose $p_T$ is more than 50\% of the original jet $p_T$;
mild trimming means keeping 0.05 $k_T$ subjets with more than 1\% of the original jet's $p_T$.
After the each jet is trimmed, the invariant mass of the pair is calculated.
}
\label{fig:trimmed_masses}
\end{center}
\end{figure}
\begin{figure}[ht]
\begin{center}
\psfrag{Higgs Significance}{\small{$\rsig$ with Trimming}}  
\psfrag{LHC ZH : Significance : }{}
\psfrag{bbbar_inv_mass_ak06}{\tiny{No Trimming $R=0.6$}}
\psfrag{bbbar_inv_mass_ak05}{\tiny{No Trimming $R=0.5$}}
\psfrag{bbbar_inv_mass_ak06_sub_ak020_trimmed_50}{\tiny{Aggressive Trimming}}
\psfrag{bbbar_inv_mass_ak06_sub_ak020_trimmed_20}{\tiny{Aggressive Trimming}}
\psfrag{bbbar_inv_mass_ak06_sub_kt005_trimmed_01}{\tiny{Mild Trimming}}
\psfrag{bbbar_inv_mass_ak06, bbbar_inv_mass_ak06_sub_kt005_trimmed_01}{\tiny{No+Mild Trimming}}
\psfrag{bbbar_inv_mass_ak06, bbbar_inv_mass_ak06_sub_ak020_trimmed_50}{\tiny{No+Agressive Trimming}}
\psfrag{bbbar_inv_mass_ak06_sub_ak020_trimmed_50, bbbar_inv_mass_ak06_sub_kt005_trimmed_01}{\tiny{Mild+Aggressive Trimming}}
\psfrag{bbbar_inv_mass_ak06, bbbar_inv_mass_ak06_sub_ak020_trimmed_50, bbbar_inv_mass_ak06_sub_kt005_trimmed_01}{\tiny{No+Mild+Aggressive Trimming}}
\includegraphics[width=0.8\textwidth]{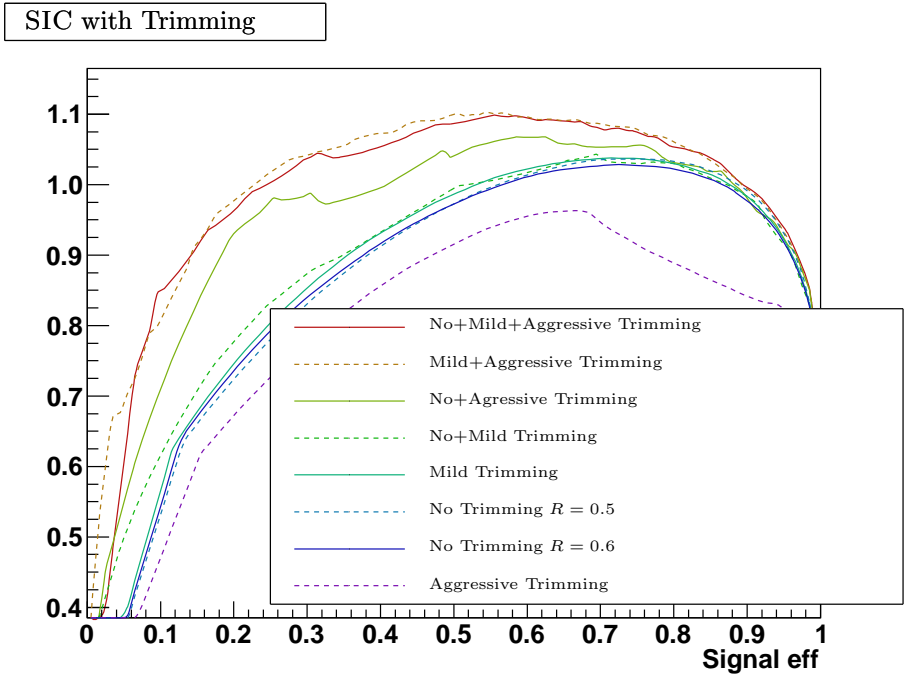}
\caption{Significance improvement $\eS/\sqrt{\eB}$ with various combinations of $\mbb$ constructed from trimmed and untrimmed jets.}
\label{fig:trimmed_rsig}
\end{center}
\end{figure}

The jet trimming starts with a jet, say one of our original anti-$k_T$ $R\!=\!0.5$ $b$-jets. Then one
reclusters the jet with a smaller jet size $r$, say $r\!=\!0.1$. If the energy of any of the smaller jets is less than a fraction $f$
of the original jets energy, the subjet is tossed. Then the remaining subjets are recombined into a trimmed jet by adding their 4-momenta.
This procedure naturally removes soft radiation, representative of underlying event contamination or soft ISR,
while keeping the hard collinear radiation from the final state shower. Trimming has two parameters $r$ and $f$, along with the
jet algorithm used to form the subjets.

First, we looked at a single $\mbb$ measure with trimming on both of the $b$-jets. We did not find that any values of the parameters were
significantly better than no trimming. On the other hand, we found that significant improvement did result from combining different
trimmed masses. We define two extreme trimmings: {\bf mild} trimming, with
$r\!=\!0.05$ and $f\!=\!1\%$ and {\bf aggressive} trimming, with $r\!=\!0.2$ and $f = 50\%$.
The two dimensional distribution of $\mbb$ with mild and aggressive trimming of both $b$-jets is shown in
Figure~\ref{fig:trimmed_masses}. One can see that $\mbb$ for the the mild and
aggressive trimmed jets are strongly correlated for the background, as evidenced by the long flat direction, but much less
correlated in the signal. By eye, one can see that drawing a contour to separate signal from background will do better than
any single line or rectangular window.

Figure~\ref{fig:trimmed_rsig} shows the significance improvement for various combinations of mild, aggressive, and no trimming. Note that
while mild trimming by itself is almost identical to no trimming, when combined with aggressive trimming, mild does
better than not trimming at all. Combining all three does not improve over mild + aggressive alone.

After concluding that multiple mass measures can improve the significance, we then included the mass measures with the other
discriminants in the multivariate analysis.
We found that for just kinematic variables, having multiple trimmed masses does
help a little. However, when the radiation variables are included, the multiple mass measures have no effect on the SIC
curves. This seems to hold for the $\zh$ and $\wh$ samples at the Tevatron or the LHC. We saw no effect either by adding the mass measures
from mild and aggressively trimmed jets on top of the top 10 variables or by incorporating  about 100 $\mbb$ measures directly
into the multivariate mix.
Thus, it seems that while multiple trimmings by themselves are useful, they share much information
with many of the showered variables.

\begin{table}\begin{center}
\begin{tabular}{|p{0.22\textwidth}|p{0.22\textwidth}|p{0.23\textwidth}|p{0.23\textwidth}|}
  \hline
  LHC ZH & LHC WH & TVT ZH & TVT WH \\
  \hline
\begin{tabular}{ll}
\scriptsize{1-10} & $|\vp_T^{\,b1}|+|\vp_T^{\,b2}|$ \\ 
\scriptsize{1-3}  & $\Delta \eta_{b \bar b}$  \\ 
\scriptsize{1-2}  & $\Delta y_{H,b2}$  \\ 
\scriptsize{2-4}  & $|\vp_T^{\,H}|-|\vp_T^{\,b2}|$  \\ 
\scriptsize{3-5}  & $\Delta y_{H,\ell 1}$  \\ 
\scriptsize{3-4}  & $\Delta y_{ZH}$  \\ 
\scriptsize{4-10} & $|\vp_T^{\,Z}|+|\vp_T^{\,b2}|$  \\ 
\scriptsize{4-10} & $m_{H,\ell 1}$  \\ 
\scriptsize{4-9}  & Sphericity${}^{\text{prim.}}$  \\ 
\scriptsize{5-9}  & $\Sigma y_{Z,b1}$  \\ 
\scriptsize{5-9}  & $\Sigma y_{Z,b1}$  \\ 
\scriptsize{5-9}  & $\Sigma y_{Z,b1}$  \\ 
\scriptsize{5-10} & $E^\text{obj.}_\mathrm{vis}$  \\ 
\scriptsize{5-10} & pull $\alpha$  \\ 
\scriptsize{6-10} & girth $g_{b2}$  \\ 
\scriptsize{6-10}  & pull $\beta_2$  \\ 
\scriptsize{8-10} & angul. $A^{-0.90}_{b1}$  \\ 
\scriptsize{9-10} & pull $\beta$  \\ 
\scriptsize{9-10} & $|\vp_T^{\,H}|+|\vp_T^{\,l2}|$  \\ 
\end{tabular}
  &
\begin{tabular}{ll}
\scriptsize{1-10} & $\Delta y_{WH}$  \\ 
\scriptsize{2-10} & $|\vp_T^{\,b1}|+|\vp_T^{\,b2}|$  \\ 
\scriptsize{1-2}  & CM  $\cos \theta_H$  \\ 
\scriptsize{2-4}  & $\Delta \phi_{\ell^+\ell^-}$  \\ 
\scriptsize{2-7}  & Twist $\tau_{b \bar b}$  \\ 
\scriptsize{3-10} & Twist $\tau_{\ell^+\ell^-}$  \\ 
\scriptsize{3-10} & $E^\text{prim.}_\mathrm{vis}$  \\ 
\scriptsize{4-10} & pull $\beta$  \\ 
\scriptsize{4-6}  & $\Delta \phi_{W,l2}$  \\ 
\scriptsize{5-10} & $m_{W,b1}$  \\ 
\scriptsize{9-10} & $m_{W,b2}$  \\ 
\scriptsize{6-10} & $\Delta R_{H,b2}$  \\ 
\scriptsize{5-10} & angul. $A^{-0.1}_{b2}$  \\ 
\scriptsize{5-10} & $|\vp_T^{\,W}|-|\vp_T^{\,\ell 1}|$\\
\scriptsize{8-9}  & pull $\alpha$  \\ 
\end{tabular}
  &
\begin{tabular}{ll}
\scriptsize{1-10} & $|\vp_T^{\,b1}|+|\vp_T^{\,b2}|$   \\ 
\scriptsize{1-10} & $\Delta \eta_{b \bar b}$  \\ 
\scriptsize{1-7}  & $\Delta y_{Z,H}$  \\ 
\scriptsize{1-10} & twist $\tau_{\ell^+\ell^-}$  \\ 
\scriptsize{2-10} & Centrality${}^{\text{prim.}}$  \\ 
\scriptsize{2-3}  & $|\vp_T^{\,b1}|$  \\ 
\scriptsize{3-10} & pull $\beta$  \\ 
\scriptsize{3-10} & pull $\alpha$  \\ 
\scriptsize{3-10} & $m_{H,\ell 2}$  \\ 
\scriptsize{3-5}  & $\cos \theta_{\ell 2}$\\
&~~($Z$ Frame)  \\ 
\scriptsize{5-8} & girth $g_{b2}$  \\ 
\scriptsize{6-10} & angul. $A^{-0.01}_{b2}$  \\ 
\scriptsize{7-9}  &  $m_{b2} / P_T^{b2}$  \\ 
\scriptsize{9-10} & angul. $A^{+0.01}_{b1}$  \\ 
\scriptsize{9-10} & $\Delta \phi_{b \bar b}$  \\ 
\end{tabular}
  &
\begin{tabular}{ll}
\scriptsize{1-10} & $\Delta y_{WH}$  \\ 
\scriptsize{1-2}  & CM $\cos \theta_H$  \\ 
\scriptsize{2-10} & $\Delta y_{\ell^+ \ell^-}$  \\ 
\scriptsize{3-10} & $H_T^\text{prim.}$  \\ 
\scriptsize{4-10} & $\Delta \eta_{b \bar b}$  \\ 
\scriptsize{4-10} & $\Delta p_T^{\ell^+ \ell^-}$  \\ 
\scriptsize{4-10} & pull $\alpha$  \\ 
\scriptsize{4-10} & pull $\beta$  \\ 
\scriptsize{7-10} & avg. subj. $p_T$\\
&~~ $(R^{k_T}_\mathrm{sub}\!\!=\!0.2)$ \\ 
\scriptsize{7-10} & $m_{b2} / p_T^{b2}$  \\ 
\scriptsize{8-9}  & $m_T^{b \bar b}$  \\ 
\scriptsize{8-10} & $m_{W,b1}$  \\ 
\end{tabular}
  \\
  \hline
\end{tabular}
\end{center}
\caption{Top variables that showed up at the stages indicated. First number is the is first stage they appeared,
second number is the last stage they appeared.
Variables that ended up in the top 10 are indicated with a 10. ``obj.'' and ``prim.'' refer to whether reconstructed objects
or the primary four objects (2 b-jets and 2 leptons) were used. ``angul.'' are the angularities.
}
\label{table:top_vars}
\end{table}

\begin{table}
\begin{center}
\begin{tabular}{|p{0.21\textwidth}|p{0.2\textwidth}|p{0.2\textwidth}|p{0.25\textwidth}|}
  \hline
  CDF ZH & \DZero~ZH & CDF WH & \DZero~ WH \\
  \hline
\begin{tabular}{l}
$\mbb$  \\ 
$p_T^{b1}$  \\  
$p_T^{b2}$  \\  
$\hat s$  \\  
$p_T^Z$  \\  
$H_T$  \\  
Sphericity${}^{\text{obj.}}$  \\  
\end{tabular}
  &
\begin{tabular}{l}
$\mbb$  \\  
$p_T^{b1}$  \\  
$p_T^{b2}$  \\  
$\Delta R_{b \bar b}$  \\  
$\Delta R_{\ell^+\ell^-}$  \\  
$\Delta R_{ZH}$  \\  
CM $\cos \theta_H$  \\  
$P_T^H$  \\  
$H_T$  \\  
\end{tabular}
  &
\begin{tabular}{l}
$\mbb$  \\  
$p_T^\mathrm{imbalance}$  \\  
$m_{W,b1}$  \\  
$m_{W,b2}$  \\  
$\eta_\ell$  \\  
$\Sigma p_T^{b \bar b}$  \\  
$p_T^W$  \\  
$H_T$  \\  
\end{tabular}
  &
\begin{tabular}{l}
$\mbb$  \\  
$p_T^{b1}$  \\  
$p_T^{b2}$  \\  
$E_{b2}$  \\  
$\Delta R_{b \bar b}$  \\  
$\Delta \phi_{b \bar b}$  \\  
$\Delta \phi_{b1,\ell}$  \\  
$p_T^H$  \\  
$p_T^W$  \\  
$\hat s$  \\  
$\Delta R_{W,H}$  \\  
$H_z$  \\  
CM  $\cos \theta_H$  \\  
\end{tabular}
  \\
  \hline
\end{tabular}
\end{center}
\caption{Subsets of variables used in recent CDF~\cite{CDF} and \DZero~\cite{D0} analyses which we use
to compare our efficiencies.}
\label{table:CDF_D0_vars}
\end{table}


\section{Final results \label{sec:money}}

\begin{figure}[t]
\begin{center}
\begin{tabular}{cc}
\psfrag{TVT HZ Significance}{\footnotesize{\qquad TVT ZH \qquad \quad $S/\sqrt{B}$ Improvement}}
\psfrag{11 TVT HZ best}{\tiny{11 TVT HZ best}}
\psfrag{9 D0 HZ}{\tiny{  9  D0 HZ}}
\psfrag{7 CDF HZ}{\tiny{  7  CDF HZ}}
\includegraphics[width=0.5\textwidth]{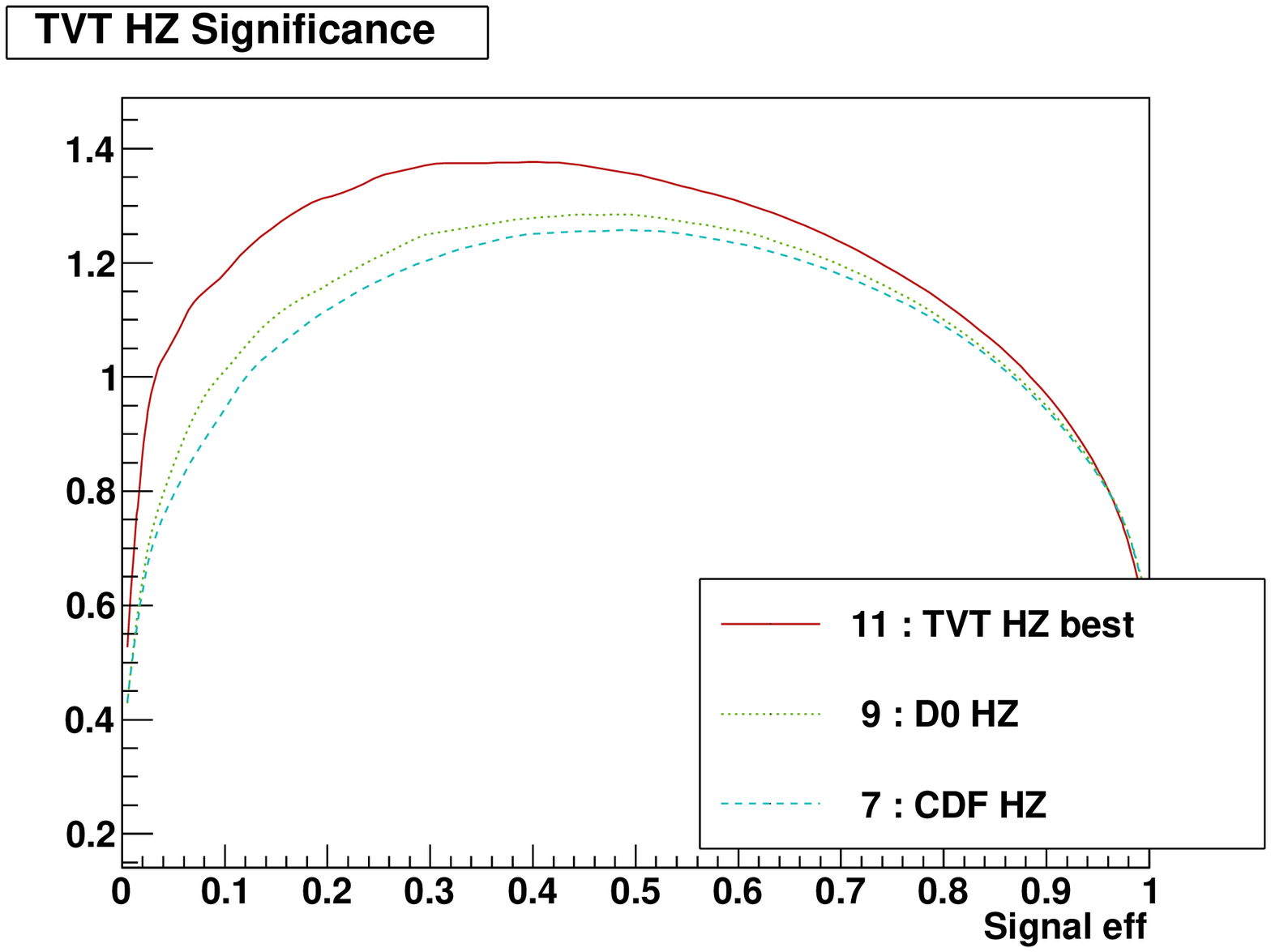} &

\psfrag{TVT HW Significance}{\footnotesize{\qquad TVT WH \qquad \quad $S/\sqrt{B}$ Improvement}}
\psfrag{ 11 TVT HW best}{\tiny{ 11 : TVT HW best}}
\psfrag{  8 CDF HW}{\tiny{  8 : CDF HW}}
\psfrag{ 13 D0 HW}{\tiny{ 13 : D0 HW}}
\includegraphics[width=0.5\textwidth]{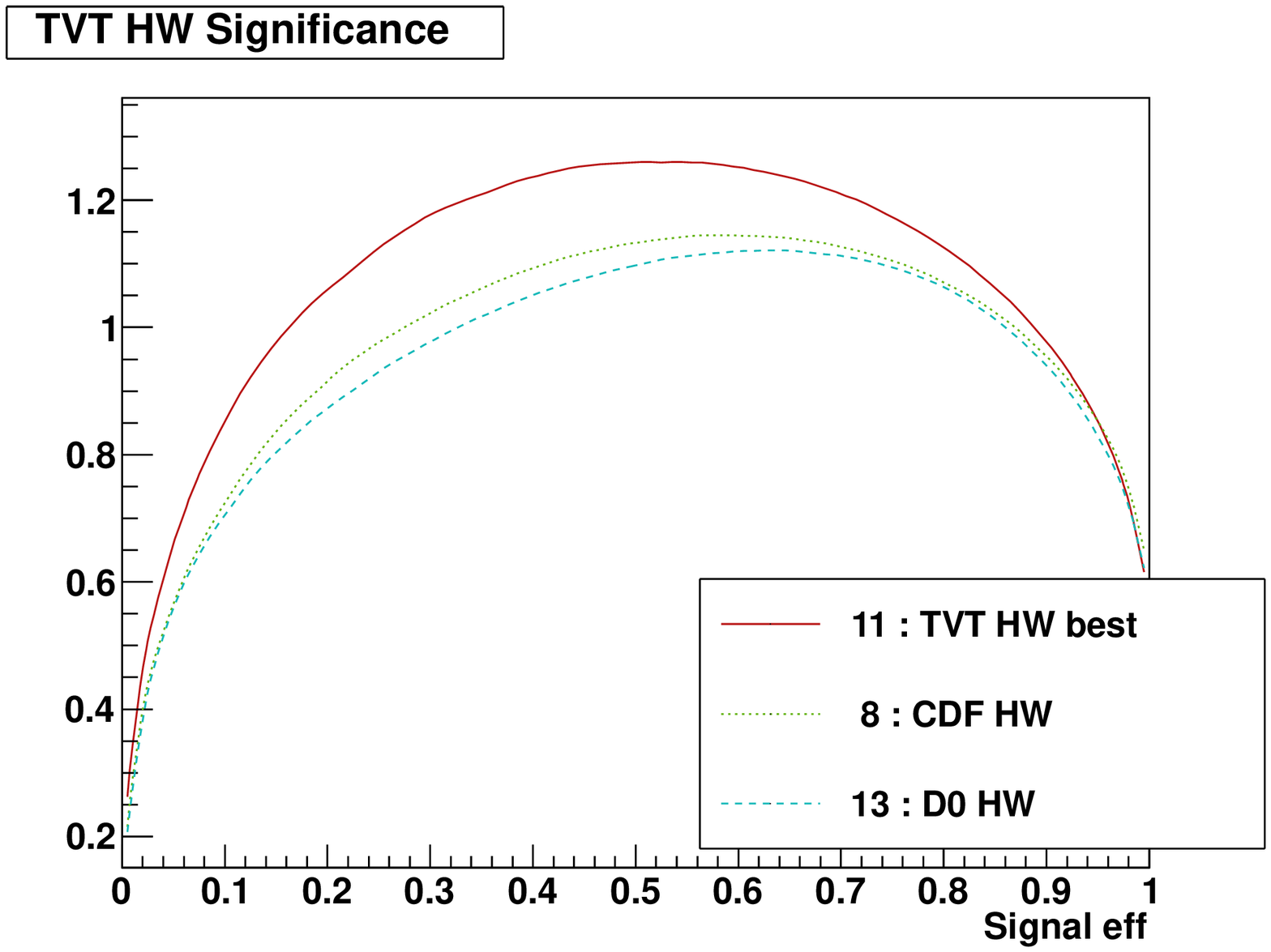} \\

\psfrag{LHC HZ Significance}{\footnotesize{\qquad LHC ZH \qquad \quad $S/\sqrt{B}$ Improvement}}
\psfrag{ 11 LHC HZ best kinematic and jet}{\tiny{ 11 : LHC HZ best kinematic and jet}}
\psfrag{  7 LHC HZ best kinematic only}{\tiny{  7 : LHC HZ best kinematic only}}
\psfrag{  2 LHC HZ Z_Pt}{\tiny{  2 : LHC HZ $P_T^Z$}}
\includegraphics[width=0.5\textwidth]{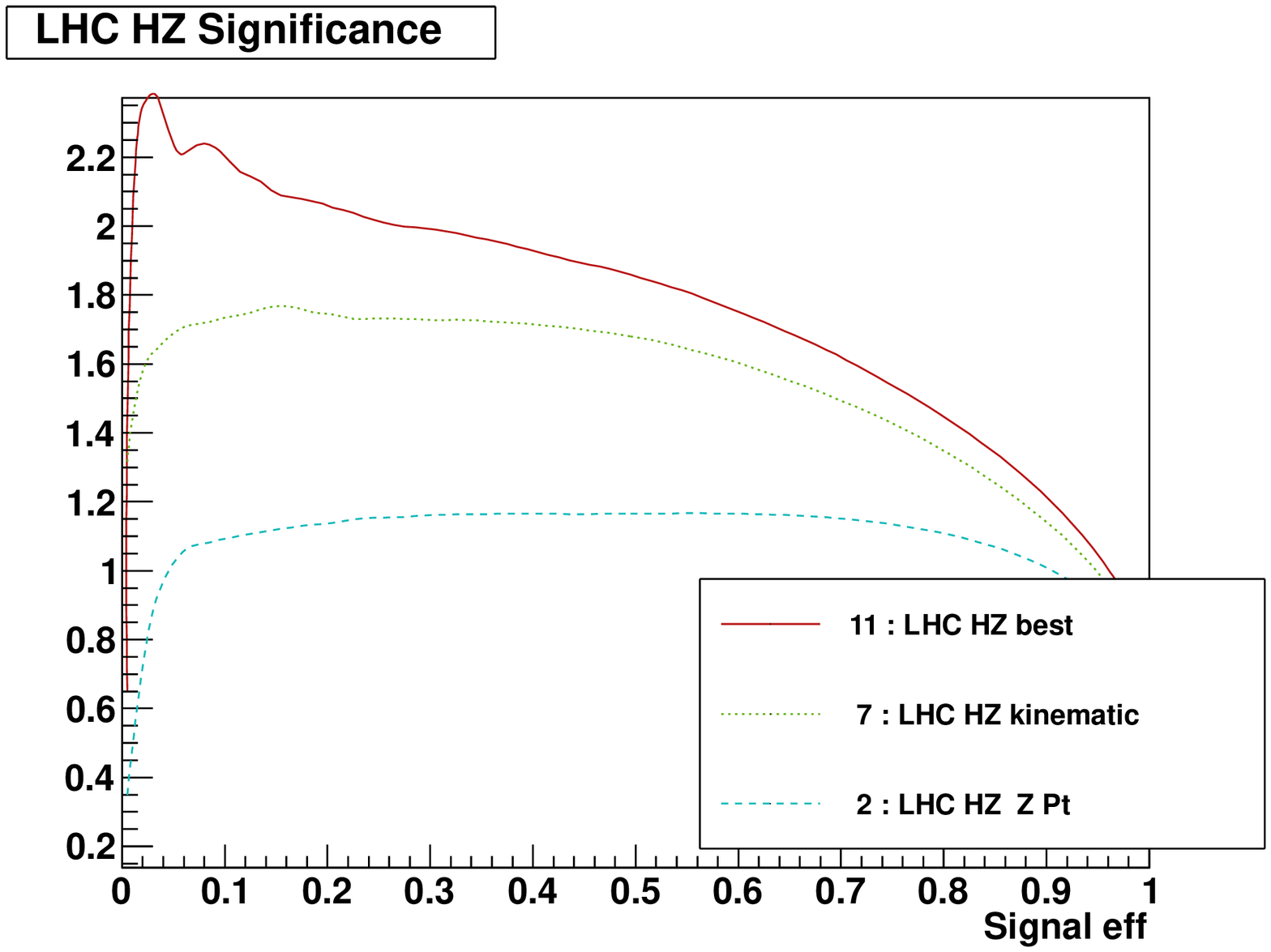} &

\psfrag{LHC HW Significance}{\footnotesize{\qquad LHC WH \qquad \quad $S/\sqrt{B}$ Improvement}}
\psfrag{ 11 LHC HW best kinematic and jet}{\tiny{ 11 : LHC HW best kinematic and jet}}
\psfrag{ 10 LHC HW best mass kinematic only}{\tiny{ 10 : LHC HW best mass kinematic only}}
\psfrag{  2 Z_Pt}{\tiny{  2 : LHC HW $P_T^W$}}
\includegraphics[width=0.5\textwidth]{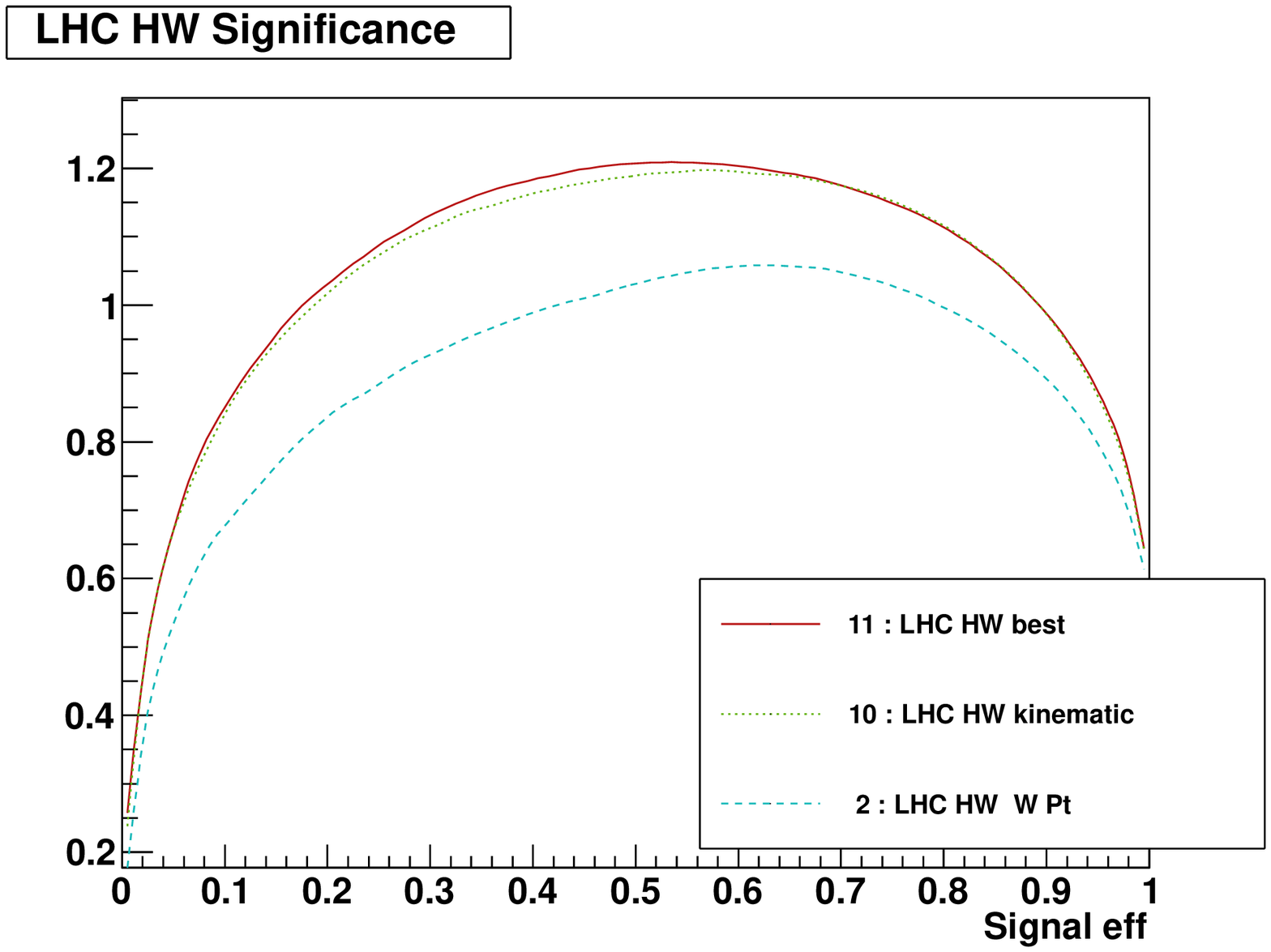}
\end{tabular}
\caption{SIC curves for sets of variables discussed in the text, along with the appropriate subset of the
variables used by CDF and \DZero, as listed in Table 5.
All curves include $\mbb$ as one of the variables, as
opposed to previous Figures which were with a fixed $\mbb$ window. The reference SIC of 1
is still with respect to the hard cut \Higgswindow.
 Numbers refer to number of variables going into the BDT (e.g. 11= top 10 + 1 for $\mbb$).
Curves labeled kinematic have the radiation variables (pull, girth, angularities) removed.
}
\end{center}
\vspace{-0.5cm}
\label{fig:moneyplot}
\end{figure}

Our final results are shown in Figure~\ref{fig:moneyplot}.
The top two panels show the significance improvement characteristics for the Tevatron, and the bottom two
panels for the LHC.  Our variables are listed in Table~\ref{table:top_vars}, and combined using BDT with 3000 trees.
For the Tevatron results, we can get a sense of how much better we do by comparing
to sets of variables
used by CDF and \DZero.
The SIC curves for our implementation
of a subset\footnote{We include all the variables used by these groups,
except for the ones which depend on missing energy. The missing energy variables
are mostly useful in the $WH$ for removing the $t\bar{t}$ background case,
which we are not considering here.
See also Section~\ref{sec:missing}.} of their variables, as listed in Table~\ref{table:CDF_D0_vars}, are also shown
in Figure~\ref{fig:moneyplot}. We see that around a 10-20\% improvement against irreducible
backgrounds is possible. There are a few important caveats associated with this conclusion:
we are only considering irreducible backgrounds, we base our study entirely on particle-level Monte Carlo, we have not
considered experimental or theoretical systematic uncertainties, and we do not know if all of these variables
can even be measured. Nevertheless, our results do imply that future study is warranted with {\it potentially} significant
gains.

In the bottom two panels, we present our results for the LHC. While we are not aware of any published multivariate approach to
this channel by either ATLAS or CMS, as a reference, we take the ATLAS search which used the high $p_T$ sample~\cite{atlas}. Comparing
the effectiveness of the $p_T$ to the multivariate approach, we can see that the multivariate approach is clearly superior.
Again, although one cannot translate this directly an improvement in sensitivity,
 it is reasonable to expect that since the boosted searches
with $p_T^H>200$ GeV has greater than $3\sigma$ discovery potential with 30 fb${}^{-1}$, a multivariate approach including the lower $p_T$
events would be at least as strong.
Part of the motivation of the hard $p_T$ cut was to kill $WH$'s $t\bar{t}$ background,
an issue we have not addressed. Nevertheless, one can
conclude from our analysis that a proper more complete multivariate study of the light Higgs boson discovery potential in $W/Z+H$ is worth pursuing.

The list of top variables is given in Table~\ref{table:top_vars}.
The first few variables tend to be $p_T$ and rapidity differences, then
other angular variables like twist or global variables like centrality or $H_T$.
Starting around the 4th variable, pull consistently proves useful, followed by
other jet properties like girth, angularities, or subjet average $p_T$'s.
Since our particle-level analysis does not take into
account detector effects, experimental jet calibration, or missing energy, the best
variables to be used experimentally will likely differ from those in this list.

\section{Conclusions \label{sec:conc}}
We explored Higgs boson production in association with a $Z$ or $W$ boson,
with $H\to\bbar$. This served as a case study in optimizing multivariate
discriminants. We attempted to systematically consider an enormous number of discriminants,
including kinematic variables natural from an experimental point of view (e.g. azimuthal angle differences),
variables natural from a physical point of view (e.g. helicity angle of the Higgs boson decay products), variables
dependent on final state radiation (e.g. pull), and even unmotivated kinematic variables (e.g. scalar sum of the $p_T$
of the harder $b$-jet and the softer lepton from the $W/Z$ decay).
Some of the variables we introduced, such as azilicity and twist
should be more generally useful.

We employed a simple algorithm to combine the variables
systematically: take the top single variables, and combine pairwise with every other variable. Then take the top pairs and combine \emph{them}
with every other variable, and so on. We found convergence at around 8 or 9 variables, with the 5th or 6th variable still
adding substantial significance improvement. Interestingly, we found that many `naive' choices of 8 good variables differ
substantially from the optimal choice. Some variables, such as pull, are not very useful by themselves, but make
a strong appearance as the $4^\textrm{th}$ or $5^\textrm{th}$ variable, leading to a relative 20\% or so $S/\sqrt{B}$ improvement. It is hard
to develop intuition for some of the more obscure variables, so the use of this algorithmic procedure provides a great advantage.
If one of many obscure variables proves powerful, it can become a focal point for new theoretical understanding and experimental study.

In order to see the marginal significance enhancements, we observed that the significance improvement $\rsig = \eS/\sqrt{\eB}$ viewed
as a function of the signal efficiency $\eS$ provides a very powerful visualization technique. The significance improvement curves
consistently
demonstrate at least three things:
\begin{enumerate}
\item They show how the improvements converge when more variables are added. We found stability at around 8 or 9.
\item They manifest instabilities when the Monte Carlo samples are inadequate.
For example, with around one million input events, the instabilities begin when $\eS$ or $\eB$ are of order $10^{-3}$ or less.
\item The curves have maxima at finite values of $\eS$ (in contrast to $\eS/\eB$, which generally diverges at small $\eS$). The $\rsig$ at this value, $\ropt$, provides a quantitative
measure of the improvement of $S/\sqrt{B}$ when the corresponding discriminant is used.
\end{enumerate}
In contrast, the ROC curves, which show $\eB$ as a function of $\eS$, and $\eS/\eB$ curves, are harder to interpret in terms
of search optimization, even though they contain the same numerical information as the $\rsig$ curves.
Nevertheless, since $S/B$ is in fact relevant
to the final analysis, one might not want to choose $\eS$ exactly to optimize $\rsig$, but rather to choose a somewhat lower value,
which increases $S/B$ at a small cost to $\rsig$.
This allows for the actual significance measure, with systematic efficiencies included, to be maximized.

We also considered various ways to measure
the $\bbar$ invariant mass, which should have a peak near $m_H$ for the signal sample. We found that the best choice of jet algorithm and size, out
of $k_T$, anti-$k_T$ and Cambridge/Aachen is anti-$k_T$ with $R\!=\!0.5$ at the LHC and anti-$k_T$ with $R\!=\!0.7$ at the Tevatron,
 although the algorithm and jet size dependence is not very strong.
We also found that using trimmed jets to construct a single $\mbb$ does not seem to help much. It seems that optimizing the jet size can
compensate for the effect of trimming. On the other hand, we found that trimming gives us an important new handle when
multiple jet measures can be combined. We found that combining $\mbb$ from aggressively trimmed jets and $\mbb$ from mildly trimmed
jets does uniformly better than a single trimming or no trimming at all. This was understood qualitatively from the 2D $\mbb$ distribution.
Nevertheless, even multiple trimmings seem not to provide marginal improvement when combined with other variables sensitive to radiation
in the event. This is important observation because, at this point, trimming has never been attempted experimentally. If
it turns out the same information is contained in, say, pull, which has already been shown to be measurable, one may be able
to avoid dealing with trimmed jets. On the other hand, if it turns out that the theoretical uncertainties on trimmed jet masses
are unusually small, or trimmed jets are less sensitive to pileup, then it may be worth using trimming instead of variables like
angularities, which are expected to degrade faster in busy events. In any case, understanding the relationship between these
many new discriminants should be a fruitful area for further investigation.

We only considered the irreducible $W/Z + \bbar$ backgrounds to $W/Z + H$ production. This is the main
reason we cannot translate our
results into a final significance estimate. Indeed, the reducible $t\bar{t}$ background is particularly important for $WH$, and
the $W/Z + jj$ background with false positive $b$-tags is important for both. We believe that the $t\bar{t}$ background can be
easily tamed with a jet veto or a more sophisticated top anti-tag, and the $jj$ background can be studied in the same way that the
we have studied the $\bbar$ background here.  The $b$-tagging quality could also be input as
an additional variable to be optimized, rather than being fixed at 60\% efficiency.

Combining multiple discriminants, we compared the significance enhancement characteristic to that coming from the set of variables
used by CDF and \DZero. We found around a 10-20\% enhancement was in principle possible. Most of the variables used by CDF and \DZero~are
kinematical, taking advantage of distinctions which are apparent in the distribution of the initial hard partons. Since we
include many variables, such as pull, which are absent at the hard parton level, it is understandable that some improvement should
result. However, our conclusion is only that enhancement  {\it may be possible}.
The same qualitative conclusion holds for the LHC light Higgs boson search.
We find that our variables work even better for $ZH$ at the LHC, partially because
initial states for the background are dominantly $gg$ in a $pp$ collider while the signal
is $q\bar{q}$ initiated at either machine.
We believe that while putting a hard cut on $p_T^Z$ reduces the problem to something more manageable, the Higgs boson can be found
without a fixed restriction on $p_T$, with potentially much larger significance.
Moreover, at the LHC, since the detectors have better resolution,
the variables related to radiation patterns and jet substructure may be more accurately measured.
Both the conclusions about the Tevatron and the LHC come with a few caveats: we have only compared to irreducible backgrounds;
we have not considered experimental or theoretical systematic uncertainty;
and we have not considered the accuracy to which all our variables can actually be measured.

In this paper, we have demonstrated that it is possible to perform comprehensive multivariate analysis using Monte Carlo simulations.
We have argued that SIC curves provide a useful visualization, and that there is room for additional discovery potential for
a light Higgs boson at both the Tevatron and the LHC. Despite this intensive effort, there remain a number of important questions which we
were forced to defer to future work. It is important to verify that many of the useful discriminants are stable with regard
to different models of the parton shower as was done for pull in \cite{Gallicchio:2010sw}, and that the set of important variables is roughly generator independent. It is also
generally useful to have a better sense of how the Boosted Decision Tree parameters should be chosen, and in which situations
other methods would be preferable. From the physics side, we believe that the dominant reducible $t\bar{t}$ background can
be removed using either a jet veto or, more importantly, a multi-variable discriminant similar to the ones we have developed here.
If that and the $W/Z+jj$ sample can be characterized, we could produce a more realistic significance estimate for the Higgs boson discovery
reach. Such a study should properly be done with fully reconstructed events. However, there is still work which can be done on the
theoretical side.

In summary, we have constructed a framework for evaluation and optimization of multivariate searches. This can form the basis for future studies in important but difficult searches at the LHC and Tevatron.

\section*{Acknowledgements}
Discussions with Michael Kirby illuminated much about the current
D\O\ multivariate Higgs boson search.
This work was supported in part by the Department of Energy
under grant DE-SC003916. The computations in this paper were
performed on the Odyssey cluster supported by the FAS
Research Computing Group at Harvard University.

\end{document}